\documentclass[12pt]{article}
\usepackage{amssymb}
\usepackage{graphics}
\usepackage{psfrag}
\usepackage{epsfig}

\DeclareFontFamily{U}{rsf}{}
\DeclareFontShape{U}{rsf}{m}{n}{
  <5> <6> rsfs5 <7> <8> <9> rsfs7 <10-> rsfs10}{}
\DeclareMathAlphabet\Scr{U}{rsf}{m}{n}

\catcode`\@=11
\def\@citex[#1]#2{%
\if@filesw \immediate \write \@auxout {\string \citation {#2}}\fi
\@tempcntb\m@ne \let\@h@ld\relax \def\@citea{}%
\@cite{%
  \@for \@citeb:=#2\do {%
    \@ifundefined {b@\@citeb}%
      {\@h@ld\@citea\@tempcntb\m@ne{\bf ?}%
      \@warning {Citation `\@citeb ' on page \thepage \space undefined}}%
      {\@tempcnta\@tempcntb \advance\@tempcnta\@ne%
      \@tempcntb\number\csname b@\@citeb \endcsname \relax%
      \ifnum\@tempcnta=\@tempcntb %Number follows previous--hold on to it
        \ifx\@h@ld\relax%
          \edef \@h@ld{\@citea\csname b@\@citeb\endcsname}%
        \else%
          \edef\@h@ld{\ifmmode{-}\else--\fi\csname b@\@citeb\endcsname}%
        \fi%
      \else%   %  non-successor--dump what's held and do this one
        \@h@ld\@citea\csname b@\@citeb \endcsname%
        \let\@h@ld\relax%
      \fi}%
    \def\@citea{,\penalty\@highpenalty\,}%
  }\@h@ld
}{#1}}

\def\@citeb#1#2{{[#1]\if@tempswa , #2\fi}}
\def\@citeu#1#2{{$^{#1}$\if@tempswa , #2\fi }}
\def\@citep#1#2{{#1\if@tempswa , #2\fi}}

\def\bcites{         % cite with []'s
        \catcode`\@=11
        \let\@cite=\@citeb
        \catcode`\@=12
}

\def\upcites{         % cite with exponents
        \catcode`\@=11
        \let\@cite=\@citeu
        \catcode`\@=12
}

\def\plaincites{      % cite without brackets
        \catcode`\@=11
        \let\@cite=\@citep
        \catcode`\@=12
}

\newcount\hour
\newcount\minute
\newtoks\amorpm
\hour=\time\divide\hour by 60
\minute=\time{\multiply\hour by 60 \global\advance\minute by-\hour}
\edef\standardtime{{\ifnum\hour<12 \global\amorpm={am}%
        \else\global\amorpm={pm}\advance\hour by-12 \fi
        \ifnum\hour=0 \hour=12 \fi
        \number\hour:\ifnum\minute<10 0\fi\number\minute\the\amorpm}}
\edef\militarytime{\number\hour:\ifnum\minute<10 0\fi\number\minute}

\def\draftlabel#1{{\@bsphack\if@filesw {\let\thepage\relax
   \xdef\@gtempa{\write\@auxout{\string
      \newlabel{#1}{{\@currentlabel}{\thepage}}}}}\@gtempa
   \if@nobreak \ifvmode\nobreak\fi\fi\fi\@esphack}
        \gdef\@eqnlabel{#1}}
\def\@eqnlabel{}
\def\@vacuum{}
\def\marginnote#1{}
\def\draftmarginnote#1{\marginpar{\raggedright\scriptsize\tt#1}}
\def\draft{
        \pagestyle{plain}
        \overfullrule=2pt
        \oddsidemargin -.5truein
        \def\@oddhead{\sl \phantom{\today\quad\militarytime} \hfil
        \smash{\Large\sl DRAFT} \hfil \today\quad\militarytime}
        \let\@evenhead\@oddhead
        \let\label=\draftlabel
        \let\marginnote=\draftmarginnote
        \def\ps@empty{\let\@mkboth\@gobbletwo
        \def\@oddfoot{\hfil \smash{\Large\sl DRAFT} \hfil}
        \let\@evenfoot\@oddhead}
        \def\@eqnnum{(\theequation)\rlap{\kern\marginparsep\tt\@eqnlabel}%
        \global\let\@eqnlabel\@vacuum}  }

\def\section{\@startsection {section}{1}{\z@}{3.ex plus 1ex minus
 .2ex}{2.ex plus .2ex}{\Large\bf}}
\def\subsection{\@startsection{subsection}{2}{\z@}{2.75ex plus 1ex minus
 .2ex}{1.5ex plus .2ex}{\large\bf}}        

\def\appendix{{\newpage\section*{Appendix}}\let\appendix\section%
        {\setcounter{section}{0}
        \gdef\thesection{\Alph{section}}}\section}

\def\abstract{\if@twocolumn
\section*{Abstract}
\else %\small
\begin{center}
{\bf Abstract\vspace{-.5em}\vspace{0pt}}
\end{center}
\quotation
\fi}

\renewcommand*\l@section[2]{%
  \ifnum \c@tocdepth >\z@
    \addpenalty\@secpenalty
    \addvspace{.0em \@plus\p@}%
    \setlength\@tempdima{1.5em}%
    \begingroup
      \parindent \z@ \rightskip \@pnumwidth
      \parfillskip -\@pnumwidth
      \leavevmode \bfseries
      \advance\leftskip\@tempdima
      \hskip -\leftskip
      #1\nobreak\hfil \nobreak\hb@xt@\@pnumwidth{\hss #2}\par
    \endgroup
  \fi}
\renewcommand*\l@subsection{\addvspace{-.5em \@plus\p@}\@dottedtocline{2}{1.5em}{2.3em}}
\renewcommand*\l@subsubsection{\addvspace{-.5em \@plus\p@}\@dottedtocline{3}{3.8em}{3.2em}}

\catcode`\@=12

\newcommand{\beq}{\begin{equation}}
\newcommand{\eeq}{\end{equation}}
\newcommand{\beqa}{\begin{eqnarray}}
\newcommand{\eeqa}{\end{eqnarray}}
\newcommand{\dd}{{\rm d}}
\newcommand{\A}{{\mathbb A}}

\newcommand{\Z}{{\mathbb Z}}
\newcommand{\ZZ}{{\mathbb Z}}

\newcommand{\R}{{\mathbb R}}
\newcommand{\C}{{\mathbb C}}
\newcommand{\CC}{{\mathbb C}}
\newcommand{\PP}{{\mathbb P}}
\newcommand{\e}{\,{\rm e}}
\newcommand{\CP}{{\CC\PP}}
\newcommand{\RP}{{\R\PP}}

\newcommand{\ts}{\textstyle}

\newcommand{\be}{\begin{equation}}
\newcommand{\ee}{\end{equation}}
\newcommand{\bea}{\begin{eqnarray}}
\newcommand{\eea}{\end{eqnarray}}

\def\to{\rightarrow}

\def\To{\longrightarrow}

\def\lae{\mathrel{\mathop{\smash{\lower .5 ex \hbox{$\stackrel<\sim$}}}}}
\def\lae{\mathrel{\mathop{\smash{\lower .5 ex \hbox{$\stackrel>\sim$}}}}}

\def\Tr{{\rm Tr}}
\def\l:{\mathopen{:}\,}
\def\r:{\,\mathclose{:}}

\catcode`\@=11
\def\theequation{\arabic{equation}}
\catcode`\@=12

\bcites

\catcode`\@=11
\def\theequation{\thesection.\arabic{equation}}
\@addtoreset{equation}{section}
\@addtoreset{footnote}{section}
\@addtoreset{footnote}{subsection}
\catcode`\@=12

\newcommand{\btheta}{\overline{\theta}}

\newcommand{\nn}{\nonumber}
\newcommand{\kket}[1]{\vert  #1\rangle\!\rangle}

\newcommand{\cket}[1]{\vert \Scr{C},#1\rangle\!\rangle}

\newcommand{\g}{{\it g}}

\newcommand{\NSNS}{{{}_{\rm NSNS}}}
\newcommand{\RR}{{{}_{\rm RR}}}
\newcommand{\NS}{{{}_{\rm NS}}}

\newcommand{\Pk}{{\rm P}_k}
\newcommand{\Mk}{{\rm M}_k}
\newcommand{\tilm}{\widetilde{m}}

\newcommand{\tilnu}{\widetilde{\nu}}
\newcommand{\tilGamma}{\widetilde{\Gamma}}
\newcommand{\tilomega}{\widetilde{\omega}}
\newcommand{\tilq}{\widetilde{q}}

\newcommand{\exact}{\longrightarrow}

\begin{document}

\begin{titlepage}

\begin{center}

\today\hfill
\hbox to 3cm
{\parbox[t]{5cm}{
hep-th/0401137\\
CERN-PH-TH/2004-006\\
KCL-MTH-04-01\\
NSF-KITP-04-01} \hss}\\

\vskip 2 cm
{\large \bf Orientifolds of Gepner Models}
\vskip 1 cm 
{Ilka Brunner$^{*}$, Kentaro Hori$^{\dagger}$, Kazuo Hosomichi$^{\dagger}$ and 
Johannes Walcher$^{\ddagger}$}\\
\vskip 0.5cm
{\it 
$^{*}$CERN Theory Division, Geneva, Switzerland}\\[0cm]
{\it and}\\[0cm]
{\it King's College, London, United Kingdom}\\[0.2cm]
{\it
$^{\dagger}$University of Toronto, Toronto, Ontario, Canada}\\[0.2cm]
{\it 
$^{\ddagger}$Kavli Institute for Theoretical Physics,
University of California, Santa Barbara, USA}
\vskip 0.2cm

\end{center}

\vskip 0.5 cm
\begin{abstract}
We systematically construct and study Type II Orientifolds
based on Gepner models which have
${\mathcal N}=1$ supersymmetry in $3+1$ dimensions.
We classify the parity symmetries and construct the crosscap states.
We write down the conditions that a configuration of rational branes
must satisfy for consistency (tadpole cancellation
and rank constraints) and spacetime supersymmetry.
For certain cases, including Type IIB orientifolds of
the quintic and a two parameter model, one can find all solutions
in this class.
Depending on the parity, the number of vacua can be
large, of the order of $10^{10}-10^{13}$.
For other models, it is hard to find all solutions
but special solutions can be found --- some of them are chiral.
We also make comparison with the large volume regime and obtain
a perfect match.
Through this study, we find a number of new features of
Type II orientifolds, including the structure of moduli space and the
change in the type of O-planes under navigation through non-geometric
phases.

\end{abstract}

\end{titlepage}

\newpage

\tableofcontents

\newpage

\section{Introduction}\label{sec:intro}

String vacua with $\mathcal N=1$ supersymmetry in $3+1$ spacetime dimensions
have been re-attracting a lot of attention in recent years. One of the 
reasons is of course that despite a lot of efforts spent on the 
heterotic string, actual connections with real world particle physics
have proven difficult to make, and that new avenues have opened up 
with our growing mastery of strings, branes, and M-theory.
But we may also wish to turn this quest around and ask for general 
lessons from exploring the duality web with four supercharges, 
which on general grounds is expected to be quite complex. Whether 
or not one will be able to make contact with phenomenology, or 
extrapolate to a situation with broken supersymmetry, it is 
natural to expect that something interesting will be learned.

Type II orientifolds with branes and fluxes
are an important class of models.
By a chain of duality,
they can be related to many other classes of models,
including the heterotic string
on Calabi-Yau 3-folds and
M-theory on $G_2$-holonomy manifolds, and therefore
may possibly provide a unifying scheme for
$4d$ ${\mathcal N}=1$ compactifications \cite{DougString}.
They provide natural set-ups for the braneworld scenario.
It should also be noted that the recent progress
in moduli stabilization is done in this framework
\cite{GKP,KKLT}.
However, most of the study in the past is done using supergravity,
or only toroidal orientifolds are given serious accounts.
This is definitely not a satisfactory state of affairs, because
the large volume or flat backgrounds are a
tiny part of the whole variety of possible theories.
What we need is a handle on the regime where supergravity is not
accessible. 

In this paper,
we study the other extreme regime where 
the internal space is very small but nevertheless
the worldsheet is extremely powerful. 
Namely, we construct and study Type II orientifolds
based on Gepner models \cite{Gepner}.
We will also try to
see how such theories are connected to the large volume regimes.

To avoid confusion, we emphasize that what we do here 
is within the framework of the perturbative NSR formalism.
We are obviously not able to include (RR) fluxes, and we are
not going to discuss the stringy quantum corrections at this stage,
except in the discussion of
the anomaly cancellation mechanism and Fayet-Iliopoulos terms.
In particular, the moduli including the dilaton remain unfixed.
However, we want to regard our work as a useful starting point 
for an explicit study of such models.
For instance, our models will have non-abelian 
gauge groups living on various RR tadpole canceling branes,
and our results may be useful also for the final step in the 
moduli stabilization \cite{KKLT}.

In fact, the roads have been
partially paved for us.
Recently, a great deal of results on D-branes in Type II string
compactifications were obtained.
They include application of Cardy's RCFT techniques
\cite{RS} and also the study of how they continue in the moduli space
to the large volume \cite{BDLR}.
There is also an orientifold version of Cardy, initiated by
Pradisi-Sagnotti-Stanev \cite{PSS} and further developed by
many people \cite{bantay,HSS,FSHSS,SaAn,BH1}.
Some of the preliminary results have been obtained in \cite{BlWi,ABPSS}
and more recently in \cite{BH2,GovMaj,HuiSch}.
In particular, we will extensively
use the results of \cite{BH2} on the minimal models
and other general properties of orientifolds of $(2,2)$ theories.

Our goal in this paper is threefold. Firstly, we want to adapt
and generalize the RCFT methods to the full string theory based on
the Gepner models.
Secondly, we want to present as unified a view as possible of the
various descriptions available for these worldsheets, such as the
Landau-Ginzburg and gauged linear sigma model pictures. In particular,
we want to generalize the relations between the Gepner point and 
large volume regimes to the situation involving unoriented strings. 
Thirdly, we want to
give rather detailed lists of explicit models that can be constructed 
within this framework. 

The ripeness of the subject and the richness of the harvest
have forced this paper to rather extended length. In order to
guide the reader towards the important results, we now give an 
overview over the organization of the presentation.

According to our global goal, we begin our discussion in section \ref{sec:CYO}
in the context of the gauged linear sigma model (GLSM), 
which provides the most global picture 
of Calabi-Yau compactifications on the worldsheet. The discussion
in subsection \ref{subsec:CYLG} is rather standard, and can safely be skipped
by experts. 

In subsection \ref{subsec:parsym}, we review the possible
orientifold projections, as discussed for example in \cite{BH2}.
As could be expected, parity symmetries of $\mathcal N=2$ supersymmetric
field theories come in two varieties, called A and B-type respectively.
The tadpoles arising from the corresponding O-planes must be canceled
by A and B-type D-branes, and the resulting $\mathcal N=1$ models can 
be thought of as Type IIA/IIB orientifolds, respectively. The associated
geometries are quite different, but are related to each other by mirror
symmetry. Of importance will be the classification of possible dressing
of the parity by various (classical and quantum) symmetries 
of the theory in such a way that the parity is involutive.

In subsection \ref{subsec:examples}, we make this discussion concrete
in the two examples which will accompany us through the rest of the
paper: the quintic hypersurface in $\PP^4$ and the degree $8$ hypersurface
in weighted projective space $\PP^4_{1,1,2,2,2}$. As we will see, many
interesting features arise in this two parameter model, which admits a
much richer set of possible orientifold projections than the quintic.
For example, we will see that with the appropriate dressing it is 
possible to project out the K\"ahler modulus corresponding to the 
overall size of the Calabi-Yau, or to select different sections of
the moduli space, corresponding to discrete fluxes in the large
volume regime.

We have organized the rest of the paper around this division into
A and B-type and the illustration in two examples, the quintic
and the two parameter model. In section \ref{sec:tad}, we discuss 
the construction of crosscap and boundary states in the full worldsheet 
theory of the Gepner model. Our approach differs slightly from the 
methods used in the literature \cite{RS,FS,BS,FSW,FKLLSW} in that we 
use a supersymmetric language throughout. Moreover, our construction
of B-type boundary states is new in the sense that it does not use 
the Greene-Plesser \cite{GreenePlesser} construction of mirror symmetry.
This approach also sheds new light on fixed-point resolution or
the appearance of so-called short-orbit branes \cite{BS,FSW,FKLLSW,MMS}.

We are then ready for discussing the consistency conditions that
constrain the possible string theory models we can build, A-
and B-type in sections \ref{sec:TCCA} and \ref{sec:TCCB}, respectively.
The discussion includes the computation of O-plane charges, the 
action of the parities on the D-branes, as well as the structure of
Chan-Paton factors. This puts us in a position to solve the consistency
conditions explicitly for our two examples. We also discuss the 
computation of the massless open string spectrum. We conclude each 
of the sections with lists of solutions to the tadpole cancellation 
conditions and open string spectra in selected cases.

The possibilities 
turn out to be extremely numerous and rich. For instance, for B-type 
models on the quintic, it turns out that there are 
$31561671503$ different supersymmetric and tadpole canceling 
configurations of rational branes at the Gepner point, all with the 
orthogonal gauge groups.
The number of vacua is similar in the two parameter model,
depending on the parity, with the the additional interesting feature
of allowing for configurations with unitary and symplectic
gauge groups.

For A-type models,
the spectrum is expected to be even richer, although we are not able
to solve the tadpole constraints completely in this case.
The number of equations and the number of branes are too many
for even the computer to find the solutions in a reasonable time.
However, special solutions can be found:
For any model with odd levels only, we always have a solution consisting
of four identical branes ---
four D6-branes on top of the O6-plane in the large volume limit.
For models including even levels, such a solution does not always exist
but one can use the recombination of branes in the
Landau-Ginzburg model to find special solutions in many cases.
Also, the size of the problem is much smaller when we consider
``intermediate'' models whose orbifold group
is not minimal (single cyclic group)
nor maximal (the mirror of single-cyclic Gepner model).

Some of the theories we obtain have chiral matter contents.
Two out of nine special solutions for the two parameter model
(A-type) are chiral.
One of them has $U(1)^8$ gauge group with chiral quiver matters, and the
other is
$Sp(1)\times Sp(1)\times U(2)$ theory with matters in
$2\times ({\bf 2},{\bf 1},{\bf 2})$,
$({\bf 1},{\bf 2},\overline{\bf 2})$ and
$({\bf 2},{\bf 2},{\bf 1})$.
We feel that there are more chiral solutions than these two, but
how many and which is not clear at the moment.
For Type IIB orientifolds on Gepner models based on a single cyclic
group, such as the quintic or the two parameter model, all the solutions
are non-chiral.
However, some of the randomly chosen solutions
of a $\Z_5$-orbifold of quintic are chiral.
Thus, we obtain the first examples of chiral supersymmetric 4d theories
out of non-toroidal orientifolds.

Section \ref{sec:FI} is an interlude, in which we make remarks on chirality,
anomaly cancellation mechanism and Fayet-Iliopoulos terms.
The bilinear identity of the Witten index, where
the only  parity-invariant closed string
ground states propagate in the tree channel, 
plays an essential role in anomaly cancellation.
We make explicit the string coupling dependence of the low energy Lagrangian
and check that it is consistent with all of the tree level results we
obtained.

Finally, in section \ref{sec:continuation}, we compare the results on
consistency conditions with the geometrical expectations in the large
volume limit, finding complete agreement.
We will here make use of 
the results of \cite{BDLR} and \cite{DiaDou} on the connection between
geometry and Gepner model boundary states (see also 
\cite{DiaRo,KLLW,Scheidegger,Mayr}), as well as the results 
on the structure of the K\"ahler moduli space of the two parameter
model \cite{Candelas} and its real sections discussed in subsection
\ref{subsec:examples}.
We find something interesting through this study:
For some Type IIB orientifolds of
the two parameter model with two large volume regions
(distinguished by the B-field), the type of O-plane changes if one
goes from one large volume region to the other,
through non-geometric domains of the K\"ahler moduli space.
We consider an example with O5-planes at a genus 9 curve
and four rational curves. Here, in one region all O-planes 
are O5$^-$ (SO-type), whereas in the other region 
the O-planes at the rational curves become O5$^+$ (Sp-type).
For Type IIA orientifolds, we find in one example
an effective description of closed and open strings that matches
the results at the Gepner point as well as large volume.
An extensive study needs more technical development
such as an A-type analog of \cite{DiaDou,DiaRo,KLLW,Scheidegger,Mayr} 
(see, however \cite{Kennaway}), geometrical study of large volume branes,
 and methods to compute superpotential in both regimes.

{\it Note:}
A part of the present work (including
Section~\ref{sec:tad} and a part of Section~\ref{sec:TCCA})
is presented in a conference in \cite{HoriString}.
While the current work was under further progress
and was being written, we noticed these papers 
by Aldazabal et al \cite{Aldazabal} and by Blumenhagen \cite{Blumenhagen}, which have some overlap with our work.
However, in these papers, only odd level Gepner models are considered.
As we will see, the rich and interesting new physics arises in 
models including even level minimal models.

\section{Calabi-Yau Orientifolds}
\label{sec:CYO}

\subsection{Calabi-Yau Sigma Models and Gepner points}\label{subsec:CYLG}

Consider a $(2,2)$ supersymmetric gauge theory in $1+1$ dimensions
with $U(1)$ gauge group and
$r+1$ fields $X_1,\ldots,X_r,P$
with tree level superpotential
\beq
W=P(X_1^{k_1+2}+\cdots+X_r^{k_r+2})
\label{Wlsm}
\eeq
and twisted superpotential
$$
\widetilde{W}=t\Sigma.
$$
$\Sigma=\overline{D}_+D_-V$ is the superfieldstrength and $t=r-i\theta$
is the Fayet-Iliopoulos-Theta parameter.
The gauge transformations act on  the fields as
$$
P\to \e^{-iH\lambda}P,\quad
X_i\to \e^{iw_i\lambda}X_i,
$$
where
\beqa
&&H:={\rm lcm}\{k_i+2\},
\nn\\
&&w_i:={H\over k_i+2}.
\nn
\eeqa

For large values of the FI parameter,
the system reduces at low energies to the sigma model on the hypersurface $M
=\{X_1^{k_1+2}+\cdots+X_r^{k_r+2}=0\}$
in a weighted projective space of dimension $r-1$.
This gauge system, introduced in \cite{phases},
is called the linear sigma model for the
manifold $M$.
The condition that $M$ is Calabi-Yau is reflected by the vanishing of 
the sum of charges
$-1+\sum_{i=1}^r{1\over k_i+2}=0$. Namely
\beq
\sum_{i=1}^r{k_i\over k_i+2}=r-2=\dim M.
\label{CY}
\eeq
In this case, the beta function for the FI parameter vanishes and therefore
$t$ is a free parameter of the system.

At large negative ${\rm Re}(t)$, the $P$ field has a vacuum expectation value
and breaks the $U(1)$ gauge symmetry to the subgroup in which
$\e^{iH\lambda}=1$. This unbroken subgroup $\Gamma$ is generated by
the one with $\lambda=2\pi/H$ which acts on the fields as
\beq
\gamma:X_i\to \e^{2\pi i\over k_i+2}X_i,
\label{gaction}
\eeq
and is a cyclic group of order $H$.
The model at $t=-\infty$ is identified as the LG orbifold
with superpotential
\beq
W_{\it G}=X_1^{k_1+2}+\cdots+X_r^{k_r+2}
\eeq
divided by the group
$\Gamma\cong\Z_{H}$
acting on fields as (\ref{gaction}).
The LG model with superpotential $W=X^{k+2}$
flows in the infra-red limit to a $(2,2)$ superconformal field theory
with central charge $c={3k\over k+2}$,
called
the (A-series) level $k$ ${\mathcal N}=2$ minimal model, $M_k$.
The infra-red limit of the above LG orbifold is thus
the $\Gamma$-orbifold of the product of the minimal models;
$$
\left(\prod_{i=1}^r M_{k_i}\right)\Biggr/\Gamma
$$
This is the {\it Gepner model}.
The generator (\ref{gaction}) of the orbifold group $\Gamma$ is identified as
\beq
\gamma=\underbrace{(\g,\ldots,\g)}_r
\label{orbgen}
\eeq
in which $\g:=\e^{-2\pi i J_0}(-1)^{\widehat{F}}$
where $J_0$ is the $U(1)$ current of the (right-moving)
${\mathcal N}=2$ superconformal algebra
and $(-1)^{\widehat{F}}$ is $1$ on NSNS sector but $-1$ on RR sector.
Note that the RR-ground states of lowest R-charge
$q=-{c\over 6}$
survives the orbifold projection, since
${c\over 6}={\dim M\over 2}={r-2\over 2}$
and thus $\gamma=\e^{\pi i (r-2)}(-1)^r=1$.
This state corresponds to the holomorphic volume form of the Calabi-Yau
manifold. We discuss more on the ground states in
Section~\ref{subsub:ground}.

Type II string theory on $M\times \R^D$ is consistent only if
$2\dim M+D=10$. If we denote the complex dimension of
the transverse space by $d=(D-2)/2$, the criticality condition is
\beq
r+d=6.
\label{crit}
\eeq
In this paper we assume both the
Calabi-Yau condition (\ref{CY}) and the criticality condition
(\ref{crit}).

\noindent
{\bf Remarks.}\\{\small
{\bf (i)}~It is possible to have some $k_i=0$. The
IR limit of $W=X^2$ is empty, but
 can be regarded as the system with
a unique (ground) state in each of R/NS-sectors, with zero energy,
zero charge.
We will regard the $k_i=0$ factor as such a quantum field theory.
The orbifold group acts on this factor non-trivially:
the generator $\gamma$ acts as
$\g=\e^{-2\pi i J_0}(-1)^{\widehat{F}}=(-1)^{\widehat{F}}$,
namely, as identity on NSNS sector but
as $(-1)$ on the RR-sector.
Thus, having this factor has a non-trivial effect.
\\
{\bf (ii)}~The behaviour of the system depends very much on
whether there is an even $k_i$. It is useful to note that
when there is at least one even $k_i$ there is actually
an even number of $i$ with largest factors of $2$ in $k_i$, under
the Calabi-Yau condition,
$\sum_{i=1}^r{H\over k_i+2}=H$.
\\
{\bf (iii)}~Let us present some examples that satisfy the Calabi-Yau
and criticality conditions.\\
$\bullet$ $(k_i+2)=(3,3,3)$; $M=$ an elliptic curve,
$D=7+1$.
\\
$\bullet$ $(k_i+2)=(4,4,4,4)$; $M=$ a K3 surface.
$D=5+1$.
\\
We will mainly consider the case with $r=5$ and $d=1$ since this corresponds 
to the string compactification to $3+1$ dimensions.
The examples of this type are\\
$\bullet$ $(k_i+2)=(5,5,5,5,5)$; $M=$ a quintic hypersurface
in $\CP^4$.
\\
$\bullet$ $(k_i+2)=(8,8,4,4,4)$.
\\
$\bullet$ $(k_i+2)=(8,8,8,8,2)$.
\\
$\bullet$ $(k_i+2)=(12,12,6,6,2)$.
\\
The first two will be our basic examples where we
examine the general story in detail.
A complete list can be found in \cite{Gepnerclassification}.
\\
{\bf (iv)}~The non-chiral GSO projection of the minimal
model $M_k$ by $(-1)^F=\e^{-\pi i (J_0-\widetilde{J}_0)}$
is the $SU(2)_k\times U(1)_2$ mod $U(1)$ gauged WZW model,
or simply $SU(2)_k\times U(1)_2/U(1)_{k+2}$ coset model.
The latter model has primaries labeled by
$(l,m,s)\in \Mk$; namely $l\in P_k=\{0,1,...,k\}$,
$m\in \Z_{2(k+2)}$, $s\in \Z_4$, with $l+m+s$ even, $(l,m,s)\equiv
(k-l,m+k+2,s+2)$.\footnote{In this paper, following the convention used by
majority of people, the $SU(2)$ spin $j$ is labeled by
 $L=2j\in\Pk=\{0,1,2,\ldots,k\}$,
rather than $j$ itself that is used in \cite{MMS,BH2}.}
The product theory $M_{k_1}\times\cdots \times M_{k_r}$
should not be confused with the tensor product of
the GSO projected models of $M_{k_1},...,M_{k_r}$.
In the latter the space of states would have mixture of NSNS and RR
factors, while in the former
 NS/R alignment is automatically imposed, as usual in
ordinary supersymmetric quantum field theories.
\\
{\bf (v)} The GSO projected model
has global symmetries $g_{n,s}$ corresponding to
simple currents $(0,n,s)$ ($n\in \Z_{2(k+2)}$, $s\in \Z_4$, with
$n+s$ even) which act on the states in
$\Scr{H}_{l',m',s'}\otimes\Scr{H}_{l',-m',-s'}$ as multiplication
by a phase $\e^{\pi i\left({nm'\over k+2}-{ss'\over 2}\right)}$.
The symmetry $\g$ above induces one of them,
$g_{2,0}$.
\\
{\bf (vi)}~``Gepner Model'' usually refers to more general models
based on orbifold of the product of minimal models.
It doesn't have to come from linear sigma models of the above types.
In Appendix~\ref{app:Gepner}, we present more general models.
In the main text of the paper, we treat only the class of models
introduced above (except Sections~\ref{sub:construction} and
\ref{sub:Cardybranes} where
the discussion is general),
in particular the case $D=3+1$ and $r=5$.
We relegate the discussion on 
the most general models
to Appendix.

}

In many cases,
$M$ has singularities that are inherited from the orbifold singularities
of the ambient space,
and their  resolution introduces extra K\"ahler parameters.
This is accommodated in the linear sigma model by extending the gauge group
and adding charged fields.
In general, the gauge group will be $U(1)^k=\prod_{a=1}^kU(1)_a$ gauge theory 
with matter fields $P,X_1,\ldots,X_{r+k-1}$ of certain charge
$Q_P^a,Q_1^a,\ldots,Q_{r+k-1}^a$ and certain (twisted) superpotential.
For example, for $(k_i+2)=(8,8,4,4,4)$,
the full system after the resolution
has $U(1)\times U(1)$ gauge group and six matter fields
of the following charges \cite{MorrisonPlesser}:
\beq
\begin{array}{cccccccc}
&P&X_1&X_2&X_3&X_4&X_5&X_6\\
U(1)_1&-4&0&0&1&1&1&1\\
U(1)_2&0&1&1&0&0&0&-2
\end{array}
\label{2pcharge}
\eeq
The system has superpotential
$$
W=P\Bigl\{\,X_6^4(X_1^8+X_2^8)+X_3^4+X_4^4+X_5^4\,\Bigr\},
$$
and twisted superpotential
\beq
\widetilde{W}=t_1\Sigma_1+t_2\Sigma_2,
\label{FIth2}
\eeq
where the $t_a=r_a-i\theta_a$ and $\Sigma_a=\overline{D}_+D_-V_a$
are the FI-Theta parameter and the fieldstrength of the $U(1)_a$ gauge
group.
In the limit $t_2\to-\infty$ with $2t_1+t_2$ fixed, $X_6$ acquires a
large absolute value and breaks the gauge group except the one
generated by $(2i,i)\in \mathfrak{u}(1)_1\oplus\mathfrak{u}(1)_2$.
We are then left with the original system with one $U(1)$ gauge symmetry
whose FI-Theta parameter is $t=2t_1+t_2$.
This corresponds to undoing the resolution.

\subsubsection{RR Ground States and Chiral Primaries}\label{subsub:ground}

Let us present the list of supersymmetric ground states of the system.
The level $k$ minimal model has
$(k+1)$ supersymmetric ground states $|l\rangle_{\RR}$ ($l=0,1,...,k$)
which
correspond to
$X^l$ and have R-charges
$q=\tilq={l+1\over k+2}-{1\over 2}$. Also, on a circle twisted by
$\e^{-2\pi i \nu J_0}$, there is a unique supersymmetric ground state
$|0\rangle_{\nu}$ which has R-charge
$q=-\tilq={l_{\nu}+1\over k+2}-{1\over 2}$ where
$l_{\nu}\in \{0,1,...,k\}$ is defined by
$l_{\nu}+1\equiv -\nu$ (mod $(k+2)$).
The RR ground states of the Gepner model are made of these states.
Since the orbifold group is generated by the tensor product of
$-\e^{-2\pi i J_0}$ for the $r=5$ factors, the condition is
that the sum of R-charges is an odd half-integer,
$\sum_iq_i\in {1\over 2}+\Z$.
Untwisted sector states are thus the products
$\otimes_{i=1}^5|l_i\rangle_{\RR}$ with the condition
$\sum_i({l_i+1\over k_i+2}-{1\over 2})=\sum_i{l_i\over k_i+2}-{3\over 2}\in
{1\over 2}+ \Z$, or
\beq
\sum_{i=1}^5{l_i\over k_i+2}=0,1,2,3.
\label{orbcond}
\eeq
They correspond to harmonic forms of degree
$(3,0)$, $(2,1)$, $(1,2)$ and $(0,3)$ respectively of the relevant
Calabi-Yau manifold.\footnote{
For non-linear sigma model on a Calabi-Yau manifold of dimension $n$,
supersymmetric ground states with R-charge
$(q,\tilq)$ correspond to harmonic $(p,\bar{p})$ forms
where $(q,\tilq)=({n\over 2}-p,\bar{p}-{n\over 2})$.}
There are also RR ground states from the twisted sectors
labeled by $\nu=1,2,...,H-1$. The orbifold condition is the same
as (\ref{orbcond}) where $l_i$ is replaced by
$l_{\nu}^{(i)}$ for such $i$ where the twist is non-trivial,
$\nu\not\equiv 0$ mod $(k_i+2)$.
For the $\nu=1$ twist, we find $l_1^{(i)}=k_i$ for all $i$ and we find a unique 
ground state with $q=-\tilq={3\over 2}$. The geometrical counterpart is the
$(0,0)$-form. For $\nu=(H-1)$, we also find a unique ground state that
corresponds to the $(3,3)$-form.
The ground states from the twisted sectors are mostly related to
$(p,p)$-forms. However, there can be states corresponding to
off-diagonal forms. For example, let us consider the case where $H$ is even
and twist by $\nu={H\over 2}$.
The twist in the $i$-th factor is non-trivial if and only if
$w_i={H\over k_i+2}$ is odd.
For such an $i$, $l_{H\over 2}^{(i)}$ is ${k_i\over 2}$ and
the ground state is $q_i=\tilq_i=0$.
For other $i$, the twist is trivial and the ground states are
ordinary ones $|l_i\rangle_{\RR}$
with R-charges $q_i=\tilq_i={l_i+1\over k_i+2}-{1\over 2}$.
They correspond to $(2,1)$ or $(1,2)$ forms.
Let us show the number of ground states in two examples.

\noindent
\underline{$(k_i+2)=(5,5,5,5,5)$}~ Untwisted ground states
 correspond to
monomials of $X_i$ with degree $0$, $5$, $10$, $15$ (with relations
$X_i^4=0$) and there are $1$, $101$, $101$, $1$ of them.
Also there is a unique ground state $\otimes_i|0\rangle_{\nu}$
from each of $\nu=1,2,3,4$ twisted sectors.
These numbers are organized into
the ``Hodge diamond''
$$
\begin{array}{ccccccc}
&&&\!\! 1\!\! &&&\\[-0.2cm]
&&\!\! 0\!\! &&\!\! 0\!\! &&\\[-0.2cm]
&\!\! 0\!\! &&\!\! 1\!\! &&\!\! 0\!\! &\\[-0.2cm]
\!\! 1\!\! &&\!\! 101\!\! &&\!\! 101\!\! &&\!\! 1\!\! \\[-0.2cm]
&\!\! 0\!\! &&\!\! 1\!\! &&\!\! 0\!\! &\\[-0.2cm]
&&\!\! 0\!\! &&\!\! 0\!\! &&\\[-0.2cm]
&&&\!\! 1\!\! &&&
\end{array}
$$

\noindent
\underline{$(k_i+2)=(8,8,4,4,4)$}~
$X_1,X_2,X_3,X_4,X_5$ have weights $w_i=1,1,2,2,2$.
Untwisted ground states correspond to
monomials of $X_i$ with total weight $0$, $8$, $16$, $24$
(with relations $X_1^7=X_2^7=X_3^3=X_4^3=X_5^3=0$).
There are $1$, $83$, $83$, $1$ of them.
There is a unique ground state $\otimes_i|0\rangle_{\nu}$
from each of $\nu=1,2,3,5,6,7$ twisted sectors.
They corresponds to diagonal forms.
For the $\nu=4$ twisted sector, ground states are
$$
\bigotimes_{i=1,2}|0\rangle_{\nu}\otimes
\bigotimes_{i=3,4,5}|l_i\rangle_{\RR}
$$
where $l_3+l_4+l_5=1$ (3 states) or $5$ (3 states).
The Hodge diamond is therefore
$$
\begin{array}{ccccccc}
&&&\!\! 1\!\! &&&\\[-0.2cm]
&&\!\! 0\!\! &&\!\! 0\!\! &&\\[-0.2cm]
&\!\! 0\!\! &&\!\! 2\!\! &&\!\! 0\!\! &\\[-0.2cm]
\!\! 1\!\! &&\!86\! &&\! 86\! &&\!\! 1\!\! \\[-0.2cm]
&\!\! 0\!\! &&\!\! 2\!\! &&\!\! 0\!\! &\\[-0.2cm]
&&\!\! 0\!\! &&\!\! 0\!\! &&\\[-0.2cm]
&&&\!\! 1\!\! &&&
\end{array}\qquad
86=83+3
$$

As usual \cite{LVW}, RR ground states are in one-to-one correspondence
with chiral primaries by a spectral flow which shifts
the R-charge as $q\to q\pm {c\over 6}$, $\tilq\to\tilq  \pm {c\over 6}$.
The spectral flow with the sign $(++)$ maps the ground states
to NSNS states corresponding to
chiral fields ($(c,c)$-fields),
and the $(-+)$-spectral flow maps them to NSNS states corresponding to
twisted chiral  fields ($(a,c)$-fields).
They are marginal operators if $q=\tilq=1$.
Marginal $(c,c)$ primaries correspond to $(2,1)$-forms and
marginal $(a,c)$ primaries correspond to $(1,1)$-forms.

\subsubsection{The Parameter Space}

\subsubsection*{\it Worldsheet Parameter Space}

The $(c,c)$ and $(a,c)$ primaries with R-charge (1,1)
are exactly marginal operators.
Parameters coupled to $(c,c)$-primaries
parametrize the complex structure of the target space.
In the linear sigma model, they are the parameters $a_i$
of the tree level superpotential $W=PG(X_i,a_i)$.
If there are twisted RR ground states corresponding to $(2,1)$-forms,
the corresponding parameters do not fit into the linear sigma model.
 Parameters coupled to $(a,c)$-primaries
parametrize
the complexified K\"ahler class $[\omega -iB]$,
where $\omega$ is the K\"ahler form and $B$ is the B-field.
In the linear sigma model, they are the FI-Theta parameters $t^a$.
In the large volume limit, the FI-Theta parameters and the complexified
K\"ahler parameters are related by
\beq
[\omega -iB]\sim \sum_{a=1}^k (t^a+\pi iQ_P^a) \omega_a
\label{kaepar}
\eeq
where $\omega_a\in H^2(M,\Z)$ is the first Chern class of the line bundle
associated with the $U(1)_a$ gauge group and $Q_P^a$
is the charge of the field $P$.

The worldsheet theory is singular
at certain loci of the parameter space.
On the complex structure moduli space, the singularity is at
the loci where $M=\{G(X_i,a_i)=0\}$ is singular 
as a complex manifold.
On the K\"ahler moduli space, the singularity is at the loci where
the linear sigma model has an unbroken gauge symmetry
and some vector multiplet
is exactly massless.
For example, in the case of quintic, the singularity is at
$$
\e^{t}=-5^5.
$$
In the example of $(k_i+2)=(8,8,4,4,4)$, there are two singular loci:
\beq
C_1=\Bigl\{e^{t_2}=4\Bigr\},
\qquad
C_{\rm con}=\Bigl\{\, \e^{t_2}(1-4^{-4}\e^{t_1})^2=4\,\Bigr\}.
\label{sing2para}
\eeq

\subsubsection*{\it Scalar Manifold of Spacetime Theory ---
Type II on Calabi-Yau}

Let us consider Type II string theory on $\R^{3+1}$ times the internal CFT
we have been discussing.
The theory has ${\mathcal N}=2$ supersymmetry on $\R^{3+1}$.
The moduli of the worldsheet theory give rise to
massless scalar fields in $3+1$ dimensions, which
are part of some ${\mathcal N}=2$ supermultiplets.
Other parts in the multiplet come from the NS-R, R-NS and R-R sectors.
In Type IIA string theory, the $h^{1,1}$ K\"ahler moduli
are the scalar components of vector multiplets, while
the $h^{2,1}$ complex structure moduli together with
the periods of the RR 3-form potential
constitute the scalar components of
hypermultiplets
For Type IIB, the complex structure moduli
are in vector multiplets, while
the K\"ahler moduli and
the periods of the RR potentials are in
hypermultiplets.
The singular loci of the worldsheet theory
are not singular in full string theory.
It is simply that there are degrees of freedom
that become massless at these loci \cite{Strominger}.

\subsubsection{Mirror Description}\label{subsub:mirror}

The mirror of the Gepner model \cite{GreenePlesser}
(see also \cite{VafaLGorb})
is
% the product of the same set of minimal models divided by a different group,
% $\widetilde{\Gamma}$, which is the subgroup of $\prod_{i=1}^r\Z_{k_i+2}$ 
% consisting of the transformations
% $\widetilde{X}_i\to \e^{2\pi i m_i\over k_i+2}\widetilde{X}_i$
% obeying $\prod_{i=1}^r\e^{2\pi i m_i\over k_i+2}=1$.
% Namely, $\widetilde{\Gamma}$-orbifold of LG model with superpotential
the IR limit of the LG orbifold with superpotential
$$
\widetilde{W}_{\it G}=\widetilde{X}_1^{k_1+2}+\cdots+\widetilde{X}_r^{k_r+2},
$$
and the group $\tilGamma\subset\prod_{i=1}^r\Z_{k_i+2}$
acting on the fields as
$$
\widetilde{X}_i\to \e^{2\pi i \tilnu_i\over k_i+2}\widetilde{X}_i,\qquad
\prod_{i=1}^r\e^{2\pi i \tilnu_i\over k_i+2}=1.
$$
The superpotential can be deformed by
polynomials of the same degree as $W$ and which are invariant under
the group $\widetilde{\Gamma}$.
The monomial $\widetilde{X}_1\cdots\widetilde{X}_r$
is an example that exists in all the cases.
In fact the model with superpotential
$\widetilde{W}_{\it G}
+\e^{t/H}\widetilde{X}_1\cdots\widetilde{X}_r$ is the mirror of
the linear sigma model with single $U(1)$ gauge group whose FI-Theta
parameter is $t$ \cite{HV}.
In the case $(k_i+2)=(5,5,5,5,5)$, this is
$$
\widetilde{W}=\widetilde{X}_1^5+\cdots+\widetilde{X}_5^5
+\e^{t/5}\widetilde{X}_1\cdots\widetilde{X}_5.
$$
In fact $\widetilde{X}_1\cdots\widetilde{X}_5$
 is the only allowed deformation for this case, which
 corresponds to the fact
that the quintic has only one K\"ahler modulus.
In more general models, there are other
$\widetilde{\Gamma}$-invariant monomials of the same degree, each
corresponding to a blow up mode.
For instance, in the case $(k_i+2)=(8,8,4,4,4)$,
the fully deformed superpotential is
\beq
\widetilde{W}=\widetilde{X}_1^8+\widetilde{X}_2^8+\widetilde{X}_3^4
+\widetilde{X}_4^4+\widetilde{X}_5^4+
\e^{t_1/4+t_2/8}\widetilde{X}_1\cdots\widetilde{X}_5
+\e^{t_2/2}\widetilde{X}_1^4\widetilde{X}_2^4,
\label{dualW2para}
\eeq
where $t_1$ and $t_2$ are the FI-Theta parameters in (\ref{FIth2}).
It indeed reduces to the one-parameter family
$\widetilde{W}_G+\e^{t/8}\widetilde{X}_1\cdots\widetilde{X}_5$
under the blow-down limit, $t_2\to-\infty$, $t=2t_1+t_2$ fixed.

\subsection{Parity Symmetries}\label{subsec:parsym}

We would like to classify involutive parity symmetries of the system
that preserves a half of the $(2,2)$ worldsheet supersymmetry.
The superfield notation we use here is introduced in \cite{BH2}:
the A and B parities on
the $(2,2)$ superspace are $\Omega_A(x^{\pm},\theta^{\pm},\btheta^{\pm})
=(x^{\mp},-\btheta^{\mp},-\theta^{\mp})$
and $\Omega_B(x^{\pm},\theta^{\pm},\btheta^{\pm})
=(x^{\mp},\theta^{\mp},\btheta^{\mp})$.

\subsubsection{Linear Sigma Model}

We first consider the parity symmetries of the linear
sigma model.

\noindent
{\bf A-parities}\\
A-parities of the single $U(1)$ gauge system with superpotential
(\ref{Wlsm}) are
$\Omega_A$ combined with $V\to V$ and
\beq
\tau^A_{{\bf m},\sigma}:\,\,
\begin{array}{l}
P\To \overline{P},\\[0.1cm]
X_i\To \e^{2\pi i m_i\over k_i+2}\overline{X_{\sigma(i)}}.
\end{array}
\label{PAlsm}
\eeq
Here, ${\bf m}$ labels the elements of the global symmetry
$(\prod_{i=1}^r\Z_{k_i+2})/\Z_H$.
Also, $i\mapsto\sigma(i)$ is an order two permutation
such that $k_{\sigma(i)}=k_i$ so that the charges are invariant.
This is involutive if and only if
$$
m_i=m_{\sigma(i)}\quad(\mbox{mod $k_i+2$}).
$$
The phase rotation can sometimes be undone by a change of
variables. For $X_i'=\e^{2\pi i n_i\over k_i+2}X_i$, the parity acts as
$X_i'\to \e^{{2\pi i\over k_i+2}(m_i+n_i+n_{\sigma(i)})}
\overline{X_{\sigma(i)}'}$. Therefore there is an equivalence relation
${\bf m}\equiv {\bf m'}$ if and only if
$$
m_i'=m_i+n_i+n_{\sigma(i)}\quad (\mbox{mod $k_i+2$}).
$$
The FI-theta parameter $t$ is unconstrained but the parameters $(a_i)$
that deforms the superpotential are constrained to be essentially real,
$G(\e^{2\pi im_i\over k_i+2}\overline{X_{\sigma(i)}},a_i)
=\overline{G(X_i,a_i)}$.

\noindent
{\bf B-parities}\\
B-parities of the single $U(1)$ gauge system (\ref{Wlsm}) are
$\Omega_B$ combined with $V\to V$ and
\beq
\tau^B_{{\bf m},\sigma}:\,\,
\begin{array}{l}
P\To -P,\\[0.1cm]
X_i\To \e^{2\pi i m_i\over k_i+2} X_{\sigma(i)}
\end{array}
\label{PBlsm}
\eeq
where $\sigma$ is an order two permutation
with $k_{\sigma(i)}=k_i$ and
$$
m_i+m_{\sigma(i)}=0\quad (\mbox{mod $k_i+2$})
$$
so that it is involutive.
For the variable $X_i'=\e^{2\pi in_i\over k_i+2}X_i$,
the parity acts as
$X_i'\to\e^{{2\pi i\over k_i+2}(m_i+n_i-n_{\sigma(i)})}X_{\sigma(i)}'$.
Thus ${\bf m}$ and ${\bf m'}$ are equivalent if and only if
$$
m_i'=m_i+n_i-n_{\sigma(i)}\quad (\mbox{mod $k_i+2$}).
$$
The FI-Theta parameter is constrained to be real
$
\overline{\e^t}=\e^t,
$
while the complex structure parameters $a_i$ are required to obey
$G(\e^{2\pi i m_i\over k_i+2}X_{\sigma(i)},a_i)
\equiv G(X_i,a_i)$.

\subsubsection{Gepner point}

The parity symmetries we have considered above,
$P^A_{{\bf m},\sigma}=\tau_{{\bf m},\sigma}^A\Omega_A$ 
and
$P^B_{{\bf m},\sigma}=\tau_{{\bf m},\sigma}^B\Omega_B$,
are of course
symmetries at the Gepner point.
Since $P$ has an expectation value $\langle P\rangle$, 
it is understood that a gauge transformation is
used so that $\langle \tau_{{\bf m},\sigma}P\rangle=\langle P\rangle$.
For A-parity, taking $\langle P\rangle$ real,
the transformation of the LG fields $X_i$ is the same as in
(\ref{PAlsm})
$$
\tau^A_{{\bf m},\sigma}:X_i\To \e^{2\pi i m_i\over k_i+2}
\overline{X_{\sigma(i)}},
$$
while for B-parity (\ref{PBlsm}) is combined
with the gauge transformation
$\e^{i\lambda}=\e^{\pi i/H}$:
$$
\tau^B_{{\bf m},\sigma}:X_i\To\e^{2\pi i m_i\over k_i+2}\e^{\pi i\over
k_i+2}X_{\sigma(i)}.
$$

At the Gepner point, there are extra symmetries
called the quantum symmetries which form a group
$\widehat{\Gamma}\cong\Z_H$. 
The quantum symmetry $g_{\omega}$ associated with an $H$-th root of unity
$\omega$ multiplies the $\gamma^{\ell}$-twisted states
by the phase $\omega^{\ell}$.
It acts on
the mirror variables $\widetilde{X}_1,\ldots,\widetilde{X}_r$ as
\beq
g_{\omega}:\widetilde{X}_i\longmapsto \omega_i\widetilde{X}_i;\qquad
\omega_i^{k_i+2}=1\,(\forall i),\quad \omega_1\cdots\omega_r=\omega.
\eeq
The monomial $\widetilde{X}_1\cdots\widetilde{X}_r$
is not invariant under $g_{\omega}$ with $\omega\ne 1$
and quantum symmetry is completely broken if $\e^t\ne 0$.
For other deformations it is broken to a subgroup.

One can use this quantum symmetry to modify the
parity symmetry.
Thus, we have a larger set of parity symmetries at the Gepner point:
\beqa
&&
P^A_{\omega;{\bf m},\sigma}=g_{\omega}\tau_{{\bf m},\sigma}^A\Omega_A,\\
&&
P^B_{\omega;{\bf m},\sigma}=g_{\omega}\tau_{{\bf m},\sigma}^B\Omega_B.
\eeqa
Actually, not all of them are involutive and not all of them
are inequivalent.
For A-type, the parity acts on the dual variables as
$\Omega_A$ combined with $\widetilde{X}_i\to \omega_i \e^{\pi i\over k_i+2}
\widetilde{X}_{\sigma(i)}$.
This is involutive if and only if $\omega^2=1$, namely
\beq
\omega=\left\{
\begin{array}{ll}
1&\mbox{$H$ odd},\\
\pm 1&\mbox{$H$ even}.
\end{array}\right.
\eeq
For B-type, the parity action is $\Omega_B$ combined with
$\widetilde{X}_i\to\omega_i^{-1}\overline{\widetilde{X}_{\sigma(i)}}$.
This is always involutive. However, some of them can be undone by a change of
variables.
Dressing by $g_{\omega}$ and $g_{\omega'}$ are equivalent
if and only if
$$
\omega'=\alpha^2\omega,
\qquad
\alpha^H=1.
$$

If $H$ is odd, there is no non-trivial involutive
dressing by quantum symmetry.
For A-type, dressed parity is not involutive unless $g_{\omega}=1$.
For B-type, any dressing is equivalent to no dressing.

If $H$ is even, there is essentially
a unique non-trivial involutive dressing by quantum symmetry.
For A-type, it is the dressing by the order 2 element $g_{-1}$.
Since $\widetilde{X}_1\cdots\widetilde{X}_r$ flips its sign
under $g_{-1}$,
the dressed parity is not a symmetry if $\e^t\ne 0$.
Thus, the K\"ahler modulus corresponding to the overall
size is frozen at $\e^t=0$ if we require this parity to be a symmetry.
For B-type, it is the dressing by the primitive element
$g_{\omega}$, $\omega=\e^{2\pi i/H}$.
It maps the monomial $\widetilde{X}_1\cdots\widetilde{X}_r$ to
$\e^{2\pi i/H}\overline{\widetilde{X}_1\cdots\widetilde{X}_r}$.
Thus, the condition of parity invariance is shifted from
$\e^{t/H}=\overline{\e^{t/H}}$ to
$\e^{t/H}\e^{2\pi i/H}=\overline{\e^{t/H}}$.
In terms of the invariant coordinate $\e^t=(\e^{t/H})^H$,
the condition is $\e^t\in \R_{\geq 0}$ if not dressed by quantum symmetry
while it is $\e^t\in \R_{\leq 0}$ if dressed by odd quantum symmetry.

\subsubsection{Type II Orientifolds}

Let us consider Type II string theory
on $\R^{3+1}$ times our internal CFT, and
gauge the worldsheet parity symmetry $P$ which acts trivially on
the $3+1$ spacetime coordinates but acts on the
internal CFT as one of the above parities (A-type or B-type).
This is the Type IIA or Type IIB orientifold.
(The original papers on more general orientifolds are
\cite{ASI,BSII,ASII,HoI,HoII}.)
To make it consistent, we need to add either D-branes or fluxes.
This is one of our main themes of this paper.
For now, let us discuss aspects that are independent of how it is done.

Since the left movers and right movers of the string modes are 
identified by the parity, ${\mathcal N}=2$ supersymmetry will be broken
to at most ${\mathcal N}=1$ supersymmetry.
Use of A-parity for Type IIA string and B-parity for Type IIB string
is the necessary condition for preserving
an ${\mathcal N}=1$ supersymmetry.
(Whether it is preserved in the full theory depends on what we add
(D-branes and fluxes), and this is another main topic of the latter part
of this paper.)

As in the case before orientifold, the worldsheet moduli
give rise to light fields of the spacetime theory.
We have seen that
these moduli are constrained by the requirement that the parity is
a symmetry of the worldsheet.
The light fields are constrained accordingly.
Together with light fields from the NS-R, R-NS and R-R sectors,
they constitute ${\mathcal N}=1$ supermultiplets.
The pattern at the large volume is analyzed in \cite{BH2}
and is summarized in Table~\ref{clmod}.
\begin{table}
\begin{center}
\begin{tabular}{|ll@{\quad\vrule width0.8pt\quad}c|c|}
\hline
&$\begin{array}{l}
\\
\\
\end{array}$
&chiral multiplets&vector multiplets
\\
\noalign{\hrule height 0.8pt}
$\begin{array}{l}
\\
\\
\end{array}$
&IIAO(6)
&$h^{1,1}_-+h^{2,1}+1$&$h^{1,1}_+$
\\
\hline
$\begin{array}{l}
\\
\\
\end{array}$
&IIBO(9,5)~~&$h^{2,1}_++h^{1,1}+1$&$h^{2,1}_-$
\\
\hline
$\begin{array}{l}
\\
\\
\end{array}$
&IIBO(7,3)&$h^{2,1}_-+h^{1,1}+1$&$h^{2,1}_+$
\\
\hline
\end{tabular}
\caption{Light Fields from Closed Strings}
\label{clmod}
\end{center}
\end{table}
Here IIAO(6) is for Type IIA orientifolds, where ``6'' is
because we generically
have orientifold 6-planes.
IIBO(9,5) is for Type IIB orientifolds with O9 or O5-planes
and IIBO(7,3) is for Type IIB orientifolds with O7 and/or O3-planes.
Also, $h^{p,\bar{p}}_{\pm}$ are the number of harmonic $(p,\bar{p})$-forms
that are invariant/anti-invariant under the involution.
Note that, even when the worldsheet
moduli receive antiholomorphic
constraints (for example,
complex structure moduli by A-parity and
K\"ahler moduli by B-parity), they combine with
periods of RR-potentials and form complex scalars of 
${\mathcal N}=1$ chiral multiplets.

\subsection{Examples}\label{subsec:examples}

Let us study the parity symmetries discussed above in
typical examples with
odd and even $H$'s 
--- the quintic $(k_i+2)=(5,5,5,5,5)$ with $H=5$ and
the two parameter model
$(k_i+2)=(8,8,4,4,4)$ with $H=8$.

\subsubsection{Quintic}\label{subsub:quintic}

This case is studied in detail in \cite{BH2}.
As we have seen above, there is no non-trivial involutive
dressing by quantum symmetry.
Also, one can show there is no non-trivial involutive dressing by the 
$\Z_5^4$ global symmetry: ${\bf m}$ that determines an
involutive parity is equivalent to ${\bf 0}$.
Thus, the parity is determined purely in terms of $\sigma\in \mathfrak{S}_5,
\sigma^2=1$.
Up to permutation of variables, there are only
three cases: $\sigma={\rm id}$, $(12)$ and $(12)(34)$.

The table shows the projected moduli as well as
 O-planes in the geometric phase,
for these six orientifolds.
\begin{table}[htb]
\caption{Six Orientifolds of Quintic}
\label{oqtable}
\begin{center}
\begin{tabular}{|l||l|l|}
\hline
parity&moduli $(K,C)$ &O-planes\\
\noalign{\hrule height 0.8pt}
$P^A_{\rm id}$&$(1_{\C},101_{\R})$&O6 at the real quintic $\cong\RP^3$\\
\hline
$P^A_{(12)}$&$(1_{\C},101_{\R})$&O6 at an $\RP^3$\\
\hline
$P^A_{(12)(34)}$&$(1_{\C},101_{\R})$&O6 at an $\RP^3$\\
\noalign{\hrule height 0.8pt}
$P^B_{\rm id}$&$(1_{\R},101_{\C})$&O9 at $M$\\
\hline
$P^B_{(12)}$&$(1_{\R},63_{\C})$&O3 at a point and O7 at a hypersurface\\
\hline
$P^B_{(12)(34)}$&$(1_{\R},53_{\C})$&O5's at a rational and a genus 6 curves
\\
\hline
\end{tabular}
\end{center}
\end{table}

\begin{figure}[htb]
\centerline{\includegraphics{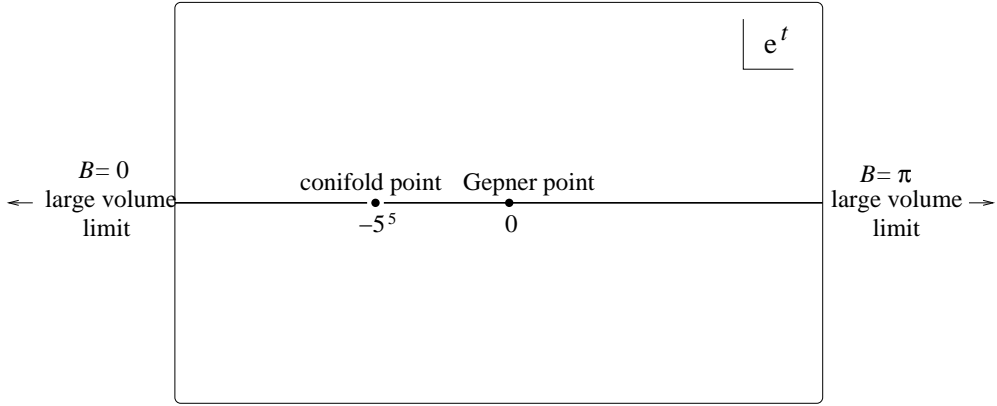}}
\caption{K\"ahler moduli space for a B-orientifold of the quintic}
\label{km}
\end{figure}

For all three B-type orientifolds, the K\"ahler moduli space
is the real line $\e^t=\overline{\e^t}$ as depicted in Figure~\ref{km}.
It passes through the Gepner point,
is broken at the conifold point and extends to the two large volume regions
--- one with $B=0$ and another with $B=\pi$.
The Gepner point is connected along $\e^t>0$ to the $B=\pi$ 
asymptotic region, as follows from (\ref{kaepar}). The path along $e^t<0$ 
is blocked by the conifold singularity.

\subsubsection{A Two parameter Model}\label{subsub:two}

In the example $(8,8,4,4,4)$, there is a unique non-trivial and
involutive dressing by quantum symmetry.
Also, there are several non-trivial and involutive 
dressing by the global symmetry $\Z_8\times\Z_4^3$.
Here, since there are already a variety of ways to choose ${\bf m}$ for a fixed
$\sigma$,
we only consider the $\sigma={\rm id}$ cases.

For A-type parity, $X_i\to \e^{2\pi i m_i\over k_i+2}\overline{X_i}$,
${\bf m}$ obey the equivalence relation
$m_i\equiv m_i+2n_i$ and $m_i\equiv m_i+1$, and there are
six independent choices
${\bf m}=(00000)$, $(00001)$, $(00011)$, $(00111)$, $(01000)$, $(01001)$.
Under the quantum symmetry $g_{-1}$, 
the term $\widetilde{X}_1\cdots\widetilde{X}_5$
in the dual superpotential (\ref{dualW2para}) flips its sign while
$\widetilde{X}_1^4\widetilde{X}_2^4$ is invariant.
Thus, if dressed by the quantum symmetry $g_{-1}$, the K\"ahler modulus
$t=2t_1+t_2$ is frozen at $\e^t=0$ but $\e^{t_2}$ is unconstrained.
If not dressed by quantum symmetry, the K\"ahler moduli are both unconstrained.
In the regime $t_1,t_2\gg 0$, one can talk about the geometry.
$\tau_{\bf m}^A$ acts as an antiholomorphic involution, and
the fixed point set is
$X_i=\e^{\pi i m_i\over k_i+2}x_i$, and $X_6=x_6$, where 
$x_i$ are all real, obey 
$$
(-1)^{m_1}x_1^8x_6^4+
(-1)^{m_2}x_2^8x_6^4+
(-1)^{m_3}x_3^4+
(-1)^{m_4}x_4^4+
(-1)^{m_5}x_5^4=0,
$$
and are subject to the gauge conditions of the GLSM preserving the reality
condition. The determination of the topology of the resulting fixed point 
sets can sometimes be a little cumbersome. This problem has been studied in 
\cite{RRW} and we review here the parity $P^A_{00001}$ as an example. The 
topology of this O-plane can be obtained by studying the solutions of the 
real equation
$$
x_6^4 (x_1^8+x_2^8) + x_3^4 + x_4^4 = x_5^4
$$
subject to the rescaling 
$$
(x_1,x_2,x_3,x_4,x_5,x_6)\equiv(\lambda x_1,\lambda x_2,\mu x_3,\mu x_4,
\mu x_5,\lambda^2\mu x_6)
$$
with $\lambda,\mu\in \R^*$. Thus, we have to require $x_5\neq 0$, 
whereupon we can set $x_5$ to one by rescaling with $\mu$.
The second rescaling can be absorbed by noting that 
$x_1^8+x_2^8>0$ in the large volume phase. After changing 
variables to $x_1^8=y_1^2$, $x_2^8=y_2^2$, $x_6^4=y_6^2$, 
$x_3^4=y_3^2$, $x_4^4=y_4^2$, the constraints become 
$y_6^2+y_3^2+y_4^2=1$, $y_1^2+y_2^2=1$, with non-trivial
$\Z_2$ identification $(y_1,y_2,y_3,y_4,y_6)\equiv
(-y_1,-y_2,y_3,y_4,y_6)$ (from $\lambda=-1$). Thus, topologically,
this O-plane is $S^2\times \RP^1=S^2\times S^1$.

We refer to \cite{RRW} for the remaining cases, and summarize the 
results in Table \ref{o2tableA}. We do not know a simple description
of the O-plane for the parity $P^A_{01001}$, except that it has Betti
numbers $b_0=1$ and $b_1=2$.
We have also indicated in table \ref{o2tableA} that even though the
number of moduli from complex structure deformations ($86$ real parameters)
is always the same, each parity selects a different real section of
the moduli space. In particular, these sections can intersect in different
ways with singular loci, as we will illustrate below.

\begin{table}[htb]
\caption{A-type Orientifolds (with $\sigma=1$) of the
Two Parameter Model}
\label{o2tableA}
\begin{center}
\begin{tabular}{|l||l|l|}
\hline
parity&moduli $(K,C)$ &O-planes\\
\noalign{\hrule height 0.8pt}
$P^A_{+;00000}$&$(2_{\C},86_{\R})$&No O-plane\\
\hline
$P^A_{-;00000}$&$(1_{\C},86_{\R})$&non-geometric\\
\hline
$P^A_{+;00001}$&$(2_{\C},86_{\R}')$&O6 at an $S^2\times S^1$\\
\hline
$P^A_{-;00001}$&$(1_{\C},86_{\R}')$&non-geometric\\
\hline
$P^A_{+;00011}$&$(2_{\C},86_{\R}'')$&O6 at a $T^3$\\
\hline
$P^A_{-;00011}$&$(1_{\C},86_{\R}'')$&non-geometric\\
\hline
$P^A_{+;00111}$&$(2_{\C},86_{\R}''')$&O6 at an $S^2\times S^1$\\
\hline
$P^A_{-;00111}$&$(1_{\C},86_{\R}''')$&non-geometric\\
\hline
$P^A_{+;01000}$&$(2_{\C},86_{\R}'''')$&O6 at an $S^3$\\
\hline
$P^A_{-;01000}$&$(1_{\C},86_{\R}'''')$&non-geometric\\
\hline
$P^A_{+;01001}$&$(2_{\C},86_{\R}''''')$&O6 at a SLAG with $b_0=1$, $b_1=2$\\
\hline
$P^A_{-;01001}$&$(1_{\C},86_{\R}''''')$&non-geometric\\
\hline
\end{tabular}
\end{center}
\end{table}

For B-type parity,
$X_i\to \e^{2\pi i (m_i+1/2)\over k_i+2}X_i$, ${\bf m}$ is constrained
by $2m_i=0$ (mod $k_i+2$) and obey the equivalence relation
$m_i\equiv m_i+1$. There are eight choices described by
the signs $\epsilon_i=\e^{2\pi i m_i\over k_i+2}$:
$(\epsilon_1\epsilon_2\epsilon_3\epsilon_4\epsilon_5)=(+++++)$,
$(++-++)$,
$(++--+)$,
$(++---)$,
$(+-+++)$,
$(+--++)$,
$(+---+)$,
$(+----)$.
If dressed with $g_{\omega}^m$ with $\omega=\e^{2\pi i /8}$,
the monomials
$\widetilde{X}_1\cdots\widetilde{X}_5$ and
$\widetilde{X}_1^4\widetilde{X}_2^4$ of the dual variables
are transformed to
$\e^{2\pi i m/8}\overline{\widetilde{X}_1\cdots\widetilde{X}_5}$
and $\e^{\pi i m}\overline{\widetilde{X}_1^4\widetilde{X}_2^4}$
respectively.
The symmetry condition $\widetilde{W}\to \overline{\widetilde{W}}$
is satisfied by the dual superpotential (\ref{dualW2para})
if $\e^{2\pi i m/8}\e^{t_1/4+t_2/8}=\overline{\e^{t_1/4+t_2/8}}$
and $\e^{\pi i m}\e^{t_2/2}=\overline{\e^{t_2/2}}$.
It follows from this that
the K\"ahler moduli are constrained by
\beqa
&&\mbox{not dressed by quantum symmetry:}\quad
\e^{t_1}\in \R,\quad\e^{t_2}\in \R_{\geq 0} 
\qquad\label{undressed}\\
&&\mbox{dressed by odd quantum symmetry:}
\quad\e^{t_1}\in \R,\quad\e^{t_2}\in\R_{\leq 0},
\qquad\label{dressed}
\eeqa
%%%%%%%%%%%%%%%%%%%%%%%%%%%%%%%%%%%%%%%%%%%%%%%%%%%%%%%%%%%%%%%%%
Each of these have two large volume regions classified by the $B$-field.
By using (\ref{kaepar}) and the charge table (\ref{2pcharge}), one learns
that in the case (\ref{undressed}) the $B$-field can be $B=0$ or 
$B=\pi\omega_1$, while in the case (\ref{dressed}), we have 
$B=\pi\omega_2$ or $B=\pi\omega_1+\pi\omega_2$.

To describe this real section of the moduli space of the two parameter 
model in somewhat more detail, we recall from \cite{Candelas} that by
introducing
$$
\xi=\e^{t_2+2t_1} \qquad \eta=\e^{t_2} \qquad \zeta=\e^{t_1+t_2} \,,
$$
we can embed the parameter space as the quadric
\beq
Q=\{ \xi \eta - \zeta^2 = 0\}
\label{quadric}
\eeq
in $\C^3$. In these coordinates, the singularities in the parameter 
space of the mirror threefold (\ref{dualW2para}) appear at the curves
\beqa
C_1&=& Q\cap \{ \eta= 4 \} \\
C_{\rm con} &=& Q\cap \{2^{-16}\xi + \eta - 2^{-7}\zeta = 4\} \,.
\eeqa
The real moduli space, $Q\cap \{\xi,\eta,\zeta\in \R\}$ is an 
ordinary double cone, which consist of the components 
$Q_+=Q\cap \{\xi,\eta>0\}$, and $Q_-=Q\cap\{\xi,\eta<0\}$, meeting 
at the tip $\xi=\eta=\zeta=0$ (Gepner point). In fact, from 
(\ref{undressed}) and (\ref{dressed}), we see that the real K\"ahler 
moduli space of the orientifold without (with) dressing by quantum 
symmetry is given by $Q_+$ ($Q_-$). Moreover, the real versions of 
$C_1$ and $C_{\rm con}$ are ordinary cone sections, and it is easy 
to check that they are parabolas and lie completely in $Q_+$. Since 
they intersect transversely and have co-dimension one, if we do not
dress by quantum symmetry, {\it the Gepner point is completely
separated from the two large volume regimes}. In that case, it is 
not possible to connect the Gepner point with a geometric interpretation 
of the orientifold without running into a singularity of the worldsheet
theory. If dressed by odd quantum symmetry, the moduli space and 
singular loci do not meet, so that Gepner point is connected to the 
corresponding two large volume regimes. 

\begin{figure}[htb]
\psfrag{Co}{$C_0$}
\psfrag{Ce}{$C_1$}
\psfrag{Cc}{$C_{\rm con}$}
\psfrag{Ci}{$C_{\infty}$}
\psfrag{a}{$a$}
\psfrag{b}{$b$}
\psfrag{c}{$c$}
\psfrag{d}{$d$}
\psfrag{e}{$e$}
\psfrag{G}{$g$} 
\centerline{\epsfig{width=6in,file=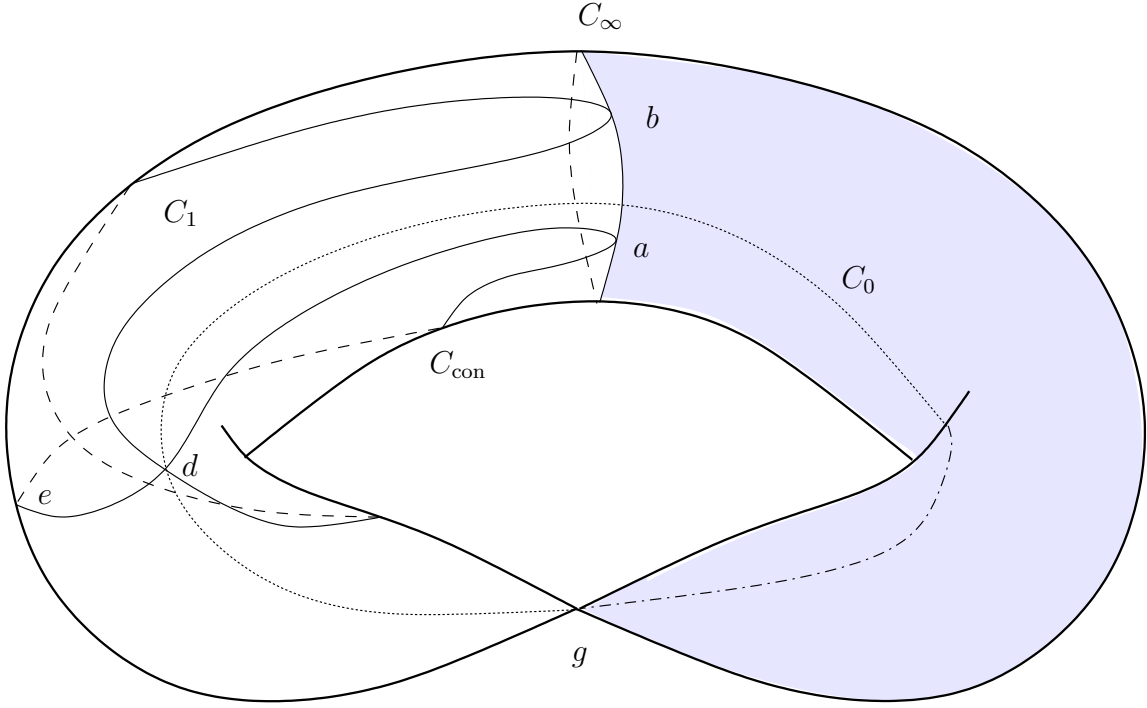}}
%\centerline{\includegraphics{twoparmod.eps}}
\caption{Real section of the K\"ahler moduli space of the two parameter
model $(k_i+2)=(8,8,4,4,4)$. The Gepner point $g$ is at the tip of
a conical singularity.  The left 
cone is the moduli space of the orientifold without dressing by 
quantum symmetry. The lines of singularity $C_1$ and $C_{\rm con}$
divide the moduli space into several perturbative regions.
The right cone (shaded region), which reaches out 
all the way to the large volume regime, is the moduli space of the 
orientifold with dressing by quantum symmetry.}
\label{kmt}
\end{figure}

In order to capture its global structure, it is convenient to compactify 
the moduli space by adding a divisor $C_\infty$ at infinity. As 
explained in \cite{Candelas}, the compactification can be achieved 
by embedding $Q$ in (\ref{quadric}) in the projective space 
$\PP^3$, which by abuse of notation we coordinatize with 
$[\xi:\eta:\zeta:\tau]$. In addition to $C_1$ and $C_{\rm con}$,
we then have the distinguished locus $C_\infty= Q\cap\{\tau=0\}$,
which also corresponds to a degeneration of (\ref{dualW2para}).
We also have the ``orbifold locus'' $C_0= \{ \xi=\zeta=0\} \subset 
Q$, which contains the Gepner point $g=[0:0:0:1]$. We show a 
picture of this compactified parameter space and the location 
of the singular loci $C_0$, $C_1$, $C_{\rm con}$, and $C_\infty$
in Figure \ref{kmt}. It is important to emphasize that in distinction 
to $C_{\rm con}$ and $C_1$, $C_0$ and $g$ do not lead to a singular 
worldsheet theory. 

\begin{figure}[htb]
\psfrag{Dd}{$\scriptstyle{D_{(0,-1)}}$}
\psfrag{D}{$\scriptstyle{D_{(-1,-1)}}$}
\psfrag{Ce}{$\scriptstyle{C_1}$}
\psfrag{Ci}{$\scriptstyle{C_{\infty}}$}
\centerline{\epsfig{width=6in,file=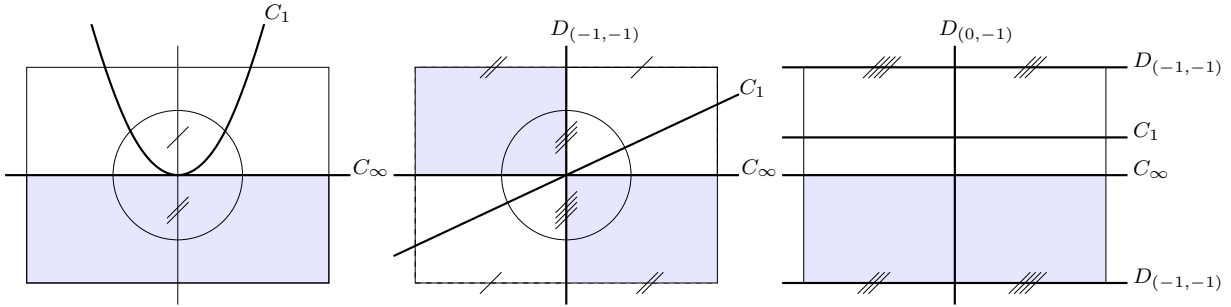}}
\caption{To see the large volume limit in the compactified moduli
space, we have to blowup the singular point $b=C_1\cap C_\infty$.
We replace twice a small neighborhood of the origin with a 
M\"obius strip, successively inserting the divisors $D_{(-1,-1)}$ 
and $D_{(0,-1)}$. The multiply stroked lines are identified. 
The shaded region can be reached smoothly from the Gepner
point.}
\label{blowups}
\end{figure}

Also, $C_\infty$ simply corresponds to the boundary of the 
uncompactified moduli space in the usual sense. In particular,
the large volume limit has been hidden inside of $C_\infty$
by the compactification process. To recover this (unique) large volume 
limit, we need to blow up the point $b$, where the two
divisors $C_1$ and $C_\infty$ intersect non transversely.
Near $C_\infty$, we can work in the patch $\xi=1$ of $\PP^3$, in which 
our real moduli space is given by the equation $\eta=\zeta^2$, 
with $\tau$ arbitrary. In this patch, $C_1$ is given by 
$\tau=\eta=\zeta^2$, while $C_\infty$ is given by $\tau=0$. A real
blowup of the origin corresponds to replacing a small disc around 
$\zeta=\tau=0$ with a M\"obius strip. The exceptional divisor (called 
$D_{(-1,-1)}$ in \cite{Candelas}) of this blowup is simply the non-trivial 
one-cycle of the M\"obius strip. In simple terms, the blowup
means that when approaching the origin along some path, we keep track 
of the first derivative $d\tau/d\zeta$, and we do not reach the
same point depending on the value of this derivative. It is easy to
see from this description that now $C_1$, $C_\infty$, and $D_{(-1,-1)}$
meet at a triple intersection, and we have to perform a second blowup,
replacing the origin by the exceptional divisor $D_{(0,-1)}$. The
large volume point is now the unique intersection point 
$D_{(0,-1)}\cap C_\infty$. In this way, we have recovered the
description of the large volume limit as a cylinder $(t_1,t_2)
\equiv (t_1+2\pi i,t_2)\equiv (t_1,t_2+2\pi i)$. We show this sequence of 
blowups in Figure \ref{blowups}. 

This description puts us in a position to illustrate geometrically
the statements on the structure of the large volume region that
we have made above. The (real) neighborhood of the large volume 
point is divided by $D_{(0,-1)}$ and $C_\infty$ into four quadrants,
which are geometrically distinguished by the value of the B-field.
By following the sequence of blowups and the global picture
in Figure \ref{kmt}, we see that starting from the Gepner point $g$, 
we can reach two of these quadrants without crossing a singularity,
but not the other two.

We will use this description of the moduli space in section 
\ref{sec:continuation} when we discuss the comparison between Gepner
model boundary and crosscap states and large volume.

Let us describe the topological structure of the
orientifold planes corresponding to each of the involutions
$\tau^B_\epsilon$ that we have defined above.
%%%%%%%%%%%%%%%%%%%%%%%%%%%%%%%%%%%%%%%%%%%%%%%%%%%%%%%%%%%%
In the large volume regime,
$\tau^B_{\epsilon}$ acts on the manifold as the
holomorphic involution $X_i\to \epsilon_i X_i$ and $X_6\to X_6$.
The fixed point set is the loci with
$\epsilon_iX_i=\lambda_2X_i$ ($i=1,2$),
$\epsilon_iX_i=\lambda_1X_i$ ($i=3,4,5$),
$X_6=\lambda_1\lambda_2^{-2}X_6$.
The solutions in the eight cases are:\\
$(+++++)$:~
 No condition (the whole manifold $M$).\\
$(++-++)$:~ $X_3=0$ (a hypersurface).\\
$(++--+)$:~ $X_3=X_4=0$ (a curve of genus 9)
and $X_5=X_6=0$ (four lines)\\
$(++---)$:~ $X_3=X_4=X_5=0$ (eight points)
and
$X_6=0$ (a hypersurface).\\
$(+-+++)$:~ $X_1=0$ or $X_2=0$ (two hypersurfaces).
The two are homologous since they are two fibres of the K3-fibrations
(with base $\{(X_1,X_2)\}$ and fibres $\{(X_3,X_4,X_5,X_6)\}$).\\
$(+--++)$:~ $X_1=X_3=0$ or $X_2=X_3=0$ (two genus 3 curves). They are
homologous to each other.\\
$(+---+)$:~ $X_1=X_5=X_6=0$ (four points),
$X_1=X_3=X_4=0$ (four points),
$X_2=X_5=X_6=0$ (four points),
$X_2=X_3=X_4=0$ (four points).\\
$(+----)$:~ $X_1=X_6=0$ or
$X_2=X_6=0$ (two genus 3 curves). They are homologous to each other.\\
These are included in Table~\ref{o2tableB}.

\newcommand{\lw}[1]{\smash{\lower2.0ex\hbox{#1}}}
\begin{table}[htb]
\caption{B-type Orientifolds (with $\sigma=1$) of the
Two Parameter Model}
\label{o2tableB}
\begin{center}
\renewcommand{\arraystretch}{1.2}
\begin{tabular}{|l||l|l|}
\hline
parity&moduli $(K,C)$ &O-planes\\
\noalign{\hrule height 0.8pt}
\multicolumn{1}{|l||}{$P^B_{0;+++++}$}&
\multicolumn{1}{l|}{$(2_{\R},86_{\C},...,83_{\C})$}&\lw{O9 at $M$}\\
\cline{1-2}
$P^B_{1;+++++}$&$(2'_{\R},86_{\C})$&\\
\hline
\multicolumn{1}{|l||}{$P^B_{0;++-++}$}&
\multicolumn{1}{l|}{$(2_{\R},57_{\C},...,56_{\C})$}&
\lw{O7 at a hypersurface}\\
\cline{1-2}
$P^B_{1;++-++}$&$(2'_{\R},57_{\C})$&\\
\hline
\multicolumn{1}{|l||}{$P^B_{0;++--+}$}&
\multicolumn{1}{l|}{$(2_{\R},46_{\C},...,47_{\C})$}&
\lw{O5's at four rational and a genus 9 curves}\\
\cline{1-2}
$P^B_{1;++--+}$&$(2'_{\R},46_{\C})$&\\
\hline
\multicolumn{1}{|l||}{$P^B_{0;++---}$}&
\multicolumn{1}{l|}{$(2_{\R},41_{\C},...,44_{\C})$}&
\lw{O7 at a hypersurface and O3's at eight points}\\
\cline{1-2}
$P^B_{1;++---}$&$(2'_{\R},41_{\C})$&\\
\hline
\multicolumn{1}{|l||}{$P^B_{0;+-+++}$}&
\multicolumn{1}{l|}{$(2_{\R},53_{\C},...,56_{\C})$}&
\lw{O7's at two homologous K3 hypersurfaces}\\
\cline{1-2}
$P^B_{1;+-+++}$&$(2'_{\R},53_{\C})$&\\
\hline
\multicolumn{1}{|l||}{$P^B_{0;+--++}$}&
\multicolumn{1}{l|}{$(2_{\R},46_{\C},...,47_{\C})$}&
\lw{O5's at two homologous genus 3 curves}\\
\cline{1-2}
$P^B_{1;+--++}$&$(2'_{\R},46_{\C})$&\\
\hline
\multicolumn{1}{|l||}{$P^B_{0;+---+}$}&
\multicolumn{1}{l|}{$(2_{\R},45_{\C},...,44_{\C})$}&
\lw{O3's at sixteen points}\\
\cline{1-2}
$P^B_{1;+---+}$&$(2'_{\R},45_{\C})$&\\
\hline
\multicolumn{1}{|l||}{$P^B_{0;+----}$}&
\multicolumn{1}{l|}{$(2_{\R},46_{\C},...,43_{\C})$}&
\lw{O5's at two homologous genus 3 curves}\\
\cline{1-2}
$P^B_{1;+----}$&$(2'_{\R},46_{\C})$&\\
\hline
\end{tabular}
\end{center}
\end{table}

To conclude this section, we count the number of complex structure
moduli in these orientifolds.
This can be done by looking at the parity action on the
corresponding chiral primary states.
To see the action, we first consider the parities $P^B_{\omega;+++++}$
that correspond to the
identity of $M$ in the large volume.
In such a case, we know that the complex structure moduli is unconstrained.
Thus the number of moduli is full $86$. Let us now go to the Gepner point
along some path in the K\"ahler moduli space.
As we have seen this can be done only for the parity $P^B_{1;+++++}$
dressed by an odd quantum symmetry.
By continuity the number of moduli at the Gepner point is still $86$.
Thus, we find that $P^B_{1;+++++}$ acts trivially
on all the marginal $(c,c)$ primaries. Since other parities are obtained from
$P^B_{1;+++++}$ by dressing global or quantum symmetries
(whose action we know), we
now know the action of all the parities $P^B_{\omega;{\bf m}}$
on the marginal $(c,c)$ primaries, at the Gepner point.
In this way, we find the number of complex moduli
at the Gepner point.\\
{\bf Remarks.}\\{\small
{\bf (i)}~By continuity the number of moduli found at the Gepner point 
applies everywhere in the K\"ahler moduli space
for the parities $P^B_{1;\epsilon_1...\epsilon_5}$.
This in particular tells us the number of moduli in the large volume limit.
The numbers are listed in Table~\ref{o2tableB}.
(It is an interesting exercise to check these numbers directly
by analyzing geometry.)
%JW: what do you mean by this comment? Doesn't this follow from the
%   action on the monomials?
\\
{\bf (ii)}~In the large volume regime, the only difference between
$P_{1;\epsilon_1...\epsilon_5}$ and $P_{0;\epsilon_1...\epsilon_5}$
is the value of the $B$-field. Thus, the number found in (i) is still
applicable for $P_{0;\epsilon_1...\epsilon_5}$, 
in the large volume regime.
\\ 
{\bf (iii)}~On the other hand, one can analyze the action of
$P_{0;\epsilon_1...\epsilon_5}$ at the Gepner point (as stated above).
The action is the same as $P_{1;\epsilon_1...\epsilon_5}$
on the untwisted sector states but differs from that by $-$ sign
on the twisted sector states. Thus, if $n$ twisted ground states survive
the $P_{1;\epsilon_1...\epsilon_5}$-projection,
then the other $(3-n)$ survive the
$P_{0;\epsilon_1...\epsilon_5}$-projection.
\\
{\bf (iv)}~Thus, the number of moduli is different between the Gepner point and
the large volume regimes for the
$P_{0,\epsilon_1...\epsilon_5}$-orientifolds.
This is not a puzzle from the worldsheet point of view, because
the two regions are separated by the singularity locus.
There are also two other regions and the number of moduli there could
be different as well.
In Table~\ref{o2tableB}, we only show
 the number at the large volume and at the Gepner point, and simply write
dots ... for the other two regions.
}

\begin{figure}[htb]
\centerline{\includegraphics{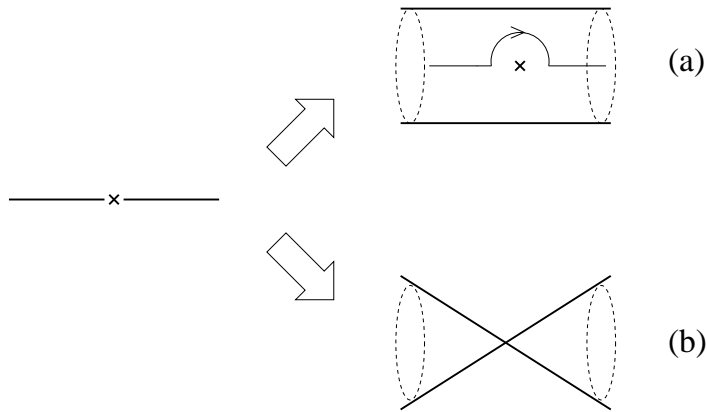}}
\caption{Two possibilities of complexifying the real moduli
space with a codimension one singularity. (a) One can go around the
singularity and the two parts are smoothly connected.
(b) The moduli space consists of branches. Any path from one region to
the other must go through the singular point.
Third possibility (not shown in Figure) would be that
the two regions are disconnected.}
\label{singul}
\end{figure}

We have seen that the complex structure moduli can jump
from from one component to another
of the real K\"ahler moduli space.
This tells us something about the full string theory.
As we have discussed, the real K\"ahler moduli
are combined with RR-potentials to form complex parameters
(which become the lowest components of ${\mathcal N}=1$
chiral superfields of the spacetime theory).
An interesting problem is to find the behaviour near the singularity.
One possibility is that
one can go around the singular loci
by turning on the RR-potential, so that
the separate regions of the
real K\"ahler moduli space are smoothly connected to each other in the full
moduli space (Figure~\ref{singul}(a)).
This happens in other situations, such as the flop transition
\cite{phases,AGM}.
This possibility is, however, eliminated in the present case
by the jump in the dimension of the complex structure moduli space.
One picture consistent with the jump is that the moduli space consists
of a number of branches (Figure~\ref{singul}(b)),
and the components of the
real moduli space belong to different branches so that
they can have different dimensions.
Another possibility is that the singularity is at infinite distance
and the two components are disconnected.
Of course, the jump does not necessarily occur (an example is the case of
quintic), and in such a case, at this stage
we do not know whether one can go around the singular
loci by turning on RR-potentials.
It is an interesting problem to find out what is the right picture
in full string theory.

\section{Tadpole States of the Gepner Model}
\label{sec:tad}

The main purpose of the present paper is to construct
consistent Type II orientifolds on Calabi-Yau manifolds and
Gepner models, with and without spacetime supersymmetry.
In the discussion of consistency and spacetime supersymmetry,
it is useful to study the ``tadpole state'' \cite{PolCai,Callan},
which is the
sum of boundary and crosscap states:
\beq
|T\rangle=|B\rangle+|C\rangle
=|B\rangle_\NSNS+i|B\rangle_\RR
+i|C\rangle_\NSNS+i|C\rangle_\RR
\label{kettot}
\eeq
and the ``bra'' version
\beq
\langle \theta T|=\langle \theta B_{\rm tot}|+\langle \theta C_{\rm tot}|
={}_\NSNS\!\langle B|
+i{}_\RR\!\langle B|
-i{}_\NSNS\!\langle C|
+i{}_\RR\!\langle C|.
\label{bratot}
\eeq
The NSNS and RR parts are
\beq
\begin{array}{l}
|B\rangle_\NSNS=|B_+\rangle_\NSNS-|B_-\rangle_\NSNS,\\
|B\rangle_\RR=|B_+\rangle_\RR-|B_-\rangle_\RR,\\
|C\rangle_\NSNS=|C_{(-1)^{F_R}P}\rangle
-|C_{(-1)^{F_L}P}\rangle,\\
|C\rangle_\RR=|C_{P}\rangle
-|C_{(-1)^FP}\rangle.
\end{array}
\label{bcdetail}
\eeq
Here $B_+$ and $B_-$ corresponds to
the boundary conditions with the opposite
spin structures, and $P$ is an involutive parity symmetry of the total
system.
Each term on the right hand sides of (\ref{bcdetail}), say
$|C_{P}\rangle$, can be written as the tensor product of the
spacetime part and internal part:
$$
|\mbox{spacetime}\rangle\otimes|\mbox{internal}\rangle.
$$
The spacetime part is associated with the Neumann boundary condition (for
boundary state)
and the standard parity $\Omega$ (for the crosscap states),
and is given by the
standard coherent state of the $D$ free bosons/fermions, the ghost and
the superghosts.
The internal part depends on the detail of D-branes and orientifold.

In this section, we construct the internal part of the crosscap
states corresponding to the orientifolds introduced in the previous
section. We also reconstruct the rational boundary states of Gepner 
models from a perspective which is somewhat different from 
the one in the literature. The Cardy-PSS construction 
\cite{Cardy,PSS} and its generalizations are usually 
formulated in the language of purely bosonic rational 
conformal field theories. In particular, in 
\cite{HSS,FS1,FSHSS}, formulas are developed for 
crosscaps and boundary states of rational conformal 
field theories with arbitrary simple-current modular 
invariants. The Gepner model, which is based on an 
$\mathcal N=2$ supersymmetric CFT, can in principle 
be formulated in this language, so that the general 
results are applicable. On the other hand, in 
\cite{BH1}, general results were derived on boundary 
and crosscap states in rational conformal field 
theories directly in the supersymmetric language, and 
using orbifold instead of simple-current techniques. 
In following this approach, we will find that it is a 
lot simpler.

\subsection{Construction of the Crosscap States}\label{sub:construction}

The crosscap states of the Gepner model
can be constructed as a straightforward application of the general
method \cite{BH1}.
Let ${\mathcal X}$ be a bosonic
conformal field theory with a finite abelian symmetry
group $G$ with which one can define an orbifold ${\mathcal X}/G$.
Suppose ${\mathcal X}$ has a parity symmetry $P$ that
commutes with the $G$-projection operator $\sum_{g\in G}g/|G|$.
Then, a parity symmetry is induced in the orbifold theory,
which is denoted again by $P$, with the crosscap state
\beq
|\Scr{C}_P\rangle^{\rm orb}
={1\over \sqrt{|G|}}\sum_{g\in G}|\Scr{C}_{gP}\rangle.
\label{sum1}
\eeq
Here $|\Scr{C}_{gP}\rangle$ are the crosscap states
for the parity symmetry $gP$ of ${\mathcal X}$, which are
supposed to obey
\beq
\langle \Scr{C}_{gP}|q_t^H|\Scr{C}_{hP}
\rangle
=\Tr\!\!\mathop{}_{{\mathcal H}_{gh^{-1}}}\Bigl[ hPq_l^H\Bigr],
\qquad \forall g,\,
\forall h,
\label{condition1}
\eeq
in which ${\mathcal H}_{gh^{-1}}$ is the space of states on the circle
with $gh^{-1}$-twist.
Indeed the Klein bottle amplitude
\beqa
{}^{\rm orb}\langle \Scr{C}_P|q_t^H|\Scr{C}_P\rangle^{\rm orb}
&=&
{1\over |G|}\sum_{g,h}\langle\Scr{C}_{gP}|q_t^H|
\Scr{C}_{hP}\rangle={1\over |G|}
\sum_{g,h}\Tr\!\!\mathop{}_{{\mathcal H}_{gh^{-1}}}\Bigl[ hPq_l^H\Bigr]\nn\\
&=&\sum_{g'\in G}\Tr\!\!\mathop{}_{{\mathcal H}_{g'}}\left[
\Bigl({1\over |G|}\sum_{h\in G}h\Bigr)Pq_l^H\right]
\nn
\eeqa
is the trace of $Pq_l^H$ over the space of states of the orbifold theory,
${\mathcal H}^{\rm orb}=\oplus_{g'\in G}{\mathcal H}_{g'}^G$.
The crosscap state for the parity
 dressed with the quantum symmetry $g_{\omega}$
associated with a character $\omega:G\to U(1)$ is
\beq
|\Scr{C}_{g_{\omega}P}\rangle^{\rm orb}
={1\over \sqrt{|G|}}\sum_{g\in G}\omega(g)^{-1}
|\Scr{C}_{gP}\rangle.
\label{sum2}
\eeq
If ${\mathcal X}$ has fermions, with mod 2 fermion number $(-1)^F$,
the above story applies, with the condition (\ref{condition1}) modified as
\beq
\langle \Scr{C}_{g(\pm)^FP}|q_t^H|\Scr{C}_{hP}
\rangle
=\Tr\!\!\mathop{}_{{\mathcal H}_{(\mp 1)^Fgh^{-1}}}
\Bigl[(-1)^FhPq_l^H\Bigr],
\qquad \forall g,\,
\forall h.
\label{condition2}
\eeq
Note that ${\mathcal H}_{(-1)^F}$ and ${\mathcal H}_{\rm id}$
are the NSNS and the RR sectors respectively.

In what follows, we apply this method to the Gepner model,
which is the orbifold of the product of the minimal models
$\prod_{i=1}^rM_{k_i}$ with respect to the group $\Gamma\cong \Z_H$.
%We treat the A-type and B-type parities separately.

\subsubsection{A-type}

We first consider A-parities of the Gepner model
$P^A_{\omega;{\bf m}}=g_{\omega}\tau^A_{\bf m}\Omega_A$, where
$\omega$ (an $H$-th root of unity) parametrizes the quantum symmetry
and ${\bf m}=(m_1,\ldots,m_r)$ labels the $\prod_{i=1}^r\Z_{k_i+2}/\Z_H$
global symmetry.
They are the ones induced from the parity symmetry of the product
theory $\prod_{i=1}^rM_{k_i}$
$$
{\bf P}^A_{\bf m}=(\g^{m_1}P_A,\ldots,\g^{m_r}P_A),
$$
where $P_A$ is the basic A-parity of the ${\mathcal N}=2$ minimal model
associated with the transformation
$X\to \overline{\Omega_A^*X}$ of the LG field.
We note that $\g P_A\g^{-1}=\g^2P_A$.

The crosscap states for the A-parities $\g^mP_A$ of the minimal model
and
their cousins $(-1)^F\g^mP_A$,
$(\pm 1)^F\g^m\widetilde{P}_A$
(where $\widetilde{P}_A=\e^{-\pi i J_0}P_A$) are
obtained in \cite{BH2} as follows
\beqa
&&|\Scr{C}_{(\pm 1)^F\g^mP_A}\rangle
=|\Scr{C}_{2m-1,-1}(\pm)\rangle,
\label{idCAmm}\\
&&|\Scr{C}_{(\pm 1)^{F}\g^m\widetilde{P}_A}\rangle
=|\Scr{C}_{2m,0}(\mp)\rangle.
\label{idCAmm2}
\eeqa
Here
\beq
|\Scr{C}_{n,s}(\pm)\rangle
=\epsilon^{\pm}_s(-1)^{n-s\over 2}
\left(\,{\sigma^{s}_{n,s}\over\sqrt{2}} 
|\Scr{C}_{n,s}\rangle
\mp {\sigma^{s}_{n,s+2}\over\sqrt{2}}
|\Scr{C}_{n,s+2}\rangle\,\right),
\label{defCns}
\eeq
where
$$
\sigma^{s}_{n,s'}
=\e^{\pi i \left(-{n^2\over 4(k+2)}+{s^2\over 8}\right)}
\e^{-\pi ih_{0ns'}},
$$
$$
\epsilon^{\pm}_{1}=\epsilon^{\pm}_{-1}=1,\qquad
\epsilon^{\pm}_0=\epsilon^{\mp}_2=
\e^{\pm{\pi i\over 4}},
$$
and
$|\Scr{C}_{n,s}\rangle$ are the PSS crosscaps
of the GSO projected model
${SU(2)_k\times U(1)_2\over U(1)_{k+2}}$,
$$
|\Scr{C}_{n,s}\rangle=\sum_{(l'm's')\in \Mk}{P_{0ns}^{\,\,\,\,l'm's'}\over
\sqrt{S_{000}^{\,\,\,\,l'm's'}}}\cket{l',m',s'}.
$$
In \cite{BH2}, the overall phase of these crosscaps are not determined.
Here, we fix the phases as above, for the following reason.

\begin{figure}[htb]
\centerline{\includegraphics{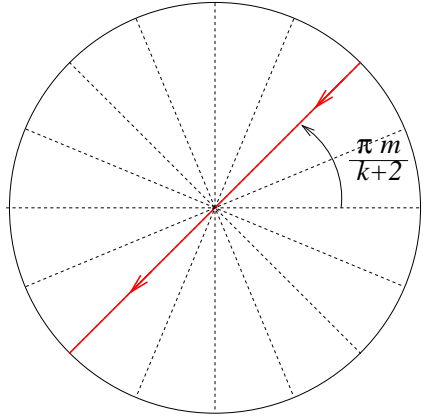}}
\caption{The O-plane corresponding to $|\Scr{C}_{2m-1,-1}(+)\rangle$.
This is the example with $k+2=8$ and $m=2$.}
\label{defOp}
\end{figure}

The $n$-dependent part of the phase is important
because this will affect the sum over the orbifold group elements,
as in (\ref{sum1}) or (\ref{sum2}).
The above choice is motivated by
the transformation property under $\g$ as well as
the periodicity under $n\to n+2(k+2)$, as we now describe.
One can show that the symmetry $\g$ acts on these states as
\beq
\g:|\Scr{C}_{n,s}(\pm)\rangle\longmapsto |\Scr{C}_{n+4,s}(\pm)\rangle.
\label{ggaction}
\eeq
This is in accord with the fact that $\g P_A\g^{-1}=\g^2P_A$
and hence that $\g|\Scr{C}_{\cdots\g^mP_A}\rangle$ is proportional to
$|\Scr{C}_{\cdots\g^{m+2}P_A}\rangle$.
What (\ref{ggaction}) means is that they are not just
proportional but {\it equal}, under the identification
(\ref{idCAmm})-(\ref{idCAmm2}) with the definition (\ref{defCns}).
Also note the periodicity
\beq
|\Scr{C}_{n+2(k+2),s}(\pm)\rangle=(-1)^s|\Scr{C}_{n,s}(\pm)\rangle.
\label{dperio}
\eeq
The crosscap states that lie in the RR-sector have a double periodicity
$n\equiv n+4(k+2)$.
Thus, with the above choice of the $n$-dependence of the phase,
these crosscaps corresponds precisely
to the oriented O-planes in the LG model;
$|\Scr{C}_{2m-1,-1}(\pm)\rangle$ corresponds to the O-plane at
$X= \e^{\pi i m\over k+2}x$, $x\in \R$, with the orientation 
that goes from positive $x$ to negative $x$.
See Figure~\ref{defOp}.
As shown in \cite{BH2}, it
has the right integral RR-charge:
overlap with the RR-boundary states
produces the correct results
for the parity-twisted open string Witten indices ---
the intersection number.

The phase factor $\epsilon_{0}^{\pm}=\epsilon_2^{\mp}=\e^{\pm {\pi i\over 4}}$
 is also added to the NSNS part of
the crosscap state, in order to simplify
the expression of the tension of the O-plane.
In fact, we need the O-plane tension to be real in the end.
Namely, we need the reality of
the overlap of the crosscap with the NSNS ground state $|0\rangle_{\NSNS}$,
in the Gepner model.
With the above choice, the overlap in the minimal model is
\beq
{}_{\NSNS}\langle 0|\Scr{C}_{2m,0}(\pm)\rangle
=\left\{
\begin{array}{ll}
\sqrt{2\over (k+2)\sin({\pi\over k+2})}
\cos\Bigl({\pi\over 2(k+2)}\Bigr)&\mbox{$k$ odd},
\\[0.2cm]
\sqrt{2\over (k+2)\sin({\pi\over k+2})}
\exp\Bigl(\pm {(-1)^m\pi i\over 2(k+2)}\Bigr)
&\mbox{$k$ even}.
\end{array}
\right.
\label{minOtens}
\eeq
If $k$ is odd, this is already real and is therefore the right choice.
If $k$ is even, this is a non-trivial phase. However, as we will see,
in the average over the orbifold group elements $\gamma^{\nu}$
($\nu\in \Z_H$),
the terms from even $\nu$ and the terms from odd $\nu$
have the opposite phase thanks to the $(-1)^m$ in
(\ref{minOtens}), and the result of the average is real or pure imaginary
depending on $\omega=1$ or $-1$.
Thus, we will simply need to multiply $\omega^{1\over 2}=\pm 1$ or $\pm i$
in the final expression. (See Section~\ref{subsec:total}.)

Applying the general method, we find that
the crosscap states
for $g_{\omega}P^A_{\bf m}=P^A_{\omega;{\bf m}}$ and
their cousins (including $\widetilde{P}^A_{\omega;{\bf m}}
=\e^{-\pi i J_0}P^A_{\omega;{\bf m}}$)
are given by
\beqa
&&
|\Scr{C}_{(\pm 1)^FP^A_{\omega;{\bf m}}}\rangle
={1\over \sqrt{H}}\sum_{\nu=1}^H\omega^{-\nu}
|\Scr{C}_{\gamma^{\nu}(\pm 1)^F{\bf P}^A_{\bf m}}\rangle^{\rm prod},
\label{CARR}
\\
&&|\Scr{C}_{(\pm 1)^{F}\widetilde{P}^A_{\omega;{\bf m}}}\rangle
={1\over \sqrt{H}}\sum_{\nu=1}^H\omega^{-\nu}
|\Scr{C}_{\gamma^{\nu}(\pm 1)^{F}
\widetilde{\bf P}^A_{\bf m}}\rangle^{\rm prod}.
\label{CANSNS}
\eeqa
in which
\beqa
&&|\Scr{C}_{\gamma^{\nu}(\pm 1)^F{\bf P}^A_{\bf m}}\rangle^{\rm prod}
=(-1)^{\sum_i{\nu\over k_i+2}}\bigotimes_{i=1}^r
|\Scr{C}_{2m_i+2\nu-1,-1}(\pm)\rangle,
\label{CAAA}\\
&&|\Scr{C}_{\gamma^{\nu}(\pm 1)^{F}
\widetilde{\bf P}^A_{\bf m}}\rangle^{\rm prod}
=\bigotimes_{i=1}^r
|\Scr{C}_{2m_i+2\nu,0}(\mp)
\label{CAAA2}
\rangle.
\eeqa
The sign factor $(-1)^{\sum_i{\nu\over k_i+2}}$
is introduced in the RR-crosscap,
in order to maintain the periodicity under $\nu\to \nu+H$,
which can be shown using (\ref{dperio}).
As mentioned above, we need to multiply
the NSNS-part of the crosscap state 
$|\Scr{C}_{(\pm 1)^{F}\widetilde{P}^A_{\omega;{\bf m}}}\rangle$ by
a phase $\omega^{1\over 2}$, in order for the O-plane tension
to be real. This will be taken care of when we discuss the
total crosscap state in string theory
in the last subsection~\ref{subsec:total}.

By the property (\ref{ggaction}) of each factor,
we find the relation
\beq
\gamma^{\nu} |\Scr{C}_{{\bf P}^A}\rangle^{\rm prod}
=|\Scr{C}_{\gamma^{2\nu}{\bf P}^A}\rangle^{\rm prod}
\label{property1}
\eeq
where ${\bf P}^A$ is one of $\gamma^{\nu'}{\bf P}^A_{\bf m}$ or any of
their cousins.
This is in accord with the relation
$\gamma^{2\nu}{\bf P}_A=\gamma^{\nu}{\bf P}^A\gamma^{-\nu}$.
%Note that $\gamma^{2\nu}{\bf P}_A=\gamma^{\nu}{\bf P}^A\gamma^{-\nu}$
%and thus we knew that $|\Scr{C}_{\gamma^{2\nu}{\bf P}^A}\rangle^{\rm prod}$
%is proportional to $\gamma^{\nu}|\Scr{C}_{{\bf P}^A}\rangle^{\rm prod}$.
%What is non-trivial here is that they are {\it equal}
%under the definitions (\ref{CAAA})-(\ref{CAAA2}).
Using this property, we can rewrite
the crosscap states in a useful way:\\[0.1cm]
{\bf\underline{$H$ odd}}\\
If $H$ is odd, the set
$\{\gamma^{\nu} {\bf P}^A\}_{\nu\in \Z_H}$ is the same as
$\{\gamma^{2\nu} {\bf P}^A\}_{\nu\in \Z_H}$.
Then, the crosscap state for the parity $P^A$
induced from ${\bf P}^A$
is simply
\beq
|\Scr{C}_{P^A}\rangle
={1\over \sqrt{H}}\sum_{\nu=1}^H\gamma^{\nu}
|\Scr{C}_{{\bf P}^A}\rangle^{\rm prod}.
\label{Cexodd}
\eeq
This state is manifestly $\Gamma$-invariant.
Note that there is no involutive dressing by quantum symmetry if $H$ is odd.
\\[0.1cm]
{\bf\underline{$H$ even}}\\
If $H$ is even, $\{\gamma^{\nu} {\bf P}^A\}_{\nu\in \Z_H}$
splits into the union of $\{\gamma^{2\nu} {\bf P}^A\}_{\nu\in \Z_{H/2}}$ and
$\{\gamma^{2\nu+1} {\bf P}^A\}_{\nu\in \Z_{H/2}}$.
The crosscap states for the parity $P^A$ or the one dressed by the
involutive quantum symmetry $g_{-1}$ are
given by
\beq
|\Scr{C}_{g_{\pm}P^A}\rangle
={1\over\sqrt{H}}\sum_{\nu=1}^{H/2}\Biggl\{
\gamma^{\nu}|\Scr{C}_{{\bf P}^A}\rangle^{\rm prod}
\pm \gamma^{\nu}|\Scr{C}_{\gamma {\bf P}^A}\rangle^{\rm prod}\Biggr\},
\label{Cexeve}
\eeq
which is also manifestly $\Gamma$-invariant.

\subsubsection{B-type}

The B-parities $P^B_{\omega;{\bf m}}$ of the Gepner model are the ones induced
from the parity symmetry of the product theory
$$
{\bf P}_{\bf m}^B=(\g^{m_1}P_B,\ldots,
\g^{m_r}P_B),
$$
where $P_B$ is the basic B-parity of the minimal model
associated with the transformation $X\to\e^{\pi i\over k+2}\Omega_B^*X$
of the LG field.

The crosscap states for the B-parities $\g^mP_B$ of the minimal model
and their cousins $(-1)^F\g^mP_B$,
$(\pm 1)^F\g^m\widetilde{P}_B$ (where
$\widetilde{P}_B=\e^{\pi i J_0}P_B$)
are
obtained in \cite{BH2} as follows
\beqa
&&|\Scr{C}_{(\pm 1)^F\g^mP_B}\rangle
=|\Scr{C}^B_{m,1}(\pm)\rangle,\\
&&|\Scr{C}_{(\pm 1)^{F}\g^m\widetilde{P}_B}\rangle
=|\Scr{C}^B_{m,0}(\mp)\rangle.
\eeqa
Here, for $r\in \Z_{k+2}$ and $p\in \{0,1\}$
\beq
|\Scr{C}^B_{r,p}(\pm)\rangle
={1\over \sqrt{k+2}}\sum_{\nu \in\Z_{k+2}}
\e^{2\pi i{\nu (r+p/2)\over k+2}}V_M|\Scr{C}_{2\nu-p,-p}(\pm)\rangle
\eeq
where $|\Scr{C}_{n,s}(\pm)\rangle$ is the A-type crosscap
defined in (\ref{defCns}) and $V_M$ is the mirror automorphism.
More explicitly,
$|\Scr{C}_{r,p}^B(\pm)\rangle
=\epsilon^{B\pm}_p\e^{\pi i p{2r+p\over k+2}}
\left({1\over \sqrt{2}}|\Scr{C}^B_{r0p}\rangle
\pm {i\over \sqrt{2}}|\Scr{C}^B_{r1p}\rangle\right)$
in which
$\epsilon_1^{B\pm}=\pm\e^{\mp{\pi i\over 4}}$, $\epsilon^{B\pm}_0=1$
and
$$
|\Scr{C}_{rqp}^B\rangle:=
{1\over 2(k+2)}\sum_{n,s\atop
{\rm even}}\e^{-\pi i\theta_{rq}(n,s)+\pi i \widehat{Q}_{(0pp)}(0ns)}
V_M|\Scr{C}_{n+p.s+p}\rangle
=\e^{-\pi i p{\widehat{2r+p}\over 2(k+2)}
+\pi i p{\widehat{2q+p}\over 4}}
|\Scr{C}_{rqp}\rangle',
$$
where
$$
|\Scr{C}_{rqp}\rangle'
= (2(k+2))^{\frac{1}{4}}
\sum_{j} \sigma_{j,2r+p,2q+p} 
\ \frac{P_{\frac{k}{2}j}}{\sqrt{S_{0j}}}
(-1)^{\frac{\widehat{2r+p}-p}{2} +q} 
\cket{j,2r+p,2q+p}_B
$$
is the state denoted by $|\Scr{C}_{rqp}\rangle$ in \cite{BH2}.
\footnote{Here $\theta_{rq}(n,s)=-{rn\over k+2}+{qs\over2}$ and
$\widehat{Q}_{a}(b)=h_a+h_b-h_{a+b}$,
and $\sigma_{j,n,s}=(-1)^{h_{j,n,s}-h_j+h_n-h_s}$
 \cite{BH2}.
Also, $\widehat{2r+p}$ is $2r+p$ (mod $2(k+2)$) brought into
 the standard range
$[-k-1,k+2]$.
Same for $\widehat{2q+p}$ (mod 4).}

Applying the general method, we find that
the crosscap states for $(\pm 1)^FP^B_{\omega;{\bf m}}$
and $(\pm 1)^{F}\widetilde{P}^B_{\omega;{\bf m}}$ are
\beqa
&&|\Scr{C}_{(\pm 1)^FP^B_{\omega;{\bf m}}}\rangle
={1\over \sqrt{H}}\sum_{\nu=1}^H\omega^{-\nu}
|\Scr{C}_{\gamma^{\nu}(\pm 1)^F{\bf P}^B_{\bf m}}\rangle^{\rm prod},
\label{CGB1}
\\
&&|\Scr{C}_{(\pm 1)^{F}\widetilde{P}^B_{\omega;{\bf m}}}\rangle
={1\over \sqrt{H}}\sum_{\nu=1}^H\omega^{-\nu}
|\Scr{C}_{\gamma^{\nu}(\pm 1)^{F}
\widetilde{\bf P}^B_{\bf m}}\rangle^{\rm prod}.
\label{CGB2}
\eeqa
in which
\beqa
&&|\Scr{C}_{\gamma^{\nu}(\pm 1)^F{\bf P}^B_{\bf m}}\rangle^{\rm prod}
=\bigotimes_{i=1}^r
|\Scr{C}^B_{m_i+\nu,1}(\pm)\rangle,\\
&&|\Scr{C}_{\gamma^{\nu}(\pm 1)^{F}
\widetilde{\bf P}^B_{\bf m}}\rangle^{\rm prod}
=\bigotimes_{i=1}^r
|\Scr{C}^B_{m_i+\nu,0}(\mp)\rangle.
\eeqa

Alternatively one could start with the A-type parities in the mirror
Gepner model, which is the orbifold of the product of minimal models with
the group $\tilGamma\subset \prod_i\Z_{k_i+2}$,
$\tilnu=(\tilnu_1,...,\tilnu_r)\in \tilGamma$ $\Leftrightarrow$
$\sum_{i=1}^r{\nu_i\over k_i+2}\in\Z$
 (see Section~\ref{subsub:mirror}).
In the mirror system, the global symmetries are parametrized by
${\bf \tilm}=(\tilm_1,...,\tilm_r)\in\prod_i\Z_{k_i+2}$ modulo
shift by $\tilnu=(\tilnu_1,...,\tilnu_r)\in \tilGamma$,
and the quantum symmetries are parametrized by
$\tilomega=(\tilomega_1,...,\tilomega_r)$, $\tilomega_i^{k_i+2}=1$,
modulo the relation $\tilomega=\tilomega'$ $\Leftrightarrow$
$\prod_i(\tilomega_i'\tilomega_i^{-1})^{\tilnu_i}=1$
($\forall\tilnu\in\tilGamma$).
They correspond respectively to the quantum symmetries $\omega$
and global symmetries ${\bf m}$
of the original Gepner model, under the map
\beqa
&&\exp\left(-2\pi i {m_i\over k_i+2}\right)=\tilomega_i.
\label{id1}\\
&&\omega=\exp\left(2\pi i \sum_{i=1}^r{\tilm_i\over k_i+2}\right),
\label{id2}
\eeqa
Then, the parity $P^B_{\omega;{\bf m}}$ of the Gepner model
corresponds to
the parity $P^A_{{\tilomega};{\bf\tilm}}$ of the mirror Gepner model,
whose crosscap states are given by
\beqa
&&
|\Scr{C}_{(\pm 1)^FP^A_{\tilomega;{\bf \tilm}}}\rangle
={1\over \sqrt{|\tilGamma|}}\sum_{\tilnu\in\tilGamma}
\tilomega^{-\tilnu}
(-1)^{\sum_i{\tilnu_i\over k_i+2}}\bigotimes_{i=1}^r
|\Scr{C}_{2\tilm_i+2\tilnu_i-1,-1}(\pm)\rangle
\\
&&|\Scr{C}_{(\pm 1)^{F}\widetilde{P}^A_{\tilomega;{\bf \tilm}}}\rangle
={1\over \sqrt{|\tilGamma|}}\sum_{\tilnu\in\tilGamma}\tilomega^{-\tilnu}
\bigotimes_{i=1}^r
|\Scr{C}_{2\tilm_i+2\tilnu_i,0}(\mp)
\rangle,
\eeqa
where $\tilomega^{-\tilnu}$ is the short-hand notation for
$\prod_i\tilomega_i^{-\tilnu_i}$.
It is straightforward to show that
\beqa
&&|\Scr{C}_{(\pm 1)^FP^B_{\omega;{\bf m}}}\rangle
=
\tilomega^{-\bf\tilm}\e^{\pi i\sum_i{\tilm_i\over k_i+2}}
V_M|\Scr{C}_{(\pm 1)^FP^A_{\tilomega;{\bf \tilm}}}\rangle
\\
&&|\Scr{C}_{(\pm 1)^{F}\widetilde{P}^B_{\omega;{\bf m}}}\rangle
=\tilomega^{-\bf\tilm}
V_M|\Scr{C}_{(\pm 1)^{F}\widetilde{P}^A_{\tilomega;{\bf \tilm}}}\rangle
\label{BvsAtil}
\eeqa
The two sets of crosscap states differ by phases.
Actually, by the condition that $P^B_{\omega;{\bf m}}$ 
is involutive, $m_i$ must be $0$ or $(k_i+2)/2$, or
$\tilomega_i=\pm 1$. Thus the overall
factor $\tilomega^{-{\bf\tilm}}$ common to all the four states
is just a sign.
On the other hand, the factor $\e^{\pi i\sum_i{\tilm_i\over k_i+2}}$
is a non-trivial phase that makes a real difference.
It turns out that the one obtained from the mirror theory
is the right choice.
This can be understood by noting that the parity twisted Witten indices
for open strings is automatically integral in the A-type crosscaps.
Thus, we modify the RR part of the crosscap in the original construction
as
\beq
(\ref{CGB1})\longrightarrow\qquad
|\Scr{C}_{(\pm 1)^FP^B_{\omega;{\bf m}}}\rangle
={1\over \sqrt{H}}\sum_{\nu=1}^H\omega^{-\nu-{1\over 2}}
|\Scr{C}_{\gamma^{\nu}(\pm 1)^F{\bf P}^B_{\bf m}}\rangle^{\rm prod},\qquad
\eeq
where $\omega^{-{1\over 2}}$ is identified as
$\e^{-\pi i\sum_i{\tilm_i\over k_i+2}}$.

\subsection{Boundary States}\label{sub:Cardybranes}

The construction of the boundary states is likewise a straightforward
application of the general method.
Let ${\mathcal X}$ and $G$ be as in the beginning of the previous subsection.
We are interested in constructing a D-brane boundary state
in ${\mathcal X}/G$ from a D-brane $\Scr{B}$ of ${\mathcal X}$.
If $\Scr{B}$ is not invariant under any element of $G$,
$\Scr{B}\mapsto g\Scr{B}\ne \Scr{B}$ ($g\ne 1$), the boundary state is
simply the sum over the images
\beq
|\Scr{B}\rangle^{\rm orb}={1\over \sqrt{|G|}}
\sum_{g\in G}g|\Scr{B}\rangle.
\label{longgeneral}
\eeq
If $\Scr{B}$ is invariant under a subgroup $H$ of $G$, 
the boundary states is the sum over images $g\Scr{B}$ ($g\in G/H$)
as well as the twist $h\in H$
\beq
|\Scr{B}\rangle^{\rm orb}
={1\over \sqrt{|G|}}\sum_{g\in G/H\atop h\in H}
g|\Scr{B}\rangle_h.
\label{shortgeneral}
\eeq
Here $|\Scr{B}\rangle_h$ is the
boundary state for $\Scr{B}$ on the $h$-twisted circle.
They are assumed to obey the relation
$$
{}_h\!\langle\Scr{B}|q_t^Hg|\Scr{B}\rangle_h
=\Tr\!\!\mathop{}_{{\mathcal H}_{\Scr{B},g\Scr{B}}}\left[ 
hq_l^H\right]
$$
where ${\mathcal H}_{\Scr{B},g\Scr{B}}$ is the space of $\Scr{B}$-$g\Scr{B}$
open string states. Indeed the cylinder amplitude
\beqa
{}^{\rm orb}\!\langle\Scr{B}|q_t^H|\Scr{B}\rangle^{\rm orb}
&=&{1\over |G|}\sum_{g_1,g_2\in G/H\atop
h\in H}
{}_h\!\langle \Scr{B}|g_1^{-1}q_t^Hg_2
|\Scr{B}\rangle_h
={1\over |H|}\sum_{g\in G/H\atop h\in H}
{}_h\!\langle \Scr{B}|q_t^Hg|\Scr{B}\rangle_h
\nn\\
&=&
\sum_{g\in G/H}\Tr\!\!\mathop{}_{{\mathcal H}_{\Scr{B},g\Scr{B}}} 
\left[\left({1\over |H|}\sum_{h\in H}h\right)q^H\right]
\nn
\eeqa
is the sum over all possible pairs $\Scr{B}$-$g\Scr{B}$
of the trace over the $H$-invariant open string
states.
If we modify the action of $H$ on the Chan-Paton factor,
we obtain a different brane.
The brane associated with the proper $H$-action given by
an $H$-character $h\mapsto \psi(h) \in U(1)$
has the boundary state
\beq
|\Scr{B}^{\psi}\rangle^{\rm orb}
={1\over \sqrt{|G|}}\sum_{g\in G/H\atop h\in H}
\psi(h) g|\Scr{B}\rangle_h.
\label{shortgeneral2}
\eeq

One may also want to include the effects of discrete torsion.
As argued in \cite{discretetorsion}, this corresponds to having
a projective representation of the orbifold group on the space
of open string states. At the level of boundary states, one uses the
alternating bihomomorphism $\epsilon:H\times H\longrightarrow
U(1)$ obtained from the discrete torsion to define a subgroup 
$K$ of $H$ for each $G$-orbit of branes as
\beq
K=\{k\in H; \epsilon(k,h)=1 \;\forall h\in H\}\,.
\label{untwisted}
\eeq
One obtains one elementary brane in the orbifold theory for
each character $k\mapsto\psi(k)\in U(1)$ of $K$. The corresponding
boundary state is a slight modification of (\ref{shortgeneral2})
\beq
|\Scr{B}^{\psi}\rangle^{\rm orb}
={1\over \sqrt{|G|}}\sqrt{\frac{|H|}{|K|}}
\sum_{g\in G/H\atop h\in H}
\psi(h) g|\Scr{B}\rangle_h,
\label{shortdiscrete}
\eeq
where the extension of $\psi$ to $H\setminus K$ is irrelevant.
We note that by general properties of group cohomology, the
factor $\sqrt{|H|/|K|}$ is always integer.

Rational boundary states of the Gepner model can be obtained
as an application of the methods described above. A-branes in 
the product theory are generically not invariant under any element 
of the orbifold group (the case $H=\{{\rm id}\}$) and thus the
boundary states in the orbifold model are simply the sum over 
images. This is how these states were first written down in 
\cite{RS}. B-branes of the product theory, on the other hand,
are all $\Gamma$-invariant (the case $H=G$) and therefore 
the boundary states are simply the sum over $\Gamma$-twists.
This way of obtaining the B-type boundary states appears to
be new, but we have found it to to be equivalent to the procedure
developed in \cite{RS,FSHSS,FKLLSW}. In particular, as we 
will see, the fixed point resolution prescription of \cite{FKLLSW} 
is correctly reproduced.

\subsubsection{A-Branes}\label{subsub:A-branes}

\begin{figure}[htb]
\centerline{\includegraphics{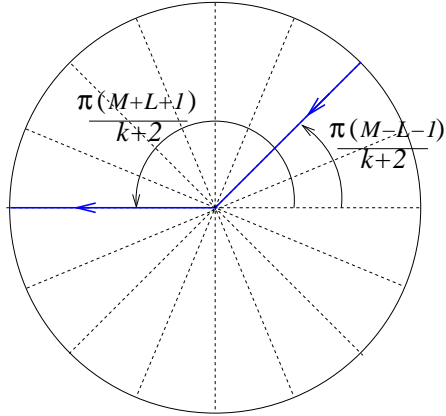}}
\caption{The A-brane $\Scr{B}^A_{L,M,S}$. This is the example
$(L,M,S)=(2,5,1)$ for $k+2=8$.}
\label{Abr}
\end{figure}

A-branes in the minimal model are denoted as
$\Scr{B}^A_{L,M,S}$ and are labeled by $(L,M,S)\in \Mk$.
 Shift of $S$-label by $2$
is simply the orientation change ---
the sign flip of RR boundary states.
$\Scr{B}^A_{L,M,S}$ preserve the combination $\overline{G}_+-(-1)^{S}G_-$
of the $(2,2)$ superconformal symmetry.
The symmetry $\g$ shifts the $M$-label by $2$.
The boundary states on the NSNS and RR sectors are given by
\beqa
&&
|\Scr{B}_{L,M}^A\rangle_\NSNS={1\over \sqrt{2}}|\Scr{B}_{L,M,S}\rangle
+{1\over\sqrt{2}}|\Scr{B}_{L,M,S+2}\rangle,
\nn\\
&&
|\Scr{B}_{L,M,S}^A\rangle_\RR={1\over \sqrt{2}}|\Scr{B}_{L,M,S}\rangle
-{1\over\sqrt{2}}|\Scr{B}_{L,M,S+2}\rangle,
\nn
\eeqa
where
$|\Scr{B}_{L,M,S}\rangle$ are the standard Cardy brane
of the GSO projected model
${SU(2)_k\times U(1)_2\over U(1)_{k+2}}$:
$$
|\Scr{B}_{L,M,S}\rangle
=\sum_{(lms)\in \Mk}{S_{LMS}^{\,\,\,\,lms}\over
\sqrt{S_{000}^{\,\,\,\,lms}}}\kket{l,m,s}.
$$
In the LG model, they correspond to the D1-brane
at the wedge-shaped lines cornering at $X=0$.
The wedge corresponding to
$\Scr{B}_{L,M,S=1}^A$ is coming in from the direction
$\arg(X)={\pi(M-L-1)\over k+2}$ and going out to the direction
$\arg(X)={\pi (M+L+1)\over k+2}$.
Replacing $S=1$ by $S=-1$ flips the orientation.
See Figure~\ref{Abr}.

We are interested in the D-brane of the Gepner model corresponding to the
product brane
$$
\Scr{B}^A_{{\bf L,M},S}=\Scr{B}^A_{L_1,M_1,S_1}\times\cdots
\times\Scr{B}^A_{L_r,M_r,S_r}.
$$
We need $S_i$ all even or all odd (i.e. $L_i+M_i$ all even or all odd),
so that either
one of $\overline{G}_+\mp G_-$ is preserved.
Orientation flip of even number of
factors does not change the total orientation.
Thus the brane depends only on the total orientation $S=[{\bf S}]$
where $[{\bf S}]\equiv [{\bf S}']$ if the number of factors with
$S_i'=S_i+2$ is even. (If $r$ is odd, $S$ can be realized as
the sum $S=\sum_iS_i\in\Z_4$.)
The orbifold group element $\gamma^{\nu}$ sends $\Scr{B}^A_{{\bf L,M},S}$
to $\Scr{B}^A_{{\bf L,M+2\nu},S}$ (where ${\bf 2\nu}=(2\nu,2\nu,...,2\nu)$).
Thus generically, $\Scr{B}^A_{{\bf L,M},S}$
is not invariant under any element of the orbifold group.
Then, the boundary state is simply the sum over images
\beqa
&&|\Scr{B}^A_{\bf L,M}\rangle_{\NSNS}
={1\over \sqrt{H}}\sum_{\nu=1}^{H}
\bigotimes_{i=1}^r
|\Scr{B}^A_{L_i,M_i+2\nu}\rangle_{\NSNS}
\label{AbrNSNS}
\\
&&|\Scr{B}^A_{{\bf L,M},S}\rangle_{\RR}
={1\over \sqrt{H}}\sum_{\nu=1}^{H}
\bigotimes_{i=1}^r
|\Scr{B}^A_{L_i,M_i+2\nu,S_i}\rangle_{\RR}.
\label{AbrRR}
\eeqa
It is a simple exercise to show that, adding
the transverse modes of the spacetime and imposing chiral GSO projection,
these lead to the boundary states obtained by Recknagel and Schomerus
\cite{RS}.

\begin{figure}[htb]
\centerline{\includegraphics{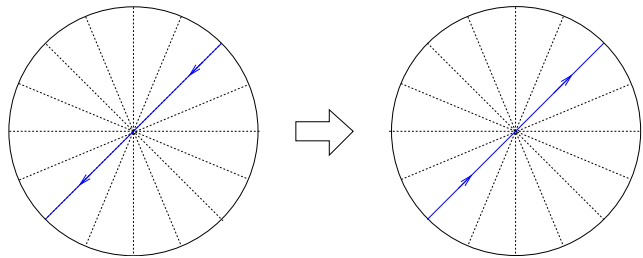}}
\caption{180${}^{\circ}$ rotation reverses the orientation of the
``straight-wedge'' branes.}
\label{rotation}
\end{figure}

The construction is different if
the product brane is invariant under a
non-trivial orbifold group element,
$\Scr{B}^A_{{\bf L,M+2\nu},S}=\Scr{B}^A_{{\bf L,M},S}$, see \cite{BS,FSW}.
Branes of different $M$ labels can be the same
because of the ``Field Identification'' (FI)
$(L,M,S)=(k-L,M+k+2,S+2)$.
Since FI is involutive we find that $2\nu=0$ mod $H$. Thus, the stabilizer
group is at most the $\Z_2$ subgroup generated by $\gamma^{H\over 2}$,
which is possible only
when $H$ is even. 
Under the symmetry $\gamma^{H/2}$, the brane $\Scr{B}^A_{{\bf L,M},S}$
transforms to $\Scr{B}^A_{{\bf L,M+{(k+2)}{{\it H}\over k+2}},S}$.
For the factor $i$ such that $w_i={H\over k_i+2}$ is even,
the brane remains the same because
$\Scr{B}_{L_i,M_i+2(k_i+2),S_i}=\Scr{B}_{L_i,M_i,S_i}$. For the
factor $i$ such that $w_i$ is odd, the brane $\Scr{B}_{L_i,M_i,S_i}$
is transformed to $\Scr{B}_{L_i,M_i+k_i+2,S_i}$.
In the LG picture, $M_i\to M_i+k_i+2$ is rotation by $\pi$, under which
a ``straight-wedge brane'' ($L_i={k_i\over 2}$)
is mapped to itself with an orientation flip. See Fig.~\ref{rotation}.
Note that there are even number of $i$'s such that $w_i$ is odd.
(Remark (ii) in Section~\ref{subsec:CYLG}.)
The total orientation is preserved even if the orientation is
reserved for each of such $i$.
Thus, the brane $\Scr{B}^A_{{\bf L,M},S}$ is invariant under
$\gamma^{H/2}$ if and only if $L_i={k_i\over 2}$ for such $i$
that $w_i$ is odd.
The boundary state is obtained as the application of the general formula
(\ref{shortgeneral2}):
\beqa
&&
|\widehat{\Scr{B}}^{(\pm)\,A}_{{\bf L,M}}\rangle_{\NSNS}
={1\over \sqrt{H}}\sum_{\nu=1}^{H\over 2}
\gamma^{\nu}|\Scr{B}^A_{{\bf L,M}}\rangle^{\rm prod}_{\NSNS}
\pm{1\over\sqrt{H}}
\sum_{\nu=1}^{H\over 2}\gamma^{\nu}
|\Scr{B}^A_{{\bf L,M}}
\rangle^{\rm prod}_{(-1)^{F}\gamma^{H/2}},
\label{shANSNS}
\\
&&
|\widehat{\Scr{B}}^{(\pm)\,A}_{{\bf L,M},S}\rangle_{\RR}
={1\over \sqrt{H}}\sum_{\nu=1}^{H\over 2}
\gamma^{\nu}|\Scr{B}^A_{{\bf L,M},S}\rangle^{\rm prod}_{\RR}
\pm{1\over\sqrt{H}}
\sum_{\nu=1}^{H\over 2}\gamma^{\nu}
|\Scr{B}^A_{{\bf L,M},S}
\rangle^{\rm prod}_{\gamma^{H/2}},
\label{shARR}
\eeqa
Note that the untwisted part is simply one half of
(\ref{AbrNSNS}) or (\ref{AbrRR}).
The boundary state
$|\Scr{B}_{{\bf L,M},S}\rangle^{\rm prod}_{(\mp 1)^{F}\gamma^{H/2}}$
is given by the product of the twisted states
$|\Scr{B}_{L_i,M_i,S_i}\rangle_{(\mp 1)^{F}\tilde{\g}^{H/2}}$
of the minimal model.
Here we replaced $\g$ by $\tilde{\g}$ 
where $\tilde{\g}=g$ if $w_i$ is even while
$\tilde{g}=a^2=\e^{-2\pi i J_0}
=\g(-1)^{\widehat{F}}$ if $w_i$ is odd.
This is because
$a^{k_i+2}$ preserves the $L_i={k_i\over 2}$
branes including the orientation.
(Since there are even number of $i$'s with odd $w_i$,
this makes no difference.)
The twist $\tilde{\g}^{H/2}$
is trivial for such $i$ that $w_i$ is even,
while it is non-trivial, $\tilde{\g}^{H/2}=a^{k_i+2}$,
for such $i$ that $w_i$ is odd.
The $a^{k+2}$-twisted boundary state
in the minimal model is given by
\beq
|\Scr{B}_{{k\over 2},M,S}^A\rangle_{(\mp 1)^Fa^{k+2}}
=\sum_{{k\over 2}+m+s\,{\rm even}\atop
s\,{\rm even/odd}}
\e^{\pi i \left(-{M+S^2+Ss+m\over 2}+{Mm+m\over k+2}\right)}
\kket{{k\over 2},m,s}.
\eeq
Since the length of the sum over images is one half of
the ordinary branes, these branes are called
{\it short orbit branes}.
``Field Identification'' is a little different on these short orbit branes.
$(L_i,M_i,S_i)\to (k_i-L_i,M_i+k_i+2,S_i+2)$ does not change the brane
if $w_i$ is even, but exchanges $+$ and $-$ label if
we do this for odd number of $i$'s with odd $w_i$.

\subsubsection{B-Branes}

B-branes in the minimal model can be obtained as the
mirror of the A-branes in the $\Z_{k+2}$-orbifold model,
see \cite{RS,FKLLSW,MMS,BH1}.
(They can also be obtained directly as an application of the methods
of \cite{FS1} by using the results of \cite{RRW} on the $\Z_2$ 
orbifold of the minimal models by mirror symmetry automorphism.)
The mirror of the brane associated with $\Scr{B}^A_{L,M,S}$ is denoted 
as $\Scr{B}^B_{L,M,S}$ and
they preserve the combination $G_+-(-1)^SG_-$ of the worldsheet
superconformal symmetry.
They are invariant under the symmetry $\g$, and
the boundary states on the various twisted circles
are given by
\beqa
|\Scr{B}_{L,M,S}^B\rangle_{(-1)^{(s'+1)F}\g^{n'}}
&=&{1\over \sqrt{k+2}}V_M\sum_{\tilnu\in \Z_{k+2}}
\e^{2\pi i{\tilnu n'\over k+2}}
|\Scr{B}^A_{L,M+2\tilnu,S}\rangle_{(-1)^{(s'+1)F}}
\nn\\
&=&(2k+4)^{{1\over 4}}\e^{-\pi i{Mn'\over k+2}+\pi i {Ss'\over 2}}
\sum_{l'\in \Pk\atop \nu_1\in \Z_2}
{S_{Ll'}\over\sqrt{S_{0l'}}}(-1)^{S\nu_1}
\kket{l',n',s'+2\nu_1}_B,
\nn\\
\eeqa
where $s'=0$ for NSNS sector and $s'=1$ for RR sector.
Note that the $M$-label appears only on the overall phase
for the boundary state with a non-trivial twist $g^{n'}\ne 1$.
Thus, the brane themselves depend only on $(L,S)$
but the $M$-label parametrizes
 the action of the global symmetry $\g$ on
the Chan-Paton factor. There is no short orbit branes
in the na\"\i ve sense,
since none of the A-branes is invariant under any
non-trivial element of $\Z_{k+2}$. However, in the GSO projected model
(i.e. in the coset model), B-branes are realized as the mirror of the
A-branes in the $\Z_{k+2}\times \Z_2$ orbifold, and for even $k$
the Cardy branes $\Scr{B}_{{k\over 2},M,S}$ are invariant under
the $\Z_2$ subgroup generated by $g_{k+2,2}$.
Thus, there are short-orbit branes in the coset model, and resolving the GSO
projection we obtain short orbit branes
$\widehat{\Scr{B}}_{{k\over 2},M,S}^B$ in the minimal model.
They are invariant under the symmetry 
$a^2=\e^{-2\pi i J_0}=\g(-1)^{\widehat{F}}$.
The boundary states on the various twisted circles are \cite{BH2}
\beqa
&&\!\!\!\!
|\widehat{\Scr{B}}_{{k\over 2},M,S}^B\rangle_{(-1)^{(n'+1)F}a^{2n'}}
={1\over\sqrt{2}}|\Scr{B}^B_{{k\over 2},M,S}\rangle_{(-1)^{(n'+1)F}g^{n'}}
\nn\\
&&\!\!\!\!
|\widehat{\Scr{B}}_{{k\over 2},M,S}^B\rangle_{(-1)^{n'F}a^{2n'}}
=\e^{-\pi i {Mn'+n'\over k+2}+\pi i {Sn'+n'\over 2}}
\sqrt{k+2\over 2}\sum_{s=\pm 1}\e^{-\pi i{S(S-s)\over 2}}
\kket{{k\over 2},{k+2\over 2}+n',s+n'}_B.
\nn
\eeqa

The long orbit brane $\Scr{B}^B_{L,M,1}$ is
described in the LG model $W=X^{k+2}$
as the one associated with the boundary
superpotential
$$
V={\mit\Gamma}X^{L+1}
$$
where ${\mit\Gamma}$ is a fermionic chiral superfield on the boundary
with constraint $\overline{D}{\mit\Gamma}=X^{k+1-L}$
and $M$ labels the action of $\g:X\to \e^{2\pi i\over k+2}X$
on the Chan-Paton ground state $|0\rangle$ (annihilated by the lowest
component of ${\mit\Gamma}$) as
$$
\g:|0\rangle\longmapsto\e^{-{2\pi i\over k+2}{M+L+1\over 2}}|0\rangle.
$$
On the other hand, short orbit branes are not realized in the LG model
with $W=X^{k+2}$ but in the model with $W=X^{k+2}-Y^2$ that also flows to
the ${\mathcal N}=2$ minimal model.
They are associated with the boundary superpotential
$$
V={\mit\Gamma}(X^{k+2\over 2}-Y), \qquad
\overline{D}{\mit\Gamma}=X^{k+2\over 2}+Y.
$$

In the open string
stretched between long and short orbit branes, there are odd number of
real fermionic zero modes \cite{BH2}.
This imposes a strong constraint in the construction of consistent set of
D-branes.

Let us first consider the product of long-orbit branes
$$
\Scr{B}^B_{{\bf L},M,S}=\Scr{B}^B_{L_1,M_1,S_1}\times\cdots\times
\Scr{B}^B_{L_r,M_r,S_r}.
$$
$S_i$ are all even or all odd and the brane
depends only on the total orientation $S=[{\bf S}]$.
Also, the $\Gamma$ action on the Chan-Paton factor depends only on
\beq
M:=H\sum_{i=1}^r{M_i\over k_i+2}\in\Z_{2H}
\eeq
which is even or odd depending on whether
$\sum_i L_i H/(k_i+2)+HrS_i$ is even or odd.
The choice of $M$ corresponds to the choice of representation of $\Gamma$
on the Chan-Paton factor.\\
The brane $\Scr{B}^B_{{\bf L},M,S}$
is invariant under all element of the orbifold group.
Thus, the boundary state in the orbifold theory is simply the
sum over the twists.
\beqa
\lefteqn{|\Scr{B}^B_{{\bf L},M,S}\rangle_{(-1)^{(s'+1)F}}
={1\over \sqrt{H}}\sum_{\nu\in\Z_{H}}
\bigotimes_{i=1}^r|\Scr{B}^B_{L_i,M_i,S_i}\rangle_{(-1)^{(s'+1)F}g^{\nu}}
}
\nn\\
&=&{1\over \sqrt{H}}\sum_{\nu\in\Z_{H}\atop
\nu_i\in \Z_2^r,l_i'\in {\rm P}_{k_i}}
\otimes_i(2k_i+2)^{1\over 4}
\e^{-\pi i {M_i\nu\over k_i+2}+\pi i {S(s'+2\nu_i)\over 2}}
{S_{L_il_i'}\over{\sqrt{S_{0l_i'}}}}\kket{l_i',\nu,s'+2\nu_i}_B,
\nn\\
\eeqa
where $s'=0$ for NSNS and $s'=1$ for RR.
This B-brane can be identified as the A-brane in the
mirror Gepner model associated with the product
$\Scr{B}_{L_1,M_1,S_1}\times\cdots\times
\Scr{B}_{L_r,M_r,S_r}$.
It is a simple exercise to reproduce the above boundary
states from this point of view.
This realization will be useful in the discussion of the
tadpole cancellation.

Next let us consider the brane involving short-orbit branes of
the minimal model.
There is one important constraint:
 the number of minimal model factors having short-orbit branes
must be even. This is to avoid the open strings to
have odd number of real fermionic zero
modes, which would be problematic upon quantization.
Thus, we will only consider product branes with even number of
$\widehat{\Scr{B}}^B_{{k_i\over 2},M_i,S_i}$ such as
$$
\widehat{\Scr{B}}_{{\bf L},M,S}
=\widehat{\Scr{B}}^B_{{k_1\over 2},M_1,S_1}\times
\widehat{\Scr{B}}^B_{{k_2\over 2},M_2,S_2}\times
\Scr{B}^B_{L_3,M_3,S_3}\times\cdots\times
\Scr{B}^B_{L_r,M_r,S_r}.
$$
The global symmetry $\g$ preserves the long-orbit brane but
reverses the orientation of the short-orbit brane.
However, since there are even number of factors with
short-orbit branes, the brane $\widehat{\Scr{B}}_{{\bf L},M,S}$ is invariant
under the orbifold group.
 Thus, again the boundary state is
the simple sum over the twists.
Note that the symmetry
$\gamma=(\g,\g,\g,...,\g)$ is the same as $(a^2,a^2,\g,...,\g)$.
The boundary state is therefore
\beqa
|\widehat{\Scr{B}}_{{\bf L},M}\rangle_\NSNS
&=&{1\over 2\sqrt{H}}\sum_{\nu\,{\rm even}}
\bigotimes_{i=1}^r|\Scr{B}^B_{L_i,M_i,S_i}\rangle_{(-1)^F\g^{\nu}}
\nn\\
&&+{1\over 2\sqrt{H}}\sum_{\nu\,{\rm odd}}
|\widetilde{\Scr{B}}^B_{{k_1\over 2},M_1,S_1}
\otimes
\widetilde{\Scr{B}}^B_{{k_2\over 2},M_2,S_2}\rangle^{}_{\nu}
\bigotimes_{i=3}^r|\Scr{B}^B_{L_i,M_i,S_i}\rangle_{(-1)^F\g^{\nu}},
\label{sslllNSNS}\\
|\widehat{\Scr{B}}_{{\bf L},M,S}\rangle_\RR
&=&{1\over 2\sqrt{H}}\sum_{\nu\,{\rm even}}
|\widetilde{\Scr{B}}^B_{{k_1\over 2},M_1,S_1}
\otimes
\widetilde{\Scr{B}}^B_{{k_2\over 2},M_2,S_2}\rangle^{}_{\nu}
\bigotimes_{i=3}^r|\Scr{B}^B_{L_i,M_i,S_i}\rangle_{\g^{\nu}}
\nn\\
&&+{1\over 2\sqrt{H}}\sum_{\nu\,{\rm odd}}
\bigotimes_{i=1}^r|\Scr{B}^B_{L_i,M_i,S_i}\rangle_{\g^{\nu}},
\label{sslllRR}
\eeqa
where
$$
|\widetilde{\Scr{B}}^B_{{k\over 2},M,S}\rangle_{\nu}=
\e^{-\pi i {M\nu+\nu\over k+2}+\pi i {S\nu+\nu\over 2}}
\sqrt{k+2}\sum_{s=\pm 1}\e^{-\pi i {S(S-s)\over 2}}
\kket{{k\over 2},{k+2\over 2}+\nu,s+\nu}_B.
$$
Let us compare this with the brane $\Scr{B}_{{\bf L},M,S}$ where
the first and the second factors are the standard ones
$\Scr{B}_{{k_1\over 2},M_1,S_1}$, $\Scr{B}_{{k_2\over 2},M_2,S_2}$.
We note that 
\beq
\begin{array}{ll}
|\Scr{B}_{{k\over 2},M,S}^B\rangle_{(-1)^F\g^{\nu}}=0&
\mbox{for odd $\nu$ and}
\\
|\Scr{B}_{{k\over 2},M,S}^B\rangle_{\g^{\nu}}=0&
\mbox{for even $\nu$.}
\end{array}
\label{rea1}
\eeq
Thus, it differs from $|\widehat{\Scr{B}}_{{\bf L},M,S}\rangle$
by the factor of $2$ and also by the absence of 
the odd $\nu$ sum in the NSNS sector
(the second line of (\ref{sslllNSNS})) and 
the even $\nu$ sum in the RR sector
(the first line of (\ref{sslllRR})).
In other words,
$$
|\Scr{B}_{{\bf L},M,S}\rangle
=|\widehat{\Scr{B}}_{{\bf L},M,S}\rangle
+|\widehat{\Scr{B}}^{(-)}_{{\bf L},M,S}\rangle
$$
where 
$|\widehat{\Scr{B}}^{(-)}_{{\bf L},M,S}\rangle$ is
obtained from 
$|\widehat{\Scr{B}}_{{\bf L},M,S}\rangle$
by flipping the sign of
the odd $\nu$ sum in the NSNS sector
and the even $\nu$ sum in the RR sector.
Thus, $\Scr{B}_{{\bf L},M,S}$ cannot be thought of as an elementary brane
but is a sum of two different branes.
The same can be said on
$\widehat{B}^B_{{\bf L},M,S}$
if two or more $L_i$ from
$L_3,...,L_r$ are the same as ${k_i\over 2}$.
If exactly one $L_i$ from $L_3,...,L_r$ is the same as ${k_i\over 2}$,
the boundary state $|\widehat{\Scr{B}}_{{\bf L},M,S}\rangle$
is simply one half of the ordinary one
$|\Scr{B}_{{\bf L},M,S}\rangle$ since the odd $\nu$ sum in NSNS
and even $\nu$ sum in RR are killed by that $i$-th factor because of
(\ref{rea1}).

By this consideration, we find that the general elementary branes are
given as follows.
For each $({\bf L},M,S)$,
things depend on the cardinality of the set
${\bf S}\subset \{1,2,...,r\}$ of $i$ for which $L_i={k_i\over 2}$.
\underline{If ${\bf S}$ is empty},
that is, if $L_i\ne {k_i\over 2}$ for all $i$,
the brane $\Scr{B}^B_{{\bf L},M,S}$
is elementary.
\underline{If $|{\bf S}|$ is even and non-zero}, the elementary branes are
\beqa
|\widehat{\Scr{B}}_{{\bf L},M,S}^{(\pm) B}\rangle_{\NSNS}
&=&{1\over 2^{|{\bf S}|\over 2}\sqrt{H}}\left\{
\sum_{\nu\,{\rm even}}
|\Scr{B}^B_{{\bf L},M,S}\rangle_{(-1)^F\gamma^{\nu}}^{\rm prod}
\pm 
\sum_{\nu\,{\rm odd}}
|\widetilde{\Scr{B}}^B_{{\bf L},M,S}\rangle^{\rm prod}_{(-1)^F\gamma^{\nu}}
\right\},
\label{fixed1}
\\
|\widehat{\Scr{B}}_{{\bf L},M,S}^{(\pm) B}\rangle_{\RR}
&=&{1\over 2^{|{\bf S}|\over 2}\sqrt{H}}
\left\{
\pm\sum_{\nu\,{\rm even}}
|\widetilde{\Scr{B}}^B_{{\bf L},M,S}\rangle^{\rm prod}_{\gamma^{\nu}}
+\sum_{\nu\,{\rm odd}}
|\Scr{B}^B_{{\bf L},M,S}\rangle_{\gamma^{\nu}}^{\rm prod}
\right\},
\label{fixed2}
\eeqa
where
\beqa
&&
|\Scr{B}^B_{{\bf L},M,S}
\rangle^{\rm prod}_{(\pm 1)^F\gamma^{\nu}}
=\bigotimes_{i=1}^r|\Scr{B}^B_{L_i,M_i,S_i}\rangle_{(\pm 1)^F\g^{\nu}},
\nn\\
&&
|\widetilde{\Scr{B}}^B_{{\bf L},M,S}
\rangle^{\rm prod}_{(\pm 1)^F\gamma^{\nu}}
=\bigotimes_{i\not\in{\bf S}}
|\Scr{B}^B_{L_i,M_i,S_i}\rangle_{(\pm 1)^F\g^{\nu}}
\otimes
\bigotimes_{i\in {\bf S}}
|\widetilde{\Scr{B}}^B_{L_i,M_i,S_i}\rangle_{\nu}.
\nn
\eeqa
\underline{If $|{\bf S}|$ is odd}, the elementary brane is
\beq
|\widehat{\Scr{B}}^B_{{\bf L},M,S}\rangle_{\NSNS\atop\RR}
={1\over 2^{|{\bf S}|-1\over 2}}
|\Scr{B}^B_{{\bf L},M,S}\rangle_{\NSNS\atop\RR}.
\label{fixed3}
\eeq

One can see that these results reproduce the fixed point
resolution prescription that is obtained by constructing the
B-type boundary states as A-type in the mirror \cite{FKLLSW}.
In this approach, one applies the Greene-Plesser orbifold
construction of mirror symmetry to the A-type brane 
$$
\Scr{B}^A_{{\bf L,M},S}=\Scr{B}^A_{L_1,M_1,S_1}\times\cdots
\times\Scr{B}^A_{L_r,M_r,S_r}.
$$
The orbifold group $G$ is the subgroup of $\prod_{i=1}^r 
\Z_{k_i+2}$ in the kernel of the elementary character of the 
diagonal subgroup $\Z_{{\rm lcm}\{k_i+2\}}$. It is then easy to see 
that the brane $\Scr{B}^A_{{\bf L,M},S}$ is invariant under the 
subgroup $H=(\Z_2)^{|{\bf S}|-1}$ generated by elements of the
form $f_{ij}=g_i^{(k_i+2)/2} g_j^{(k_j+2)/2}$ for all pairs $i, j\in\bf 
S\neq\emptyset$. The discrete torsion on $H$ was computed in 
\cite{FKLLSW}, and shown to be maximal in the sense that the size of
$K$ (see eq.\ (\ref{untwisted}). $K$ is called ``untwisted stabilizer'' 
in \cite{FKLLSW}) is the minimal compatible with the constraint 
that $|H|/|K|$ be the square of an integer. Explicitly, one finds
$$
\epsilon(f_{1i},f_{1j})= (-1)^{1+\delta_{ij}} \,.
$$
It is easy to see that this implies $K=\{{\rm id}\}$ if $|{\bf S}|-1$
is even, while $K=\Z_2$ if $|{\bf S}|-1$ is odd. Applying the general
theory of \cite{FSHSS} explained around (\ref{shortdiscrete}), this 
gives the same results for the structure of elementary short orbit 
B-branes that we have obtained in eqs.\ (\ref{fixed1}), 
(\ref{fixed2}),(\ref{fixed3}), including the normalization factor.

\subsection{Boundary/Crosscap States
in String Theory}\label{subsec:total}

We have constructed the internal parts of the 
boundary and crosscap states.
We now use them to construct the ones in full string theory
relevant for compactifications to $3+1$ dimensions
--- we add the spacetime part ($D=3+1$ free bosons and fermions
as well as ghost and superghost), and also make sure
that the states obey the chiral GSO projection condition:
$$
\mbox{IIA}:\,
\left\{
\begin{array}{l}
(-1)^{F_L}=-1\\
(-1)^{F_R}=-(-1)^{s}
\end{array}
\right.
\qquad
\mbox{IIB}:\,
\left\{
\begin{array}{l}
(-1)^{F_L}=-1\\
(-1)^{F_R}=-1
\end{array}
\right.
$$
where $s=0$ for NS-sector and $s=1$ on R-sector.
We are interested in branes filling the $D$-dimensional spacetime
and the ordinary worldsheet orientation reversal $\Omega$
that acts trivially on these $D$ coordinates.
Thus, boundary and crosscap states in the spacetime part are independent
of IIA or IIB, and are the standard coherent state
$|\Scr{B}_{\pm}^{\rm \,st}\rangle$,
$|\Scr{C}_{\pm}^{\rm \,st}\rangle$.
They are related to each other by
$$
|\Scr{B}_{-}^{\rm \,st}\rangle
=(-1)^{F_R^{\rm \,st}}|\Scr{B}_+^{\rm \,st}\rangle,
\qquad
|\Scr{C}_{-}^{\rm \,st}\rangle
=(-1)^{F_R^{\rm \,st}}|\Scr{C}_+^{\rm \,st}\rangle.
$$
Here $(-1)^{F_R^{\rm \,st}}$ is the spacetime part of the
right-moving mod 2 fermion number, which is defined so that
\beq
(-1)^{F_R}=(-)^{F_R^{\rm \,st}}\e^{\pi i J_0},
\eeq
where $J_0$ is the $U(1)$-charge of
the right-moving ${\mathcal N}=2$ superconformal algebra.
Finally, we also need to make sure that the O-plane tension is real.
This requires us to multiply the NSNS-part of the crosscap state
by a suitable phase.

In what follows in the main part of the paper,
we assume $D=3+1$, $r=5$ and
$$
\sum_{i=1}^r{1\over k_i+2}=1.
$$
Since $r=5$ is odd, the $S$ label can be represented
by $S_1=S_2=\cdots=S_5=:S$.
More general models are treated in Appendix.

\subsubsection{Type IIA Orientifolds}

To find the combination obeying
 the chiral GSO projection condition, we need to know the action of
$\e^{\pi i J_0}$ on the boundary and crosscap states we have determined.
In the individual minimal model, the action is as follows:
\beqa
&&\e^{\pi i J_0}|\Scr{B}_{L,M,S}\rangle_{\NSNS\atop\RR}
=|\Scr{B}_{L,M-1,S-1}\rangle_{\NSNS\atop\RR},
\nn\\
&&\e^{\pi i J_0}|\Scr{C}_{n,s}(\pm)\rangle
=\left\{\begin{array}{ll}
|\Scr{C}_{n-2,s}(\mp)\rangle&\mbox{$s$ even}\\
\pm \e^{-\pi i {s+1\over 2}}|\Scr{C}_{n-2,s}(\mp)\rangle&\mbox{$s$ odd}
\end{array}\right.
\nn
\eeqa
%This essentially follows from the fact that $\e^{\pi i J_0}$
%corresponds to the simple current symmetry $g_{-1,-1}$
%of the coset model ${SU(2)_k\times U(1)_2\over U(1)_{k+2}}$.
%Also, the preserved worldsheet supersymmetry is flipped
%($S\to S+1$ (mod 2) and $(\pm)\to (\mp)$ corresponds to
%the flip $\overline{Q}_++Q_-\leftrightarrow
%\overline{Q}_+-Q_-$), because
%$\e^{\pi i J_0}$ anticommutes with the right-moving supercharges
%but commutes with the left-moving supercharges.
Using this, we find that the boundary and crosscap states of the Gepner model
are transformed as
\beqa
&&\e^{\pi i J_0}|\Scr{C}_{\widetilde{P}^A_{\omega;{\bf m}}}\rangle
=\omega|\Scr{C}_{(-1)^F\widetilde{P}^A_{{\omega;{\bf m}}}}\rangle,
\nn\\
&&\e^{\pi i J_0}|\Scr{C}_{P^A_{\omega;{\bf m}}}\rangle
=-\omega|\Scr{C}_{(-1)^FP^A_{{\omega;{\bf m}}}}\rangle,
\nn\\
&&\e^{\pi i J_0}|\Scr{B}_{{\bf L,M},S}\rangle_{\NSNS\atop\RR}
=|\Scr{B}_{{\bf L,M-1},S-1}\rangle_{\NSNS\atop\RR}.
\nn
\eeqa
The appearance of $\omega$ is because of the shift in the
summation index $\nu$, and the appearance of the minus sign
in the RR-part of
the crosscap state is from the prefactor
$(-1)^{\sum_i{\nu\over k_i+2}}$ in the summand (\ref{CAAA})
of the $\nu$-summation.

We also need to make sure that the tension of the D-branes are real positive,
and the tension of the O-planes are real. We know that
${}_{\NSNS}\langle 0|\Scr{B}_{L,M,S}\rangle$ is real positive,
and thus we can use the
NSNS boundary state without modification.
As for the crosscap states, using the formula (\ref{minOtens})
for the minimal model, we find that for $H$ odd
(all $k_i$ odd)
$$
{}_{\NSNS}\langle 0|\Scr{C}_{(\pm 1)^F\widetilde{P}_{{\bf m}}}\rangle
=\sqrt{H}\prod_{i=1}^r\mbox{$\sqrt{2\over (k_i+2)\sin({\pi\over k_i+2})}$}
\cos\Bigl({\pi\over
2(k_i+2)}\Bigr),
$$
and for $H$ even (some $k_i$ even)
$$
{}_{\NSNS}\langle 0|\Scr{C}_{(\pm 1)^F\widetilde{P}_{\omega;{\bf m}}}\rangle
={\sqrt{H}\over 2}\prod_{i=1}^r
\mbox{$\sqrt{2\over (k_i+2)\sin({\pi\over k_i+2})}$}
\prod_{k_i\,{\rm odd}}\cos\Bigl({\pi\over
2(k_i+2)}\Bigr)\cdot\Bigl(
\e^{\mp i\Theta}+\omega\e^{\pm i\Theta}\Bigr),
$$
$$
\Theta=\sum_{k_i\,{\rm even}}{(-1)^{m_i}\pi\over 2(k_i+2)}.
$$
We see that it is real if $H$ is odd and also for the $\omega=1$ case
if $H$ is even. However, for the $\omega=-1$ case ($H$ even),
it is pure imaginary. To make it real, me must multiply
the state by $i$.
In general, multiplication by $\omega^{1\over 2}$ will do the job.

Collecting all these items, we find that the total crosscap and
boundary states are given by
\beqa
&&|C_{\omega;{\bf m}}\rangle_{\NSNS}
=\omega^{1\over 2}|\Scr{C}_{\widetilde{P}^A_{\omega;{\bf m}}}\rangle\otimes 
|\Scr{C}_+^{\rm \,st}\rangle_{\NSNS}
-\omega^{-{1\over 2}}
|\Scr{C}_{(-1)^{F}\widetilde{P}^A_{\omega;{\bf m}}}\rangle\otimes 
|\Scr{C}_-^{\rm \,st}\rangle_{\NSNS}\\
&&|C_{\omega;{\bf m}}\rangle_{\RR}
=|\Scr{C}_{P^A_{\omega;{\bf m}}}\rangle\otimes
|\Scr{C}_+^{\rm \,st}\rangle_{\RR}
-\omega
|\Scr{C}_{(-1)^FP^A_{\omega;{\bf m}}}\rangle\otimes
|\Scr{C}_-^{\rm \,st}\rangle_{\RR},
\label{CARtotal}
\eeqa
and
\beqa
&&|B_{\bf L,M}\rangle_{\NSNS}=
|\Scr{B}_{{\bf L,M+1},1}\otimes\Scr{B}_+^{\rm \,st}\rangle_{\NSNS}
-|\Scr{B}_{{\bf L,M},0}\otimes\Scr{B}_-^{\rm \,st}\rangle_{\NSNS},
\\
&&|B_{{\bf L,M}}\rangle_{\RR}
=|\Scr{B}_{{\bf L,M+1},1}\otimes\Scr{B}_+^{\rm \,st}\rangle_{\RR}
+|\Scr{B}_{{\bf L,M},0}\otimes\Scr{B}_-^{\rm \,st}\rangle_{\RR}.
\eeqa
We note that there are still a freedom to flip the sign of
them except the NSNS part of the boundary state.
The sign flip of
the RR-parts of the boundary/crosscap states corresponds to
orientation flip, and the sign flip of the
NSNS part of the crosscap state corresponds to the flip in the type of
the orientifold.
The choice of this sign for the NSNS crosscap can be made by the choice of
the phase $\omega^{1\over 2}$ (that is, $1$ or $-1$ for $\omega=1$,
and $i$ or $-i$ for $\omega=-1$).

\subsubsection{Type IIB Orientifolds}

The action of $\e^{\pi i J_0}$ on B-type boundary and crosscap states
can be found either directly or by using the mirror description.
Here, we present the latter way.
We first note that $\e^{\pi i J_0}$ and mirror automorphism obey the
following relation
$$
\e^{\pi i J_0}V_M=V_M\e^{-\pi i J_0}
=\left\{\begin{array}{ll}
V_M\e^{\pi i J_0}&\mbox{on NSNS sector}\\
(-1)^rV_M\e^{\pi i J_0}&\mbox{on RR sector}.
\end{array}\right.
$$
Using this and using the mirror realization of the crosscap and boundary
state we find
\beqa
&&\e^{\pi i J_0}|\Scr{C}_{\widetilde{P}^B_{\omega;{\bf m}}}\rangle
=\e^{2\pi i\sum_i{m_i\over
k_i+2}}|\Scr{C}_{(-1)^F\widetilde{P}^B_{\omega;{\bf m}}}\rangle,
\nn\\
&&\e^{\pi i J_0}|\Scr{C}_{P^B_{\omega;{\bf m}}}\rangle
=\e^{2\pi i\sum_i{m_i\over
k_i+2}}|\Scr{C}_{(-1)^FP^B_{\omega;{\bf m}}}\rangle,
\nn\\
&&\e^{\pi i J_0}|\Scr{B}_{{\bf L},M,S}\rangle_{\NSNS\atop\RR}
=|\Scr{B}_{{\bf L},M+\sum_i{H\over k_i+2},S+1}\rangle_{\NSNS\atop\RR}
\nn
\eeqa
We also find, by direct computation, that the B-brane including short-orbit
brane factors are transformed in the same way as
$|\Scr{B}_{{\bf L},M,S}\rangle$.

The next item is the reality of the overlap of the crosscap states
with the NSNS ground state. Here again, the mirror description is useful.
We have just experienced what to do for the A-type crosscaps.
This tells us that for the reality of the overlap with
$|0\rangle_{\NSNS}$ we need to multiply
$|\Scr{C}_{(\pm 1)^F\widetilde{P}^A_{\tilomega;{\bf\tilm}}}\rangle$
by the phase
$$
\tilomega^{1\over 2}=\prod_i\tilomega_i^{1\over 2}
=\exp\left(-\pi i\sum_{i=1}^r{m_i\over k_i+2}\right).
$$

The total crosscap and boundary states are given by
\beqa
&&|C^B_{\omega;{\bf m}}\rangle_{\NSNS}
=\tilomega^{1\over 2}|\Scr{C}_{\widetilde{P}^B_{\omega;{\bf m}}}\rangle\otimes 
|\Scr{C}_+^{\rm \,st}\rangle_{\NSNS}
-\tilomega^{-{1\over 2}}
|\Scr{C}_{(-1)^{F}\widetilde{P}^B_{\omega;{\bf m}}}\rangle\otimes 
|\Scr{C}_-^{\rm \,st}\rangle_{\NSNS},
\label{CBNStotal}\\
&&|C^B_{\omega;{\bf m}}\rangle_{\RR}
=|\Scr{C}_{P^B_{\omega;{\bf m}}}\rangle\otimes
|\Scr{C}_+^{\rm \,st}\rangle_{\RR}
-\tilomega^{-1}
|\Scr{C}_{(-1)^FP^B_{\omega;{\bf m}}}\rangle\otimes
|\Scr{C}_-^{\rm \,st}\rangle_{\RR},
\label{CBRtotal}
\eeqa
and
\beqa
&&|B^B_{{\bf L},M}\rangle_{\NSNS}=
|\Scr{B}^B_{{\bf L},M+\sum_i{H\over k_i+2},1}
\otimes\Scr{B}_+^{\rm \,st}\rangle_{\NSNS}
-|\Scr{B}^B_{{\bf L},M,0}\otimes\Scr{B}_-^{\rm \,st}\rangle_{\NSNS},\\
&&|B^B_{{\bf L},M}\rangle_{\RR}=
|\Scr{B}^B_{{\bf L},M+\sum_i{H\over k_i+2},1}
\otimes\Scr{B}_+^{\rm \,st}\rangle_{\RR}
+|\Scr{B}^B_{{\bf L},M,0}\otimes\Scr{B}_-^{\rm \,st}\rangle_{\RR},
\eeqa
The sign of the second term of the RR boundary state is $+$ because
$|\Scr{B}^B_{{\bf L},M+\sum_i{2H\over k_i+2},2}\rangle_{\RR}
=(-1)^r|\Scr{B}^B_{{\bf L},M,0}\rangle_{\RR}
=-|\Scr{B}^B_{{\bf L},M,0}\rangle_{\RR}$, where $r=5$ is used.
The choice of the sign for the NSNS crosscap can be made by the choice of
the phase $\tilomega^{1\over 2}$
($1$ or $-1$ for $\tilomega=1$,
and $i$ or $-i$ for $\tilomega=-1$).

\section{Consistency Conditions and Supersymmetry --- A}
\label{sec:TCCA}

In this and the next sections,
we determine the conditions of consistency and
spacetime supersymmetry of 
Type II orientifolds on Gepner model with rational D-branes.
We focus on compactification down to $3+1$ dimensions.

The main part of consistency conditions is the
RR tadpole cancellation \cite{PolCai,Callan}
$$
\langle \mbox{massless RR scalar}|T\rangle =0.
$$
In terms of the internal CFT, this can be written as
\beq
{}_\RR\langle i |\Scr{C}_P\rangle\,+\,
{1\over 4}\,
{}_\RR\langle i |\Scr{B}_+\rangle_\RR\,=\,0,
\label{internal}
\eeq
for any RR ground states $|i\rangle_\RR$ of the internal theory responsible
for RR scalars.
The factor of $1/4$ is from the $3+1$ dimensional spacetime part.
The other condition is when there are D-branes invariant under
the orientifold action. If that is of $Sp$-type, the number of such branes
must be even.

Spacetime supersummetry is conserved by a set of branes $\Scr{B}^a$
($a=1,...,N$) when
the overlaps ${}_\NSNS\langle 0|\Scr{B}^a_+\rangle_\NSNS$ and 
${}_\RR\langle 0|\Scr{B}^a_+\rangle_\RR$ differ by a phase common to all
$a$.
This phase determines the conserved combination of supercharges.
A spacetime supersymmetry exists in the orientifold model when
the supersymmetries preserved by the D-branes and the O-planes are the
same;
\beq
{{}_\RR\langle 0|\Scr{B}^a_+\rangle_\RR
\over {}_\NSNS\langle 0|\Scr{B}^a_+\rangle_\NSNS}
={{}_\RR\langle 0|\Scr{C}_P\rangle_\RR
\over {}_\NSNS\langle 0|\Scr{C}_P\rangle_\NSNS}.
\label{ssusy}
\eeq

In the rest of this section,
we write down these conditions for Type IIA orientifolds.
We will also find a very simple class of solutions to these conditions, 
and compute the particle spectrum in selected examples. Finding the most 
general solution is a rather hard problem, about which we will also make 
some comments towards the end of this section. In section 
\ref{sec:TCCB}, we will present complete solutions of the tadpole conditions 
for Type IIB orientifolds of Gepner models, which are a lot simpler. 
To be sure, we do not mean to say that A-type tadpole conditions are 
intrinsically harder to solve than B-type. Indeed, A and B-type are 
identified under mirror symmetry. The solutions in the Gepner model we 
seek in this section are interpreted in the large volume limit as A-type 
on the quintic or B-type on the mirror quintic (and vice-versa in section 
\ref{sec:TCCB}). There are tadpole cancellation problems in Gepner models 
which are of intermediate difficulty, such as in certain orbifolds of the 
quintic. We discuss one of them in the appendix.

\subsection{Charge and Supersymmetry of O-planes}

Let us first review the description of RR-charge of
the A-type D-branes and O-planes in a general LG model
(see \cite{HIV,HKKPTVVZ,BH2} for more details).
Let us consider a LG model on
a non-compact K\"ahler manifold $X$ of dimension $n$
with superpotential $W$.
A-branes are D$n$-branes wrapped on an oriented
Lagrangian submanifolds of $X$ that lie in level sets of
${\rm Im}(W)$. An A-type orientifold is associated with
an antiholomorphic involution $\tau$ of $X$ that maps $W$ to
its complex conjugate $\overline{W}$ up to a constant shift.
The O-plane $O_{\tau \Omega}$, the fixed point set of $\tau$, 
is also a Lagrangian submanifold in a level set
of ${\rm Im}(W)$ and we assume that an orientation is chosen.
To describe their charges, we introduce
the subspaces $B^{\pm}\subset X$ which are the set of points with large
values of $\pm {\rm Im}(W)$, say, $B^{\pm}=\{x\in X|
\pm {\rm Im}(W(x))\geq R\}$ for a sufficiently large $R>0$.
For an A-brane $\gamma$, we deform its asymptotics so that their
$W$-images are deformed to $\pm{\rm Im}(W)>R$. Let us denote the resulting
submanifolds by $\gamma^{\pm}$.
The submanifold $\gamma^{+}$ has its boundaries in $B^+$,
and defines a homology class relative to $B^+$:
\beq
[\gamma^+]\in H_n(X,B^+).
\eeq
This is the one that represents the RR-charge of the A-brane.
To be precise, this is the charge
at the in-coming boundary
preserving the supercharge $\overline{Q}_++Q_-$.
The charge at the out-going boundary preserving the same
supercharge (or at the
in-coming boundary preserving the opposite combination $\overline{Q}_+-Q_-$)
is given by the other class $[\gamma^-]\in H_n(X,B^-)$.
The Witten index for the open string
stretched from $\gamma_1$ and $\gamma_2$ is given by the intersection
number
$\#(\gamma_1^-\cap\gamma_2^+).$
The RR-charge of the O-plane at the in-coming crosscap
for the parity commuting with the supercharge $\overline{Q}_++Q_-$
is similarly given by
\beq
[O_{\tau\Omega}^+]\in H_n(X,B^+).
\eeq

\begin{figure}[htb]
\centerline{\includegraphics{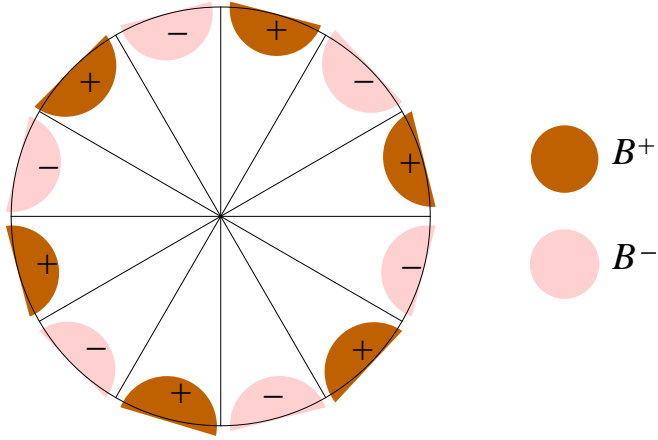}}
\caption{The regions $B^{\pm}$ for the case $k=4$.}
\label{B}
\end{figure}

Let us apply this to the LG model with $W=X^{k+2}$
that flows to the ${\mathcal N}=2$ minimal model at level $k$.
The $X$-plane is separated into $2(k+2)$ regions by the inverse images of
the real line of the $W$-plane, and
$B^+$ and $B^-$ consist of the asymptotic regions that appears alternately,
as depicted in Fig.~\ref{B}.

\noindent
As mentioned in section \ref{subsub:A-branes}, the A-brane $\Scr{B}_{L,M,S}$
corresponds to the D1-brane at the wedge-shaped line $\gamma_{L,M,S}$
 coming in from the direction
$\arg(X)=\pi {M-L-1\over k+2}$, cornering at $X=0$, and going out to
the direction $\arg(X)=\pi {M+L+1\over 2}$
if $S=0$ or $1$ ($S=2$ or $-1$ are their orientation reversals).
The branes with $S=\pm 1$ preserve the supercharge
$\overline{Q}_++Q_-$ while those with $S=0,2$ preserves the opposite
combination $\overline{Q}_+-Q_-$.
The cycle $\gamma_{L,M,1}^+$ is obtained 
by slightly rotating $\gamma_{L,M,1}$, counter-clock-wise.
This correspondence $\gamma_{L,M,S}\leftrightarrow |\Scr{B}_{L,M,S}\rangle$
is at the in-coming boundary.
At the out-going boundary, the correspondence is slightly different:
$\gamma_{L,M,1}$ (resp. $\gamma_{L,M,0}$)
corresponds to $\langle \Scr{B}_{L,M-1,0}|$ (resp. 
$\langle \Scr{B}_{L,M-1,1}|$). This can be understood by comparing
the RR-charges as well as the conserved worldsheet
supersymmetries.

\begin{figure}[htb]
\centerline{\includegraphics{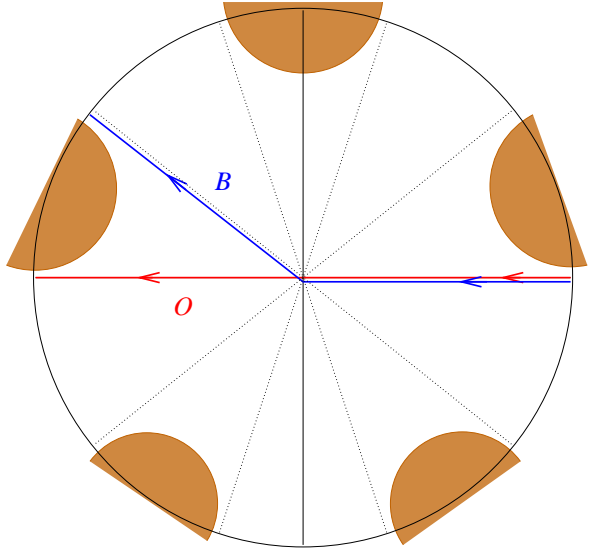}}
\caption{The O-plane $O=O_{P_A}$ and the brane
$B=\gamma_{1,2,1}$ in the $k=3$ minimal model.
They have the same in-coming RR-charge}
\label{k=3}
\end{figure}

The parity $\g^mP_A$ commutes with the worldsheet supercharge
$\overline{Q}_++Q_-$ which is preserved by branes $\gamma_{L,M,S}$ with
odd $S$. It acts on the LG field as
$X\to \e^{2\pi i m\over k+2}\overline{X}$ and
the O-plane $O_{\g^mP_A}$ is the straight line
at $X\in\e^{\pi i m\over k+2}\R$. We assume the orientation
that goes from $\arg(X)={\pi m\over k+2}$ to the opposite direction.
Note that $m\to m+(k+2)$ is the orientation flip.
The cycle $O_{\g^mP_A}^+$ is obtained by deforming it so
that both of the two asymptotics are in the region $B^+$.
This involves bending when $k$ is odd while it is a small 
rotation when $k$ is even.
To see this, let us first consider the basic A-parity
$P_A$ whose O-plane $O_{P_A}$ is the real line that goes from
$+\infty$ to $-\infty$.
If $k$ is even, $O_{P_A}^+$ is the slight counter-clockwise rotation of
$\R$. In fact there is an A-brane that does the same ---
$\gamma_{{k\over 2},{k+2\over 2},1}$.
Thus, the O-plane $O_{P_A}$ and $\gamma_{{k\over 2},{k+2\over 2},1}$ has
the same location and the same charge.
If $k$ is odd, $O_{P_A}^+$ is obtained by small counter-clockwise rotation
of the real-positive half and small clockwise rotation of
the real-negative half.
There is no A-brane at the same location, but the brane
$\gamma_{{k-1\over 2},{k+1\over 2},1}$ has the same in-coming charge.
(Another brane $\gamma_{{k-1\over 2},{k+1\over 2}+1,0}$
may appear to have the same charge, but it preserves a different
combination $\overline{Q}_+-Q_-$ of the supercharge
--- $P_A$ preserves the combination $\overline{Q}_++Q_-$
and thus must be compared to the branes
with odd $S$.)
Figure~\ref{k=3} depicts the example of $k=3$.
Repeating this consideration in the general case, we find that the
O-plane $O_{\g^mP_A}$ has the same RR-charge as one of the A-branes.
The result is
\beqa
\mbox{$k$ even}
&&[O_{\g^mP_A}^+]=
\left\{\begin{array}{ll}
[\gamma_{{k\over 2},{k+2\over 2}+m,1}^+]&\mbox{$m$ even}\\
{[}\gamma_{{k\over 2},{k+2\over 2}+m-1,1}^+]&\mbox{$m$ odd},
\end{array}\right.
\label{Ochevmin}
\\
\mbox{$k$ odd}
&&[O_{\g^mP_A}^+]=
\left\{\begin{array}{ll}
[\gamma_{{k-1\over 2},{k+1\over 2}+m,1}^+]&\mbox{$m$ even}\\
{[}\gamma_{{k+1\over 2},{k+1\over 2}+m,1}^+]&\mbox{$m$ odd}.
\end{array}\right.
\label{Ochodmin}
\eeqa
This can also be checked by showing
${}_{\RR}\langle i|\Scr{C}_{\g^mP_A}\rangle={}_{\RR}\langle
i|\Scr{B}_{L,M,S}\rangle_{\RR}$ for any RR-ground state
$|i\rangle_{\RR}$
with $(L,M,S)$ as indicated in
(\ref{Ochevmin})-(\ref{Ochodmin}).
Note that $m\to m+(k+2)$ indeed corresponds to orientation flip
since RR-part of the corresponding boundary states flips its sign.

Having learned the RR-charge of the O-plane in the minimal model,
we can now compute the charge in the Gepner model.
For this purpose, the expressions (\ref{Cexodd}) and (\ref{Cexeve})
of the crosscap states are useful. These expressions simply says that
the O-plane charge in the Gepner model is given by
the same type of average formula for the A-brane charge.

If $H$ is odd, the average formula (\ref{Cexodd})
is identical to the one for an A-brane.
Note that we only have to consider the basic parity
$P^A=P^A_{1;{\bf 0}}$ since there is no involutive dressing by quantum
symmetry and dressing by global symmetry ${\bf m}$ is equivalent to
no-dressing.
By the relation (\ref{Ochodmin}) for each minimal model
we find that the O-plane charge is the same as the charge of the
D-brane associated with the product
$\prod_{i=1}^r\Scr{B}_{{k_i-1\over 2},{k_i+1\over 2},1}$.
Namely,
\beq
\Bigl[O_{P^A}\Bigr]=4\Bigl[B_{\bf {k-1\over 2},{k-1\over 2}}\Bigr],
\label{Ochod}
\eeq
where the factor of $4$ comes from the spacetime part.

If $H$ is even, the sum splits into two parts (\ref{Cexeve})
and each part is the same as the untwisted part of the sum for an 
A-brane with $\Z_2$ stabilizer group.
The charge for $|\Scr{C}_{{\bf P}^A_{\bf m}}\rangle$
is the same as
the charge for the product brane
$\prod_i\Scr{B}_{{k_i\over 2},{k_i+2\over 2}+m_i-\delta_{m_i},1}$
where 
$$
\delta_{m_i}=\left\{\begin{array}{ll}
0&\mbox{$m_i$ even}\\
1&\mbox{$m_i$ odd}.
\end{array}\right.
$$
The charge for $|\Scr{C}_{\gamma{\bf P}^A_{\bf m}}\rangle$ is the same as
the charge for
$-\prod_i\Scr{B}_{{k_i\over 2},{k_i+2\over 2}+m_i+\delta_{m_i},1}$,
where the minus sign is from the
factor $(-1)^{\sum_i{\nu\over k_i+2}}$ in (\ref{CAAA}).
Thus, the charge is
\beq
\Bigl[O_{P^A_{\pm;{\bf m}}}\Bigr]=
2\Bigl[B_{\bf {k\over 2},{k\over 2}+m-\delta_m}\Bigr]
\mp
2\Bigl[B_{\bf {k\over 2},{k\over 2}+m+\delta_m}\Bigr],
\label{Ochev}
\eeq
where $[B_{\bf {k\over 2},M}]$ is the sum
of the two short-orbit brane charges
$[\widehat{B}_{\bf {k\over 2},M}^{(+)}]
+[\widehat{B}_{\bf {k\over 2},M}^{(-)}]$
which has no twisted state component.

\begin{figure}[htb]
\centerline{\includegraphics{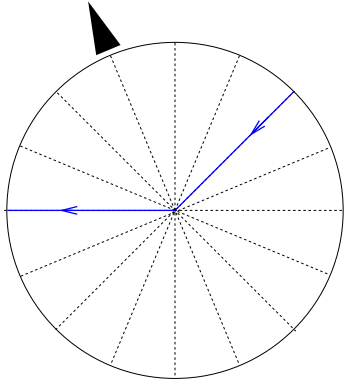}}
\caption{The mean direction of a brane.}
\label{mean}
\end{figure}

Let us discuss the spacetime supersymmetry preserved by
D-branes and the O-plane. This is to compute the ratio of the 
overlap of the boundary/crosscap state with RR-ground state $|0\rangle_{\RR}$
and the brane/plane tension.
Here $|0\rangle_{\RR}$ is the RR ground state of the internal
system with the lowest R-charge.
Let us first present the RR-overlap for the A-brane in the minimal model.
This has been computed in many ways.
In the LG description, it is realized as the integral over
$\gamma_{L,M,1}^+$ of the 1-form $\overline{c_0}\e^{-i\overline{X}^{k+2}}
*\dd \overline{X}$ where $c_0$ is a certain normalization factor
\cite{HIV,HKKPTVVZ,BH2}. The result is
\beqa
{}_{\RR}\langle0|\Scr{B}_{L,M,1}\rangle_{\RR}
&=&
i\sqrt{2\over (k+2)\sin({\pi\over k+2})}\e^{-\pi i {M\over k+2}}
\sin\Bigl({\pi (L+1)\over k+2}\Bigr)
\nn\\
&=&i\e^{-\pi i {M\over k+2}}{}_{\NSNS}\langle 0|
\Scr{B}_{L,M,1}\rangle_{\NSNS}.
\nn
\eeqa
Using this, we find that the overlap in the Gepner model is
$$
{}_{\RR}\langle0|\Scr{B}_{{\bf L,M},1}\rangle_{\RR}
=i^r\e^{-\pi i \sum_i{M_i\over k_i+2}}
{}_{\NSNS}\langle 0|\Scr{B}_{{\bf L,M},1}\rangle_{\NSNS}.
$$
The phase determining the spacetime supersymmetry is the ratio
\beq
\exp\left(i\theta_{\bf L,M}\right)
=-i
\exp\left(-\pi i\sum_{i=1}^r{M_i\over
k_i+2}
\right).
\label{SUSYAbrane}
\eeq
We find that the phase is determined by the sum over the angles
${M_i\over k_i+2}$ of the ``mean-direction'' of the wedge in the LG
realization. See Figure~\ref{mean}. The result is applicable also to short
orbit branes.

Let us next compute the RR-overlap for the crosscap states.
In the minimal model,
this is essentially computed in \cite{BH2}, in both using PSS
crosscap and also using LG model.
Here one could also use the relation of the O-plane charge and D-brane
charge given in (\ref{Ochevmin}) and (\ref{Ochodmin}) and the above
expression for the brane overlaps.
In any way, we find
\beqa
{}_{\RR}\langle 0|\Scr{C}_{\g^mP_A}\rangle
&=&\left\{\begin{array}{ll}
i\sqrt{2\over (k+2)\sin({\pi\over k+2})}
\cos\Bigl({\pi\over 2(k+2)}\Bigr)\e^{-\pi i {{k+1\over 2}+m\over k+2}}
&\mbox{$k$ odd},
\\[0.2cm]
i\sqrt{2\over (k+2)\sin({\pi\over k+2})}
\e^{-\pi i {{k+2\over 2}+m-\delta_m\over k+2}}
&\mbox{$k$ even}
\end{array}
\right.
\nn\\
&=&i\e^{-\pi i {m+{k+1\over 2}\over k+2}}\times {}_{\NSNS}\langle 0|
\Scr{C}_{\g^m\widetilde{P}_A}\rangle.
\nn
\eeqa
It follows from this that in the Gepner model
$$
{}_{\RR}\langle 0|\Scr{C}_{P_{\omega;{\bf m}}^A}\rangle
=i^r\e^{-\pi i\sum_i{m_i+{k+1\over 2}\over k_i+2}}\times
{}_{\NSNS}\langle 0|\Scr{C}_{\widetilde{P}_{\omega;{\bf m}}^A}\rangle
$$
Since the NSNS crosscap in string theory is obtained by multiplying
$\omega^{1\over 2}$ to $|\Scr{C}_{\widetilde{P}_{\omega;{\bf
m}}^A}\rangle$,
we find that the ratio is
\beq
\exp\left(i\theta_{P^A_{\omega;{\bf m}}}\right)
=-i
\omega^{-{1\over 2}}\exp\left(-\pi i\sum_{i=1}^r{m_i+{k_i-1\over 2}\over
k_i+2}
\right).
\label{SUSYAOplane}
\eeq
The phase is essentially the sum over the slopes 
${m_i+{k_i-1\over 2}\over k_i+2}$ of the direction perpendicular
to the O-planes if $\omega=1$, but it differs from that sum by
right angle if $\omega=-1$.

In Table \ref{tableAOch}, we describe the
RR-charge, the tension, and the phase determining the conserved
supersymmetry of the twelve A-type orientifolds of the
two parameter model $(k_i+2)=(8,8,4,4,4)$.
\begin{table}[htb]
\begin{center}
\renewcommand{\arraystretch}{1.2}
\begin{tabular}{|l||l|l|l|}
\hline
parity&RR-charge&Tension&SUSY\\
\noalign{\hrule height 0.8pt}
\multicolumn{1}{|l||}{$P^B_{+;00000}$}
&
\multicolumn{1}{l|}{$0$}
&
\multicolumn{1}{l|}{$0$}&\lw{$i\omega^{-{1\over 2}}$}\\
\cline{1-3}
$P^B_{-;00000}$
&
$4[B_{{\bf k\over 2},{\bf k\over 2}}]$
&
$-i\omega^{1\over 2}2\sqrt{2\sqrt{2}+2}$
&
\\
\hline
\multicolumn{1}{|l||}{$P^B_{+;00001}$}
&
\multicolumn{1}{l|}{$2[B_{{\bf k\over 2},{\bf k\over 2}}
-B_{{\bf k\over 2},(33113)}]$}
&
\multicolumn{1}{l|}{$\omega^{1\over 2}2\sqrt{\sqrt{2}+1}$}
&\lw{$i\omega^{-{1\over 2}}\e^{-{\pi i\over 4}}$}\\
\cline{1-3}
$P^B_{-;00001}$
&$2[B_{{\bf k\over 2},{\bf k\over 2}}
+B_{{\bf k\over 2},(33113)}]$
&$-i\omega^{1\over 2}2\sqrt{\sqrt{2}+1}$
&
\\
\hline
\multicolumn{1}{|l||}{$P^B_{+;01000}$}
&
\multicolumn{1}{l|}{$2[B_{{\bf k\over 2},{\bf k\over 2}}
-B_{{\bf k\over 2},(35111)}]$}
&
\multicolumn{1}{l|}{$2\omega^{1\over 2}$}
&\lw{$i\omega^{-{1\over 2}}\e^{-{\pi i\over 8}}$}\\
\cline{1-3}
$P^B_{-;01000}$
&$2[B_{{\bf k\over 2},{\bf k\over 2}}
+B_{{\bf k\over 2},(35111)}]$
&$-2i\omega^{1\over 2}(\sqrt{2}+1)$
&
\\
\hline
\multicolumn{1}{|l||}{$P^B_{+;00011}$}
&
\multicolumn{1}{l|}{$2[B_{{\bf k\over 2},{\bf k\over 2}}
-B_{{\bf k\over 2},(33133)}]$}
&
\multicolumn{1}{l|}{$\omega^{1\over 2}2\sqrt{2}\sqrt{\sqrt{2}+1}$}
&\lw{$\omega^{-{1\over 2}}$}\\
\cline{1-3}
$P^B_{-;00011}$
&$2[B_{{\bf k\over 2},{\bf k\over 2}}
+B_{{\bf k\over 2},(33133)}]$
&$0$
&
\\
\hline
\multicolumn{1}{|l||}{$P^B_{+;01001}$}
&
\multicolumn{1}{l|}{$2[B_{{\bf k\over 2},{\bf k\over 2}}
-B_{{\bf k\over 2},(35113)}]$}
&
\multicolumn{1}{l|}{$\omega^{1\over 2}2(\sqrt{2}+1)$}
&\lw{$i\omega^{-{1\over 2}}\e^{-{3\pi i\over 8}}$}\\
\cline{1-3}
$P^B_{-;01001}$
&$2[B_{{\bf k\over 2},{\bf k\over 2}}
+B_{{\bf k\over 2},(35113)}]$
&$-2i\omega^{1\over 2}$
&
\\
\hline
\multicolumn{1}{|l||}{$P^B_{+;11000}$}
&
\multicolumn{1}{l|}{$2[B_{{\bf k\over 2},{\bf k\over 2}}
-B_{{\bf k\over 2},(55111)}]$}
&
\multicolumn{1}{l|}{$\omega^{1\over 2}2\sqrt{\sqrt{2}+1}$}
&\lw{$i\omega^{-{1\over 2}}\e^{-{\pi i\over 4}}$}\\
\cline{1-3}
$P^B_{-;11000}$
&$2[B_{{\bf k\over 2},{\bf k\over 2}}
+B_{{\bf k\over 2},(55111)}]$
&$-i\omega^{1\over 2}2\sqrt{\sqrt{2}+1}$
&
\\
\hline
\end{tabular}
\caption{Charge and Tension of O-planes in the two parameter model
($\omega=1$)}
\label{tableAOch}
\end{center}
\end{table}

\subsection{Parity Action on D-branes}

The next task is to find out how the parities act on the D-branes.

Let us first consider the action in the minimal model, which is studied
\cite{BH2}.
The action is encoded in the formulae
\beqa
&&\langle \Scr{C}_{\g^mP_A}|q_t^H|\Scr{B}_{L,M,S}\rangle_{\RR}
={}_{\RR}\langle\Scr{B}_{L,2m-M-1,-S-1}|q_t^H|
\Scr{C}_{(-1)^F\g^mP_A}\rangle,
\label{PonDmin1}
\\
&&\langle \Scr{C}_{\g^m\widetilde{P}_A}|q_t^H|\Scr{B}_{L,M}\rangle_{\NSNS}
={}_{\NSNS}\langle\Scr{B}_{L,2m-M}|q_t^H|
\Scr{C}_{(-1)^F\g^m\widetilde{P}_A}\rangle.
\label{PonDmin2}
\eeqa
They can be shown using the properties of
the P-matrix $\e^{2\pi i Q_g(j)}P_{gj}^*=P_{gj}$.
They can also be geometrically understood in the LG model as follows.
Under the basic parity $P_A$ that acts on the LG field as
$X\to\overline{X}$, the wedge $\gamma_{L,M,1}$ is mapped to its
complex conjugate. The initial and final angles 
$({\pi(M-L-1)\over k+2},{\pi (M+L+1)\over k+2})$ of the wedge
are mapped to $({\pi(-M+L+1)\over k+2},{\pi (-M-L-1)\over k+2})$
which are the initial and final angles of the wedge $\gamma_{L,-M,1}$ with
the opposite orientation.
Thus the brane $\gamma_{L,M,1}$ is mapped under $P_A$ to
$-\gamma_{L,-M,1}=\gamma_{L,-M,-1}$.
More general parity maps the brane as
$$
\g^mP_A:\gamma_{L,M,1}\to\gamma_{L,2m-M,-1}.
$$
Note that the parity exchanges in-coming and out-going
boundaries. Thus, if $\gamma_{L,M,1}$ is at the in-coming boundary and
corresponds to $|\Scr{B}_{L,M,1}\rangle$, then
$\gamma_{L,2m-M,-1}$ is at the out-going boundary
and corresponds to $\langle \Scr{B}_{L,2m-M-1,-2}|$.
Thus we find that the parity acts as
$$
\g^mP_A:|\Scr{B}_{L,M,1}\rangle\to\langle\Scr{B}_{L,2m-M-1,-2}|.
$$
This is nothing but the $S=1$ case of (\ref{PonDmin1}).
The one with other values of $S$ and the other
relation (\ref{PonDmin2}) can also be understood in a similar way.

Let us now discuss the orientifold
action on D-branes in the full string theory.
The relations (\ref{PonDmin1}) and (\ref{PonDmin2})
readily imply
\beqa
&&\langle \Scr{C}_{P^A_{\omega;{\bf m}}}|q_t^H|
\Scr{B}_{{\bf L,M},S}\rangle_{\RR}
={}_{\RR}\langle\Scr{B}_{{\bf L,2m-M-1},-S-1}|q_t^H|
\Scr{C}_{(-1)^FP^A_{\omega;{\bf m}}}\rangle,
\label{PonDG1}
\\
&&\langle \Scr{C}_{\widetilde{P}^A_{\omega;{\bf m}}}|q_t^H
|\Scr{B}_{\bf L,M}\rangle_{\NSNS}
={}_{\NSNS}\langle\Scr{B}_{\bf L,2m-M}|q_t^H|
\Scr{C}_{(-1)^F\widetilde{P}^A_{\omega;{\bf m}}}\rangle.
\label{PonDG2}
\eeqa
Applying these to the total crosscap states,
we find
\beqa
&&{}_{\RR}\langle C_{\omega;{\bf m}}|q_t^H|B_{\bf L,M}\rangle_{\RR}
=
\omega\times {}_{\RR}\langle
B_{\bf L,2m-M}|q_t^H|C_{\omega;{\bf m}}\rangle_{\RR},\\
&&{}_{\NSNS}\langle C_{\omega;{\bf m}}|q_t^H|B_{\bf L,M}\rangle_{\NSNS}
=-{}_{\NSNS}\langle
B_{\bf L,2m-M}|q_t^H|C_{\omega;{\bf m}}\rangle_{\NSNS}.
\label{secon}
\eeqa
Recall that the overlaps appears in the one-loop diagram
as the combination
$$
i{}_{\NSNS}\langle B|q_t^H|C\rangle_{\NSNS}
-i{}_{\NSNS}\langle C|q_t^H|B\rangle_{\NSNS}
-{}_{\RR}\langle B|q_t^H|C\rangle_{\RR}
-{}_{\RR}\langle C|q_t^H|B\rangle_{\RR}.
$$
Thus the equation (\ref{secon}) shows
that the brane $B_{\bf L,M}$ is mapped to $B_{\bf L,2m-M}$ if the 
brane orientations are ignored. The first equation includes the
information on the orientations.
It shows that the branes are mapped under
the orientifold action as
\beq
P^A_{\omega;{\bf m}}:\,
B_{\bf L,M}\longmapsto 
\omega B_{\bf L,2m-M},
\label{PAonD}
\eeq
where $-B$ stands for the orientation reversal of $B$.
We see that dressing by quantum symmetry affects the action on orientation.

Let us see how the short-orbit branes are transformed.
We recall that if $L_i={k_i\over 2}$ for each
$i$ with odd $w_i$ (possible only when $H$ is even),
the brane $B_{\bf L,M}$ must be regarded as the sum of
two short-orbit branes $\widehat{B}_{\bf L,M}^{(+)}$
and $\widehat{B}_{\bf L,M}^{(-)}$.
The boundary states are given in (\ref{shANSNS}) and (\ref{shARR}).
The overlap $\langle \widehat{B}^{(\pm)}|C\rangle$
is simply one-half of $\langle B|C\rangle$ for both $\pm$, since
the crosscap state $|C\rangle$ does not have twisted state components.
Thus we see that the 
$({\bf L,M})$-label is transformed in the same way as
the long-orbit branes.
To see how the $\pm$ label is transformed, we need to compare with
the $\langle\widehat{B}^{(\varepsilon)}|\widehat{B}^{(\varepsilon')}\rangle$
overlaps.
By the loop channel expansion of the latter overlaps,
one can read the spectrum of open string states between two short-orbit
branes: the states labeled by
$\otimes_{i=1}^r(l_i,m_i,s_i)$ are subject to the projection
\beq
  \frac12\left(\,1+\varepsilon\varepsilon' 
\prod_{w_i~{\rm odd}}(-1)^{\frac{1}{2}(l_i+m_i-s)}\,\right),
\label{AshProj}
\eeq
where $s=0$ for NS states and $1$ for R ones.
%It follows that the short-orbit branes also support $U(1)$ gauge
%symmetry as the massless vector states between the branes with the
%same sign are not projected out by (\ref{AshProj}).
Let us compare this with the loop-channel expansion of
the overlaps $\langle\widehat{\Scr{B}}|q_t^H|\Scr{C}\rangle$. 
It turns out that the open string states are subject to the projection
\beq
  \frac12\left(\,1+\omega^{\frac{H}{2}}(-1)^{\frac{\sigma}{2}}
  \prod_{w_i~{\rm odd}}(-1)^{\frac{1}{2}(l_i+m_i-s)}\,\right),
\label{AshPA}
\eeq
where $\sigma$ is the number of $i$'s such that $w_i$ is odd.
The parity action on short-orbit branes is therefore summarized as
\beq
P^A_{\omega;{\bf m}}~:~
\widehat{B}_{\bf L,M}^{(\varepsilon)}\longmapsto
\omega
\widehat{B}_{\bf L,2m-M}^{(\varepsilon')},~~~
\varepsilon'=\omega^{\frac{H}{2}}(-1)^{\frac{\sigma}{2}}\varepsilon.
\label{PAonDshort}
\eeq
For the computations that leads to (\ref{AshProj})
and (\ref{AshPA}), see Appendix~\ref{app:detail1}.

\subsubsection{Invariant Branes}

Let us see which of the branes are invariant under the parity symmetries.
By (\ref{PAonD}), the long-orbit brane $B_{\bf L,M}$ is invariant under
$P^A_{\omega;\bf m}$ when $\omega B_{\bf L,2m-M}=B_{\bf L,M}$.
This requires that, for each $i$, $\Scr{B}_{L_i,2m_i-M_i,0}$ is equal to
$\Scr{B}_{L_i,M_i,0}$ up to orientation (and up to the uniform
shift in $M_i$'s).
One possibility is $L_i$ arbitrary and
$2m_i-M_i=M_i$ (mod $2k_i+4$) which is the case with the positive
orientation, and another is $L_i=k_i/2$ and
$2m_i-M_i=M_i+k_i+2$ (mod $2k_i+4$) which is the case with the reversed
the orientation.
For $\omega=1$, we need the total orientation to be positive, and thus
the case ``$L_i=k_i/2$ and
$2m_i-M_i=M_i+k_i+2$'' must occur for even number of $i$'s:
\beq
P^A_{+;{\bf m}}\mbox{-fixed}:\left\{\begin{array}{ll}
L_i={k_i\over 2},\,\,\,
M_i=m_i+{k_i+2\over 2}\,\,(\mbox{mod $k_i+2$}),&
\mbox{for even \# of $i$'s}\\
L_i\,\,\mbox{arbitrary,}\,\,\,
M_i=m_i\,\,\,(\mbox{mod $k_i+2$}),&
\mbox{for other $i$}
\end{array}\right.
\eeq
For $\omega=-1$ (which is possible only when some $k_i$ are even),
we need the total orientation to be negative,
and thus
the case ``$L_i=k_i/2$ and
$2m_i-M_i=M_i+k_i+2$'' must occur for odd number of $i$'s:
\beq
P^A_{-;{\bf m}}\mbox{-fixed}:\left\{\begin{array}{ll}
L_i={k_i\over 2},\,\,\,
M_i=m_i+{k_i+2\over 2}\,\,(\mbox{mod $k_i+2$}),&
\mbox{for odd \# of $i$'s}\\
L_i\,\,\mbox{arbitrary,}\,\,\,
M_i=m_i\,\,\,(\mbox{mod $k_i+2$}),&
\mbox{for other $i$}
\end{array}\right.
\eeq

Let us next consider the short orbit branes
$\widehat{B}_{\bf L,M}^{(\varepsilon)}$.
For this case, the ``Brane Identification'' involves the change in
$\varepsilon$-label:
$M_i\to M_i+k_i+2$ for $i$ with odd $w_i$
does the flip of $\varepsilon$ in addition to the flip of orientation.
Also, the parity acts on the $\varepsilon$ label as
$$
\varepsilon\to\omega^{\frac{H}{2}}(-1)^{\frac{\sigma}{2}}\varepsilon.
$$
Thus, the invariant branes are those with
$$
\left\{\begin{array}{ll}
L_i={k_i\over 2},\,\,\,
M_i=m_i+{k_i+2\over 2}\,\,(\mbox{mod $k_i+2$}),&
i\in I\\
L_i\,\,\mbox{arbitrary,}\,\,\,
M_i=m_i\,\,\,(\mbox{mod $k_i+2$}),&
i\not\in I
\end{array}\right.
$$
where $I$ is a subset of $\{1,...,r=5\}$ obeying 
some condition that depends on the parity and case as described in the table.
\begin{table}[htb]
\begin{center}
\renewcommand{\arraystretch}{1.2}
\begin{tabular}{|c|c|c|c|c|}
\hline
&$P^A_{+;{\bf m}}$, ${\sigma\over 2}$ even&
$P^A_{+;{\bf m}}$, ${\sigma\over 2}$ odd&
$P^A_{-;{\bf m}}$, ${\sigma+H\over 2}$ even&
$P^A_{-;{\bf m}}$, ${\sigma+H\over 2}$ odd
\\
\noalign{\hrule height 0.8pt}
$\# I$&even&even&odd&odd
\\
\hline
$\# (I\cap\Sigma)$&even&odd&even&odd
\\
\hline
\end{tabular}
\end{center}
\end{table}
Here $\Sigma$ is the set of $i$'s with odd $w_i$ ($\sigma=\#\Sigma$).

\subsection{Structure of Chan-Paton Factor}

Let us now determine the structure of Chan-Paton factor on the
D-branes. If not with orientifolds, $N$ D-branes on top of
each other, i.e., $N$ copies of a D-brane, support $U(N)$ gauge group.
In the orientifold model, this is modified.
If the D-brane $B_a$ is not invariant under the parity,
$P:B_a\to B_a'\ne B_a$,
the gauge group is still $U(N)$ since $B_a$-$B_a$ string is simply related to
$B_a'$-$B_a'$ string under the orientifold projection.
However, for an invariant D-brane, a non-trivial projection is imposed on
the open string ending on it, and the gauge group is usually either
$O(N)$ or $USp(N)=Sp(N/2)$.
In the latter case $N$ must be even, which
is one of the consistency requirement.

Thus, we would like to find the orientifold projections on
invariant D-branes.
Let $\psi^{\mu}_{-{1\over 2}}|IJ\rangle$ ($1\leq I,J\leq N$) be
the open string states corresponding to the massless gauge bosons.
The parity action is
$$
\psi^{\mu}_{-{1\over 2}}|IJ\rangle\mapsto
-\sum_{I'J'}\gamma_{II'}\psi^{\mu}_{-{1\over 2}}|J'I'\rangle
\gamma^{-1}_{J'J}
$$
where $\gamma^T=\pm \gamma$ is required for the parity to be involutive.
The gauge group is $O(N)$ if $\gamma^T=\gamma$ (solved by $\gamma={\bf 1}_N$)
and $Sp({N\over 2})$ if
$\gamma=-\gamma^T$ (solved by $\gamma={{\bf 0}\,-{\bf 1}\choose
{\bf 1}\,\,\,\,{\bf 0}}$).
One consequence is that
$$
\Tr\Bigl|_{\rm gauge\,\,boson}Pq^H
=-2{\rm tr}(\gamma^T\gamma^{-1})q^{1\over 2}=-2N\sigma q^{1\over 2},
$$
where
\beq
\sigma=\left\{\begin{array}{ll}
1&\mbox{if $O(N)$}\\
-1&\mbox{if $Sp({N\over 2})$}.
\end{array}
\right.
\eeq
Thus, to find out the type of the Chan-Paton factor, we want to look at
the sign in front of the gauge boson part $q^0$
in the twisted partition function
$\Tr Pq^H=i\langle B|q_t^H| C\rangle$.

It is straightforward to compute the $\langle B|C\rangle$
overlaps of the minimal model and their loop-channel expansions.
In particular, the ground state contribution is
\beq
{}_{\NSNS}\langle \Scr{B}_{L,M,S}|q_t^H|\Scr{C}_{2m,0}(\mp)\rangle
=\left\{
\begin{array}{l}
\e^{\mp{\pi i\over 4}}\delta_{M,m}\delta_{S,0}\\
\e^{\pm{\pi i\over 4}}\delta_{M,m}\delta_{S,1}\\
\e^{\pm{\pi i\over 4}}\delta_{L,{k\over 2}}
\delta_{M,m+{k+2\over 2}}\delta_{S,0}\\
\e^{\mp{\pi i\over 4}}\delta_{L,{k\over 2}}
\delta_{M,m+{k+2\over 2}}\delta_{S,1}
\end{array}\right\}\times\widehat{\chi}_{0,0,0}(q_l)+\cdots
\label{PonGSmin}
\eeq
where the delta functions are mod $k+2$ for $M$-indices and 
mod 2 for $S$-indices. 
For the universal part, we find
\beqa
&&\langle\Scr{B}_+|q_t^H|\Scr{C}_+\rangle_{\NSNS}
=\langle\Scr{B}_-|q_t^H|\Scr{C}_-\rangle_{\NSNS}
=\e^{-{\pi i\over 4}}q^{-{1\over 2}}
\prod_{n=1}^{\infty}(1-i(-1)^nq_l^{n-{1\over 2}})^2,
\nn\\
&&\langle\Scr{B}_+|q_t^H|\Scr{C}_-\rangle_{\NSNS}
=\langle\Scr{B}_-|q_t^H|\Scr{C}_+\rangle_{\NSNS}
=\e^{{\pi i\over 4}}q^{-{1\over 2}}
\prod_{n=1}^{\infty}(1+i(-1)^nq_l^{n-{1\over 2}})^2,
\nn
\eeqa
up to the factors from
bosonic transverse oscillators, longitudinal modes, and ghost/super\-ghost
sectors.
Combining the above equations, we find
\beqa
\lefteqn{\langle B_{\bf L,M}|q_t^H|C_{\omega;{\bf m}}\rangle_{\NSNS}}
\nn\\
&=&\!\!\omega^{-{1\over 2}}\e^{{\pi i\over 4}(-1+r_1-r_2)}
q^{-{1\over 2}}\prod_{n=1}^{\infty}(1-i(-1)^nq_l^{n-{1\over 2}})^2
-\omega^{1\over 2}\e^{{\pi i\over 4}(1-r_1+r_2)}
q^{-{1\over 2}}\prod_{n=1}^{\infty}(1+i(-1)^nq_l^{n-{1\over 2}})^2
\nn\\
&&+\cdots
\eeqa
up to the universal factor, where we have decomposed $r$ as
$r_1+r_2$;
$$
\begin{array}{l}
r_1=\#\{i|M_i=m_i\}\\
r_2=\#\{i|L_i={k_i\over 2},M_i=m_i+{k_i+2\over 2}\}
\end{array}
$$
Note that
$$
\omega^{1\over 2}\e^{{\pi i \over 4}(1-r_1+r_2)}
=\e^{{\pi i \over 4}(1-r)}\omega^{1\over 2}\e^{{\pi i\over 2}r_2}
=-\omega^{1\over 2}\e^{{\pi i\over 2}r_2}
$$
where $r=5$ is used. It is a sign factor since
$r_2$ is even
for $\omega=1$, and $r_2$ is odd if $\omega=-1$.
This is the sign $\sigma$ that determines the structure of Chan-Paton factor.
$N$ D-branes support $O(N)$ gauge group if it is $+1$
while they support $Sp(N/2)$ gauge group if it is $-1$:
$$
-\omega^{1\over 2}\e^{{\pi i\over 2}r_2}=
\left\{\begin{array}{lcl}
1&\Longrightarrow&O(N),\\[0.2cm]
-1&\Longrightarrow&Sp({N\over 2}).
\end{array}\right.
\label{ACPf}
$$
For example, consider the case where $\omega^{1\over 2}=-1$.
Then, the branes with $M_i=m_i$ for all $i$ support $O(N)$ gauge group,
those with two $i$'s with $L_i={k_i\over 2},M_i=m_i+{k_i+2\over 2}$ support
$Sp(N/2)$ gauge group, and those with
four $i$'s with $L_i={k_i\over 2},M_i=m_i+{k_i+2\over 2}$ support
$O(N)$ gauge group.

To see the LG image of this rule, let us look at the two kinds
of invariant branes in the minimal model, one with
$M=m$ another with $L={k\over 2}$, $M=m+{k+2\over 2}$.
\begin{figure}[htb]
\centerline{\includegraphics{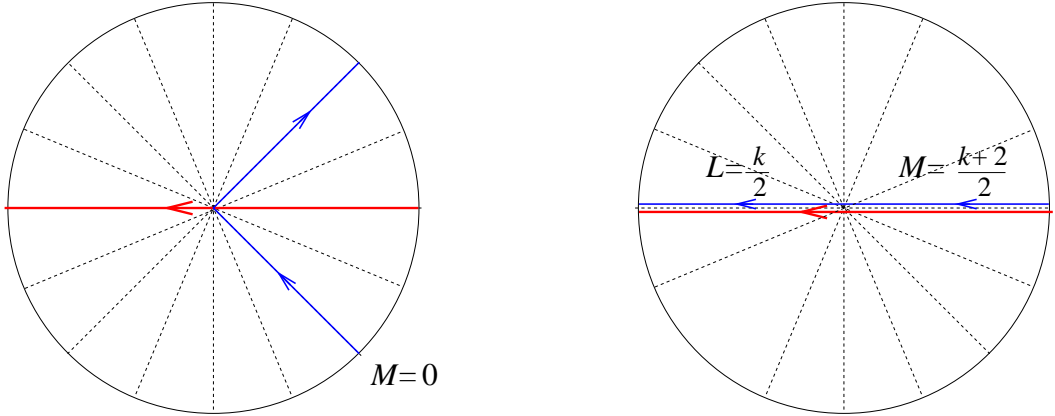}}
\caption{Two branes invariant under the parity with $m=0$.}
\label{CPfig}
\end{figure}
The $M=m$ branes intersect transversely with the O-plane and the
$L={k\over 2}$, $M={k+2\over 2}$ branes are parallel to the O-plane.
We have seen above that replacing transverse branes
by parallel brane in two factors flips the type of the CP factor
from $O(N)$ to $Sp({N\over 2})$ and vice versa.
Note that two factors means real four-dimensions.
This is very much reminiscent of what happens in 
the standard superstring in flat space.
For example, consider a Type II orientifold with an O7-plane.
If D7-branes parallel to O7-plane support $O(N)$,
D7-branes intersecting orthogonally to O7-plane in real four-dimensions
support $Sp({N\over 2})$.
(This is what happens if we obtain this system by T-duality from Type I
and decompactification \cite{GimPol}, but we could also consider
the opposite case --- parallel branes support $Sp({N\over 2})$
and orthogonal branes support $O(N)$.)

The result (\ref{ACPf})
applies also to short-orbit branes,
since the overlap with the crosscap state does not
receive contribution from the twisted sector.

\subsection{A Class of Consistent and Supersymmetric D-brane
Configurations}

We have obtained the expression of the charge and the supersymmetry
of the O-plane, and we also described the orientifold action on rational
D-branes and the structure of CP-factor of the invariant D-branes.
Thus we have obtained the condition of consistency as well as the
spacetime supersymmetry on the D-brane configurations.
Now, we are interested in finding solutions.
It is not an easy task to classify all solutions
because the rank of the charge lattice is very large (typically 100)
and also there are many D-branes preserving the unbroken supersymmetry.
In this subsection, we present special solutions.
In the case with odd $H$, we find one solution in each case.
In the case with even $H$, we present an algorithm to find a solution.
It works in most of the cases but sometimes it fails.

To simplify the notation, we consider the O-plane of the reversed
orientation. Namely, we use $-|C_{\omega;{\bf m}}\rangle_{\RR}$
in place of the RR-crosscap state. For this choice the
RR-charge and the phase for the spacetime supersymmetry (e.g.
the ones in the table of page~\pageref{tableAOch}) have extra minus sign.

\subsubsection{Odd $H$}

If $H$ is odd, we have seen that the RR-charge of the crosscap state
is equal to the RR-charge of one of the D-branes, which is
$B_{\bf {k-1\over 2},{k-1\over 2}}$.
One can also see that this brane preserves the same spacetime supersymmetry
unbroken by the orientifold --- the phases (\ref{SUSYAbrane}) and
(\ref{SUSYAOplane}) are both $-i^r\exp(-\pi i\sum_i{k_i-1\over 2(k_i+2)})$.
Furthermore, this brane is invariant under the orientifold action.
This can be shown as follows.
As we found above, the brane
is mapped to $B_{\bf {k-1\over 2},-{k-1\over 2}}$.
Here, we note that
$H=(k_i+2)w_i$ where $w_i$ is an integer --- in fact $w_i$ is an odd
integer in the present case where $H$ is odd.
This means that $H\equiv k_i+2$ mod $2(k_i+2)$, or
$$
(H-3)\equiv k_i-1 \quad\mbox{mod $2(k_i+2)$}.
$$
Using this we find that the orientifold action is
$$
P^A:B_{\bf {k-1\over 2},{k-1\over 2}}\longrightarrow
B_{\bf {k-1\over 2},-{k-1\over 2}}=
B_{{\bf {k-1\over 2},-{k-1\over 2}}+(H-3){\bf 1}}
=B_{\bf {k-1\over 2},{k-1\over 2}}.
$$
Namely the brane $B_{\bf {k-1\over 2},{k-1\over 2}}$
is mapped to itself under the
orientifold action.

Thus, we find that 
a consistent and spacetime supersymmetric configuration is given by
four $B_{\bf {k-1\over 2},{k-1\over 2}}$'s. 
This may be regarded as the configuration of four D-branes ``on top of'' the
O-plane, although we do not have a geometrical picture.

One consequence of this result is that, when continued on the
K\"ahler moduli space from the Gepner point to a large volume limit,
the branes $B_{\bf {k-1\over 2},{k-1\over 2}}$  becomes the
D-brane wrapped on the D6-brane wrapped on the fixed point set
of the involution $\tau:X_i\to \overline{X}_i$.
For example, the D-brane wrapped on the real quintic in the quintic
hypersurface in $\CP^4$ is the continuation of
 $B_{{\bf 1,1}}$.

\subsubsection{Even $H$}

If $H$ is even, the RR-charge of the O-plane is expressed
as in (\ref{Ochev}) as the sum or the difference of the
untwisted part of the RR-charge of two kinds of branes,
$B_{\bf {k\over 2},{k\over 2}+m-\delta_m}$ and
$B_{\bf {k\over 2},{k\over 2}+m+\delta_m}$.
However, generically the two preserve different combinations of
supersymmetry as one can see from their phases, and in particular, neither
one of them preserve the supersymmetry unbroken by the orientifold.
In order to preserve spacetime supersymmetry, one has
to find the set of D-branes all with the same phase whose RR-charge
in total equals that of the O-plane.
We study the example $(k_i+2)=(8,8,4,4,4)$
in detail.

\noindent
{\underline{$\bullet$ $P^A_{\pm;00000}$}}\\[0.1cm]
This is a special case in which the two D-branes are the same.
For the $P^A_{+;00000}$-orientifold,
the O-plane has no RR charge and hence it gives
a consistent supersymmetric configuration 
\underline{without adding any
D-branes}.
For the $P^A_{-;00000}$-orientifold,
the O-plane charge
is equal to $-4[B_{\bf {k\over 2},{k\over 2}}]
=-4[\widehat{B}_{\bf {k\over 2},{k\over 2}}^{(+)}
+\widehat{B}_{\bf {k\over 2},{k\over 2}}^{(-)}]$.
Thus,
\underline{four $\widehat{B}^{(+)}_{\bf {k\over 2},{k\over 2}}$ and
four $\widehat{B}^{(-)}_{\bf {k\over 2},{k\over 2}}$}
provide a tadpole canceling brane configuration.
Note that the twisted part of the RR-charge carried
by the $(+)$-brane and the $(-)$-brane cancel against each other.
One can also see that
they preserve the same supersymmetry unbroken by the orientifold
with $\omega^{1\over 2}=-i$.
The $+$ branes and $-$ branes are exchanged with each other under the
orientifold. Hence
the gauge group supported by the branes is $U(4)$.

\begin{figure}[htb]
\centerline{\includegraphics{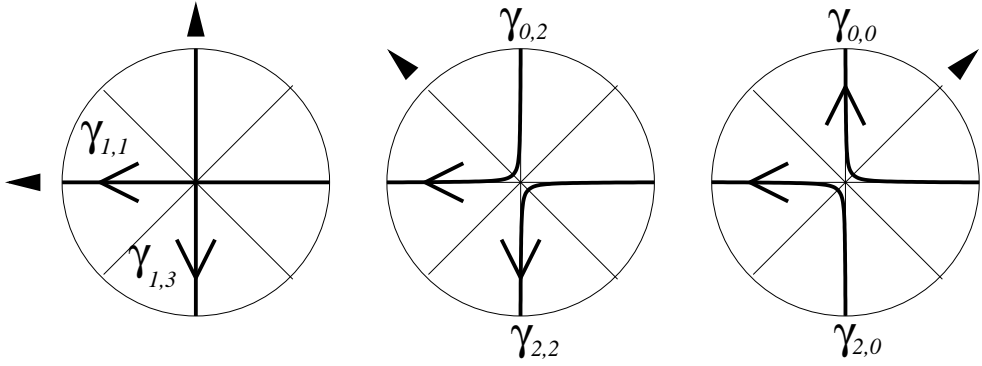}}
\caption{The recombination of the branes for $\gamma_{1,1}+\gamma_{1,3}$
(middle) and $\gamma_{1,1}-\gamma_{1,3}$ (right).}
\label{keq2}
\end{figure}

\noindent
{\underline{$\bullet$ $P^A_{\pm;00001}$}}\\[0.1cm]
The O-plane charge is given by
$$
 [O_{P^A_{\pm;00001}}] ~=~
    -2[\widehat{B}_{{\bf k\over 2},(33111)}]
\pm 2[\widehat{B}_{{\bf k\over 2},(33113)}].
$$
The two D-branes indeed preserve different combinations of supersymmetry
since the $M$-labels of the 5-th factor are different.
Let us now focus on this factor.
As one can see from Fig.~\ref{keq2},
the sum and the difference of the two charges
can be recombined as follows (to simplify the notation,
we denote $\gamma_{L,M+1,1}$ by $\gamma_{L,M}$):
$$
  [\gamma^+_{1,1}]-[\gamma^+_{1,3}] ~=~
  [\gamma^+_{2,0}]+[\gamma^+_{0,0}],\quad
  [\gamma^+_{1,1}]+[\gamma^+_{1,3}] ~=~
  [\gamma^+_{2,2}]+[\gamma^+_{0,2}].
$$
In either case, the two resulting
wedges have a common ``mean-direction''.
This shows that the two new D-branes that result from the recombination
preserve the same supersymmetry.
The new D-branes are
$B_{\bf L_1,M_1}$ and $B_{\bf L_2,M_1}$
for $\omega=1$
and $\widehat{B}_{\bf L_1,M_2}$ and $\widehat{B}_{\bf L_2,M_2}$
for $\omega=-1$, where
$$
{\bf L_1}=(33110),\,\,{\bf L_2}=(33112),\,\,
{\bf M_1}=(33110),\,\,{\bf M_2}=(33112).
$$
They split into the sum of the
$(+)$ and the $(-)$ short orbit branes.
Thus, we find
supersymmetric and tadpole-canceling configurations are given by:
\beqa
&&\mbox{\underline{two each of $\widehat{B}^{(+)}_{\bf L_1,M_1}$,
$\widehat{B}^{(-)}_{\bf L_1,M_1}$,
$\widehat{B}^{(+)}_{\bf L_2,M_1}$,
$\widehat{B}^{(-)}_{\bf L_2,M_1}$ for $P^A_{+;00001}$-orientifold
($\omega^{1\over 2}=-1$)}},
\nn\\
&&\mbox{\underline{two each of
$\widehat{B}^{(+)}_{\bf L_1,M_2}$,
$\widehat{B}^{(-)}_{\bf L_1,M_2}$,
$\widehat{B}^{(+)}_{\bf L_2,M_2}$,
$\widehat{B}^{(-)}_{\bf L_2,M_2}$
for $P^A_{-;00001}$-orientifold ($\omega^{1\over 2}=-i$).}}
\nn
\eeqa 
Again, we need the same number of $+$ branes and $-$ branes to cancel
the twisted part of the RR-tadpole.
In both cases,
the $+$ branes and the $-$ branes are exchanged by the orientifold
action (${\bf L_i}$ fixed for $P^A_{+;00001}$
and exchanged for $P^A_{-;00001}$).
Therefore the gauge group is $U(2)\times U(2)$.

\begin{figure}[htb]
\centerline{\includegraphics{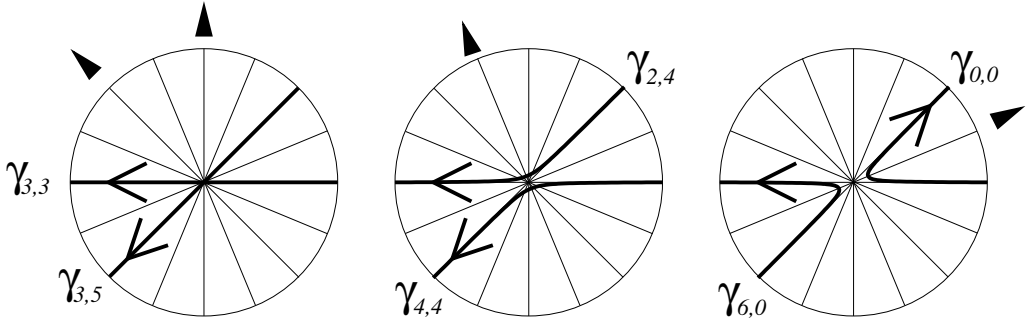}}
\caption{The recombination of branes for
$\gamma_{3,3}+\gamma_{3,5}$ (middle) and 
$\gamma_{3,3}-\gamma_{3,5}$ (right).}
\label{keq6}
\end{figure}

\noindent
{\underline{$\bullet$ $P^A_{\pm;01000}$}}\\[0.1cm]
O-plane charge is proportional to the difference or the sum of 
$B_{{\bf k\over 2},(33111)}$ and
$B_{{\bf k\over 2},(35111)}$.
For this case, we focus of the second factor.
The recombination relevant for this is depicted in
Fig.~\ref{keq6}.
In each case, the ``mean-directions'' of the two wedges are aligned
after the recombination.
We also note that, in each of these cases, the two branes
are mapped into each other by rotation of four steps.
To be precise, the orientation is revered but that is compensated by
the orientation reversal for the $L={k_1\over 2}$ brane of the first factor.
Thus, they are in the same (long) orbit of the $\Z_8$ orbifold group.
We therefore found a supersymmetric and tadpole-canceling
configuration:
\underline{four $B_{\bf L_3,M_3}$ for the $P^A_{+;01000}$-orientifold}
($\omega^{1\over 2}=-1$)
and
\underline{four $B_{\bf L_4,M_4}$ for the $P^A_{-;01000}$-orientifold}
($\omega^{1\over 2}=-i$),
where
$$
{\bf L_3}=(30111),\quad
{\bf M_3}=(30111),\quad
{\bf L_4}=(32111),\quad
{\bf M_4}=(34111).
$$
These branes are invariant under the respective orientifolds.
The gauge group is $O(4)$ for both
$P^A_{+;01000}$ ($\omega^{1\over 2}=-1$)
and $P^A_{-;01000}$ ($\omega^{1\over 2}=-i$).

\noindent
{\underline{$\bullet$ $P^A_{\pm;00011}$}}\\[0.1cm]
O-plane charge is proportional to the difference or the sum of
$B_{{\bf k\over 2},(33111)}$ and
$B_{{\bf k\over 2},(33133)}$.
For this case, we need to focus on the two factors, 4-th and 5-th.
The recombination to align the ``mean-directions'' is not obvious, but we
can use the following trick. It is to use the identity
$$
A_1A_2-B_1B_2
={1\over 2}(A_1+B_1)(A_2-B_2)
+{1\over 2}(A_1-B_1)(A_2+B_2).
$$
For the $P^A_{+;00011}$-orientifold,
we find
\beqa
(\gamma_{1,1})^2-(\gamma_{1,3})^2
&=&{1\over 2}(\gamma_{1,1}+\gamma_{1,3})(\gamma_{1,1}-\gamma_{1,3})
+{1\over 2}(\gamma_{1,1}-\gamma_{1,3})(\gamma_{1,1}+\gamma_{1,3})
\nn\\
&=&{1\over 2}(\gamma_{0,2}+\gamma_{2,2})(\gamma_{0,0}+\gamma_{2,0})
+{1\over 2}(\gamma_{0,0}+\gamma_{2,0})(\gamma_{0,2}+\gamma_{2,2})
\nn
\eeqa
where we have used the recombination used in the case of
$P^A_{\pm;00001}$.
This indeed aligns the sum of ``mean-directions'', and thus
a supersymmetry is preserved ---
it is the one preserved by the $\omega^{1\over 2}=-1$ orientifold.
The resulting brane configuration is the collection of \underline
{sixteen branes $\widehat{B}^{(\pm)}_{\bf L_i,M_j}$, $i=5,6,7,8$, $j=5,6$}
where
\beqa
&{\bf L_5}=(33100),\,\,
{\bf L_6}=(33102),\,\,
{\bf L_7}=(33120),\,\,
{\bf L_8}=(33122),
\nn\\
&{\bf M_5}=(33120),\,\,
{\bf M_6}=(33102).
\nn
\eeqa
The orientifold exchanges the $+$ and $-$ labels 
(and acts on the ${\bf L_i}$ labels in a certain way).
Thus the gauge group is $U(1)^8$.

For the $P^A_{-;00011}$-orientifold,
the same procedure on the 4-th and 5-th factors
gives
$$
(\gamma_{1,1})^2+(\gamma_{1,3})^2
={1\over 2}(\gamma_{0,0}+\gamma_{2,0})^2
+{1\over 2}(\gamma_{0,2}+\gamma_{2,2})^2
$$
and the sum of the ``mean-directions'' are not aligned.
(The two terms have opposite phases.)
Thus, this recipe of recombination
does not work to find a supersymmetric configuration.
In fact, in this case, the O-plane tension is vanishing
(see the table in page~\ref{tableAOch}),
and there is no supersymmetric brane
configuration that cancels the RR-tadpole.

\noindent
{\underline{$\bullet$ $P^A_{\pm;01001}$}}\\[0.1cm]
This case is similar to the above.
For the $P^A_{+;01001}$-orientifold,
the recombination successfully aligns the ``mean-direction''
and we find that a supersymmetric and tadpole canceling
configuration is given by \underline{two each of
$B_{\bf L_9,M_7}$, $B_{\bf L_{10},M_7}$, $B_{\bf L_{11},M_8}$
and $B_{\bf L_{12},M_8}$}, where
\beqa
&{\bf L_9}=(32110),\,\,
{\bf L_{10}}=(32112),\,\,
{\bf L_{11}}=(30110),\,\,
{\bf L_{12}}=(30112),
\nn\\
&{\bf M_7}=(34110),\,\,
{\bf M_8}=(30112).
\nn
\eeqa
The preserved supersymmetry is that of $\omega^{1\over 2}=-1$.
${\bf M}={\bf M_7}$ branes are invariant under the orientifold action
and are of $Sp$-type, whereas the two ${\bf M}={\bf M_8}$ branes are
mapped to each other.
Thus the gauge group is $Sp(1)\times Sp(1)\times U(2)$.

For the $P^A_{-;01001}$-orientifold, recombination does not work.

\noindent
{\underline{$\bullet$ $P^A_{\pm;11000}$}}\\[0.1cm]
For the $P^A_{+;11000}$-orientifold,
a supersymmetric and tadpole canceling configuration is given by
\underline{two each of
$B_{\bf L_{13},M_9}$, $B_{\bf L_{14},M_9}$, $B_{\bf L_{15},M_{10}}$
and $B_{\bf L_{16},M_{10}}$}, where
\beqa
&{\bf L_{13}}=(20111),\,\,
{\bf L_{14}}=(26111),\,\,
{\bf L_{15}}=(02111),\,\,
{\bf L_{16}}=(04111),
\nn\\
&{\bf M_9}=(40111),\,\,
{\bf M_{10}}=(04111).
\nn
\eeqa
The preserved supersymmetry is that of $\omega^{1\over 2}=-1$.
Orientifold action preserves the ${\bf M}$-label
but exchanges the ${\bf L}$-labels as
${\bf L_{13}}\leftrightarrow {\bf L_{14}}$
and ${\bf L_{15}}\leftrightarrow {\bf L_{16}}$.
Thus the gauge group is $U(2)\times U(2)$.

For the $P^A_{-;11000}$-orientifold, 
recombination does not work.

\subsection{Particle Spectra in Some Supersymmetric Models}

Let us find out the spectrum of massless particles for the
configurations obtained in the previous subsection.
The problem here is to count the numbers of
scalar fields in various open string sectors
and study the action of parity on them.
They are read off from the annulus and M\"{o}bius strip amplitudes.
Here are some essential facts:
\begin{itemize}
\item The open string states $\otimes_i(l_i,n_i,s_i)$ between
  two D-branes $B_{\bf L,M}$ and $B_{\bf L',M'}$ satisfy
  $n_i=M'_i-M_i+2\nu$ mod $(2k_i+4)$, and $l_i$'s are also
   constrained from the $SU(2)$ fusion rule.
\item Massless
  scalars correspond to chiral or antichiral primary states
  $\otimes_i(l_i,l_i,0)$ or $\otimes_i(l_i,-l_i,0)$
  with $\sum_i\frac{l_i}{k_i+2}=1$.
  They are the lowest components of four-dimensional ${\cal N}=1$
%Kazuo
  chiral or antichiral multiplets, and are related to each other
  by the worldsheet orientation reversal.
  Namely, chiral primary states on $B-B'$ string and antichiral
  primary states on $B'-B$ string are related to each other.
\item For the open string states on the parity invariant D-branes
  we have to study the action of parity.
  If the brane $B_{\bf L,M}$ is invariant under the parity
  $P^A_{\omega,{\bf m}}$, the open strings $\otimes_i(l_i,n_i,s_i)$
  on the M\"{o}bius strip satisfy $n_i = 2M_i-2m_i+2\nu$ mod $(2k_i+4)$,
  and the constraint on $l_i$ from the $SU(2)$ fusion rule.
  For chiral or antichiral states satisfying the above two conditions,
  the parity eigenvalue is then given by
\beq
  P = -i\omega^{\nu-\frac12}(-i)^{\{\#{\rm of}\,(s_i=2)\}},
\label{pareig}
\eeq
 where we have to put $\omega^{\frac12}=-1$ or $-i$ for $\omega=\pm1$
 as before.
\end{itemize}

We present here the relevant amplitudes
that lead to the above conclusions. (We describe them for general $r$ and $d$
but we are interested in the case $r=5$ and $d=1$.)
The NS part of the annulus amplitude between the A-branes
$B_{\bf L,M}$ and $B_{\bf L',M'}$ is given by
\begin{equation}
  \frac12\sum_{\nu=1}^H\sum_{l_i}\prod_{i=1}^rN_{L_iL'_i}^{~l_i}\times
  \left\{
   \chi^{\rm (st)NS+}\prod_{i=1}^r\chi^{\rm NS+}_{l_i,M'_i-M_i+2\nu}
  -\chi^{\rm (st)NS-}\prod_{i=1}^r\chi^{\rm NS-}_{l_i,M'_i-M_i+2\nu}
  \right\}
\end{equation}
where $\chi^{\rm NS\pm}_{l,n}=\chi_{l,n,0}\pm\chi_{l,n,2}$ are
linear combinations of minimal model characters and
$\chi^{\rm (st)NS\pm}$ represent the non-compact spacetime $\R^{2d+2}$
plus ghost contribution
\begin{equation}
  \chi^{\rm (st)NS\pm} = q^{-\frac{d}{8}}(1\pm 2d\cdot q^{\frac12}+\cdots).
\end{equation}
For pairs of short-orbit branes, the sum over the open string states
is subject to the projection
\begin{equation}
  \frac14(1+\varepsilon\bar\varepsilon\prod_{w_i~{\rm odd}}
  (-1)^{\frac12(l_i+M'_i-M_i)})
\end{equation}
The NS part of M\"{o}bius strip amplitude between the A-brane
$B_{\bf L,M}$ and its image under the parity $P^A_{\omega;\bf m}$ 
is given by
\begin{equation}
  {\rm Re}\left\{ ie^{\frac{\pi i(r-d)}{4}}
     \sum_{\nu=1}^H\sum_{l_i}
     \omega^{-\frac12-\nu}
     \hat{\chi}^{\rm (st)NS+}
     \prod_{i=1}^rN_{L_iL_i}^{~l_i}\hat{\chi}^{\rm NS+}_{l,2M_i-2m_i+2\nu}
 \right\},
\end{equation}
where $\hat\chi^{\rm (st)NS\pm}$ represent the spacetime and ghost
contributions
\begin{equation}
  \hat\chi^{\rm (st)NS\pm} =
  q^{-\frac{d}{8}}(1\pm 2id\cdot q^{\frac12}+\cdots)
\end{equation}
and $\hat{\chi}^{\rm NS\pm}_{l,n}$ are defined by
\begin{equation}
  \hat{\chi}^{\rm NS\pm}_{l,n} = (-1)^{\frac12(l+n)}
  (\hat\chi_{l,n,0}\pm i\hat\chi_{l,n,2}),~~~
   \hat\chi_{l,n,s}(\tau) =
   \e^{-\pi i(\frac{l(l+2)-n^2}{4k+8}+\frac{s^2}{8}-\frac{c}{24})}
   \chi_{l,n,s}(\tau+1/2).
\end{equation}
For short-orbit branes the amplitude gets an extra factor of $\frac12$.
It is easy to read off the parity eigenvalue (\ref{pareig}) for
(anti)chiral primary states from this formula.

\subsubsection{Odd $H$}\label{subsubsec:oddH}

We have seen that the configuration of
four $B_{\bf \frac{k-1}{2},\frac{k-1}{2}}$'s
in the $P^A_{+,{\bf 0}}$-orientifold is
supersymmetric and free of tadpoles, and the gauge symmetry
is $O(4)$ in all cases.
However, the spectrum of massless matters depends on the model.
Let us illustrate here our analysis in some examples.

\noindent
{\underline{$k_i=(33333)$}}\\[0.1cm]
On the annulus and on the M\"{o}bius strip we find one
chiral primary state $\otimes_{i=1}^5(2,6,2)$ which is
equivalent to $\otimes_{i=1}^5(1,1,0)$.
Since this has $P=1$, there is one chiral multiplet
belonging to the symmetric tensor representation {\bf 10}
of $O(4)$.

\noindent
{\underline{$k_i=(11777)$}}\\[0.1cm]
There are two chiral multiplets, which appear on the M\"{o}bius
strip as chiral primary states
$\otimes_{i=1,2}(0,4,2)\otimes_{i=3,4,5}(6,10,2)$ and
$\otimes_{i=1,2}(0,0,0)\otimes_{i=3,4,5}(4,12,2)$ respectively.
The former has $P=1$ while the latter has $P=-1$, so they belong
to one symmetric and one antisymmetric tensor representations
of $O(4)$.

The analysis of other models goes in much the same way.
The result is summarized in the table below.

\begin{table}[htb]
\begin{center}
\renewcommand{\arraystretch}{1.2}
\begin{tabular}{|c|c|c|c|c|c|c|}
\hline
($k_i$)& (3,3,3,3,3)  &(1,1,7,7,7) &(1,3,3,3,13)
       & (1,1,3,13,13)&(1,1,5,5,19)&(1,1,3,7,43) \\
\noalign{\hrule height 0.8pt}
$\#{\bf 10}$ & 1& 1& 2& 2& 2& 8  \\\hline
$\#{\bf 6} $ & 0& 1& 1& 1& 3& 3  \\\hline
\end{tabular}
\label{tableAHOM}
\end{center}
\end{table}

\subsubsection{Even $H$ -- two parameter model $k_i=(66222)$ in detail}

\noindent
{\underline{$\bullet$ $P^A_{-;00000}$}}\\[0.1cm]
Four short-orbit branes $\widehat{B}^{(+)}_{\bf\frac{k}{2},\frac{k}{2}}$
and their parity images support $U(4)$ gauge symmetry,
and there is a single adjoint matter.

\noindent
{\underline{$\bullet$ $P^A_{\pm;00001}$}}\\[0.1cm]
The structure of massless spectrum are the same for the two examples
\beqa
  P^A_{+;00001}:&& 2\widehat{B}^{(+)}_{\bf L_1,M_1}
 +2\widehat{B}^{(-)}_{\bf L_1,M_1}
 +2\widehat{B}^{(+)}_{\bf L_2,M_1}
 +2\widehat{B}^{(-)}_{\bf L_2,M_1},
\nn\\
  P^A_{-;00001}:&& 2\widehat{B}^{(+)}_{\bf L_1,M_2}
 +2\widehat{B}^{(-)}_{\bf L_1,M_2}
 +2\widehat{B}^{(+)}_{\bf L_2,M_2}
 +2\widehat{B}^{(-)}_{\bf L_2,M_2},
\nn
\eeqa
with ${\bf L_1}=(33110),\,\,{\bf L_2}=(33112),\,\,
      {\bf M_1}=(33110),\,\,{\bf M_2}=(33112).$
In both cases, the gauge group is $U(2)\times U(2)$ and
there are matters in the
representation $({\bf 2,\bar{2}})\oplus({\bf \bar{2},2})$.

\noindent
{\underline{$\bullet$ $P^A_{\pm;01000}$}}\\[0.1cm]
Here we found the configurations with four long-orbit branes
which are invariant under the parity.
The $P^A_{+;01000}$-orientifold with four
$B_{(30111),(30111)}$
gives
$$
\mbox{$O(4)$ pure Super-Yang-Mills.}
$$
The $P^A_{-;01000}$-orientifold with four
$B_{(32111),(34111)}$ gives
$$
\mbox{$O(4)$ with one symmetric and one
antisymmetric matters.}
$$

\noindent
{\underline{$\bullet$ $P^A_{+;00011}$}}\\[0.1cm]
Here we find a very interesting situation.
The tadpole canceling configuration we have found is
eight short-orbit branes $B_{\rm I},\cdots,B_{\rm VIII}$
and their parity images, where
$$
\begin{array}{rclrclrclrcl}
B_{\rm I}   &=& \widehat{B}^{(+)}_{{\bf L_5, M_5}},&
B_{\rm II}  &=& \widehat{B}^{(+)}_{{\bf L_6, M_5}},&
B_{\rm III} &=& \widehat{B}^{(+)}_{{\bf L_7, M_5}},&
B_{\rm IV}  &=& \widehat{B}^{(+)}_{{\bf L_8, M_5}},\\
B_{\rm V}   &=& \widehat{B}^{(+)}_{{\bf L_5, M_6}},&
B_{\rm VI}  &=& \widehat{B}^{(+)}_{{\bf L_6, M_6}},&
B_{\rm VII} &=& \widehat{B}^{(+)}_{{\bf L_7, M_6}},&
B_{\rm VIII}&=& \widehat{B}^{(+)}_{{\bf L_8, M_6}},
\end{array}
$$
with
${\bf L_5}=(33100)$,
${\bf L_6}=(33102)$,
${\bf L_7}=(33120)$,
${\bf L_8}=(33122)$,
${\bf M_5}=(33120)$,
and ${\bf M_6}=(33102)$.
The gauge group is $U(1)^8$, and we have quite a few matter
fields which are charged under two of $U(1)$'s.
The spectrum are the most neatly expressed in terms of the
quiver diagram, where each arrow represents a chiral multiplet
charged $+1$ and $-1$ under the $U(1)$'s on its head
and tail.
\begin{figure}[htb]
\centerline{\includegraphics{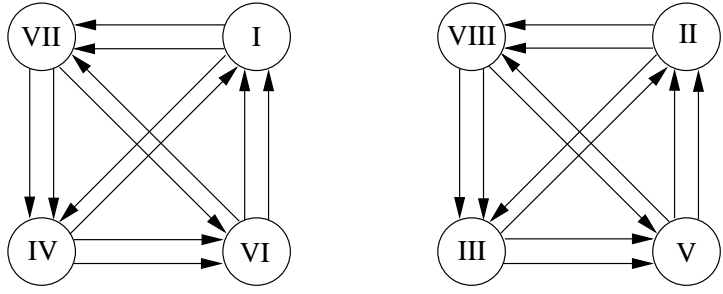}}
\caption{quiver diagram representing a D-brane
         configuration with $P^A_{+;00011}$}
\label{quiver1}
\end{figure}
Note that the gauge theory is chiral.
There is a mixed $U(1)_aU(1)_b^2$ anomaly for each pair
$(a,b)$ of neighboring groups of the quiver
(i.e. VII-I,IV-VII,VI-IV,VI-I for the first square,
and similarly for the second square).
Anomaly cancellation mechanism will be discussed in
Section~\ref{subsec:anomaly}.

\noindent
{\underline{$\bullet$ $P^A_{+;01001}$}}\\[0.1cm]
For this parity we found a tadpole canceling configuration
$$
 2B_{\bf L_9,   M_7}
+2B_{\bf L_{10},M_7}
+2B_{\bf L_{11},M_8}
+2B_{\bf L_{12},M_8}
$$
with
${\bf L_9}=(32110)$,
${\bf L_{10}}=(32112)$,
${\bf L_{11}}=(30110)$,
${\bf L_{12}}=(30112)$,
${\bf M_7}=(34110)$,
and ${\bf M_8}=(30112)$.
The gauge group is $Sp(1)\times Sp(1)\times U(2)$
and there are matters in the representations
$$
 \bf {\rm 2\times}(2,2,1)\oplus(2,1,2)\oplus(1,2,\bar{2}).
$$
This system is also chiral.
There are mixed $U(1)Sp(1)_a^2$ anomalies.
 Anomaly cancellation mechanism will be discussed in
Section~\ref{subsec:anomaly}.

\noindent
{\underline{$\bullet$ $P^A_{+;11000}$}}\\[0.1cm]
We found a D-brane configuration
$$
2B_{\bf L_{13},M_9}+
2B_{\bf L_{14},M_9}+
2B_{\bf L_{15},M_{10}}+
2B_{\bf L_{16},M_{10}}
$$
with
${\bf L_{13}}=(20111)$,
${\bf L_{14}}=(26111)$,
${\bf L_{15}}=(02111)$,
${\bf L_{16}}=(04111)$,
${\bf M_9}=(40111)$, and
${\bf M_{10}}=(04111)$.
The gauge group is $U(2)\times U(2)$, and the matter
belongs to $\bf (2,\bar{2})\oplus(\bar{2},2)$.

\subsection{More general tadpole canceling configurations}
\label{subsec:general}

In the previous subsection, we have seen that it is generically 
rather easy to find a supersymmetric tadpole canceling brane 
configuration for Type IIA orientifolds of Gepner models. When 
all levels are odd, these configurations corresponds to placing 
$4$ D-branes on top of the O-plane, and leads to $O(4)$ 
gauge group with some matter content which depends on the 
particular model. On the other hand, we have seen that when 
some levels are even, we can have somewhat more interesting 
configurations which support unitary gauge groups and chiral 
matter.

It would be interesting to know whether these are all solutions and 
if not, how to describe the set of all possibilities. Let us recall
the general nature of the problem. First of all, we emphasize once
again that we have only been considering certain rational boundary states,
which are just a subset of all possible branes that could be used to
cancel the tadpoles. It would be interesting to see if one can
obtain more interesting possibilities by using for example the
boundary states constructed in \cite{recknagel} (which are still
rational, but more general than the ones we have considered here). 
Secondly, we wish to point out that the problem of finding tadpole 
canceling configurations does not actually depend on whether we 
are considering Type IIA or Type IIB (the two are just exchanged 
by mirror symmetry). In other words, it is sufficient to discuss 
the conditions (\ref{internal}) in the internal CFT.

\noindent
With these comments in mind, we are looking for sets of rational
D-branes which

(i) have the same 
RR-charge as the O-plane,

(ii) are invariant under the parity, 

(iii) allow a consistent assignment of Chan-Paton factor,

\noindent
and, if we are interested in spacetime supersymmetric configurations,

(iv) preserve a common ${\mathcal N}=1$ supersymmetry.

\noindent
These conditions are solved in steps.\\
\underline{Step 1:} Choose the parity $P$. Compute the
RR-charge $[O_P]$ and the preserved spacetime supersymmetry 
$M_O$ of the corresponding O-plane.\\
\underline{Step 2:} Make a list of rational branes $B_i$ 
preserving the same spacetime supersymmetry as the O-plane, and 
compute their RR-charges $[B_i]$. We note that we 
have to distinguish branes even if they have the same RR charge.\\
\underline{Step 3:} Determine for each brane its image 
$B_{P(i)}$ under the parity. If a brane is fixed under the parity,
determine whether the gauge group is of $O$ or $Sp$-type.
We will use an indicator $\sigma_i$ to concisely denote this 
gauge group. If the gauge group supported on $n_i$ branes 
$[B_i]$ is $O(n_i)$, we will set $\sigma_i=+1$,
if it is of type $Sp(n_i/2)$ we set $\sigma_i=-1$. If
the brane is not invariant under $P$ (so the gauge group is
$U(n_i)$), we will set $\sigma_i=0$.\\
\underline{Step 4:} Solve the equation
\beq
\sum_i n_i [B_i] = [O_P]
\label{tadcan}
\eeq
for positive integers $n_i$ under the condition that
$n_i=n_{P(i)}$ and that $n_i$ is even if $\sigma_i=-1$.

Steps 1-3 are of course just those that we have been taking above.
The hard part is solving (\ref{tadcan}). Indeed, while this
is a linear equation, there are a large number of equations
to solve (on the order of $100$ RR charges) and a large number
of variables (on the order of several thousands branes preserving
the same supersymmetry as any given O-plane). The number of
solutions to this Diophantine problem is finite when restricted
to positive integers $n_i$, because there is always one
equation in which all $n_i$ appear with a positive 
coefficient. The simplest way to obtain this equation is 
to take the overlap with the RR ground state $|0\rangle_{\rm RR}$. 
By eq.\ (\ref{ssusy}), this is proportional to the overlap 
with the NSNS ground state, so what we are saying is 
simply that the tensions of the branes are all positive 
and must cancel the tension of the O-plane. 

Knowing that the number of solutions is finite, one
would like to count or even enumerate the solutions. A
priori, it is not even clear that there is a single one
(besides the somewhat trivial ones we have already found).
To estimate the difficulty of the problem, we notice that
the tension of the A-type orientifold in the quintic is about 
(minus) 20 times the typical tension of A-branes.
So if we try to scan all the configurations in which the tension
of D-branes cancel the negative tension of orientifold, there
are roughly $\left({2000\atop20}\right)\sim10^{50}$ of them,
which is too large a number to look through even with the help
of a computer. One has to resort to a more direct method.

The problem becomes dramatically simpler when the number of
equations and the number of possible branes is smaller.
For B-type on the quintic, for example, there are $2$ linearly
independent equations and $32$ variables. This problem 
(and its analogs for the two parameter model) can be solved 
completely, as we do in section \ref{sec:TCCB}. If we consider
the orbifold of the quintic by a certain $\Z_5$ phase symmetry
(see \cite{GreenePlesser}), it turns out that there
are $6$ equations in $96$ variables. This problem is still 
tractable, and we present some solutions in appendix 
\ref{orbifold}.

A purely technical difficulty is to find the right basis in
which to write the equations (\ref{internal}). The simplest
basis might seem to be the basis of RR ground states
which are products of the minimal model ground states in 
the form
\beq
|l_i\rangle = \prod_{i=1}^5|l_i,l_i+1,1\rangle\times|l_i,-l_i-1,-1\rangle
\qquad
(1\le l_i+1\le k_i+1,~\sum_{i}{\textstyle\frac{l_i+1}{k_i+2}}\in\Z)
\label{ground}
\eeq
However, the problem is that when written in this basis,
the Diophantine equations (\ref{tadcan}) are not manifestly
integral --- the coefficients are certain combinations of
trigonometric and exponential functions. To remedy this
situation, one can use the fact that (at the level of 
charges) some branes can be written as integral linear
combinations of other branes. A convenient choice of
reference branes --- for any Gepner model --- are the 
branes with ${\bf L}={\bf 0}$ and varying ${\bf M}$.
This ``basis'' has often been used in previous works on the RS 
boundary states in Gepner models. We note three important facts. 
\\ (i) The ${\bf L}={\bf 0}$ branes in general only generate
a sublattice of the full BPS charge lattice. As we have mentioned
before, the generic Gepner model has chiral ring elements from 
twisted sectors, corresponding to non-toric blowups in the geometry.
The corresponding RR fields do not couple to the RS branes, which
preserve a diagonal chiral algebra in each minimal model.
\\ (ii) The ${\bf L}={\bf 0}$ branes are not in general primitive
generators of the charge lattice. It is an outstanding problem
to find boundary states which are integral generators of the
charge lattice.
\\ (iii) The charges of the ${\bf L}={\bf 0}$ branes are not
linearly independent. It can sometimes be a little cumbersome 
to eliminate these relations in (iii) in order to find the linearly 
independent conditions. We discuss how this can be done in appendix 
\ref{integral}.

\subsubsection*{O(4) Configuration is Not Always Possible}

To conclude this section, we answer (in the negative) the following
question: At large volume, the tadpole of the O6-plane is always
exactly canceled by the four D-branes all wrapping on it. Can
we always find a corresponding solution at the Gepner point?
We have found such a configuration for each of odd $H$ Gepner models.
However, in the model with even $H$ one cannot always
find such a configuration.
In the two parameter model, it turns out that the only such solutions
are those for $P^A_{\pm,01000}$ found already.

To see this, let us first look at the tensions of the O-planes.
For the
$P^A_{\omega;{\bf m}}$-orientifold, it is related to
the tension of the D-brane $B_{(33111),{\bf M}}$
by
\beqa
T_{O^A_{+;{\bf m}}}&=&
 -4\sin{\ts\frac{\pi n}{8}}T_{(33111),{\bf M}},\nn\\
T_{O^A_{-;{\bf m}}} &=&
 -4\left|\cos{\ts\frac{\pi n}{8}}\right|T_{(33111),{\bf M}}
 \nn
\eeqa
where $n:=\sum_iw_im_i$ if we assume $m_i=0$ or $1$.
One can show that there is no $O(4)$ configurations
when $n\!=\!0$ or $4$, in the following way.
In these cases the O-plane tension, if nonvanishing,
is equal to $-4T_{(33111),{\bf M}}$,
but the brane $B_{(33111),{\bf M}}$ is sum of two short orbit branes.
Moreover, any elementary brane has tension less that
$T_{(33111),{\bf M}}$.
Thus, the O-plane tension cannot be canceled with just four identical
elementary branes.
For $n=1,2$ or $3$ the tensions are rewritten as
\beqa
T_{O^A_{+;{\bf m}}} &=&
 -4T_{(3,n-1,1,1,1),{\bf M}}~=~
- 4T_{(n-1,3,1,1,1),{\bf M}},\nn\\
T_{O^A_{-;{\bf m}}} &=&
 -4T_{(3,3-n,1,1,1),{\bf M}} ~=~
 -4T_{(3-n,3,1,1,1),{\bf M}}
\nn
\eeqa
and similarly for $n=5,6,7$.
There are no other branes with the same tension.
Thus the only configurations with four identical D-branes
that can cancel the O-plane tensions are
$$
\begin{array}{c}
P^A_{+;00001}\Rightarrow 4B_{(31111),{\bf M}}~,\\
P^A_{+;01000}\Rightarrow 4B_{(30111),{\bf M}}~,\\
P^A_{+;01001}\Rightarrow 4B_{(32111),{\bf M}}~,\\
P^A_{+;11000}\Rightarrow 4B_{(31111),{\bf M}}~,
\end{array}
~~
\begin{array}{c}
P^A_{-;00001}\Rightarrow 4B_{(31111),{\bf M}}~,\\
P^A_{-;01000}\Rightarrow 4B_{(32111),{\bf M}}~,\\
P^A_{-;01001}\Rightarrow 4B_{(30111),{\bf M}}~,\\
P^A_{-;11000}\Rightarrow 4B_{(31111),{\bf M}}~
\end{array}
$$
and those obtained by exchanging $L_1$ and $L_2$.

Let us then see whether any of the above relations are lifted
to the full equality between the RR-charges.
The possibilities in the first and the fourth rows
can be easily excluded by looking at their transformation
property under the permutations of the first two minimal models.
Also, those in the first and the third rows are excluded because
the branes cannot be parity invariant under any choice of ${\bf M}$ label.
This is easily seen by noting that, for the branes $B_{\bf L,M}$
in the two parameter model to be invariant under
$P^A_{\omega,{\bf m}}$, $M_i-m_i$ have to be all even or all add.
Thus we are left with the ones in the second row, for which
the permutation of minimal models, parity invariance and $O(4)$
gauge symmetry reduce the possibilities to
$$
P^A_{+;01000}\Rightarrow 4B_{(30111),(30111)}\,,~~
P^A_{+;01000}\Rightarrow 4B_{(30111),(12111)}\,,~~
P^A_{-;01000}\Rightarrow 4B_{(32111),(34111)}\,.
$$
Two of them are the configurations that was already found before.
The remaining (second) case does not satisfy the full tadpole condition,
as can be guessed from the comparison with the first one and
confirmed by a more detailed analysis of the tadpole condition.

\section{Chirality, Anomaly Cancellation, and Fayet-Ilio\-poulos Terms}
\label{sec:FI}

In this section, we take a break and make some general remarks
on chirality, anomaly cancellation mechanism, and
Fayet-Iliopoulos terms.

\subsection{Chirality and Witten Indices}

Chirality of the theory can be measured by
the open string Witten indices of the internal CFT,
\beqa
&&
I(B,B')=\Tr\!\!\mathop{}_{{\mathcal H}_{\Scr{B},\Scr{B}'}}(-1)^{F_{\rm int}},
\nn\\
&&
I(B,O_P)=\Tr\!\!\mathop{}_{{\mathcal H}_{\Scr{B},P(\Scr{B})}}
(-1)^{F_{\rm int}}P.
\nn
\eeqa
To see this, note that the GSO operator can be written as the product 
$(-1)^F=(-1)^{F_{\rm int}}(-1)^{F_{\rm st}}$.
The spacetime part $(-1)^{F_{\rm st}}$ acts on the RR sector states,
that is, on the spacetime fermions, as the gamma-five matrix,
$\Gamma^5=\Gamma^0\Gamma^1\Gamma^2\Gamma^3$.
Thus, by the GSO projection, the Witten index of the internal part
$\Tr (-1)^{F_{\rm int}}$ is proportional to
$\Tr_{\rm massless}\Gamma^5$, which measures the chirality of the theory.

Let $B$ and $B'$ be branes that are {\it not} the orientifold images
of each other, $P(B)\ne B'$. The parity maps the $B$-$B'$ string to
the $P(B')$-$P(B)$ string, and hence
the orientifold projection simply relates the two
string sectors. The chirality of the bifundamental representation
is given by the index
\beq
\#(\overline{\bf n}_B,{\bf n}_{B'})-
\#({\bf n}_B,\overline{\bf n}_{B'})
=I(B,B').
\eeq
By definition, it must be antisymmetric with respect to
the exchange of $B$ and $B'$. This is guaranteed by the antisymmetry
of the index in the internal CFT, $I(B',B)=-I(B,B')$.

The string stretched from a brane $B$
to its parity image $P(B)$ is invariant under the parity action.
Its Chan-Paton factor is 
$\overline{\bf n}_B\otimes {\bf n}_{P(B)}
=\overline{\bf n}_B\otimes\overline{\bf n}_B$,
the second rank tensor product of $\overline{\bf n}_B$.
The orientifold projection selects the symmetric tensor
times the $P=1$ states
as well as the antisymmetric tensor times the $P=-1$ states.
Note that the ordinary $B$-$P(B)$ index
is the sum of the index in the $P=1$ subspace and the one in the
$P=-1$ subspace,
%$I(B,P(B))=I^{P=1}(B,P(B))+I^{P=-1}(B,P(B))$,
while the twisted index is the difference.
%,
%$I(B,O_P)=I^{P=1}(B,P(B))-I^{P=-1}(B,P(B))$.
Thus, the chirality of the symmetric and antisymmetric representation
is
\beqa
&&\#{\rm S}^2\overline{\bf n}_B
-\#{\rm S}^2{\bf n}_B={1\over 2}\Bigl(I(B,P(B))+I(B,O_P)\Bigr)
\\
&&\#{\rm A}^2\overline{\bf n}_B
-\#{\rm A}^2{\bf n}_B={1\over 2}\Bigl(I(B,P(B))-I(B,O_P)\Bigr)
\eeqa
They must vanish if the brane is parity invariant,
$P(B)=B$, and the representation $\overline{\bf n}_B$ is real or pseudo-real.
This is again guaranteed by the antisymmetry of the indices
$I(P(B),B)=-I(B,P(B))$, $I(O_P,P(B))=-I(B,O_P)$.

\subsubsection{Examples in IIA Gepner Models}

Let us study the Witten index of branes
in Type IIA orientifolds of Gepner model.
Long-orbit A-branes are simply the sum over images
under the orbifold group $\Z_H$. Accordingly, the index is also given by the
sum of the product of the minimal model indices
\beq
I(B_{\bf L,M},B_{\bf L',M'})
=\sum_{\nu=0}^{H-1}\prod_{i=1}^5I(B_{L_i,M_i},
B_{L_i',M_i'+2\nu}).
\label{sumoo}
\eeq
Furthermore, the index for each minimal model is given by
the intersection number of the wedge cycle, $I(B_{L_i,M_i},B_{L_i',M_i'})
=\#(\gamma_{L_i,M_i}^-\cap \gamma_{L_i',M_i'}^+)$.
In many supersymmetric brane configurations,
there are two or more branes with the same ${\bf M}$-label.
So let us examine such a pair, $B_{\bf L,M}$ and $B_{\bf L',M}$.
In the LG picture, the two product 
branes $\prod_i\gamma_{L_i,M_i}$ and $\prod_i\gamma_{L'_i,M_i}$
 have the common mean-direction at each minimal model factor.
After rotation by the orbifold group they no longer have the same
mean-direction, but there is an interesting relation between
different steps of rotation:
$\gamma_{L'_i,M_i}^+$ rotated by a step $\nu$
and $\gamma_{L'_i,M_i}^+$ rotated by the opposite step $-(\nu+1)$
have the opposite intersection number with
any $\gamma_{L_i,M_i}$:
$$
\#(\gamma_{L_i,M_i}^-\cap \gamma_{L_i',M_i'+2\nu}^+)
=-\#(\gamma_{L_i,M_i}^-\cap \gamma_{L_i',M_i'-2\nu-2}^+).
$$
It follows that
the $\nu$-th term in (\ref{sumoo}) is opposite to the $(H-1-\nu)$-th
term, and the sum vanishes.
Thus, we find the

\noindent
{\bf Vanishing Theorem:}\\
{\it The index between two long-orbit branes of the same ${\bf M}$-label
vanishes.}

\noindent
This helps us in finding chiral pairs of branes in a given model:
If two D-branes have the same ${\bf M}$-labels,
we find that they are non-chiral before analyzing the spectrum.
We will see a similar vanishing theorem in Type IIB orientifolds.

Supersymmetry does not require that the branes to have the same
${\bf M}$-label but only that $\sum_i{M_i\over k_i+2}$ to be
the same (modulo $H$). Indeed, in the examples we have studied,
there are many configurations with various ${\bf M}$-labels.
For example, consider the $P^A_{+;01001}$-orientifold with
the branes $B_{\bf L_9,M_7}$,
$B_{\bf L_{10},M_7}$,
$B_{\bf L_{11},M_8}$ and
$B_{\bf L_{12},M_8}$ (two each).
We find
$$
I(B_{\bf L_9,M_7},B_{\bf L_{11},M_8})=1,\quad
I(B_{\bf L_9,M_7},B_{\bf L_{12},M_8})=-1.
$$
Indeed, we have seen that this system is chiral by an explicit
spectrum analysis.

\subsection{Anomaly Cancellation Mechanism}\label{subsec:anomaly}

Let us consider a tadpole canceling brane configuration $\{n_aB_a\}$
in a Type II orientifold with respect to a worldsheet
parity symmetry $P$.
The gauge group $G_a$ supported by
the $n_a$ branes $B_a$ is
$U(n_a)$ if $B_a$ is not invariant under the parity
while it is $G_a=O(n_a)$ or $Sp(n_a/2)$ if the brane is invariant.
The tadpole cancellation condition is
$$
\sum_an_a[B_a]=[O_P].
$$
The standard triangle anomalies in the low energy field theory
are proportional to
\beqa
&&
A_{U(n_a)U(n_b)^2}=I(B_a,B_b)+I(B_a,B_{P(b)})
\label{Ananb}\\
&&
A_{U(n_a){\rm Gravi}^2}=\sum_bI(B_a,B_b)n_b=I(B_a,O_P).
\eeqa
where tadpole cancellation condition is used in the second equation.
For the $U(n_a)G_b^2$ anomaly with $G_b=O(n_b)$ or $Sp(n_b/2)$,
it is simply $I(B_a,B_b)$ (no
 extra term $I(B_a,B_{P(b)})$ as in (\ref{Ananb})).
Note that only the $U(1)_a$ subgroup of $U(n_a)$ is anomalous.

This field theory anomaly is canceled by the Green-Schwarz mechanism
\cite{GS,GSW,DSW,DM}. The relevant
 Green-Schwarz terms are obtained from
the disc diagrams with bulk insertion of a RR axion $\vartheta^{\bf i}$
and one or two boundary insertion of the gauge bosons,
or from the $\RP^2$ diagrams with insertion of a $\vartheta^{\bf i}$
and two gravitons.
They are proportional to the overlaps
$$
\Pi_{{\bf i}}^a=\langle\Scr{B}_a|\e^{-\pi i J_0}|{\bf i}\rangle_{\RR},
\quad
\widetilde{\Pi}^b_{\bf i}={}_{\RR}\langle {\bf i}|\Scr{B}_b\rangle,
\quad
\widetilde{\Pi}^P_{\bf i}={}_{\RR}\langle {\bf i}|\Scr{C}_P\rangle,
$$
and are given by
\beq
\Pi_{\bf i}^aA^{U(1)_a}_{\mu}\partial^{\mu}\vartheta^{\bf i}
\label{axion}
\eeq
and
$$
(\widetilde{\Pi}^b_{\bf i}+\widetilde{\Pi}^{P(b)}_{\bf i})
\vartheta^{\bf i}\Tr_{\bf n_b}F^b\wedge F^b,\qquad
\widetilde{\Pi}^P_{\bf i}\vartheta^{\bf i}\Tr R\wedge R.
$$
Note that the axion $\vartheta^{\bf i}$ corresponds to the RR ground state
$|{\bf i}\rangle_{\RR}$ that survives the orientifold projection.
If there are $2(h+1)$ RR ground states obeying the same
R-charge selection rule as the boundary states, the number of
such $|{\bf i}\rangle_{\RR}$ is $(h+1)$.
Also, we have chosen a basis such that the overlaps
$\Pi^a_{\bf i}$, $\widetilde{\Pi}^b_{\bf i}$ are real.
The coupling (\ref{axion}) induces an anomalous
$U(1)_a$ gauge transformation of the axion
$$
\vartheta^{\bf i}\longrightarrow 
\vartheta^{\bf i}+g^{\bf ij}\Pi_{\bf j}^a\lambda_a,
$$
where $g^{\bf ij}$ is the inverse matrix of
$g_{\bf ij}={}_{\RR}\langle {\bf j}|{\bf i}\rangle_{\RR}$
which determines the axion kinetic term,
$g_{\bf ij}\partial^{\mu}\vartheta^{\bf i}\partial_{\mu}\vartheta^{\bf j}$.
Then, the triangle anomalies are canceled
as a consequence of the bilinear identity
\footnote{Alternatively, one can
use the $\theta^{\bf i}$-equations of motion (instead of the
anomalous transformation).
This again introduces $g^{\bf ij}$ here from the axion kinetic term.}
$$
\sum_{\bf i,j}\Pi^a_{\bf i}g^{\bf ij}(\widetilde{\Pi}^b_{\bf j}
+\widetilde{\Pi}^{P(b)}_{\bf j})
=I(B_a,B_b)+I(B_a,B_{P(b)}),
\qquad
\sum_{\bf i,j}\Pi^a_{\bf i}g^{\bf ij}\widetilde{\Pi}^P_{\bf j}=I(B_a,O_P).
$$
The bilinear identity holds for the sum over {\it all}
RR ground states \cite{BH2}. However, in the present case, the sum
can be restricted to the orientifold-invariant states
$|{\bf i}\rangle_{\RR}$,
because $(\widetilde{\Pi}^b_i+\widetilde{\Pi}^{P(b)}_i)$
and $\widetilde{\Pi}^P_i$ are non-vanishing only for such
$|{\bf i}\rangle_{\RR}$.
Same is true on the overlap $\widetilde{\Pi}_i^b$ for a parity invariant
brane, $P(b)=b$.

For Type IIA orientifolds of a Calabi-Yau manifold $M$,
the RR-ground states $|i\rangle_{\RR}$ contributing to the overlap
$\Pi_i^a$ with A-branes corresponds to middle dimensional forms,
and the axions are the KK reduction of the RR 3-form
on $H^3(M)$. At the Gepner point,
the rational A-branes have overlap only with the untwisted states
since the boundary states are sum over images.
Thus the Green-Schwarz mechanism works with the untwisted
RR-fields, as long as rational A-branes are concerned.
Similar situations are encountered in the context of
toroidal orbifold in \cite{Alda1,CSU}.

For Type IIB orientifolds of a Calabi-Yau manifold $M$,
the RR-ground states $|i\rangle_{\RR}$ contributing to the overlap
$\Pi_i^a$ with B-branes corresponds to diagonal forms, $H^{p,p}(M)$,
and the axions are the KK reduction of the RR 0, 2, 4-forms.
At the Gepner point, the rational B-branes
generically have overlap with the twisted sector states
since the boundary state is sum over twists.
Thus the Green-Schwarz mechanism works with the twisted
RR-fields, just as in Type I orbifolds studied in \cite{DM}
(see also \cite{IRU1}).

\subsection{Fayet-Iliopoulos Terms}

The coupling (\ref{axion})
is extended to the full kinetic term
$(\partial_{\mu}\vartheta^{\bf i}+g^{\bf ij}\Pi_{\bf j}^aA_{\mu}^{U(1)_a})^2$,
and its supersymmetric completion is
\beq
\int\dd^4\theta\,\, K\Bigl(Y^{\bf i}+\overline{Y^{\bf i}}
+g^{\bf ij}\Pi_{\bf i}^aV_a\Bigr).
\label{FIterm}
\eeq
Here $Y^{\bf i}$ is a chiral superfield whose lowest component
is a complex scalar whose imaginary part is the axion
$y^{\bf i}=c^{\bf i}-i\vartheta^{\bf i}$.
This means that the real part $c^{\bf i}$
enters into the Fayet-Iliopoulos parameter \cite{DSW}
$$
\zeta^a=\sum_{\bf i}c^{\bf i}\Pi^a_{\bf i}.
$$
This is true as long as the gauge group includes $U(1)$ factors,
independently of whether the particle spectrum is chiral.

In Type IIA orientifolds, the superpartner of RR axions are
the complex structure moduli fields which are constrained
to the ``real section'' by the parity invariance. 
Thus, the ``real'' complex structure moduli fields
can enter into the FI parameters.
In the previous section, we have constructed many supersymmetric
(and tadpole canceling) brane configurations at the Gepner point.
As we move away from the Gepner point in the complex structure moduli space,
 the phases $\Pi^a_0$ may no longer align and
 the branes preserve different combinations of the
spacetime supersymmetry. In such a situation, we expect either
the branes recombine into other branes so that the supersymmetry is
restored, or there is no such configuration and the supersymmetry is broken.
This is exactly the situation
described by the above low energy field theory:
Under the deformation of $c^{\bf i}$ such that the FI parameter $\zeta^a$
becomes non-zero,
some charged scalar fields become tachyonic and condense
to find a supersymmetric vacua, or supersymmetry is broken.
A local model of such phenomenon was in fact constructed by
Kachru and McGreevy \cite{KM}.

The $U(1)$ gauge boson with non-zero $\Pi^a_{\bf i}$
acquires a mass by eating a combination of
the moduli fields $Y^{\bf i}$.
This must be a string loop effect to be consistent with the
tree level spectrum at the Gepner point which says that the gauge bosons
are all massless.
On the other hand, the tachyonic mass term of some charged open string fields 
after deformation of complex structure
must be at string tree level.
How can these be consistent?
To see this, let us be careful in the factors of
$g_{\it st}\propto g^2$.
The term (\ref{FIterm}) is correct provided that the gauge kinetic term
is normalized as ${1\over g^2}(F_{\mu\nu})^2$ and that 
the $y$-field is written as $y={c\over g^2}-i\vartheta$ so that
the complex structure fields $c$ have the standard NSNS kinetic term
${1\over g_{\it st}^2}(\partial_{\mu}c)^2$.
Then the FI parameter behaves as
$\zeta={c\over g^2}$.
Therefore the relevant terms in the effective Lagrangian depend on
the open string coupling $g$ as follows
$$
-{1\over g^2}(F_{\mu\nu})^2-(A_{\mu}+\partial_{\mu}\vartheta)^2
-|D_{\mu} Q_I|^2-{g^2\over 2}\left(
\pm |Q_I|^2-{c\over g^2}\right)^2,
$$
where $Q_I$ are open string fields charged under the $U(1)$.
We indeed see that the gauge boson mass is of open string one-loop level
(at the vacuum with $c=0$), which is
consistent with the tree level spectrum at the Gepner point.
Also, we find that the (sometimes tachyonic) mass term
for the charged open string fields is
$\pm c |Q_I|^2$ which is indeed tree level.

If all the branes are invariant under the parity, the gauge group
has no $U(1)$ factor and there is no room for FI term.
Thus, in such a case, we do not expect the brane-recombination nor
supersymmetry breaking as we move away from the Gepner point, or any
supersymmetric point, as long as each brane remains parity invariant.
This is indeed the case.
To be specific, let us show this in the large volume limit
(the same can be said near the Gepner point).
The supersymmetry preserved by the brane $W$ is measured by the phase of the
period integral $\int_W{\mit\Omega}$ where
${\mit\Omega}$ is the holomorphic 3-form of the Calabi-Yau manifold.
The supersymmetry preserved by the O-plane $O_P$ is
the phase of $\int_{O_P}{\mit\Omega}$.
As we change the complex structure, these phases vary.
We are considering the parity $P=\tau\Omega$ associated with
the antiholomorphic involution $\tau$, and we have
$$
\tau^*{\mit\Omega}=\e^{i\theta_{\tau}}\overline{\mit\Omega}.
$$
If we use the invariance $\tau W=W$, $\tau O_P=O_P$,
we find that the phases for $\int_W{\mit\Omega}$ and 
$\int_{O_P}{\mit\Omega}$ are both $\e^{i\theta_{\tau}/2}$
up to sign. But they have the same sign since
we started with the point where the phases are the same.
Thus, the phases of $\int_W{\mit\Omega}$ and 
$\int_{O_P}{\mit\Omega}$ are always aligned.
Therefore, as long as the branes are parity invariant, 
they preserve the same supersymmetry as the O-plane,
under any deformation of the complex structure compatible with the
parity.

What is said here can be repeated for Type IIB orientifolds:
This time K\"ahler moduli enter into the FI terms,
corresponding to the fact that the stability of B-branes is
controlled by the K\"ahler moduli \cite{DFRI,DFRII,Dougcat}.
In fact, the direct computation of the FI term is done in
similar contexts in \cite{DM,poppitz}.
See also \cite{Cvetic,Lalak,IRU2,IbaQue} for discussions.

\section{Consistency Conditions and Supersymmetry --- B}
\label{sec:TCCB}

In this section,
we write down the conditions of consistency and supersymmetry
and count the number of
solutions, for Type IIB orientifolds.
We will follow the general strategy outlined in subsection
\ref{subsec:general} and solve the tadpole constraints
completely. We find, for example, that the IIB orientifold 
of the Gepner model for quintic with respect to the parity 
without exchange has one the order of 30 billion supersymmetric 
and exactly solvable brane configurations.

\subsection{Charge and Supersymmetry of O-plane}\label{sec:chSB}

The first step is to study the charge of the O-plane.
To this end, it is useful to express it
in terms of the charges of the B-branes which
have been studied a lot in the past.

Here again, mirror A-type picture is convenient.
We know that the B-parity $P^B_{\omega;{\bf m}}$ is
the mirror of the A-parity $P^A_{\tilomega;{\bf \tilm}}$
in the model with the orbifold group $\tilGamma$ of order
$H^{-1}\prod_i(k_i+2)$.
Dressing by global symmetry ${\bf m}$ (resp. quantum symmetry
$\omega$) corresponds to
dressing by quantum symmetry $(\tilomega_i)$ (resp. global symmetry
${\bf \tilm}$): %(\ref{id1})-(\ref{id2}):
$$\e^{-2\pi i{m_i\over k_i+2}}=\tilomega_i,
\qquad
\omega=\e^{2\pi i \sum_i{\tilm_i\over k_i+2}}=:
\exp\left(2\pi i {M_{\omega}\over H}\right).
$$
We discuss
the odd $H$ and even $H$ cases separately.

\newcommand{\tilgamma}{\widetilde{\gamma}}

If $H$ is odd, we only have to consider the basic one
$P^B$ without dressing --- dressing by global symmetry
is not involutive and dressing by quantum symmetry is equivalent to
no dressing.
The structure of the crosscap state for the mirror
A-parity $P^A$ is just like
(\ref{Cexodd}), where the group $\Gamma$ is replaced by
the mirror orbifold group $\tilGamma$:
$$
|\Scr{C}_{P^A}\rangle
={1\over \sqrt{|\tilGamma|}}\sum_{\tilgamma\in \tilGamma}\tilgamma
|\Scr{C}_{{\bf P}^A}\rangle^{\rm prod}.
$$
This has the same structure as the sum-over-image 
formula for the boundary state, and we know that
$|\Scr{C}_{{\bf P}^A}\rangle^{\rm prod}$
has the same RR-charge as
the product brane $\Scr{B}_{{k_1-1\over 2},{k_1+1\over 2},1}\times
\cdots\times\Scr{B}_{{k_r-1\over 2},{k_r+1\over 2},1}$.
Thus, we find that $|\Scr{C}_{P^A}\rangle$ has the same charge as the
the brane $\Scr{B}_{{\bf {k-1\over 2},{k+1\over 2}},1}$.
Taking the mirror, we find that
$|\Scr{C}_{P^B}\rangle$ has the same RR-charge as
the B-brane $\Scr{B}_{{\bf k-1\over 2},H\sum_i{k_i+1\over 2(k_i+2)},1}$.
Including the spacetime part, we find the following relation of RR-charges
\beq
\Bigl[O_{P^B}\Bigr]=4\Bigl[
B^B_{{\bf k-1\over 2},H\sum_i{k_i-1\over 2(k_i+2)}}\Bigr]
\label{OBchodd}
\eeq

If $H$ is even, the structure of the 
crosscap state for the mirror
A-parity $P^A_{\tilomega;{\bf\tilm}}$ is analogous to
(\ref{Cexeve}). As in that case, we classify
the orbit of parity symmetries
$\{\tilgamma{\bf P}^A_{\bf m}\}_{\tilgamma\in\tilGamma}$
with respect to the subgroup
$$
\tilGamma^2=\{\tilgamma^2|\tilgamma\in \tilGamma\}\subset\tilGamma.
$$
This is a proper subgroup of $\tilGamma$ if $H$ is even (if $H$ is odd,
this agrees with $\tilGamma$ and hence the orbit sum has a
simple structure as we have discussed above).
The orbit $\{\tilgamma{\bf P}^A_{\bf m}\}_{\tilgamma\in\tilGamma}$
decomposes into blocks
$\{\tilgamma^2
{\bf P}^A_{\bf m+\tilnu}\}_{\tilgamma^2\in \tilGamma^2}$
parametrized by the coset $\tilnu\in \tilGamma/\tilGamma^2$.
Thus, the crosscap state has the following structure
$$
|\Scr{C}_{P^A_{\tilomega;{\bf \tilm}}}\rangle
={1\over\sqrt{|\tilGamma|}}
\sum_{\tilgamma\in\tilGamma}\tilomega^{-\tilgamma}
|\Scr{C}_{\tilgamma{\bf P}^A_{\bf \tilm}}\rangle
=
{1\over\sqrt{|\tilGamma|}}
\sum_{\tilnu\in\tilGamma/\tilGamma^2}\tilomega^{-\tilnu}
\sum_{\tilgamma^2\in \tilGamma^2}
|\Scr{C}_{\tilgamma^2{\bf P}^A_{\bf \tilm+\tilnu}}\rangle
$$
where we have used the fact that $\tilomega_i=\pm 1$ and
hence $\tilomega^{-\tilgamma^2}=1$.
At this stage we use the relation
$|\Scr{C}_{\tilgamma^2{\bf P}^A_{\bf \tilm+\tilnu}}\rangle
=\tilgamma|\Scr{C}_{{\bf P}^A_{\bf \tilm+\tilnu}}\rangle$,
and also replace the sum over $\tilgamma^2\in \tilGamma^2$
by the sum over $\tilgamma\in \tilGamma$ times the ratio of the orders
$|\tilGamma^2|/|\tilGamma|$:
\beq
|\Scr{C}_{P^A_{\tilomega;{\bf \tilm}}}\rangle
={1\over\sqrt{|\tilGamma|}}
\sum_{\tilnu\in\tilGamma/\tilGamma^2}\tilomega^{-\tilnu}
{|\tilGamma^2|\over |\tilGamma|}
\sum_{\tilgamma\in \tilGamma}
\tilgamma|\Scr{C}_{{\bf P}^A_{\bf \tilm+\tilnu}}\rangle
=
{1\over |\tilGamma/\tilGamma^2|}
\sum_{\tilnu\in\tilGamma/\tilGamma^2}\tilomega^{-\tilnu}
\left({1\over\sqrt{|\tilGamma|}}\sum_{\tilgamma\in \tilGamma}
\tilgamma|\Scr{C}_{{\bf P}^A_{\bf \tilm+\tilnu}}\rangle\right).
\eeq
The expression in the parenthesis of the right hand side has the same
structure as the sum-over-image formula for the boundary states.
If $k_i$ are all even, this has the same RR-charge as
the brane $\Scr{B}_{{\bf {k\over 2},{k+2\over 2}+\tilm+\tilnu}
-\delta_{\bf \tilm+\tilnu},1}$ times the possible orientation flip
$(-1)^{\sum_i{\tilnu_i\over k_i+2}}$. Bringing this mirror relation
back into the original side and adding the spacetime part, we find
%Kazuo: $\tilm$ inserted
\beq
\Bigl[O_{P^B_{\omega;{\bf m}}}\Bigr]
={4\over |\tilGamma/\tilGamma^2|}\sum_{\tilnu\in\tilGamma/\tilGamma^2}
\tilomega^{-\tilm-\tilnu}(-1)^{\sum_i{\tilnu_i\over k_i+2}}
\Bigl[B^B_{{\bf k\over 2},M_{{\bf \tilm}+\tilnu}}\Bigr]
\label{OBcheve}
\eeq
where
\beq
M_{{\bf \tilm}+\tilnu}=
H\sum_{i=1}^r{{k_i\over 2}+
\tilm_i+\tilnu_i-\delta_{\tilm_i+\tilnu_i}\over k_i+2}.
\eeq
If there are both even and odd $k_i$, the expression is the obvious
mixture of (\ref{OBcheve}) and (\ref{OBchodd}).
An alternative approach to find the O-plane charge
directly in the B-type picture will be
outlined in section \ref{subsec:Dbranecharges}.

The next thing to find is
the phase determining the spacetime supersymmetry
preserved by the branes and the orientifold.
For the branes, we find
$$
{}_{\RR}\langle 0|\Scr{B}_{{\bf L},M,S}\rangle_{\RR}
=\e^{\pi i \sum_i({M_i\over k_i+2}-{S\over 2})}
{}_{\NSNS}\langle 0|\Scr{B}_{{\bf L},M,S}\rangle_{\NSNS}
$$
and thus the phase is
\beq
\exp\Bigl(i\theta^B_{{\bf L},M}\Bigr)
=i
\exp\left(\pi i {M \over H}\right)
\eeq
For the crosscap, one can see that
\beq
{}_{\RR}\langle 0|\Scr{C}_{P^B_{\omega;{\bf m}}}\rangle
=\tilomega\e^{\pi i \sum_i{\tilm_i-{1\over 2}\over k_i+2}}
{}_{\NSNS}\langle 0|\Scr{C}_{\widetilde{P}^B_{\omega;{\bf m}}}\rangle,
\eeq
where $\tilm_i$ parametrizes the global symmetry in the mirror which is
the quantum symmetry $\omega=\e^{2\pi i \sum_i{\tilm_i\over k_i+2}}$
of the original side.
We note here that the NSNS part of the total crosscap state
has the factor $\tilomega^{1\over 2}$,
see Eqn~(\ref{CBNStotal}).
Thus, the ratio is
\beq
\exp\Bigl(i\theta^B_{\omega;{\bf m}}\Bigr)
=-i\tilomega^{{1\over 2}}
\exp\left(\pi i \sum_i{\tilm_i\over k_i+2}\right)
%=-i^{-r}
%\omega^{1\over 2}(\pm 1)\exp\left(\pi i \sum_{i=1}^r{m_i+{k_i-1\over 2}\over
%k_i+2}\right).
\eeq
For completeness, we record here the expression of the O-plane tension;
$$
4\tilomega^{1\over 2}
{}_{\NSNS}\langle 0|\Scr{C}_{\widetilde{P}^B_{\omega,{\bf m}}}
\rangle
={4\over\sqrt{H}}
\prod_{i=1}^r\sqrt{2\over\sin({\pi\over k_i+2})}
\prod_{k_i:\,{\rm odd}}
\cos\Bigl({\pi\over 2(k_i\!+\!2)}\Bigr)\cdot
{1\over |\tilGamma/\tilGamma^2|}
\sum_{\tilnu\in \tilGamma/\tilGamma^2}
\tilomega^{-\tilnu-{1\over 2}}
\e^{i\Theta_{\bf\tilm+\tilnu}},
$$
$$\Theta_{\bf \tilm+\tilnu}=\sum_{k_i\,{\rm even}}
{\pi (-1)^{\tilm_i+\tilnu_i}\over 2(k_i+2)}.
$$

\subsubsection{Example --- Quintic}\label{subsub:qex}

For the model $(k_i+2)=(5,5,5,5,5)$ corresponding to the quintic,
the charge of the O-plane for the parity $P^B=P^B_{+;{\bf 0}}$
is four times that of the B-brane
$B_{{\bf L},M}^B$ with ${\bf L}=(1,1,1,1,1)$ and $M=5$.
\beq
\Bigl[O_{P^B}\Bigr]
=4\Bigl[B_{{\bf 1},5}\Bigr].
\label{qcross}
\eeq
The tension of the O-plane is
$$
4\tilomega^{1\over 2}
{}_{\NSNS}\langle 0|\Scr{C}_{\widetilde{P}^B}\rangle
= {4\tilomega^{1\over 2}\over \sqrt{5}}
\sqrt{2\over \sin({\pi\over 5})}^5\cos^5\left({\pi\over 10}\right);\qquad
\tilomega^{1\over 2}=\pm 1.
$$
The phase determining the spacetime supersymmetry is
\beq
\e^{i\theta_{\omega;{\bf 0}}^B}=-i\tilomega^{{1\over 2}}
\eeq
while the one for the brane
$B^B_{{\bf L},M}$ is $\e^{i\theta^B_{{\bf L},M}}
=i\e^{\pi i M/5}$.
For the orientifold with
$\tilomega^{1\over 2}=-1$, branes preserving the same
supersymmetry are $B^B_{{\bf L},M=0}$
and $\overline{B^B_{{\bf L},M=5}}$. More explicitly, they are
$B^B_{(00000),0}$, $\overline{B^B_{(10000),5}}$ and permutations,
$B^B_{(11000),0}$ and permutations,
$\overline{B^B_{(11100),5}}$ and permutations,
$B^B_{(11110),5}$ and permutations, and
$\overline{B^B_{(11111),5}}$. In total, there are
$1+5+10+10+5+1=32$ of them.

\subsubsection{Example --- The Two Parameter Model}\label{sec:oplanec}

Let us consider the model $(k_i+2)=(8,8,4,4,4)$.
The parity symmetries $P^B_{\omega;{\bf m}}$ are denoted as in
Section~\ref{subsub:two} as
$P^B_{\mu;\epsilon_1...\epsilon_5}$ where $\omega=\e^{2\pi i \mu/8}$ and
$\epsilon_i=\e^{2\pi i{m_i\over k_i+2}}=\pm$.
We only have to consider $P^B_{0;\epsilon_1...\epsilon_5}$
and $P^B_{1;\epsilon_1...\epsilon_5}$ since
others are related to these by symmetry conjugations.
Also, there are eight inequivalent choices for $(\epsilon_1...\epsilon_5)$:
$(+++++)$, $(++-++)$, $(++--+)$, $(++---)$,
$(+-+++)$, $(+--++)$, $(+---+)$, $(+----)$.

The mirror orbifold group $\tilGamma$ is the set of
$\tilnu=(\nu_1,\nu_2,\nu_3,\nu_4,\nu_5)\in
\Z_8\times \Z_8\times\Z_4\times\Z_4\times\Z_4$ with
${\nu_1+\nu_2\over 8}+{\nu_3+\nu_4+\nu_5\over 4}\in \Z$.
One may solve for $\nu_1$ as
$\nu_1=-\nu_2-2(\nu_3+\nu_4+\nu_5)$, and thus
the group is isomorphic to $\Z_8\times (\Z_4)^3$.
This also shows that the element of $\tilGamma/\tilGamma^2$ is labeled
by the mod 2 reduction of $(\nu_2,\nu_3,\nu_4,\nu_5)$ and hence
$\tilGamma/\tilGamma^2\cong (\Z_2)^4$.

For parities $P^B_{0;\epsilon_1...\epsilon_5}$
without dressing by quantum symmetry,
we have ${\bf\tilm}={\bf 0}$.
We find
 $M_{\bf\tilm+\tilnu}=M_{\tilnu}=12-2(\nu_2+\nu_3+\nu_4+\nu_5)$
if $\nu_2,...,\nu_5$ are assumed to take values in $\{0,1\}$.
The charge and the tension of the O-plane
for the eight cases are summarized in the table \ref{Otwopar1}.
\begin{table}[htb]
\begin{center}
\renewcommand{\arraystretch}{1.2}
\begin{tabular}{|l||l|l|}
\hline
parity&RR-charge&Tension\\
\noalign{\hrule height 0.8pt}
$P^B_{0;+++++}$
&
${1\over 4}\left([B_{{\bf k\over 2},12}]
+4[B_{{\bf k\over 2},10}]
+6[B_{{\bf k\over 2},8}]
+4[B_{{\bf k\over 2},6}]
+[B_{{\bf k\over 2},4}]\right)$
&$\begin{array}{l}
\\
\\
\end{array}
\!\!\!\!
4\tilomega^{-{1\over 2}}{3+2\sqrt{2}\over \sqrt{2\sqrt{2}-2}}$
\\
\hline
\multicolumn{1}{|l||}{$P^B_{0;++-++}$}&
\lw{$
{1\over 4}\left([B_{{\bf k\over 2},12}]
+2[B_{{\bf k\over 2},10}]
-2[B_{{\bf k\over 2},6}]
-[B_{{\bf k\over 2},4}]\right)
$}&\lw{$4i\tilomega^{-{1\over 2}}{1+\sqrt{2}\over \sqrt{2\sqrt{2}-2}}
$}\\
\cline{1-1}
$P^B_{0;+-+++}$&&\\
\hline
\multicolumn{1}{|l||}{$P^B_{0;++--+}$}&
\lw{$
{1\over 4}\left([B_{{\bf k\over 2},12}]
-2[B_{{\bf k\over 2},8}]
+[B_{{\bf k\over 2},4}]\right)
$}&\lw{$
-4\tilomega^{-{1\over 2}}{1\over \sqrt{2\sqrt{2}-2}}
$}\\
\cline{1-1}
$P^B_{0;+--++}$&&\\
\hline
\multicolumn{1}{|l||}{$P^B_{0;++---}$}&
\lw{$
{1\over 4}\left([B_{{\bf k\over 2},12}]
-2[B_{{\bf k\over 2},10}]
+2[B_{{\bf k\over 2},6}]
-[B_{{\bf k\over 2},4}]\right)
$}&\lw{$
-4i\tilomega^{-{1\over 2}}{\sqrt{2}-1\over \sqrt{2\sqrt{2}-2}}
$}\\
\cline{1-1}
$P^B_{0;+---+}$&&\\
\hline
$P^B_{0;+----}$&
$
{1\over 4}\left([B_{{\bf k\over 2},12}]
-4[B_{{\bf k\over 2},10}]
+6[B_{{\bf k\over 2},8}]
-4[B_{{\bf k\over 2},6}]
+[B_{{\bf k\over 2},4}]\right)
$
&$\begin{array}{l}
\\
\\
\end{array}
\!\!\!\!
4\tilomega^{-{1\over 2}}{3-2\sqrt{2}\over \sqrt{2\sqrt{2}-2}}
$
\\
\hline
\end{tabular}
\caption{Charge and Tension of the O-plane ($\omega=1$)}
\label{Otwopar1}
\end{center}
\end{table}
The spacetime supersymmetry preserved by the orientifold is
\beq
\e^{i\theta_O}=-i\tilomega^{1\over 2},
\eeq
where we note that $\tilomega=\epsilon_1\cdots\epsilon_5$.
Branes preserving the same supersymmetries are 
$B_{{\bf L},8}$,$\overline{B_{{\bf L},0}}$ for
$\tilomega^{1\over 2}=1$;
$B_{{\bf L},0}$,$\overline{B_{{\bf L},8}}$ for
$\tilomega^{1\over 2}=-1$;
$B_{{\bf L},12}$,$\overline{B_{{\bf L},4}}$ for
$\tilomega^{1\over 2}=i$; and
$B_{{\bf L},4}$,$\overline{B_{{\bf L},12}}$ for
$\tilomega^{1\over 2}=-i$.

For parities $P^B_{1;\epsilon_1...\epsilon_5}$
dressed by the quantum symmetry $\omega=\e^{2\pi i/8}$,
we have ${\bf\tilm}=(1,0,0,0,0)$.
We find
 $M_{\bf\tilm+\tilnu}=12-2(\nu_3+\nu_4+\nu_5)$
if $\nu_2,...,\nu_5$ are assumed to take values in $\{0,1\}$.
The charge and the tension of the O-plane
for the eight cases are summarized in the table \ref{Otwopar2}.
\begin{table}[htb]
\begin{center}
\renewcommand{\arraystretch}{1.2}
\begin{tabular}{|l||l|l|}
\hline
parity&RR-charge&Tension
\\
\noalign{\hrule height 0.8pt}
$P^B_{1;+-***}$
&$0$
&$\begin{array}{l}
\\
\\
\end{array}
\!\!\!\!
0$
\\
\hline
$P^B_{1;+++++}$
&$
{1\over 2}\left([B_{{\bf k\over 2},12}]
+3[B_{{\bf k\over 2},10}]
+3[B_{{\bf k\over 2},8}]
+[B_{{\bf k\over 2},6}]\right)
$
&$\begin{array}{l}
\\
\\
\end{array}
\!\!\!\!
4\tilomega^{-{1\over 2}}\sqrt{10+7\sqrt{2}\over \sqrt{2}-1}
$
\\
\hline
$P^B_{1;++-++}$
&$
{1\over 2}\left([B_{{\bf k\over 2},12}]
+[B_{{\bf k\over 2},10}]
-[B_{{\bf k\over 2},8}]
-[B_{{\bf k\over 2},6}]\right)
$
&$\begin{array}{l}
\\
\\
\end{array}
\!\!\!\!
4i\tilomega^{-{1\over 2}}\sqrt{2+\sqrt{2}\over \sqrt{2}-1}
$
\\
\hline
$P^B_{1;++--+}$
&$
{1\over 2}\left([B_{{\bf k\over 2},12}]
-[B_{{\bf k\over 2},10}]
-[B_{{\bf k\over 2},8}]
+[B_{{\bf k\over 2},6}]\right)
$
&$\begin{array}{l}
\\
\\
\end{array}
\!\!\!\!\!\!
-4\tilomega^{-{1\over 2}}\sqrt{2-\sqrt{2}\over \sqrt{2}-1}
$
\\
\hline
$P^B_{1;++---}$
&$
{1\over 2}\left([B_{{\bf k\over 2},12}]
-3[B_{{\bf k\over 2},10}]
+3[B_{{\bf k\over 2},8}]
-[B_{{\bf k\over 2},6}]\right)
$
&$\begin{array}{l}
\\
\\
\end{array}
\!\!\!\!\!\!
-4i\tilomega^{-{1\over 2}}\sqrt{10-7\sqrt{2}\over \sqrt{2}-1}
$
\\
\hline
\end{tabular}
\caption{Charge and Tension of the O-plane ($\omega=\e^{2\pi i/8}$)}
\label{Otwopar2}
\end{center}
\end{table}
The spacetime supersymmetry preserved by the orientifold is
\beq
\e^{i\theta_O}=-i\tilomega^{1\over 2}\exp\left({\pi i\over 8}\right).
\eeq
Branes preserving the same supersymmetries are 
$B_{{\bf L},9}$,$\overline{B_{{\bf L},1}}$ for
$\tilomega^{1\over 2}=1$;
$B_{{\bf L},1}$,$\overline{B_{{\bf L},9}}$ for
$\tilomega^{1\over 2}=-1$;
$B_{{\bf L},13}$,$\overline{B_{{\bf L},5}}$ for
$\tilomega^{1\over 2}=i$; and
$B_{{\bf L},5}$,$\overline{B_{{\bf L},13}}$ for
$\tilomega^{1\over 2}=-i$.

\subsection{D-branes in the Orientifold Models}

\subsubsection{Parity Action on D-branes}

\newcommand{\bfS}{{\bf S}}

Let us now find how the B-type orientifold acts on the B-branes.
We first consider long-orbit branes. 
To see the action, we compare the $\langle B|C\rangle$
and $\langle C|B\rangle$ M\"obius strips.
We find
\beqa
&&{}_{\RR}\langle C^B_{\omega;{\bf m}}|q_t^H|B^B_{{\bf L},M}\rangle_{\RR}
=\tilomega^{-1}\times {}_{\RR}\langle
B^B_{{\bf L},2M_{\omega}-M}|q_t^H
|C^B_{\omega;{\bf m}}\rangle_{\RR},\\
&&{}_{\NSNS}\langle C^B_{\omega;{\bf m}}|q_t^H|
B^B_{{\bf L},M}\rangle_{\NSNS}
=-{}_{\NSNS}\langle
B^B_{{\bf L},2M_{\omega}-M}|q_t^H
|C^B_{\omega;{\bf m}}\rangle_{\NSNS}.
\label{seconB}
\eeqa
This can again be shown 
using the mirror description.
Thus, the parity acts on the branes as
\beq
P^B_{\omega;{\bf m}}:\,B^B_{{\bf L},M}
\longmapsto
\tilomega^{-1}
B^B_{{\bf L},2M_{\omega}-M},
\label{PBonBb}
\eeq
where we recall that
$\tilomega^{-1}=\e^{2\pi i \sum_i{m_i\over k_i+2}}$
and $\e^{2\pi i M_{\omega}/H}=\omega$.
 
Let us now consider short-orbit branes.
We denote by $\bfS$ the set of $i$ such that $L_i={k_i\over 2}$
If the number of elements $|\bfS|$ is odd, there is no difference
from the above result. Thus we focus on the branes
 $\widehat{B}^{(\varepsilon)}$ with even $|\bfS|$.
The action on the $({\bf L},M)$-label is the same as above, and
the difference appears in the action on
the $\varepsilon$-label.
We find that the result is
\beq
P^B_{\omega;{\bf m}}:\varepsilon\longmapsto \varepsilon'
=(-1)^{|\bfS|\over 2}\prod_{i\in\bfS}\widetilde{\omega}_i^{\frac{k_i+2}{2}}
\cdot \varepsilon.
\label{PBonBshort}
\eeq

\subsubsection{Invariant Branes}

Let us find out which of the B-branes are invariant under the
orientifold action.
By (\ref{PBonBb}), the condition is
$B^B_{{\bf L},2M_{\omega}-M}=B^B_{{\bf L},M}$.
Here it is useful to note the ``brane identification'':
$B^B_{{\bf L}',M'}=B^B_{{\bf L},M}$ if and only if 
$M'=M$ and $L_i'=L_i$ except for even number of
$i$'s with $L_i'=k_i-L_i$.
Also, $B^B_{{\bf L}',M'}=\overline{B^B_{{\bf L},M}}$
if and only if $M'=M+H$ and $L_i'=L_i$ except for odd number of $i$'s with
$L_i'=k_i-L_i$. Using this we find that invariant branes are
\beqa
\tilomega=1:&B^B_{{\bf L},M_{\omega}},B^B_{{\bf L},M_{\omega}+H},
\,\,\mbox{${\bf L}$ arbitrary},
\\
\tilomega=-1:&B^B_{{\bf L},M_{\omega}\pm {H\over 2}},\,\,\,L_i={k_i\over 2}\,\,
\mbox{for a single $i$}.
\eeqa
This applies also to short orbit branes with odd $|\bfS|$.

For short-orbit branes $\widehat{B}^{(\varepsilon)}_{{\bf L},M}$
with even $|\bfS|$, this is modified because of the
new type of ``Brane identification'' where
$M\to M+H$ does the flip of $\varepsilon$ as well as the orientation.
The invariant branes are those with
$M=M_{\omega}$ (mod $H$) if $\tilomega=1$
and $M=M_{\omega}+{H\over 2}$ (mod $H$) if $\tilomega=-1$, just as above
but there is an extra condition
on the number $|\bfS|$:
\beq
(-1)^{|\bfS|\over 2}
=\tilomega
  \prod_{i\in\bfS}\tilomega_i^{\frac{k_i+2}{2}}.
\eeq

\subsubsection{Structure of Chan-Paton Factor}

Let us find the gauge group supported by $N$ of the
invariant D-branes by computing
 the $\langle B|C\rangle$ overlap in the NSNS sector.
The computation can be done most easily in the mirror picture,
but we have to be careful for the factor $\tilomega^{-\tilm}$
appeared in (\ref{BvsAtil}).
Using the formula
(\ref{PonGSmin}) for the minimal model and the ones for the universal sector,
we find (up to the standard factor)
$$
\langle B^B_{{\bf L},M}|q_t^H|C^B_{\omega;{\bf m}}\rangle_{\NSNS}
=\overline{\epsilon_{{\bf L},M}^{\omega;{\bf m}}}
\prod_{n=1}^{\infty}(1-i(-1)^nq_l^{n-{1\over 2}})^2
-\epsilon_{{\bf L},M}^{\omega;{\bf m}}
\prod_{n=1}^{\infty}(1+i(-1)^nq_l^{n-{1\over 2}})^2
+\cdots
$$
where $\epsilon_{{\bf L},M}^{\omega;{\bf m}}$
is given as follows;
\beqa
\epsilon_{{\bf L},M}^{\omega;{\bf m}}
&=&\e^{\pi i\over 4}
\sum_{\tilnu\in \tilGamma}
\tilomega^{\tilm+\tilnu+{1\over 2}}
\prod_i\left(
\e^{-{\pi i\over 4}}\delta_{M_i,\tilm_i+\tilnu_i}
+\e^{\pi i\over 4}\delta_{L_i,{k_i\over 2}}
\delta_{M_i,\tilm_i+\tilnu_i+{k_i+2\over 2}}\right)
\nn\\
&=&
-\tilomega^{{\bf M}+{1\over 2}}
\sum_{\tilnu\in \tilGamma}
\prod_{L_i\ne {k_i\over 2}}\delta_{M_i,\tilm_i+\tilnu_i}
\prod_{L_i={k_i\over 2}}\left(\delta_{M_i,\tilm_i+\tilnu_i}
+i\tilomega_i^{k_i+2\over 2}
\delta_{M_i,\tilm_i+\tilnu_i+{k_i+2\over 2}}\right)
\nn\\
&=&-\tilomega^{{\bf L}+{1\over 2}}
\sum_{p_i\in\{0,1\}}\delta^{(H)}_{M,M_{\omega}+{\sum_ip_i\over 2}H}
\prod_{L_i={k_i\over 2}}\left(i\tilomega_i^{k_i+2\over 2}\right)^{p_i}.
\eeqa
This is indeed a sign factor for the long-orbit branes
for which $L_i={k_i\over 2}$ at most for one $i$, and
$M\equiv M_{\omega}$ (mod $H$) if $\tilomega=1$
and $M\equiv M_{\omega}+{H\over 2}$ (mod $H$) if $\tilomega=-1$.
More concretely,
\beq
\epsilon_{{\bf L},M}^{\omega;{\bf m}}
=\left\{
\begin{array}{ll}
-\tilomega^{{\bf L}+{1\over 2}}&\tilomega=1
\\
-i\tilomega^{{\bf L}+{1\over 2}}\tilomega_{i_*}^{k_{i_*}+2\over 2}
&\tilomega=-1.
\end{array}\right.
\label{Ggroup}
\eeq
where $i_*$ is the one that has $L_{i_*}={k_{i_*}\over 2}$.
The invariant branes with $\epsilon_{{\bf L},M}^{\omega;{\bf m}}=1$
or $-1$ support the $O$ or $Sp$-type gauge symmetries.

One can do the same computation for short-orbit branes
satisfying $L_i=\frac{k_i}{2}$ for $i\in \bfS~ (|\bfS|\ge 2)$.
Taking into account the correct normalization factor, one finds
\beqa
\epsilon_{{\bf L},M}^{\omega;{\bf m}}
&=&
 -2^{-\left[|\bfS|/2\right]}\tilomega^{{\bf L}+{1\over 2}}
\sum_{p_i\in\{0,1\}}\delta^{(H)}_{M,M_{\omega}+{\sum_ip_i\over 2}H}
\prod_{i\in\bfS}\left(i\tilomega_i^{k_i+2\over 2}\right)^{p_i}.
\nn \\ &=&
 -{\rm Re}\left(2^{-\left[|\bfS|/2\right]}\tilomega^{{\bf L}+{1\over 2}}
 \prod_{i\in\bfS}(1+i\tilomega_i^{k_i+2\over 2})\right).
\label{Ggroup2}
\eeqa
This is indeed a sign factor again: for odd $|\bfS|$ the quantity in the large
parenthesis is of the form $(\pm1\!\pm'\!i)$, and for even $|\bfS|$
it squares to
\beq
  \tilomega(-1)^{\frac{|\bfS|}{2}}\prod_{i\in\bfS}\tilomega_i^{k_i+2\over 2}
\eeq
which is unity for parity-invariant short-orbit branes.

\subsubsection{Examples}

{\bf Quintic}

The branes in the $(5,5,5,5,5)$-model are transformed by the B-parity $P^B$
as $B^B_{{\bf L},M}\longmapsto B^B_{{\bf L},-M}$. Invariant branes are those with
$M=0$ and $M=5$. All of them support $O(N)$ (resp. $Sp(N/2)$)
gauge group for the choice
$\tilomega^{1\over 2}=-1$ (resp. $\tilomega^{1\over 2}=1$).

\bigskip
\noindent
{\bf The Two Parameter Model}

The branes in the $(8,8,4,4,4)$-model are transformed by B-parities as
\beqa
&&P^B_{0;\epsilon_1...\epsilon_5}:
B_{{\bf L},M}\longmapsto \epsilon_1\cdots\epsilon_5 B_{{\bf L},-M},\nn\\
&&P^B_{1;\epsilon_1...\epsilon_5}:
B_{{\bf L},M}\longmapsto \epsilon_1\cdots\epsilon_5 B_{{\bf L},2-M}.\nn
\eeqa
Invariant branes are
\beqa
&&P^B_{0;\epsilon_1...\epsilon_5},\,\epsilon_1\cdots\epsilon_5=1:
B^B_{{\bf L},0},B^B_{{\bf L},8};
\nn\\
&&P^B_{0;\epsilon_1...\epsilon_5},\,\epsilon_1\cdots\epsilon_5=-1:
B^B_{{\bf L^*},4},B^B_{{\bf L^*},12};
\nn\\
&&P^B_{1;\epsilon_1...\epsilon_5},\,\epsilon_1\cdots\epsilon_5=1:
B^B_{{\bf L},1},B^B_{{\bf L},9};
\nn\\
&&P^B_{1;\epsilon_1...\epsilon_5},\,\epsilon_1\cdots\epsilon_5=-1:
B^B_{{\bf L^*},5},B^B_{{\bf L^*},13}.
\nn
\eeqa
Here, ${\bf L^*}$ is such that $L_i={k_i\over 2}$ for a single $i$.
Invariant short-orbit branes with even $|\bfS|$ also satisfy
$(-1)^{\frac{|\bfS|}{2}}=\epsilon_1\cdots\epsilon_5$.
The gauge group depends on $\prod_i\epsilon_i^{L_i}$ as in
(\ref{Ggroup}) and (\ref{Ggroup2}).

\subsection{D-brane charges}\label{subsec:Dbranecharges}

\def\II{\bf S}
Recall: Rational branes in the Gepner model 
$(k_1,k_2,\ldots,k_r)$ with $H={\rm lcm}\{k_i+2\}$
are in bijective correspondence with the following labeling system.
We need:\\
(i) A label ${\bf L}=(L_1,\ldots,L_r)$ with $0\leq L_i \leq k_i/2$.
We denote by $\II$ the set of $i$ for which $L_i=k_i/2$. \\
(ii) A label $M\in\ZZ_{2H}$ with $M=\sum w_i L_i\bmod 2$.
($w_i=H/(k_i+2)$)\\
(iii) If $d+r$ is even AND all $L_i<k_i/2$ (ie, $\II=\emptyset$),
a label $S=0,2$.\\
(iv) If $\II\neq\emptyset$ AND $d+r+|\II|$ is even,
a label $\psi=\pm$.\\
In this paper, we denote such a brane by $B^B_{{\bf L},M,S,\psi}$,
where it is understood that $S$ and $\psi$ can be omitted or
neglected if they are unnecessary. Again, we emphasize that any 
brane has a unique label of this type and that any label of this 
type uniquely specifies a brane. This labeling system assigns
different labels to a brane and its antibrane, but we will 
sometimes take the freedom to indicate the antibrane with a
minus-sign. For branes with an $S=0$ label, the antibrane has
$S=2$, while for branes without an $S$ label, the antibrane
is obtained by sending $M\mapsto M+H$. The spacetime supersymmetry 
preserved by such a brane is a phase given by $\e^{\pi i (M/H+S/2)}$.

To summarize the RR charge of these branes, we find it convenient
to introduce generators for the charge lattice. For B-type branes,
such a generating set is conveniently obtained from the charges of 
the $H$ branes with ${\bf L}= {\bf 0}$, and $M=0,2,\ldots, 2H-2$. 
The relations they satisfy can be understood quite easily from 
divisibility properties of the weights and we will make this more 
explicit below. We will denote the linear operator mapping the $H$ 
${\bf L}={\bf 0}$ branes onto a linearly independent set generically 
by $T$. To expand the charges of the other branes in terms of the 
${\bf L}={\bf 0}$ ones, we shall use as before brackets $[\Scr B]$ 
to denote the RR charge vector of a brane $\Scr B$. Neglecting for 
a very short moment the $S$ and $\psi$ labels, we denote by 
$([B^B_{{\bf L},M}])$ the $\ZZ_H$ orbit of rational branes with 
definite ${\bf L}$ label and $M$ label running over $M,M+2,\ldots, 
(M+2H-2)(\bmod 2H)$. We can then write
\beq
\left(\bigl[ B^B_{{\bf L},M}\bigr]\right) =
\left(\bigl[B^B_{{\bf 0},0}\bigr]\right)
Q_{{\bf L},M}(g)
\label{run}
\eeq
where the $H\times H$-dimensional matrix $Q_{{\bf L},M}(g)$ is a simple
polynomial expression in the ``shift generator'' $g$, which is 
the matrix with entry $1$ on the first lower diagonal and in the 
upper right corner, and zeros elsewhere. Explicitly
\beq
Q_{{\bf L},M}(g) = g^{M/2} \prod_{i=1}^r \Bigl( \sum_{k_i=0}^{L_i} 
\bigl(g_i\bigr)^{L_i/2-k_i}\Bigr) \,,
\label{branec}
\eeq
where $g_i=g^{w_i}$ and $w_i=H/(k_i+2)$. Even more explicitly,
the components of the charge vector of the $M$-th brane on the
orbit (\ref{run}) are given by the $M$-th column of the matrix
(\ref{branec}).

If the brane carries an $S$ label, the formula (\ref{branec}) is 
simply multiplied by $(-1)^{S/2}$. If the brane has ${\bf S}\neq
\emptyset$, the formula (\ref{branec}) gets corrected by a fixed 
point resolution factor $f=1/2^{[\nu/2]}$, where $\nu={\bf S}+1$ if 
$d+r$ is odd and $\nu=|{\bf S}|$ if $d+r$ is even (see subsection
\ref{sub:Cardybranes}). We realize that this factor gets some time 
to get accustomed to, so we write it out explicitly
for the canonical case $r=5$, $d=1$, all levels even, such as
our two parameter model. Then if the number of $L_i$ which 
are equal to $k_i/2$ is $1,2,3,4,5$, we have $f=1,2,2,4,4$, 
respectively. 
Let us also note here that in can happen under exceptional 
circumstance that the brane charge depends on the $\psi$
label, because the non-toric K\"ahler parameters sit in the
$\Z_{H/2}$ twisted sector. (An example of this is the Gepner 
model $(k_i)=(2,2,4,4,4)$.)

In Section~\ref{sec:chSB}, we obtained the RR charge of
the O-plane using the mirror picture and expressed it in terms of
the D-brane charge.
Here we comment on an alternative
way, directly in the B-type picture, to find and express it
in terms of the generators of the charge lattice introduced in this
section.
We look for a stack of ${\bf L=0}$ branes that have
the same intersection numbers with
any other set of branes as the orientifold. Since the ${\bf L=0}$ branes
form a basis of the charge lattice, it is sufficient to check this
for a general configuration of ${\bf L=0}$ branes. So all we have to
do is to solve the linear system ``intersection matrix of the
${\bf L}={\bf 0}$ branes times a charge vector equals 
the twisted Witten indices of the ${\bf L=0}$ branes with the orientifold
plane''.
The charge vector of this linear system yields the orientifold
charge in terms of the ${\bf L}={\bf 0}$ branes.

We are now ready to explicitly write down the tadpole cancellation
conditions and find supersymmetric solutions at the Gepner point.

\subsection{Solutions of the Tadpole Conditions --- Quintic Case}

The problem simplifies somewhat for the case of the quintic because
there is only a single possible parity and because all branes
preserving the same spacetime supersymmetry as the O-plane are
invariant under the parity. As explained above, we will study
tadpole cancellation using the ${\bf L}={\bf 0}$ RS branes as a
``basis'' for the charge lattice. The charges of these branes
satisfy one linear relation
\beq
\bigl[B^B_{{\bf 0},0}\bigr]+
\bigl[B^B_{{\bf 0},2}\bigr]+
\bigl[B^B_{{\bf 0},4}\bigr]+
\bigl[B^B_{{\bf 0},6}\bigr]+
\bigl[B^B_{{\bf 0},8}\bigr] =0  \,.
\label{quinrel}
\eeq
In conjunction with the invariance of the tadpole canceling brane 
configuration under the parity this linear
relation implies that the equation (\ref{tadcan}) will reduce to
two linearly independent equations on the $n_i$.

\subsubsection{O-plane charge}

The charge of the O-plane is given in (\ref{qcross}) as
$\bigl[O_{P^B}\bigr] = 4\bigl[B_{{\bf 1},5}\bigr]$
and thus is expressed in terms of ${\bf L=0}$ branes as
\beq
\bigl[O_{P^B}\bigr]
=\left(
\bigl[B^B_{{\bf 0},0}\bigr],
\bigl[B^B_{{\bf 0},2}\bigr],
\bigl[B^B_{{\bf 0},4}\bigr],
\bigl[B^B_{{\bf 0},6}\bigr],
\bigl[B^B_{{\bf 0},8}\bigr]
\right) \cdot
\left(
\begin{array}{c}
8 \\
20 \\
40\\
40\\
20
\end{array}
\right)
\eeq

\subsubsection{Supersymmetry preserving branes}
 
As studied in Section~\ref{subsub:qex},
for any given parity, the set 
of rational branes contains $32$ branes preserving the same 
supersymmetry as the O-planes, and these $32$ branes have $6$ 
different charges. Representatives of these $6$ charges are
the branes
\beq
\bigl[B^B_{(00000),0} \Bigr],
\bigl[\overline{B^B_{(10000),5}} \bigr],
\bigl[B^B_{(11000),0} \bigr],
\bigl[\overline{B^B_{(11100),5}} \bigr],
\bigl[B^B_{(11110),5} \bigr] ,
\bigl[\overline{B^B_{(11111),5}} \bigr] .
\label{qlist}
\eeq
The other branes are obtained by permuting the ${\bf L}$ label,
leading to multiplicities $m_i=1,5,10,10,5,1$ for the $i$-th
charge, respectively.  To simplify the enumeration of solutions,
we will then consider tadpole canceling brane configurations 
containing $n_i$ branes with charge of each type and subsequently 
multiply by the combinatorial factor $\left(\!\!\begin{array}{c} 
n_i+m_i-1\\[-.2cm] n_i\end{array}\!\!\right)$ counting the number 
of ways of distributing the charge.

Using the formulas of the previous subsection, we obtain the following 
expression for these $6$ charges in terms of those of the
${\bf L}={\bf 0}$ branes.
\beqa
&&\left(
\bigl[B^B_{(00000),0} \Bigr],
\bigl[\overline{B^B_{(10000),5}} \bigr],
\bigl[B^B_{(11000),0} \bigr],
\bigl[\overline{B^B_{(11100),5}} \bigr],
\bigl[B^B_{(11110),5} \bigr] ,
\bigl[\overline{B^B_{(11111),5}} \bigr] 
\right) \qquad\qquad
\nn\\
&&\,\,\,\,\,\,\qquad=\left(
\bigl[B^B_{{\bf 0},0}\bigr],
\bigl[B^B_{{\bf 0},2}\bigr],
\bigl[B^B_{{\bf 0},4}\bigr],
\bigl[B^B_{{\bf 0},6}\bigr],
\bigl[B^B_{{\bf 0},8}\bigr] 
\right) 
Q
\eeqa
where $Q$ is the matrix
\beq
Q= \left(
\begin{array}{cccccc}
1&0&2&0&6&-2\\
0&0&1&-1&4&-5\\
0&-1&0&-3&1&-10\\
0&-1&0&-3&1&-10\\
0&0&1&-1&4&-5
\end{array}
\right)\,.
\eeq
We can take the linear relation (\ref{quinrel}) into account by
multiplying from the left with the matrix
\beq
T=\left(
\begin{array}{ccccc}
1&0&0&0&-1\\
0&1&0&0&-1\\
0&0&1&0&-1\\
0&0&0&1&-1
\end{array}
\right)
\label{inter}
\eeq

\subsubsection{Action of parity on D-branes}

As shown above, all branes are invariant under the parity and
support an orthogonal gauge group, ie, we have $\sigma=+1$ for
all branes in the list (\ref{qlist}).

\subsubsection{Solutions}

The positive integers $n_i$ must then satisfy the tadpole condition
(\ref{tadcan}) in the explicit form
\beq
T\; Q \;(n_1,n_2,n_3,n_4,n_5,n_5)^t = T\;(8,20,40,40,20)^t \,,
\eeq
which indeed reduces to two linearly independent equations,
\beq
\begin{array}{rcl}
n_1+ n_3+n_4+2n_5+3n_6&=&12 \\
n_2+n_3+2 n_4+ 3n_5 + 5n_6 &=&20 \,.
\end{array}
\label{condition}
\eeq
Obviously, there is only a finite number of solutions to the equations 
(\ref{condition}) with positive $n_i$ (negative $n_i$ means using 
the antibrane and this breaks supersymmetry). A simple computer 
aided count shows that there are in fact $417$ solutions. For
each such solution, the number  of ways of distributing the charge 
among the $32$ branes with the same supersymmetry is then given
by
\beq
\# (n_1,n_2,n_3,n_4,n_5,n_6) = 
\prod_{i=1}^6 \left(\!\!
\begin{array}{c} 
n_i+m_i-1 \\[-.2cm] n_i
\end{array}\!\!\right)
\label{distribute}
\eeq
where the multiplicities $m_i$ are given by $1$, $5$, $10$, $10$, 
$5$, and $1$. Again using a computer, one then finds that the 
grand total number of tadpole canceling brane configurations 
using only the rational branes at the Gepner point is equal to 
$31561671503$, as advertised.

\noindent
{\bf Remarks.}\\{\small
{\bf (i)}~All these solutions of the tadpole conditions we have
constructed above have a worldsheet description in terms of 
rational conformal field theories based on orbifolds of 
${\mathcal N}=2$ minimal models, and are ``in principle 
exactly solvable'' perturbative string vacua with 
${\mathcal N}=1$ spacetime supersymmetry in $3+1$ dimensions. 
This impressive number is much larger than a comparable number 
in heterotic string constructions (see, e.g., \cite{Gepnerclassification}).
Of course, all these vacua and their moduli spaces are potentially 
unstable to non-perturbative effects.
\\ 
{\bf (ii)}~We emphasize that in counting the number of solutions,
we have {\it not} divided out by the symmetry group $\mathfrak{S}^5$
which exchanges the various minimal model factors. It might
seem that this is overcounting, since the solutions mapped onto
each other under such a symmetry must lead to the same 
perturbative low energy physics. However, one also has to 
take into account that the Gepner point is a special point in
the (closed string) K\"ahler moduli space. There are perturbation
away from that point which break the exchange symmetries, also
after orientifold projection. Once such a perturbation has
been turned on (or if the corresponding moduli are fixed away
from the Gepner point by some mechanism), the various solutions
will no longer lead to the same physics at low energies, so
we count them as distinct ``vacua''. (But, of course, they 
are still exchanged if we act on all fields (closed and open 
strings) simultaneously.)
\\ 
{\bf (iii)}~The solutions we have constructed are valid right at the
Gepner point.
One can ask what happens to these solutions when one moves
away from the Gepner point. 
On general grounds, the branes we
have used to cancel the tadpoles might disappear at lines
of marginal stability. When this happens, then as discussed 
in section \ref{sec:FI}, we expect that either there is
a new supersymmetric brane configuration obtained by
condensing some open string tachyon, or there is none, in
which case the K\"ahler moduli is lifted at string loop level.
On top of this there could also be string non-perturbative effect
that may fix some of the K\"ahler moduli.
}

%JW: new subsubsection
\subsubsection{Distribution of gauge group rank}

The solutions to the tadpole conditions we have found are certainly
too numerous to make a complete list. But we can gain a qualitative 
overview over the possibilities by looking at the distribution of a 
certain property over the set of all models, for example the rank of
the unbroken gauge group or the number of massless chiral fields. 
Such a statistical approach to exploring the string theory vacua
has recently been advocated in particular in 
\cite{DougString,statistics}. Let us here present the result of 
this type of counting for a very simple quantity, the total rank 
of the gauge group.

\begin{figure}[ht]
\centerline{\scriptsize $\begin{array}[b]{|l|r|}
\hline
{\rm Rank} & {\rm Number\; of\; solutions}\\
\hline\hline
0&   41100850    \\
1&   410137435    \\ 
2&   1767975754   \\ 
3&   4320652050   \\ 
4&   6758910800   \\ 
5&   7251800650   \\ 
6&   5593308703   \\ 
7&   3227024877   \\ 
8&   1450260204   \\ 
9&   527957402    \\ 
10&  161242450    \\ 
11&  41130702     \\ 
12&  8534850      \\ 
13&  1460250      \\ 
14&  159225       \\ 
15&  14300        \\ 
16&  1001         \\
\hline
\end{array}$
\qquad\qquad
\psfrag{r}{rank}
\psfrag{n}{}
\psfrag{0}{$0$}
\psfrag{1}{$1$}
\psfrag{2}{$2$}
\psfrag{3}{$3$}
\psfrag{4}{$4$}
\psfrag{5}{$5$}
\psfrag{6}{$6$}
\psfrag{7}{$7$}
\psfrag{8}{$8$}
\psfrag{9}{$9$}
\psfrag{10}{$10$}
\psfrag{11}{$11$}
\psfrag{12}{$12$}
\psfrag{13}{$13$}
\psfrag{14}{$14$}
\psfrag{15}{$15$}
\psfrag{16}{$16$}
\psfrag{100}{$\!\!\!\!\!\! 1\times 10^9$}
\psfrag{200}{$\!\!\!\!\!\! 2\times 10^9$}
\psfrag{300}{$\!\!\!\!\!\! 3\times 10^9$}
\psfrag{400}{$\!\!\!\!\!\! 4\times 10^9$}
\psfrag{500}{$\!\!\!\!\!\! 5\times 10^9$}
\psfrag{600}{$\!\!\!\!\!\! 6\times 10^9$}
\psfrag{700}{$\!\!\!\!\!\! 7\times 10^9$}
\epsfig{width=4in,file=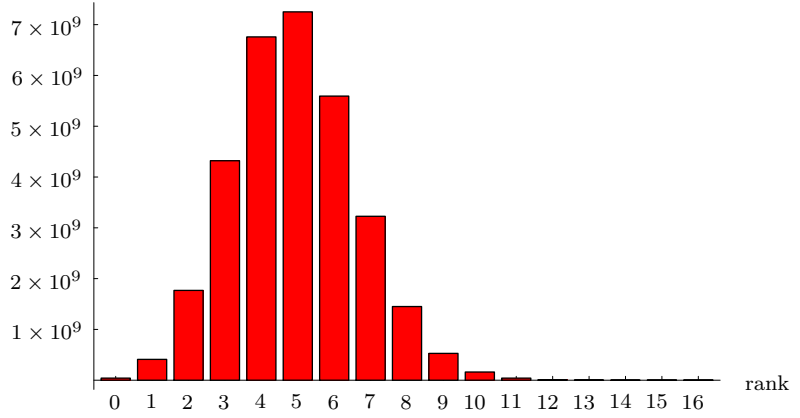}}
\caption{Distribution of total rank of gauge group over the 
solutions of tadpole conditions for Type IIB orientifold of 
quintic Gepner model.}
\label{fig:ranks}
\end{figure}

In all type IIB orientifolds of the quintic we have found, the gauge 
group is a product of orthogonal groups $G=\prod_{j} O(N_j)$, where 
$N_j$ is a positive integer. By  slight abuse of terminology, we will 
call the maximal number of $U(1)$ subgroups of $G$ the rank of $G$. In 
particular, the ranks of $O(1)$ and $O(2)$ are defined to be $0$ and 
$1$, respectively. Then, for each solution of the equations
(\ref{condition}), we have to distribute the $n_i$ (the number of 
times a given charge appears) among the various branes with that 
given charge. This is similar to what we did in (\ref{distribute})
to count the total number of solutions, but we have to take into
account that the ranks depends on how we distribute the $n_i$'s.
In any event, the rank of a given solution is then computed as
$$
\sum_j \biggl[\frac{N_j}{2}\biggr]  \,.
$$
The maximal possible rank is $16$ \cite{Blumenhagen}. It is 
obtained from the solution $n_1=12$, $n_2=20$ of (\ref{condition})
by choosing $12$ times the brane $B_{(0,0,0,0,0),0}$ and
$20$ branes $\overline{B_{{\bf L},5}}$ with one $L_i=1$, an 
even number of each. The number of possibilities is 
$\left({24\atop4}\right)=1001$. The results for the
other ranks between $0$ and $16$ are shown in Figure \ref{fig:ranks}.

It is interesting that the peak of the distribution 
lies around rank $5$, which is rather close to the
value of the Standard Model. Let us also note that the 
distribution goes to zero very rapidly for large ranks, 
and --- somewhat surprisingly --- has a rather large 
support at small ranks. In particular, rank $0$, which 
corresponds in our language to having only distinct 
parity-invariant branes supporting $O(1)$ gauge groups, 
appears with appreciable frequency.

\subsection{Solutions of the tadpole conditions --- Two parameter model}

Similarly to the quintic, we use the $8$ ${\bf L}={\bf 0}$ branes to express 
the charges of the other branes and the O-planes. The charges of these $8$
branes satisfy the two relations
\beqa
\bigl[B^B_{(00000),0} \bigr] + 
\bigl[B^B_{(00000),4} \bigr] + 
\bigl[B^B_{(00000),8} \bigr] + 
\bigl[B^B_{(00000),12} \bigr] &=&0\nn\\
\bigl[B^B_{(00000),2} \bigr] + 
\bigl[B^B_{(00000),6} \bigr] + 
\bigl[B^B_{(00000),10} \bigr] + 
\bigl[B^B_{(00000),14} \bigr] &=&0
\label{tprels}
\eeqa
For the practical computations, we 
will use the relations (\ref{tprels}) to project to a linearly 
independent set of $6$ charges with the matrix
\beq
T=
\left(
\begin{array}{cccccccc}
1&0&0&0&0&0&-1&0\\
0&1&0&0&0&0&0&-1\\
0&0&1&0&0&0&-1&0\\
0&0&0&1&0&0&0&-1\\
0&0&0&0&1&0&-1&0\\
0&0&0&0&0&1&0&-1\\
\end{array}
\right)
\label{Ttwo}
\eeq

\subsubsection{Parities and O-planes}

In the two parameter model, we have various parities $P^B_{(\omega;
\epsilon_1,\ldots,\epsilon_5)}$ to consider. Here $\omega=0,1$ 
and $(\epsilon_1,\ldots,\epsilon_5)$, $\epsilon_i=\pm$, denote the 
twisting by quantum and classical symmetries, respectively. This is 
a slightly redundant labeling because as we recall  $(\epsilon_1,
\epsilon_2,\epsilon_3,\epsilon_4,\epsilon_5)$ is the same parity 
as $(-\epsilon_1,-\epsilon_2,\epsilon_3,\epsilon_4,\epsilon_5)$.
The charges of the corresponding O-planes are given in table
\ref{Otwopar1} and table \ref{Otwopar2} in subsection 
\ref{sec:oplanec}. 

\subsubsection{Supersymmetry preserving branes}

\subsubsection*{Even quantum symmetry}

According to the discussion in subsection \ref{sec:oplanec}, O-planes
corresponding to parities with even quantum symmetry dressing
preserve a spacetime supersymmetry with even $M$-label, $M_O=0,8,4,12$,
depending on the $\epsilon_i$'s. Branes preserving the same
supersymmetry must have $M=M_O$ or $M=M_O+8$, and this restricts
the possible ${\bf L}$ labels of the branes to $L_1+L-2={\rm even}$.
In order to get familiar with the use of the formula (\ref{branec}), 
we give here a list of branes preserving the same supersymmetry as 
any given O-plane as well as their charges. We only list ${\bf L}$ 
labels up to permutation of factors with equal levels. In the last
column, we give the expansion of the brane charge (for $M_O=0$) in 
terms of the ${\bf L}={\bf 0}$ branes, modulo the relations
(\ref{tprels}). As one can see, there are $15$ different charges.
The eight-component vectors give, for example, the following
equation:
\beq
[\widehat{B}^B_{{\bf\frac{k}{2}},M}]~=~
\frac14[B^B_{{\bf\frac{k}{2}},M}]~=~
  [B^B_{{\bf 0},M+4}]+2[B^B_{{\bf 0},M+2}]+2[B^B_{{\bf 0},M  }]
+2[B^B_{{\bf 0},M-2}] +[B^B_{{\bf 0},M-4}].
\label{chBk/2}
\eeq
It is also a useful exercise to check the ``multiplicity'' or number 
of inequivalent branes with the same charge.

{\scriptsize
\begin{center}
\begin{tabular}{|l||lll|l|l|}
\hline
charge \# & ${\bf L}$ & $M$    &  $S$ & $g^{-M_O/2} Q$ & multiplicity $m_i$ \\
\hline
$1$     & $(00000)$ &  $M_O$   &  $0$ & $(1, 0, 0, 0, 0, 0, 0, 0)$        & $1$\\  
$2$     & $(00000)$ &  $M_O+8$ &  $2$ & $(1, 0, 1, 0, 0, 0, 1, 0)$       & $1$\\
$3$     & $(11000)$ &  $M_O$   &  $0$ & $(2, 1, 0, 0, 0, 0, 0, 1)$        & $1$\\
$4$     & $(11000)$ &  $M_O+8$ &  $2$ & $(2, 1, 2, 0, 0, 0, 2, 1)$       & $1$\\  
$5$     & $(20000)$ &  $M_O$   &  $0$ & $(1, 1, 0, 0, 0, 0, 0, 1)$        & $2$\\
$6$     & $(20000)$ &  $M_O+8$ &  $2$ & $(1, 1, 1, 0, 0, 0, 1, 1)$       & $2$\\
$7$     & $(22000)$ &  $M_O$   &  $0$ & $(3, 2, 1, 0, 0, 0, 1, 2)$        & $1$\\
$8$     & $(22000)$ &  $M_O+8$ &  $2$ & $(3, 2, 2, 0, 0, 0, 2, 2)$       & $1$\\
$9$     & $(00100)$, $(00111)$ 
                    &  $M_O$   &      & $(0, 1, 0, 0, 0, 0, 0, 1)$        & $4$\\
$10$    & $(00110)$ &  $M_O$   &      & $(1, 0, \frac12, 0, 0, 0, \frac12, 0)$ & $6$\\
$11$    & \parbox[t]{2.5cm}{$(11100)$,  $(11111)$, \\[-.1cm] 
$(31000)$, $(31110)$,\\[-.1cm]
$(33100)$, $(33111)$} 
                    &  $M_O$   &      & $(2, 2, 1, 0, 0, 0, 1, 2)$        & $16$\\
$12$    & \parbox[t]{2.5cm}{$(11110)$, $(20100)$,\\[-.1cm]
 $(20111)$, $(31100)$,\\[-.1cm]
$(31111)$, $(33000)$, \\[-.1cm]
$(33110)$} 
                    &  $M_O$   &      & $(2, 1, 1, 0, 0, 0, 1, 1)$        & $38$\\        
$13$    & $(20110)$ &  $M_O$   &      & $(1, 1, \frac12, 0, 0, 0, \frac12, 1)$ & $12$\\
$14$    & $(22100)$, $(22111)$ 
                    &  $M_O$   &      & $(4, 3, 2, 0, 0, 0, 2, 3)$        & $4$ \\
$15$    & $(22110)$ &  $M_O$   &      & $(3, 2, \frac32, 0, 0, 0, \frac32, 2)$ &$6$\\
\hline
\end{tabular}
\end{center}
}

\subsubsection*{Odd quantum symmetry}

O-planes corresponding to dressing with odd quantum symmetry
preserve a spacetime supersymmetry with odd $M$-label $M=1,5,9,13$,
depending on the $\epsilon_i$'s. We are then restricted to 
branes with $L_1+L_2$ odd. The list of branes and charges
is

{\scriptsize
\begin{center}
\begin{tabular}{|l||lll|l|l|}
\hline
charge \# & ${\bf L}$ & $M$    &  $S$ & $g^{-(M_O-1)/2} Q$ & multiplicity $m_i$ \\
\hline
$1$     & $(10000)$            &  $M_O$   &  $0$ & $(1, 1, 0, 0, 0, 0, 0, 0)$    &$2$\\
$2$     & $(10000)$            &  $M_O+8$ &  $2$ & $(1, 1, 1, 1, 0, 0, 1, 1)$   &$2$\\
$3$     & $(21000)$            &  $M_O$   &  $0$ & $(2, 2, 1, 0, 0, 0, 0, 1)$    &$2$\\
$4$     & $(21000)$            & $M_O+8$  &  $2$ & $(2, 2, 2, 1, 0, 0, 1, 2)$   &$2$\\
$5$     & \parbox[t]{2.5cm}{$(10100)$, $(10111)$\\[-.1cm]
         $(30000)$, $(30110)$} & $M_O$    &      & $(1, 1, 1, 0, 0, 0, 0, 1)$    &$16$\\
$6$     & \parbox[t]{2.5cm}{$(10110)$, $(30100)$\\[-.1cm]
        $(30111)$}             & $M_O$    &      & $(1, 1, \frac12, \frac12, 0, 0, \frac12, \frac12)$&$28$\\
$7$     & \parbox[t]{2.5cm}{$(21100)$, $(21111)$\\[-.1cm]  
         $(32000)$, $(32110)$} & $M_O$    &      & $(3, 3, 2, 1, 0, 0, 1, 2)$   &$16$\\
$8$     & \parbox[t]{2.5cm}{$(21110)$, $(32100)$\\[-.1cm]
         $(32111)$}            & $M_O$    &      & $(2, 2, \frac32,
 \frac12, 0, 0, \frac12, \frac32)$ &$28$\\
\hline
\end{tabular}
\end{center}
}

\subsubsection{Action of parities on D-branes}

We summarize the action of the B-type parities $P^B_{\omega;
\epsilon_1,\ldots,\epsilon_5}$ on the branes $B^B_{{\bf L},M,S,\psi}$
for the two parameter model $(6,6,2,2,2)$.

Neglecting again for a very short moment the $S$ and $\psi$ labels,
the branes are mapped under parity as follows:
\beqa
&&P^B_{0;\epsilon_1...\epsilon_5}:
B_{{\bf L},M}\longmapsto \epsilon_1\cdots\epsilon_5 
B_{{\bf L},-M},\nn\\
&&P^B_{1;\epsilon_1...\epsilon_5}:
B_{{\bf L},M}\longmapsto \epsilon_1\cdots\epsilon_5 
B_{{\bf L},2-M}.\nn
\eeqa
For branes carrying a $\psi$ label, this $\psi$ label is
transformed according to
\beq
\psi\longmapsto (-1)^{|\II|/2}\;\psi.
\eeq
If the sign $\epsilon_1\cdots\epsilon_5$ is negative, this means
that the branes are mapped to antibranes. This minus signs can be
absorbed according to the rule
\beq
-B_{{\bf L},M,S}= B_{{\bf L},M,S+2},~~
-B_{{\bf L},M}  = B_{{\bf L},M+H},~~
-\widehat{B}_{{\bf L},M}^\psi  = \widehat{B}_{{\bf L},M+H}^{-\psi}.
\eeq
Restricting ourselves to the supersymmetry-preserving branes,
we find that the branes with $|{\bf S}|=$ odd are all parity invariant.
As for the branes with $|{\bf S}|=$ even, those with $|{\bf S}|=0$ or 
$4$ are invariant under parities with $\epsilon_1\cdots\epsilon_5=1$
and not under parities with $\epsilon_1\cdots\epsilon_5=-1$.
The branes with $|{\bf S}|=2$ behave in an opposite way.

The gauge group on parity invariant branes is either $O$ or
$Sp$ according to the sign (\ref{Ggroup}) or (\ref{Ggroup2}).
Other branes support the unitary gauge groups.
A complete list of supersymmetry-preserving branes together
with the gauge groups is given in tables \ref{cp-2pb1},
\ref{cp-2pb2} in subsection \ref{spectrum}.

We have presented the list of D-branes preserving the same
supersymmetry as any given orientifold, and they formed 15 or 8
groups according to the RR-charge.
The number of groups depends on the dressings with quantum symmetries.
Each group contains branes having different signature $\sigma$ and
therefore supporting different gauge groups,
and it is necessary for counting the supersymmetric vacua
to know the numbers $m_i^{\sigma}$ of branes with definite
signature in a given group.
They are defined to satisfy
\beq
  m_i = m_i^{+}+2m_i^{0}+m_i^{-},
\eeq
where $m_i$ is the total multiplicity of branes in the $i$-th group.
For later convenience we give a table of these numbers below.

\begin{table}[tbh]
{\scriptsize\begin{center}
\begin{tabular}{|c||c|c|c|c|c|c|c|c|}
\hline
$\#$
 &$\!\!P_{0;+++++}\!\!$&$\!\!P_{0;+-+++}\!\!$
 &$\!\!P_{0;++-++}\!\!$&$\!\!P_{0;+--++}\!\!$
 &$\!\!P_{0;++--+}\!\!$&$\!\!P_{0;+---+}\!\!$
 &$\!\!P_{0;++---}\!\!$&$\!\!P_{0;+----}\!\!$\\
\hline
1,2,7,8&(1,0,0)&(0,1,0)&(0,1,0)&(0,0,1)&(0,0,1)&(0,1,0)&(0,1,0)&(1,0,0)\\
 3,4 &(1,0,0)&(0,1,0)&(0,1,0)&(1,0,0)&(0,0,1)&(0,1,0)&(0,1,0)&(0,0,1)\\
 5,6 &(2,0,0)&(0,2,0)&(0,2,0)&(0,0,2)&(0,0,2)&(0,2,0)&(0,2,0)&(2,0,0)\\
 9,14&(3,0,1)&(0,0,4)&(2,0,2)&(1,0,3)&(3,0,1)&(2,0,2)&(0,0,4)&(1,0,3)\\
10,15&(0,6,0)&(0,0,6)&(4,0,2)&(0,6,0)&(0,6,0)&(2,0,4)&(6,0,0)&(0,6,0)\\
11   &(5,0,11)&(15,0,1)&(7,0,9)&(11,0,5)&(7,0,9)&(9,0,7)&(9,0,7)&(9,0,7)\\
12   &(6,20,12)&(20,10,8)&(12,10,16)&(10,20,8)
     &(12,20,6)&(16,10,12)&(8,10,20)&(8,20,10)\\
13   &(0,12,0)&(0,0,12)&(8,0,4)&(0,12,0)&(0,12,0)&(4,0,8)&(12,0,0)&(0,12,0)\\
\hline
\end{tabular}
\vskip5mm
\begin{tabular}{|c||c|c|c|c|c|c|c|c|}
\hline $\#$
 &$\!\!P_{1;+++++}\!\!$&$\!\!P_{1;+-+++}\!\!$
 &$\!\!P_{1;++-++}\!\!$&$\!\!P_{1;+--++}\!\!$
 &$\!\!P_{1;++--+}\!\!$&$\!\!P_{1;+---+}\!\!$
 &$\!\!P_{1;++---}\!\!$&$\!\!P_{1;+----}\!\!$\\
\hline
1,2,3,4&(2,0,0)&(0,2,0)&(0,2,0)&(1,0,1)&(0,0,2)&(0,2,0)&(0,2,0)&(1,0,1)\\
5,7    &(8,0,8)&(8,0,8)&(8,0,8)&(8,0,8)&(8,0,8)&(8,0,8)&(8,0,8)&(8,0,8)\\
6,8    &(0,24,4)&(12,4,12)&(12,4,12)&(2,24,2)
       &(4,24,0)&(12,4,12)&(12,4,12)&(2,24,2)\\
\hline
\end{tabular}
\end{center}}
\caption{Multiplicities $(m_i^+,2m_i^0,m_i^-)$ of branes
         with different gauge groups.}
\end{table}

\subsubsection{Counting the Solutions}

Let us now count the number of supersymmetric vacua in various
orientifolds of two parameter model.
As compared to the case with quintic, there arise a new
complication due to the presence of $Sp$ or $U$-type branes.
We explain the detail of the counting in the cases $P^B_{0,+++++}$
and $P^B_{0,+-+++}$, and state the results for other cases briefly.

\subsubsection*{Parity $P^B_{0;+++++}$}

Using (\ref{chBk/2}) and the expression in the table {\ref{Otwopar1}},
this O-plane is shown to have the RR-charge
\beq
\bigl[O_P\bigr] ~=~ ( 2,  6, 16, 26, 30, 26, 16,  6)^t
          ~\simeq~ -(28, 20, 14,  0,  0,  0, 14, 20)^t.
\eeq
Under the choice $\tilomega^{\frac12}=-1$, it preserves a
spacetime supersymmetry corresponding to $M_O=0$.
We denote by $n_i$, $i=1,2,\ldots,15$ the number of times
a given charge appears in a tadpole canceling configuration.
Computing the charges, projecting onto $6$ independent ones
and equating brane and O-plane charges leads to the following
three linearly independent equations on the $n_i$'s.
\beq
\left(
\begin{array}{ccccccccccccccc}
1& 1& 2& 2& 1& 1& 3& 3& 0& 1& 2& 2& 1& 4& 3\\
0& 0& 1& 1& 1& 1& 2& 2& 1& 0& 2& 1& 1& 3& 2\\
0& 1& 0& 2& 0& 1& 1& 2& 0& \frac12& 1& 1& \frac12& 2& \frac32
\end{array}
\right)
\left(
\begin{array}{c}
n_1\\
n_2\\
\vdots
\\
n_{15}
\end{array}
\right)
=\left(
\begin{array}{c}
28\\
20\\
14
\end{array}
\right)
\label{btcc2p1}
\eeq
Note that again, there is a finite number of solutions
to these equations, and we can count the total number of
tadpole canceling D-brane configurations.

The counting goes in the following way.
Suppose that the numbers $\{n_i\}$~($i=1,\cdots,15$ or $8$) give
a solution to (\ref{btcc2p1}).
For each solution, we have to distribute the
charge over the various branes with fixed charge,
taking into account their signature $\sigma$.
The first 8 charges are carried only by $O$-type branes,
but the other 7 are carried by branes with various signatures.
We note that even though the charges $i>8$ are invariant
under the parity, the branes themselves need not be
because of the action on the $\psi$-label.
Consider all the possible decompositions of each $n_i$~($i=9,\cdots,15$)
into
\beq
  n_i = n_i^++n_i^0+n_i^-
\eeq
with the condition that $n_i^0$ and $n_i^-$ be even.
The combinatorial factor associated to a solution $\{n_i\}$
is given by the sum over all the possible decompositions
\beq
\#(n_i)=
  (n_5+1)(n_6+1)
  \prod_{i=9}^{15}\sum_{n_i=\Sigma_\sigma n_i^\sigma}
\left(\!\!
\begin{array}{c} 
\frac{n_i^-}2 + m^-_i-1 \\[0cm] \frac{n_i^-}2
\end{array}\!\!\right)
\left(\!\!
\begin{array}{c} 
\frac{n_i^0}2 + m_i^0-1 \\ \frac{n_i^0}2
\end{array}\!\!\right)
\left(\!\!
\begin{array}{c} 
n_i^++m_i^+-1 \\ n_i^+
\end{array}\!\!\right).
\label{comb1}
\eeq
The number of vacua is therefore the sum of this over
all the solutions $\{n_i\}$.
The total number turns out to be $13213511375147\approx 10^{13}$.

\subsubsection*{Parity $P^B_{0;+-+++}$}

Under the choice $\tilomega^{\frac12}=-i$, this O-plane preserves
a spacetime supersymmetry corresponding to $M_O=4$. Its RR-charge is
\beq
\bigl[O_P\bigr]~=~  (0,-4,-6,-4, 0, 4, 6, 4)^t
          ~\simeq~ -(6, 8,12, 8, 6, 0, 0, 0)^t.
\eeq
One can easily see that the first $8$ branes on the list are
mapped onto each other under this parity, and give rise to
a unitary gauge group.
We have to require $n_i=n_{i+1}$ for $i=1,3,5,7$.
Equating crosscap and brane charge then leads to the
two independent conditions
\beq
\left(
\begin{array}{ccccccccccc}
2& 4& 2& 6&  0& 1& 2& 2& 1& 4& 3 \\
0& 2& 2& 4&  1& 0& 2& 1& 1& 3& 2
\end{array}
\right)
\left(
\begin{array}{c}
n_1\\
n_3\\
\vdots
\\
n_{15}
\end{array}
\right)
=\left(
\begin{array}{c}
12\\
8
\end{array}
\right).
\label{btcc2p2}
\eeq
The number of vacua is given by the sum of the combinatoric factors
\beq
\# (n_i) = \bigl(n_5+1\bigr)\;
\prod_{i=9}^{15} 
\sum
\left(\!\!
\begin{array}{c} 
\frac{n_i^-}2 + m^-_i-1 \\[0cm] \frac{n_i^-}2
\end{array}\!\!\right)
\left(\!\!
\begin{array}{c} 
\frac{n_i^0}2 + m_i^0-1 \\ \frac{n_i^0}2
\end{array}\!\!\right)
\left(\!\!
\begin{array}{c} 
n_i^++m_i^+-1 \\ n_i^+
\end{array}\!\!\right),
\label{comb2}
\eeq
over all the solutions of (\ref{btcc2p2}).
The total number is $47803952$.

\subsubsection*{Other parities (with even quantum symmetry dressings)}
One can analyze the cases with other parities in a similar way.
Let us denote by $|\epsilon|$ the number of minus signs in $\epsilon_i$.
Choosing $\tilomega^{\frac12}=-i^{|\epsilon|}$ for the parity
$P^B_{0;\epsilon_1\epsilon_2\epsilon_3\epsilon_4\epsilon_5}$,
the O-plane preserves a spacetime supersymmetry corresponding
to $M_O=4|\epsilon|$.
The O-planes have the RR-charges
\beqa
\bigl[O_{P^B_{0;+++++}}\bigr]
 &\simeq& -(28,20,14, 0, 0, 0,14,20)^t, \nn\\
\bigl[O_{P^B_{0;+-+++}}\bigr] ~=~
\bigl[O_{P^B_{0;++-++}}\bigr]
 &\simeq& -(6, 8,12, 8, 6, 0, 0, 0)^t, \nn\\
\bigl[O_{P^B_{0;+--++}}\bigr] ~=~
\bigl[O_{P^B_{0;++--+}}\bigr]
 &\simeq& -( 0, 0, 2, 4, 4, 4, 2, 0)^t, \nn\\
\bigl[O_{P^B_{0;+---+}}\bigr] ~=~
\bigl[O_{P^B_{0;++---}}\bigr]
 &\simeq& -( 2, 0, 0, 0, 2, 0, 4, 0)^t, \nn\\
\bigl[O_{P^B_{0;+----}}\bigr]
 &\simeq& -(-4, 4,-2, 0, 0, 0,-2, 4)^t.
\label{ochvec1}
\eeqa
So the tadpole cancellation conditions for other O-planes are
given by the replacements $(28,20,14)\to (x,y,z)$ in (\ref{btcc2p1}),
where $x,y$ and $z$ are the $(2|\epsilon|+1)$-st, $(2|\epsilon|+2)$-nd
and $(2|\epsilon|+3)$-rd components of the above vectors.
When $|\epsilon|$ is odd, the branes in the groups $i=1,\cdots,8$
are all U-type and the parity maps the charges $1\leftrightarrow2$,
$3\leftrightarrow4$ and so on.
So we have to put $n_i=n_{i+1}$ for $i={1,3,5,7}$ in these cases,
and there remain only two linearly independent equations
for $n_1,n_3,\cdots,n_{15}$.

The total number of vacua is given by the sum of combinatoric
factors $\#(n_i)$, defined in a similar way as (\ref{comb1})
or (\ref{comb2}), over all the solutions $\{n_i\}$ of tadpole
cancellation condition.
The result is summarized below.
\begin{center}
{\footnotesize
\begin{tabular}{|c|c|c|c|c|c|c|c|}
\hline
$\!\!B_{0;+++++}\!\!$&
$\!\!B_{0;+-+++}\!\!$&
$\!\!B_{0;++-++}\!\!$&
$\!\!B_{0;+--++}\!\!$&
$\!\!B_{0;++--+}\!\!$&
$\!\!B_{0;+---+}\!\!$&
$\!\!B_{0;++---}\!\!$&
$\!\!B_{0;+----}\!\!$ \\
\hline
13213511375147 &
47803952 &
434841441 &
1051 &
2162 &
35 &
148 &
0
\\ \hline
\end{tabular}}
\end{center}

\subsubsection*{Other parities (with odd quantum symmetry dressings)}

For parities $P^B_{1;\epsilon_1\epsilon_2\epsilon_3\epsilon_4\epsilon_5}$
with odd quantum symmetry dressings,
we choose $\tilomega^{\frac12}=-i^{|\epsilon|}$ and consider the
systems of an O-plane and D-branes with $M_O=1+4|\epsilon|$.
In solving the tadpole cancellation condition, note first that
for the parities $P^B_{+-***}$ the O-planes have no RR-charge so that
the O-planes by themselves give the unique consistent superstring
backgrounds.
For other parities, the O-planes have the RR-charges
\beqa
\bigl[O_{P^B_{1;+++++}}\bigr]
 &\simeq& -(28,28,20, 8, 0, 0, 8,20)^t, \nn\\
\bigl[O_{P^B_{1;++-++}}\bigr]
 &\simeq& -( 4, 8,12,12, 8, 4, 0, 0)^t, \nn\\
\bigl[O_{P^B_{1;++--+}}\bigr]
 &\simeq& -( 0, 0, 0, 4, 4, 4, 4, 0)^t, \nn\\
\bigl[O_{P^B_{1;++---}}\bigr]
 &\simeq& -( 0, 4, 0, 0, 4, 0, 4, 4)^t.
\label{ochvec2}
\eeqa

Tadpole cancellation condition is then given by three linearly
independent equations on 8 numbers.
For the parity $P^B_{1;+++++}$ it becomes
\beq
\left(\begin{array}{cccccccc}
1& 1& 2& 2& 1& 1     & 3& 2 \\
0& 1& 1& 2& 1& \frac12& 2& \frac32 \\
0& 1& 0& 1& 0& \frac12& 1& \frac12
\end{array}\right)
\left(\begin{array}{c} n_1\\n_2\\\vdots\\n_8\end{array}\right)
~=~ 
\left(\begin{array}{c} 28\\20\\8\end{array}\right),
\eeq
and for other parities we only have to replace the numbers $(28,20,8)$
on the right hand side with
$(2|\epsilon|+2,2|\epsilon|+3,2|\epsilon|+4)$-th components of the
vectors (\ref{ochvec2}).
When $|\epsilon|$ is odd, the branes in the group 1 and 2, 3 and 4
are mapped to each other.
So we have to put $n_1=n_2, n_3=n_4$ for such cases,
and the number of independent equations on $n_{1,3,5,6,7,8}$
turns out to be reduced by one.

The total number of vacua is then calculated as a sum of a certain
combinatoric factor over all the solutions $n_{i}$ of the above
equations.
The results are summarized below.
\begin{center}
{\footnotesize
\begin{tabular}{|c|c|c|c|c|}
\hline
$\!\!B_{1;+++++}\!\!$&
$\!\!B_{1;++-++}\!\!$&
$\!\!B_{1;++--+}\!\!$&
$\!\!B_{1;++---}\!\!$&
$\!\!B_{1;+-***}\!\!$ \\
\hline
28956442028638 & 1093287843 & 654 & 0 & 1
\\ \hline
\end{tabular}}
\end{center}

\subsubsection{Distribution of gauge group rank}

Let us see the distribution of the rank of gauge group over
the supersymmetric vacua.
For the brane configuration with gauge group
$G=\prod O(N_{j^+})\times\prod U(N_{j^0})\times\prod Sp(N_{j^-})$,
the rank is counted as
$$
 \sum_{j^+}\left[\frac{N_{j^+}}{2}\right]
+\sum_{j^0}N_{j^0}
+\sum_{j^-}N_{j^-}.
$$
The results are summarized in the table below.

As the table shows, in the type I cases the rank of the gauge group
is peaked around $9$, differently from the quintic case.
Also, the number of vacua with low ranks are more strongly suppressed
because of the presence of $U$ or $Sp$ type branes.
For other orientifolds the maximum allowed rank is reduced, and
the distributions are peaked at lower values of rank.

{\scriptsize
\begin{center}
\begin{tabular}{|c||r|r|r|r|r|r|r|}
\hline
\lw{rank}
 &$P_{0;+++++}$
 &$P_{0;+-+++}$
 &$P_{0;++-++}$
 &$P_{0;+--++}$
 &$P_{0;++--+}$
 &$P_{0;+---+}$
 &$P_{0;++---}$ \\
&13213511375147
&47803952
&434841441
&1051
&2162
&35
&148
\\ \hline\hline
 0
&646540
&508725
&4926687
&166
&480
&0
&15
\\
 1
&44771470
&2554170
&29783246
&330
&721
&7
&105
\\
 2
&1031791551
&7173709
&76613078
&397
&719
&28
&28
\\
 3
&11643923756
&11188898
&113881856
&88
&172
&&\\
 4
&75080785790
&11195422
&102828964
&70
&70
&&\\
 5
&302754231919
&8532104
&66661000
&&&&\\
 6
&816375589073
&4126724
&28541380
&&&&\\
 7
&1555478380691
&1860600
&9347940
&&&&\\
 8
&2202010164391
&501900
&1980090
&&&&\\
 9
&2424675084374
&129360
&244860
&&&&\\
10
&2159636846181
&32340
&32340
&&&&\\
11
&1607633137394
&&&&&&\\
12
&1023393658328
&&&&&&\\
13
&567624907070
&&&&&&\\
14
&277143210040
&&&&&&\\
15
&120183191993
&&&&&&\\
16
&46373508969
&&&&&&\\
17
&15919273033
&&&&&&\\
18
&4851273490
&&&&&&\\
19
&1288731061
&&&&&&\\
20
&300818948
&&&&&&\\
21
&56875115
&&&&&&\\
22
&9505650
&&&&&&\\
23
&961928
&&&&&&\\
24
&106392
&&&&&&\\
\hline
\end{tabular}
\end{center}
}

{\scriptsize
\begin{center}
\begin{tabular}{|c||r|r|r|}
\hline
\lw{rank}
 &$P_{1;+++++}$
 &$P_{1;++-++}$
 &$P_{1;++--+}$ \\
&28956442028638
&1093287843
&654
\\ \hline\hline
 0
&131418
&10073409
&70
\\
 1
&12060448
&98476432
&448
\\
 2
&355100152
&281607952
&136
\\
 3
&5191991568
&398177360
&\\
 4
&44410530386
&248315690
&\\
 5
&245511738472
&52357760
&\\
 6
&928376315288
&4279240
&\\
 7
&2485035106608
&&\\
 8
&4801648669394
&&\\
 9
&6693313716784
&&\\
10
&6689330341632
&&\\
11
&4555609978656
&&\\
12
&2009612464368
&&\\
13
&457636777344
&&\\
14
&40397106120
&&\\
\hline
\end{tabular}
\end{center}
}

\subsection{Particle Spectrum in Some Supersymmetric Models}
\label{spectrum}

A closer look at the annulus and the M\"{o}bius strip
amplitudes gives us a more detailed information on the spectrum
of open strings.
Let us now turn to count the number of matter fields between
the same branes.
There are many branes and each has quite a few scalar fields, so
the best way to count them is again to use computers.
As in the analysis of A-branes, the massless chiral fields in the
spacetime theory are in one to one correspondence with the
chiral primary states in the internal theory.
Let us summarize the necessary materials.
\begin{itemize}
\item
The open string NS state $\otimes_i(l_i,n_i,s_i)$ between the branes
$B^B_{{\bf L},M}$ and $B^B_{{\bf L}',M'}$ obey
$\sum_in_iw_i = M'-M~({\rm mod}~ 2H)$, as well
as the usual selection rule $l_i+n_i\in 2\Z$ and the $SU(2)$ fusion
rule constraint.
The states between short-orbit branes with the same $\II$
and additional labels $\psi, \psi'=\pm1$ are subject to a projection
$\prod_{i\in \bfS}(-1)^{\frac12(l_i+n_i)}=\psi\psi'$.
\item
The open string states on parity-invariant D-branes have definite
eigenvalues of parity. The M\"{o}bius strip amplitude between
$B_{{\bf L},M}$ and its image under $P^B_{M_\omega;\epsilon_i}$
contains NS states $\otimes_i(l_i,2\nu_i,s_i)$ satisfying
$\sum_i\nu_iw_i=M_\omega-M$ (mod $H$), as well as
the $SU(2)$ fusion constraint on $l_i$.
The contribution of massless states to M\"{o}bius strip amplitude
is given by the sum of chiral primary states satisfying these condition,
with the phase
$$
 i^{-1+|\epsilon|+\{\#{\rm of}(s_i=2)\}}\prod_i\epsilon_i^{L_i+\nu_i}.
$$
\end{itemize}

We present here the relevant amplitudes from which the result follows.
The annulus amplitudes between two long-orbit B-branes
$B^B_{{\bf L},M}$ and $B^B_{{\bf L}',M'}$ has the following 
NS part:
\beqa
\lefteqn{\langle B^B_{{\bf L},M}|q^H|B^B_{{\bf L}',M'}\rangle|_{\NS}}
 \nn\\ &=&
\frac12\sum_{n_i}
\delta^{(2H)}_{\sum_iw_in_i+M-M'}
\sum_{l_i}\prod_{i=1}^rN_{L_iL'_i}^{~l_i}
\times \left\{
 \chi^{(\rm st)NS+}\prod_{i=1}^r\chi^{\rm NS+}_{l_i,n_i}
-\chi^{(\rm st)NS-}\prod_{i=1}^r\chi^{\rm NS-}_{l_i,n_i}\right\}
\eeqa
Here $\chi^{(\rm st)NS\pm}$ and
$\chi^{\rm NS\pm}_{l,n}=\chi_{l,n,0}\pm\chi_{l,n,2}$ are the same
as those used in the discussion of A-branes.
The delta symbol represents that the sum over $n_i$ is
taken over the $\tilGamma$-orbit but is shifted by $M$ and $M'$.
For two short-orbit branes $\widehat{B}^B_{{\bf L},M}$ and
$\widehat{B}^B_{{\bf L}',M'}$ the above amplitude has to be divided
by $2^{[|\bfS|/2]}2^{[|\bfS'|/2]}$, where $\bfS$ and $\bfS'$ are
the sets of $i$'s such that $L_i$ or $L'_i$ coincide with $\frac{k_i}{2}$.
When $\bfS=\bfS'$ and $|\bfS|=2$ or $4$, the twisted parts of boundary
states yield
\beqa
&&
\pm\frac{1}{2^{1+|\bfS|}}\sum_{n_i}
\delta^{(2H)}_{\sum_iw_in_i+M-M'}
\sum_{l_i}\prod_{i=1}^rN_{L_iL'_i}^{~l_i}\times
\prod_{i\in\bfS}(-1)^{\frac12(l_i+n_i)}
\nn\\ && \hskip20mm
\times \left\{
 \chi^{(\rm st)NS+}\prod_{i=1}^r\chi^{\rm NS+}_{l_i,n_i}
-\chi^{(\rm st)NS-}\prod_{i=1}^r\chi^{\rm NS-}_{l_i,n_i}\right\},
\eeqa
so the states with $\prod_{i\in\bfS}(-1)^{\frac12(l_i+n_i)}=1\,(-1)$
propagates between the branes with the same (opposite) signs.
All these maintain the integrality of the open string spectrum, and
ensure that every single B-brane supports a $U(1)$ gauge symmetry
in the absence of orientifolds.

For parity-invariant branes, we have to find the
action of parity on the matter fields on their worldvolume.
The NS part of M\"{o}bius strip amplitude between a long-orbit B-brane
$B^B_{{\bf L},M}$ and its image under the parity $P^B_{\omega,{\bf m}}$
reads (recall $\omega = \e^{\frac{2\pi i M_\omega}{H}}$ and
       $\tilomega_i= \e^{-\frac{2\pi i m_i}{k_i+2}}=\pm1$)
\beqa
\lefteqn{\langle B^B_{{\bf L},M}|q^H|C^B_{\omega,{\bf m}}\rangle|_{\NS}}
 \nn\\ &=&
  {\rm Re}\left\{
   i\e^{\frac{\pi i(r-d)}{4}}\tilomega^{{\bf L}-\frac12}
   \sum_{\tilnu_i\in\Z_{k_i+2}}
   \delta^{(1)}_{\sum_i\frac{\tilnu_i}{k_i+2},\frac{M_\omega-M}{H}}
   \sum_{l_i}\tilomega^{\tilnu}\hat\chi^{(\rm st)NS+}
   \prod_{i=1}^rN_{L_iL_i}^{~l_i}\hat
   \chi^{\rm NS+}_{l_i,2\tilnu_i}
  \right\}
\eeqa
where the characters $\hat\chi^{(\rm st)NS\pm}$, $\hat\chi^{\rm NS\pm}_{l,n}$
are the same as those appeared in the discussion of A-branes.
The expressions are the same for short-orbit branes except for obvious
change of normalizations.
The states contributing to the above M\"{o}bius strip amplitude
with $+(-)$ signs are the eigenstates of the parity
with eigenvalues $P^B_{\omega;{\bf m}}=+1(-1)$.
We also see that, in the $q_l$-expansion of this amplitude,
the terms of order $q_l^0$ corresponding to gauge fields
appear with the sign $-\epsilon^{\omega;{\bf m}}_{{\bf L},M}$.

\subsubsection{Quintic}

%Ilka
For the quintic, there exists only a single parity of interest. 
We are considering  supersymmetry preserving branes,
which in the case of the quintic are invariant under the parity.
The following
table lists the number of massless scalars on these branes
and their transformation properties
under parity: $(n_1,n_2)$ denotes the number of (symmetric, antisymmetric)
massless scalars.
\begin{center}
{\scriptsize
\begin{tabular}{|c||c|}
\hline
($L_i$)&$(n_1,n_2)$\\
\hline
(00000)&$(0,0)$\\
(10000)&$(4,0)$\\
(11000)&$(8,3)$\\
(11100)&$(15,9)$\\
(11110)&$(28,22)$\\
(11111)&$(51,50)$\\
\hline
\end{tabular}}
\end{center}
It is now straightforward
to find the matter content of supersymmetric tadpole canceling
configurations. Two such solutions have been given in \cite{Blumenhagen}, 
the standard solution with $4$ branes of type $(11111)$
and the one with $12$ branes of type $(00000)$ and $20$ of type $(10000)$,
which is the configuration with the highest possible rank in this example.
Just for the purposes of illustration, we give a 
third configuration, which is chosen completely randomly. We consider a
setup consisting of $4$ branes of type $(11000)$ and $8$ of $(00111)$. 
The matter content
under the gauge group $O(4) \times O(8)$ is
\beq
8({\bf 10},{\bf 1} ) \oplus 3({\bf 6},\bf{1}) 
\oplus 15 ({\bf 1},{\bf 36}) \oplus 9({\bf 1},{\bf 28}) \oplus 101 ({\bf 4}, {\bf 8}). 
\eeq
The part of the spectrum involving only one type of brane can be directly
read off from the above table. For those strings that connect one type
of brane to another, note that there is always a linear combination of
any open string operator and its parity image that survives the projection.
As a consequence,this part of the spectrum can be determined using the results
on the open string spectrum without orientifolds.

%end Ilka

\subsubsection{Two parameter model}

The tables \ref{cp-2pb1} and \ref{cp-2pb2} list the  gauge groups
$G$ and the number of matters on the D-branes for various choices
of orientifolds of the two parameter model.
For $G=O$ or $Sp$, the two numbers in $G_{(n_1,n_2)}$ mean there
are massless scalars in $n_1$ symmetric and $n_2$ antisymmetric
tensor representations of $G$.
For $G=U$ the three numbers in $U_{(n_1,n_2,n_3)}$ mean
there are $n_1$ adjoint, $n_2$ symmetric tensor and $n_3$ antisymmetric
tensor representations.

%\begin{table}
\begin{center}
{\scriptsize
\begin{tabular}{|c||l|l|l|l|l|l|}
\hline
($L_i$)&$P_{0;+++++}$&$P_{0;++-++}$&$P_{0;++++-}$
       &$P_{0;++--+}$&$P_{0;+++--}$&$P_{0;++---}$\\
\hline
(00000)&$O_{(0,0)}$&$U_{(0,2,1)}$&$U_{(0,2,1)}$&$Sp_{(0,0)}$&$Sp_{(0,0)}$&$U_{(0,3,0)}$\\
(00100)&$O_{(2,1)}$&$O_{(1,2)}$&$Sp_{(2,1)}$&$O_{(2,1)}$&$Sp_{(3,0)}$&$Sp_{(0,3)}$\\
(00110)&$U_{(2,1,0)}$&$O_{(1,1)}$&$Sp_{(2,0)}$&$U_{(2,0,1)}$&$U_{(2,0,1)}$&$O_{(2,0)}$\\
(00111)&$Sp_{(3,0)}$&$O_{(1,2)}$&$O_{(1,2)}$&$O_{(2,1)}$&$O_{(2,1)}$&$Sp_{(0,3)}$\\
\hline
(11000)&$O_{(5,0)}$&$U_{(5,4,6)}$&$U_{(5,4,6)}$&$Sp_{(2,3)}$&$Sp_{(2,3)}$&$U_{(5,4,6)}$\\
(11100)&$O_{(9,6)}$&$O_{(11,4)}$&$Sp_{(6,9)}$&$O_{(7,8)}$&$Sp_{(6,9)}$&$Sp_{(8,7)}$\\
(11110)&$U_{(9,3,3)}$&$O_{(6,3)}$&$Sp_{(3,6)}$&$U_{(9,5,1)}$&$U_{(9,3,3)}$&$O_{(4,5)}$\\
(11111)&$Sp_{(6,9)}$&$O_{(9,6)}$&$O_{(9,6)}$&$O_{(9,6)}$&$O_{(9,6)}$&$Sp_{(10,5)}$\\
\hline
(20000)&$O_{(4,0)}$&$U_{(4,3,4)}$&$U_{(4,3,4)}$&$Sp_{(2,2)}$&$Sp_{(2,2)}$&$U_{(4,4,3)}$\\
(20100)&$O_{(7,4)}$&$O_{(8,3)}$&$Sp_{(5,6)}$&$O_{(5,6)}$&$Sp_{(6,5)}$&$Sp_{(5,6)}$\\
(20110)&$U_{(6,3,2)}$&$O_{(4,2)}$&$Sp_{(3,3)}$&$U_{(6,4,1)}$&$U_{(6,2,3)}$&$O_{(3,3)}$\\
(20111)&$Sp_{(6,5)}$&$O_{(6,5)}$&$O_{(6,5)}$&$O_{(7,4)}$&$O_{(7,4)}$&$Sp_{(7,4)}$\\
\hline
(22000)&$O_{(13,3)}$&$U_{(16,7,12)}$&$U_{(16,7,12)}$&$Sp_{(7,9)}$&$Sp_{(7,9)}$&$U_{(16,6,13)}$\\
(22100)&$O_{(20,15)}$&$O_{(25,10)}$&$Sp_{(14,21)}$&$O_{(16,19)}$&$Sp_{(13,22)}$&$Sp_{(20,15)}$\\
(22110)&$U_{(18,8,9)}$&$O_{(12,6)}$&$Sp_{(5,13)}$&$U_{(18,13,4)}$&$U_{(18,9,8)}$&$O_{(7,11)}$\\
(22111)&$Sp_{(13,22)}$&$O_{(21,14)}$&$O_{(21,14)}$&$O_{(20,15)}$&$O_{(20,15)}$&$Sp_{(24,11)}$\\
\hline
(31000)&$O_{(9,6)}$&$Sp_{(6,9)}$&$Sp_{(6,9)}$&$Sp_{(6,9)}$&$Sp_{(6,9)}$&$O_{(5,10)}$\\
(31100)&$U_{(9,3,3)}$&$O_{(6,3)}$&$Sp_{(3,6)}$&$U_{(9,3,3)}$&$U_{(9,1,5)}$&$Sp_{(5,4)}$\\
(31110)&$Sp_{(6,9)}$&$O_{(9,6)}$&$Sp_{(4,11)}$&$O_{(9,6)}$&$Sp_{(8,7)}$&$O_{(7,8)}$\\
(31111)&$Sp_{(0,5)}$&$U_{(5,6,4)}$&$U_{(5,6,4)}$&$O_{(3,2)}$&$O_{(3,2)}$&$U_{(5,6,4)}$\\
\hline
(33000)&$U_{(9,3,3)}$&$Sp_{(3,6)}$&$Sp_{(3,6)}$&$U_{(9,1,5)}$&$U_{(9,1,5)}$&$O_{(0,9)}$\\
(33100)&$Sp_{(6,9)}$&$O_{(9,6)}$&$Sp_{(4,11)}$&$Sp_{(8,7)}$&$O_{(1,14)}$&$Sp_{(10,5)}$\\
(33110)&$Sp_{(0,5)}$&$U_{(5,6,4)}$&$U_{(5,0,10)}$&$O_{(3,2)}$&$Sp_{(4,1)}$&$U_{(5,4,6)}$\\
(33111)&$Sp_{(0,15)}$&$Sp_{(10,5)}$&$Sp_{(10,5)}$&$O_{(7,8)}$&$O_{(7,8)}$&$O_{(9,6)}$\\
\hline
\end{tabular}}
\end{center}
%\end{table}

\begin{table}[htb]
\begin{center}
{\scriptsize
\begin{tabular}{|c||l|l|l|l|l|l|}
\hline
($L_i$)&$P_{1;+++++}$&$P_{1;++-++}$&$P_{1;++++-}$
       &$P_{1;++--+}$&$P_{1;+++--}$&$P_{1;++---}$\\
\hline
(10000)&$O_{(1,0)}$&$U_{(1,3,3)}$&$U_{(1,3,3)}$&$Sp_{(0,1)}$&$Sp_{(0,1)}$&$U_{(1,3,3)}$\\
(10100)&$O_{(4,3)}$&$O_{(4,3)}$&$Sp_{(3,4)}$&$O_{(4,3)}$&$Sp_{(3,4)}$&$Sp_{(3,4)}$\\
(10110)&$U_{(5,1,1)}$&$O_{(3,2)}$&$Sp_{(2,3)}$&$U_{(5,1,1)}$&$U_{(5,1,1)}$&$O_{(3,2)}$\\
(10111)&$Sp_{(3,4)}$&$O_{(4,3)}$&$O_{(4,3)}$&$O_{(4,3)}$&$O_{(4,3)}$&$Sp_{(3,4)}$\\
\hline
(21000)&$O_{(9,0)}$&$U_{(9,5,9)}$&$U_{(9,5,9)}$&$Sp_{(4,5)}$&$Sp_{(4,5)}$&$U_{(9,5,9)}$\\
(21100)&$O_{(14,9)}$&$O_{(18,5)}$&$Sp_{(9,14)}$&$O_{(10,13)}$&$Sp_{(9,14)}$&$Sp_{(13,10)}$\\
(21110)&$U_{(13,5,5)}$&$O_{(9,4)}$&$Sp_{(4,9)}$&$U_{(13,9,1)}$&$U_{(13,5,5)}$&$O_{(5,8)}$\\
(21111)&$Sp_{(9,14)}$&$O_{(14,9)}$&$O_{(14,9)}$&$O_{(14,9)}$&$O_{(14,9)}$&$Sp_{(17,6)}$\\
\hline
(30000)&$O_{(4,3)}$&$Sp_{(3,4)}$&$Sp_{(3,4)}$&$Sp_{(3,4)}$&$Sp_{(3,4)}$&$O_{(4,3)}$\\
(30100)&$U_{(5,1,1)}$&$O_{(3,2)}$&$Sp_{(2,3)}$&$U_{(5,1,1)}$&$U_{(5,1,1)}$&$Sp_{(2,3)}$\\
(30110)&$Sp_{(3,4)}$&$O_{(4,3)}$&$Sp_{(3,4)}$&$O_{(4,3)}$&$Sp_{(3,4)}$&$O_{(4,3)}$\\
(30111)&$Sp_{(0,1)}$&$U_{(1,3,3)}$&$U_{(1,3,3)}$&$O_{(1,0)}$&$O_{(1,0)}$&$U_{(1,3,3)}$\\
\hline
(32000)&$O_{(14,9)}$&$Sp_{(9,14)}$&$Sp_{(9,14)}$&$Sp_{(9,14)}$&$Sp_{(9,14)}$&$O_{(6,17)}$\\
(32100)&$U_{(13,5,5)}$&$O_{(9,4)}$&$Sp_{(4,9)}$&$U_{(13,5,5)}$&$U_{(13,1,9)}$&$Sp_{(8,5)}$\\
(32110)&$Sp_{(9,14)}$&$O_{(14,9)}$&$Sp_{(5,18)}$&$O_{(14,9)}$&$Sp_{(13,10)}$&$O_{(10,13)}$\\
(32111)&$Sp_{(0,9)}$&$U_{(9,9,5)}$&$U_{(9,9,5)}$&$O_{(5,4)}$&$O_{(5,4)}$&$U_{(9,9,5)}$\\
\hline
\end{tabular}}
\end{center}
\caption{Gauge group and number of massless scalar fields}
\label{cp-2pb1}
\end{table}

%\begin{table}
\begin{center}
{\scriptsize
\begin{tabular}{|c||l|l|l|l|l|l|}
\hline
($L_i$)&$P_{0;+-+++}$&$P_{0;+--++}$&$P_{0;+-++-}$
       &$P_{0;+---+}$&$P_{0;+-+--}$&$P_{0;+----}$\\
\hline
(00000)&$U_{(0,0,3)}$&$Sp_{(0,0)}$&$Sp_{(0,0)}$&$U_{(0,1,2)}$&$U_{(0,1,2)}$&$O_{(0,0)}$\\
(00100)&$Sp_{(0,3)}$&$O_{(0,3)}$&$Sp_{(1,2)}$&$Sp_{(2,1)}$&$O_{(1,2)}$&$Sp_{(1,2)}$\\
(00110)&$Sp_{(0,2)}$&$U_{(2,0,1)}$&$U_{(2,0,1)}$&$O_{(0,2)}$&$Sp_{(1,1)}$&$U_{(2,1,0)}$\\
(00111)&$Sp_{(0,3)}$&$Sp_{(1,2)}$&$Sp_{(1,2)}$&$O_{(1,2)}$&$O_{(1,2)}$&$O_{(0,3)}$\\
\hline
(11000)&$U_{(5,10,0)}$&$O_{(1,4)}$&$O_{(1,4)}$&$U_{(5,6,4)}$&$U_{(5,6,4)}$&$Sp_{(2,3)}$\\
(11100)&$O_{(11,4)}$&$Sp_{(14,1)}$&$O_{(7,8)}$&$O_{(5,10)}$&$Sp_{(8,7)}$&$O_{(9,6)}$\\
(11110)&$O_{(6,3)}$&$U_{(9,5,1)}$&$U_{(9,3,3)}$&$Sp_{(9,0)}$&$O_{(4,5)}$&$U_{(9,1,5)}$\\
(11111)&$O_{(9,6)}$&$O_{(9,6)}$&$O_{(9,6)}$&$Sp_{(10,5)}$&$Sp_{(10,5)}$&$Sp_{(14,1)}$\\
\hline
(20000)&$U_{(4,1,6)}$&$Sp_{(2,2)}$&$Sp_{(2,2)}$&$U_{(4,2,5)}$&$U_{(4,2,5)}$&$O_{(0,4)}$\\
(20100)&$Sp_{(3,8)}$&$O_{(3,8)}$&$Sp_{(4,7)}$&$Sp_{(7,4)}$&$O_{(2,9)}$&$Sp_{(6,5)}$\\
(20110)&$Sp_{(1,5)}$&$U_{(6,2,3)}$&$U_{(6,0,5)}$&$O_{(1,5)}$&$Sp_{(4,2)}$&$U_{(6,3,2)}$\\
(20111)&$Sp_{(1,10)}$&$Sp_{(6,5)}$&$Sp_{(6,5)}$&$O_{(4,7)}$&$O_{(4,7)}$&$O_{(3,8)}$\\
\hline
(22000)&$U_{(16,9,10)}$&$Sp_{(7,9)}$&$Sp_{(7,9)}$&$U_{(16,8,11)}$&$U_{(16,8,11)}$&$O_{(5,11)}$\\
(22100)&$Sp_{(16,19)}$&$O_{(18,17)}$&$Sp_{(15,20)}$&$Sp_{(18,17)}$&$O_{(13,22)}$&$Sp_{(19,16)}$\\
(22110)&$Sp_{(7,11)}$&$U_{(18,9,8)}$&$U_{(18,5,12)}$&$O_{(9,9)}$&$Sp_{(10,8)}$&$U_{(18,8,9)}$\\
(22111)&$Sp_{(12,23)}$&$Sp_{(19,16)}$&$Sp_{(19,16)}$&$O_{(17,18)}$&$O_{(17,18)}$&$O_{(18,17)}$\\
\hline
(31000)&$O_{(11,4)}$&$O_{(7,8)}$&$O_{(7,8)}$&$Sp_{(8,7)}$&$Sp_{(8,7)}$&$Sp_{(8,7)}$\\
(31100)&$O_{(6,3)}$&$U_{(9,5,1)}$&$U_{(9,3,3)}$&$O_{(4,5)}$&$Sp_{(5,4)}$&$U_{(9,3,3)}$\\
(31110)&$O_{(9,6)}$&$O_{(9,6)}$&$Sp_{(8,7)}$&$Sp_{(10,5)}$&$O_{(7,8)}$&$O_{(7,8)}$\\
(31111)&$U_{(5,6,4)}$&$O_{(3,2)}$&$O_{(3,2)}$&$U_{(5,6,4)}$&$U_{(5,6,4)}$&$Sp_{(4,1)}$\\
\hline
(33000)&$O_{(6,3)}$&$U_{(9,3,3)}$&$U_{(9,3,3)}$&$Sp_{(5,4)}$&$Sp_{(5,4)}$&$U_{(9,5,1)}$\\
(33100)&$O_{(9,6)}$&$O_{(9,6)}$&$Sp_{(8,7)}$&$O_{(7,8)}$&$Sp_{(10,5)}$&$Sp_{(6,9)}$\\
(33110)&$U_{(5,6,4)}$&$O_{(3,2)}$&$Sp_{(4,1)}$&$U_{(5,6,4)}$&$U_{(5,4,6)}$&$O_{(3,2)}$\\
(33111)&$Sp_{(10,5)}$&$O_{(7,8)}$&$O_{(7,8)}$&$O_{(9,6)}$&$O_{(9,6)}$&$Sp_{(8,7)}$\\
\hline
\end{tabular} }
\end{center}
%\end{table}
\begin{table}[htb]
\begin{center}
{\scriptsize
\begin{tabular}{|c||l|l|l|l|l|l|}
\hline
($L_i$)&$P_{1;+-+++}$&$P_{1;+--++}$&$P_{1;+-++-}$
       &$P_{1;+---+}$&$P_{1;+-+--}$&$P_{1;+----}$\\
\hline
(10000)&$U_{(1,0,6)}$&$Sp_{(1,0)}$&$Sp_{(1,0)}$&$U_{(1,2,4)}$&$U_{(1,2,4)}$&$O_{(0,1)}$\\
(10100)&$Sp_{(1,6)}$&$O_{(0,7)}$&$Sp_{(3,4)}$&$Sp_{(5,2)}$&$O_{(2,5)}$&$Sp_{(3,4)}$\\
(10110)&$Sp_{(1,4)}$&$U_{(5,0,2)}$&$U_{(5,0,2)}$&$O_{(0,5)}$&$Sp_{(3,2)}$&$U_{(5,2,0)}$\\
(10111)&$Sp_{(1,6)}$&$Sp_{(3,4)}$&$Sp_{(3,4)}$&$O_{(2,5)}$&$O_{(2,5)}$&$O_{(0,7)}$\\
\hline
(21000)&$U_{(9,10,4)}$&$O_{(4,5)}$&$O_{(4,5)}$&$U_{(9,8,6)}$&$U_{(9,8,6)}$&$Sp_{(5,4)}$\\
(21100)&$O_{(14,9)}$&$Sp_{(15,8)}$&$O_{(12,11)}$&$O_{(10,13)}$&$Sp_{(13,10)}$&$O_{(12,11)}$\\
(21110)&$O_{(8,5)}$&$U_{(13,6,4)}$&$U_{(13,6,4)}$&$Sp_{(9,4)}$&$O_{(6,7)}$&$U_{(13,4,6)}$\\
(21111)&$O_{(14,9)}$&$O_{(12,11)}$&$O_{(12,11)}$&$Sp_{(13,10)}$&$Sp_{(13,10)}$&$Sp_{(15,8)}$\\
\hline
(30000)&$Sp_{(1,6)}$&$Sp_{(3,4)}$&$Sp_{(3,4)}$&$O_{(2,5)}$&$O_{(2,5)}$&$O_{(0,7)}$\\
(30100)&$Sp_{(1,4)}$&$U_{(5,0,2)}$&$U_{(5,0,2)}$&$Sp_{(3,2)}$&$O_{(0,5)}$&$U_{(5,2,0)}$\\
(30110)&$Sp_{(1,6)}$&$Sp_{(3,4)}$&$O_{(0,7)}$&$O_{(2,5)}$&$Sp_{(5,2)}$&$Sp_{(3,4)}$\\
(30111)&$U_{(1,0,6)}$&$Sp_{(1,0)}$&$Sp_{(1,0)}$&$U_{(1,2,4)}$&$U_{(1,2,4)}$&$O_{(0,1)}$\\
\hline
(32000)&$Sp_{(9,14)}$&$Sp_{(11,12)}$&$Sp_{(11,12)}$&$O_{(10,13)}$&$O_{(10,13)}$&$O_{(8,15)}$\\
(32100)&$Sp_{(5,8)}$&$U_{(13,4,6)}$&$U_{(13,4,6)}$&$Sp_{(7,6)}$&$O_{(4,9)}$&$U_{(13,6,4)}$\\
(32110)&$Sp_{(9,14)}$&$Sp_{(11,12)}$&$O_{(8,15)}$&$O_{(10,13)}$&$Sp_{(13,10)}$&$Sp_{(11,12)}$\\
(32111)&$U_{(9,4,10)}$&$Sp_{(5,4)}$&$Sp_{(5,4)}$&$U_{(9,6,8)}$&$U_{(9,6,8)}$&$O_{(4,5)}$\\
\hline
\end{tabular}}
\end{center}
\caption{Gauge group and number of massless scalar fields (continued)}
\label{cp-2pb2}
\end{table}

Note that the symmetric and antisymmetric tensors of $U(n)$ are
complex representations, and are supported on strings between a
$U$-type brane and its parity image.
The analysis of the spectrum also shows that all the $U$-type branes
in the tables support equal number of (anti)symmetric tensors and
their conjugates, namely, all of them support non-chiral matters.
We will extend this observation later and show the non-chirality
of the spectrum for general supersymmetric brane configurations
in any type IIB orientifolds of Gepner model.

\subsubsection{Spectrum in Sample Examples}

It is straightforward to determine the matter contents
in sample tadpole canceling configurations.
Let us consider a few examples as an exercise.

As the first example, let us take the parity $P^B_{0;+++++}$
and take six branes from the group $\#14$ and two from $\#12$ to
cancel the tadpole.
There are still many ways to do so.
For example, there are the following two inequivalent configurations
supporting $O(6)\times O(2)$ gauge group:
\beqa
 6(22100)+2(20100) &:&
20({\bf 21,1})\oplus 15({\bf 15,1})\oplus
 7({\bf  1,3})\oplus  4({\bf  1,1})\oplus 12({\bf  6,2}), \nn\\
 6(22100)+2(20010) &:&
20({\bf 21,1})\oplus 15({\bf 15,1})\oplus
 7({\bf  1,3})\oplus  4({\bf  1,1})\oplus  6({\bf  6,2}).
\eeqa
Here ${\bf 2}$ and ${\bf 3}$ of $O(2)$ mean the reducible
representations ${\bf [1]\oplus[-1]}$ and
${\bf [2]\oplus[0]\oplus[-2]}$ of $U(1)$.
The configurations supporting $O(6)\times U(1)$ are
\beqa
 6(22100)+(11110)^++(11110)^- &:&
20({\bf 21})\oplus 15({\bf 15})
\oplus 9({\bf 1})\oplus 3({\bf 1})^{\pm\pm}\oplus 8({\bf 6})^{\pm}
\nn\\
 6(22100)+(11011)^++(11011)^- &:&
20({\bf 21})\oplus 15({\bf 15})
\oplus 9({\bf 1})\oplus 3({\bf 1})^{\pm\pm}\oplus 2({\bf 6})^{\pm}
\nn\\
 6(22100)+(31100)^++(31100)^- &:&
20({\bf 21})\oplus 15({\bf 15})
\oplus 9({\bf 1})\oplus 3({\bf 1})^{\pm\pm}\oplus 12({\bf 6})^{\pm}
\nn\\
 6(22100)+(31010)^++(31010)^- &:&
20({\bf 21})\oplus 15({\bf 15})
\oplus 9({\bf 1})\oplus 3({\bf 1})^{\pm\pm}\oplus 6({\bf 6})^{\pm}
\nn\\
 6(22100)+(33000)^++(33000)^- &:&
20({\bf 21})\oplus 15({\bf 15})
\oplus 9({\bf 1})\oplus 3({\bf 1})^{\pm\pm}\oplus 10({\bf 6})^{\pm}
\eeqa
Here the $\pm$ signs represent $U(1)$ charge.
$18({\bf 1})$ are neutral scalars corresponding to open strings
with both ends on the same short-orbit brane, while
$({\bf 1})^{\pm\pm}$ correspond to strings stretching between a short-orbit
brane and its parity image.
There are six configurations supporting $O(6)\times Sp(1)$:
\beqa
 6(22100)+2(20111) &:&
20({\bf 21,1})\oplus 15({\bf 15,1})
\oplus 6({\bf 1,3})\oplus 5({\bf 1,1})\oplus 6({\bf 6,2})
\nn\\
 6(22100)+2(31111) &:&
20({\bf 21,1})\oplus 15({\bf 15,1})
\oplus 5({\bf 1,1})\oplus 6({\bf 6,2})
\nn\\
 6(22100)+2(33110)^\pm &:&
20({\bf 21,1})\oplus 15({\bf 15,1})
\oplus 5({\bf 1,1})\oplus 10({\bf 6,2})
\nn\\
 6(22100)+2(33011)^\pm &:&
20({\bf 21,1})\oplus 15({\bf 15,1})
\oplus 5({\bf 1,1})\oplus 4({\bf 6,2})
\eeqa
There are indeed a lot more tadpole-canceling configurations
with various choices of orientifold and D-branes, and the spectrum
of massless states can be obtained in the same way.

\subsection{Chirality --- Vanishing Theorem}

As was explained before, chirality of the theory is measured
by the Witten index.
Given a tadpole-free set of an O-plane and D-branes, the theory
is chiral if there is a pair of D-branes with nonzero open string Witten
index, or any D-brane and its parity image with nonzero twisted Witten index.

The index between two long-orbit B-branes in Gepner model can be easily
computed as the diagonal elements\footnote
  {All the diagonal elements take the same value because $g$ is
   a $H$-dimensional shift matrix.}
of the following polynomial of the $H$-dimensional shift matrix $g$,
\beq
  Q_{{\bf L},M}(g)Q_{{\bf L}',M'}(g^{-1})\prod_{i=1}^r(1-g^{w_i}),
\label{bindex}
\eeq
where $Q_{{\bf L},M}(g)$ is the polynomial defined in (\ref{branec}).
The parity twisted Witten index is given by replacing one of the two
polynomials with the one representing the O-plane charge, which are
given in (\ref{ochvec1}) and (\ref{ochvec2}).

Using the index formula (\ref{bindex}), one can show that any
tadpole-free configurations of an O-plane and long-orbit D-branes
are non-chiral.
To do this, notice first that the polynomial $Q_{{\bf L},M}(g)$
is symmetric under $g^i\to g^{M-i}$.
Similarly, the polynomials representing the O-plane charges
are symmetric under $g^i\to g^{M_O-i}$, where
$M_O=M_\omega+4|\epsilon|$ characterize the spacetime supersymmetry
preserved by the O-planes.
On the other hand, under the assumption $\sum_iw_i=H$ the last factor
in (\ref{bindex}) is transformed to $(-1)^r$ times itself under
$g\to g^{-1}$.
Using all these one finds that, in standard four-dimensional models
with $r=5$, the polynomial (\ref{bindex}) has no $g^0$ term
for any susy-preserving pairs of long-orbit D-branes and the O-plane.
Thus the index vanishes for all such pairs.

There is still a possibility of having chiral models with B-type
orientifolds of Gepner models.
The point is that, in some Gepner models with even $H$, there are
RR-charges carried by some short-orbit B-branes and none of
long-orbit B-branes.
In general, the number of RR-charges in type IIB orientifolds
is $2h_{1,1}+2$, and the number of RR-charges carried by
B-branes is fewer than this:
\beqa
2h_{1,1}+2 &\ge&
\#\mbox{(charges carried by all the short- and long-orbit B-branes)}
\nn\\ &\ge&
\#\mbox{(charges carried by $L_i=0$ B-branes)}.
\eeqa
The first inequality shows that there can be RR-charges carried
by none of rational B-type boundary states constructed in this paper.
It is expected that such RR-charges are associated with non-toric
blowups, as there are non-polynomial deformations in the IIA case.
In the $k=(66222)$ model both of the above equalities hold,
so there is no chiral brane configurations.

As an example where neither of the two equalities hold, let us
consider the $k=(22444)$ model which is known to have $h_{1,1}=6$.
Let us first work out the $14=2h_{1,1}+2$ RR ground states.
First, take the RR ground states
$|0\rangle_\nu$ ($\nu=1,\cdots,k+1$) in the level $k$ minimal model
\beq
 |0\rangle_\nu = |\nu-1,\nu,1\rangle \times |\nu-1,\nu,1\rangle,
\eeq
and construct the ground states of the form
$|0\rangle_{(\nu_i)}=\prod_i |0\rangle_{\nu_i}$,
with $\nu_i= \nu$ (mod $k_i+2$) for all $i$.
There are only eight such states:
\beqa
&
|0\rangle_{(11111)},~~
|0\rangle_{(22222)},~~
|0\rangle_{(33333)},~~
|0\rangle_{(33111)},\nn\\ &
|0\rangle_{(11555)},~~
|0\rangle_{(11333)},~~
|0\rangle_{(22444)},~~
|0\rangle_{(33555)}.
\eeqa
Other states are obtained by looking for mixed products
of $|0\rangle_\nu$ and $|l\rangle_\RR$, where
\beq
 |l\rangle_\RR = |l,l+1,1\rangle \times |l,-l-1,-1\rangle.
\eeq
One finds six additional states of the form
$|l_1,l_2\rangle_\RR\times|0\rangle_{(\nu_3\nu_4\nu_5)}$:
\beqa
&
|2,0\rangle_\RR|0\rangle_{(222)},~~
|1,1\rangle_\RR|0\rangle_{(222)},~~
|0,2\rangle_\RR|0\rangle_{(222)},\nn\\
&
|2,0\rangle_\RR|0\rangle_{(444)},~~
|1,1\rangle_\RR|0\rangle_{(444)},~~
|0,2\rangle_\RR|0\rangle_{(444)}.
\eeqa
(Note that the state $|1,1\rangle_\RR|0\rangle_{(222)}$ is different
 from $|0\rangle_{(22222)}$, although they are labeled by the same
 quantum numbers. Recall that in Gepner model certain closed string
 states appear more than once in the toroidal partition function,
 and we should distinguish them as they are sitting in
 different twisted sectors.)
The $L_i=0$ B-branes can only couple to the first eight states,
and the short-orbit branes with $L_1=L_2=1$ couple also to the
the two states with $l_1=l_2=1$ in the second group.
The remaining four RR ground states have no overlaps
with any B-branes.

Unfortunately, one can also show that the index vanishes
for pairs of short-orbit branes with these extra RR charges,
using a similar index formula as before.
Let us take two short-orbit B-branes with the same $\bfS$
of even order.
As was given in (\ref{fixed1}) and (\ref{fixed2}), the
boundary states are sums of two terms orthogonal to each other.
So the index is also a sum of two terms,
one of which is $2^{-|\bfS|}$ times the expression for long-orbit
branes (\ref{bindex}) and the other represents the new contribution
\beq
 2^{-|\bfS|}
 \widetilde{Q}_{{\bf L},M}(g)\widetilde{Q}_{{\bf L}',M'}(g^{-1})
 \prod_{i\;/\!\!\!\!\in\bfS}(1-g^{w_i})
 \prod_{i\in\bfS}(1+g^{w_i}+\cdots+g^{w_i(k_i+1)}),
\eeq
where
\beq
 \widetilde{Q}_{{\bf L},M}(g)= g^{M/2}\prod_{i\;/\!\!\!\!\in\bfS}
 \left(\sum_{j_i=0}^{L_i}g^{w_i(\frac{L_i}{2}-j_i)}\right).
\eeq
Using the symmetry or antisymmetry of each factor under the
inversion $g\to g^{-1}$ one finds that no supersymmetry-preserving
pair of short-orbit B-branes can have non-zero index.

Thus, we find

\noindent
{\bf Theorem:}\\
{\it The index of any pair
of branes in a tadpole canceling and supersymmetric rational brane
configuration vanishes in Type IIB orientifolds of Gepner models.
In particular, there is no chiral and supersymmetric theory
in this class of solutions.}

\noindent
{\bf Remarks.}\\{\small
{\bf (i)}~This theorem applies only to the Gepner model obtained as
the orbifold of the product of minimal models by a
{\it single} cyclic group $\Z_H$, and may not hold for
orbifolds with more than one cyclic group factors.
For example, Type IIA models we considered in
Section~\ref{sec:TCCA} is nothing but Type IIB models on
orbifolds with four cyclic group factors \cite{GreenePlesser},
and we indeed found
chiral supersymmetric models there.
Actually there is an existence proof of chiral model
if the orbifold group has two cyclic factors (next to minimal):
In Appendix~\ref{orbifold}, we analyze
the condition for a Type IIB orientifold of
the model $M_3^5/\Z_5\times\Z_5$ corresponding to
the $\Z_5$-orbifold of the quintic.
There we find some chiral solutions.\\
{\bf (ii)}~The theorem applies
to more general models with $r\geq 5$, as long as
the orbifold group is $\Z_H$.
The essential point we have used is that $r$ is odd.
In our supersymmetric formulation, we indeed need
$r$ to be odd as discussed in Appendix~\ref{app:Gepner}
(if $r=6$ or $8$ in the formulation
as in \cite{Gepnerclassification} we need to add $k=0$ factor(s)
to make $r$ odd).}

\section{Continuation to Geometry}\label{sec:continuation}

In this section, we compare the results at the Gepner point
and what is expected at the large volume regions.
Namely we compare two different domains of the K\"ahler moduli space.
The story is very much different between Type IIA and
Type IIB cases since the role of the K\"ahler moduli
are different.

In Type IIA orientifolds, K\"ahler class and B-field form complex moduli
fields. The large volume region, if consistent with orientifold,
is always smoothly connected to the Gepner point and the comparison
makes sense.
The K\"ahler moduli can enter into the tree level
superpotential.
The comparison of the two regions may be useful to find
out the set of low energy fields and the global determination of the
tree level superpotential.

In Type IIB orientifolds, K\"ahler moduli are real and
are complexified by RR potentials.
In some cases the large volume regions are separated from the Gepner point,
but in some other cases they are smoothly connected.
It is only in the latter case where the comparison makes sense.
The K\"ahler moduli do not enter into the
tree level superpotential, though they may enter into FI parameters
as well as non-perturbative superpotential.

The main focus of this section will  be on the Type IIB cases.
One technical advantage in these cases is that
the large volume interpretation of the branes at the Gepner
model has been worked out in detail.
Thus, Sections~\ref{subsec:LV} through \ref{subsec:twoparaLV}
are about Type IIB orientifolds.
However, in the last subsection, we make some remarks
on the Type IIA cases.

\subsection{Consistency Condition at Large Volume}
\label{subsec:LV}

\newcommand{\sfX}{{\bf X}}

Let us first
present the tadpole cancellation condition in the large volume regime.
We consider a spacetime manifold $\sfX$ with an involution $\tau$, and a
D-brane supporting a complex vector bundle $E$ with an antilinear map
that descends to $\tau$. 
The tadpole cancellation condition for the $\tau$-orientifold
of this system is 
\beq
{\rm ch}(E)\e^{-B}\sqrt{\widehat{\rm A}(\sfX)}
=2^{2\dim_c\sfX^{\tau}-\dim_c \sfX}
\epsilon [\sfX^{\tau}] \sqrt{{L({1\over 4}T\sfX^{\tau})\over
L({1\over 4}N\sfX^{\tau})}}.
\label{tccLV}
\eeq
This is found by comparing the formulae for the RR-overlaps
with the boundary state
$\widetilde{\Pi}^E_i$ and 
the crosscap state $\widetilde{\Pi}^{\tau\Omega}_i$
computed in the non-linear sigma models (see e.g. page 27 of \cite{BH2}).
Some remarks are in order:\\
$\bullet$~ $B$ is the B-field. In this section, we normalize it
so that $B$ is trivial for closed strings if and only if
$B\in H^2(\sfX,\Z)$.\\
$\bullet$~ $\sfX^{\tau}$ is the O-plane, the fixed point set of
$\tau$. $\sfX^{\tau}$ may consist of
several connected components. In such a case the right hand side
 is regarded as the sum over
components.\\
$\bullet$~ In the power of $2$, $\dim_c\sfX^{\tau}$ and $\dim_c\sfX$
include the $\R^4$ directions (counted as 2 complex dimensions) as well as
the internal dimensions.
So, the power is $32$ for O9-plane, $8$ for O7,
$2$ for O5, and $1/2$ for O3.\\
$\bullet$~ $[\sfX^{\tau}]$ is the Poincar\'e dual of
(the component of) the O-plane.
``$\epsilon$'' stands for a sign which is determined by the orientation
of O-plane.\\
$\bullet$~
Useful identities to be remembered (on a Calabi-Yau three-fold $M$) are
\beqa
&&
\widehat{\rm A}(M)={\rm td}(M)=1+{c_2(M)\over 12},
\nn\\
&&
L({1\over 4}V)=1+{p_1(V)\over 48}
=1-{c_2(V\otimes \C)\over 48},\quad \mbox{for a real vector bundle $V$}
\nn
\eeqa

Let us apply this to Type I string theory compactified
on a Calabi-Yau 3-fold $M$ --- Type IIB orientifold
of $\sfX=M\times \R^4$ associated with $\tau={\rm id}_{\sfX}$.
In this case, $\sfX^{\tau}=\sfX$ and $[\sfX^{\tau}]=1$.
Applying the useful formula we find
 $\sqrt{{\rm td}(M)}=1+{1\over 24}c_2(M)$
and $\sqrt{L({1\over 4}TM)}=1-{1\over 48}c_2(M)$.
The condition is therefore
$$
{\rm ch}(E)\e^{-B}=32+2{\rm ch}_2(M),
$$
which is the rank and the anomaly cancellation condition in the
standard form.

We will examine whether the condition (\ref{tccLV})
is satisfied
for the D-brane configuration at the Gepner model
continued to the large volume, whenever the continuation is
possible. We work in two examples  --- the quintic case $(5,5,5,5,5)$
and the two parameter model $(8,8,4,4,4)$.

\subsection{Quintic}\label{subsec:quinticLV}

Let us first discuss the model $(k_i+2)=(5,5,5,5,5)$
that continues to the sigma model on the quintic hypersurface 
$M$ of $\CP^4$.
As we have seen in Section~\ref{subsub:quintic},
the moduli space of the orientifold model is real,
$\e^t\in \R$: The Gepner point $\e^t=0$ is separated from the $B=0$
large volume ($\e^t\ll -1$)
by the conifold point $\e^t=-5^5$, but is connected
to the large volume region with $B={H\over 2}$ ($\e^t\gg 1$),
where $H=c_1({\mathcal O}(1))|_M$ is the integral
generator of $H^2(M,\Z)$.
Thus, we expect the match of the condition only with
the large volume with $B={H\over 2}$.

Let us first write down the tadpole cancellation condition at the
large volume. The Chern character of $M$ can be read from
the exact sequence $0\to T_M\to T_{\CP^4}\to N_{M/\CP^4}\to 0$
as ${\rm ch}(T_M)={\rm ch}(T_{\CP^4})|_M
-{\rm ch}(N_{M/\CP^4})$. We know that $N_{M/\CP^4}={\mathcal O}(5)|_M$
since $M$ is quintic, and
also that
${\rm ch}(T_{\CP^4})={\rm ch}({\mathcal O}(1)^5)-{\rm ch}({\mathcal O})$
from the tautological sequence.
Thus, ${\rm ch}(T_M)=5\e^H-1-\e^{5H}
=3-10H^2-20H^3$, and in particular
${\rm ch}_2(M)=-10 H^2$.
Thus, the tadpole cancellation condition in the large volume region
is
\beq
{\rm ch}(E)\e^{-B}=32-20 H^2.
\eeq

Now we would like to compare this with the condition we obtained in
Section~\ref{sec:TCCB}.
In order to make the comparison, we need to know the relation of
the basis of the D-brane charges at the Gepner model
and the basis at the large volume region.
This has been studied in \cite{BDLR}, and the result is
\beq
B_{{\bf L},M}=B_{(00000),2m+2n}\longleftrightarrow V_{m}
\label{iden}
\eeq
for some $n\in \Z_5$ where
\beqa
V_0={\mathcal O},&& {\rm ch}(V_0)=1,
\nn\\
V_1=\overline{T^*_{\CP^4}(1)},&&
{\rm ch}(V_1)=-4+H+{1\over 2}H^2+{1\over 6}H^3,
\nn\\
V_2=\wedge^2T^*_{\CP^4}(2),&&
{\rm ch}(V_2)=6-3H-{1\over 2}H^2+{1\over 2}H^3,
\nn\\
V_3=\overline{\wedge^3T^*_{\CP^4}(3)},&&
{\rm ch}(V_3)=-4+3H-{1\over 2}H^2-{1\over 2}H^3,
\nn\\
V_4=\wedge^4T^*_{\CP^4}(4),&&
{\rm ch}(V_4)=1-H+{1\over 2}H^2-{1\over 6}H^3.
\nn
\eeqa
We found in Section~\ref{sec:TCCB}
that the O-plane has the D-brane charge
$4[B_{{\bf 1},5}]=4(2[B_{{\bf 0},0}]
+5[B_{{\bf 0},2}]
+10[B_{{\bf 0},4}]
+10[B_{{\bf 0},6}]
+5[B_{{\bf 0},8}])$
We try all the 5 possible identifications (\ref{iden}) to compute
the rank of the tadpole canceling brane:
\beqa
V_{m}\leftrightarrow M=2m&\Longrightarrow&
{\rm rank}=28,
\nn\\
V_{m}\leftrightarrow M=2m+2&\Longrightarrow&
{\rm rank}=28,
\nn\\
V_{m}\leftrightarrow M=2m+4&\Longrightarrow&
{\rm rank}=-32,
\nn\\
V_{m}\leftrightarrow M=2m+6&\Longrightarrow&
{\rm rank}=-12,
\nn\\
V_{m}\leftrightarrow M=2m+8&\Longrightarrow&
{\rm rank}=-12,
\nn
\eeqa
Thus, we find that the identification
$V_{m}\leftrightarrow M=2m+4$ may work.
Indeed under this identification the full charge of the tadpole canceling
D-brane is
$$
{\rm ch}(E)=4\left(-8+4H+4H^2-{7\over 3}H^3\right),
$$
and for the choice $B=-{H\over 2}$, we find
\beq
{\rm ch}(E)\e^{-B}=-32+20 H^2,
\eeq
which is nothing but the large volume condition.

\subsection{The Two Parameter Model}\label{subsec:twoparaLV}

Let us now discuss the two parameter model that includes the Gepner model
with $(k_i+2)=(8,8,4,4,4)$. As we have seen in Section~\ref{subsub:two},
the Gepner point and the large volume regions are separated in
the K\"ahler moduli space of the orientifold models by the parities
$P^B_{0,\epsilon_1...\epsilon_5}$.
Thus in this case, we do not expect that the tadpole cancellation condition
at the Gepner point matches with that in the large volume.
On the other hand, for the orientifolds by $P_{1;\epsilon_1...\epsilon_5}^B$,
the Gepner point is connected to the large volume regions
in the moduli space.
In fact, there are two separate large volume regions --- one with
$B={L\over 2}$ and another with $B={H\over 2}+{L\over 2}$.
Thus, the condition at the Gepner point must match with the conditions
at {\it both} of the large volume region.
We will check this in what follows.

\subsubsection{Topology of the manifold and O-planes}

We first describe the topology of the Calabi-Yau manifold $M$ itself.
Let $X$ be the toric manifold associated to the $U(1)^2$ gauge theory
with six matter fields of the following charge
$$
\begin{array}{ccccccc}
&X_1&X_2&X_3&X_4&X_5&X_6\\
U(1)_1&0&0&1&1&1&1\\
U(1)_2&1&1&0&0&0&-2
\end{array}
$$
Our Calabi-Yau manifold $M$ is a hypersurface of $X$ given by
$X_6^4(X_1^8+X_2^8)+X_3^4+X_4^4+X_5^4=0$.
The cohomology ring of $X$ is generated by the divisor class
$H=(X_3=0)=(X_4=0)=(X_5=0)$ and $L=(X_1=0)=(X_2=0)$ that obey the relations
\beqa
&&L^2=H^3(H-2L)=0,\nn\\
&&\int_XH^3L=1.\nn
\eeqa
Holomorphic tangent bundle of $X$ fits into an exact sequence
$$
0\to {\mathcal O}\oplus {\cal O}\to 
{\mathcal L}_1\oplus
{\mathcal L}_2\oplus
{\mathcal L}_3\oplus
{\mathcal L}_4\oplus
{\mathcal L}_5\oplus
{\mathcal L}_6
\to T_X\to 0
$$
where ${\mathcal L}_i$ is the line bundle with section $X_i$.
Chern lass of $X$ is therefore given by
$$
c(X)=(1+L)^2(1+H)^3(1+H-2L).
$$
The hypersurface $M$ yields the divisor class $[M]=4H$ and the normal
bundle has $c_1(N_{M/X})=4H|_M$. We shall hereafter denote
$H|_M,L|_M$ simply by $H,L$. They obey
$$
\int_MH^2L=4,\quad
\int_MH^3=8.
$$
Chern class of $M$ is given by $c(M)=c(X)|_Mc(N_{M/X})^{-1}$
namely,
\beqa
&&c_1(M)=0\\
&&c_2(M)=2HL+6H^2\\
&&c_3(M)=-21 H^3
\eeqa

Now, we write down the tadpole cancellation condition 
(\ref{tccLV}) for the various involutions we discussed in
Section~\ref{subsub:two}.

\subsubsection*{\underline{$(+++++)$}}

When $\tau:M\to M$ is identity (the case for Type I string theory),
the consistency condition for the
bundle $E$ is
\beq
{\rm ch}(E)\e^{-B}
=32-4HL-12 H^2.
\eeq

\subsubsection*{\underline{$(++-++)$ etc}}

The fixed point set of $\tau:
(X_1,...,X_6)\mapsto (X_1,X_2,-X_3,X_4,X_5,X_6)$ is the divisor
$X_3=0$. For this we have
\beqa
[M^{\tau}]&=&H,
\nn\\
N_{M^{\tau}/M}&=&{\mathcal L}_3|_{M^{\tau}},\quad
c(N_{M^{\tau}/M})=1+H
\nn\\
c(T_{M^{\tau}})&=&c(M)|_{M^{\tau}}c(N_{M^{\tau}/M})^{-1}
=1-H+7H^2+2HL
\nn\\
p_1(TM^{\tau})
&=&-c_2(T_{M^{\tau}}\oplus \overline{T}_{M^{\tau}})
=-13 H^2-4HL,
\nn\\
p_1(NM^{\tau})
&=&-c_2(N_{M^{\tau}}\oplus \overline{N}_{M^{\tau}})
=H^2
\nn\\
{\rm td}(M)|_{M^{\tau}}
&=&\left.1+{1\over 6}HL+{1\over 2}H^2\right|_{M^{\tau}}
=1+{7\over 12}H^2
\nn\\
L({1\over 4}TM^{\tau})&=&
1-{1\over 48}(13H^2+4HL)=1-{15\over 48}H^2,
\nn\\
L({1\over 4}NM^{\tau})&=&
1+{1\over 48}H^2,
\nn\\
{L({1\over 4}TM^{\tau})\over
L({1\over 4}NM^{\tau})}
{\rm td}(M)^{-1}&=&1-{11\over 12}H^2.
\nn
\eeqa
Thus, consistency condition for this orientifold is
\beq
{\rm ch}(E)\e^{-B}
=\pm \left(8 H -{11\over 3} H^3\right).
\eeq

\subsubsection*{\underline{$(++--+)$ etc}}

For
$\tau:(X_1,...,X_6)\mapsto (X_1,X_2,-X_3,-X_4,X_5,X_6)$,
the fixed point sets are the curve
$C=\{X_3=X_4=0\}$ and four lines
$\ell_a=\{X_5=X_6=0,X_3=\e^{{\pi i\over 4}+{\pi i a\over 2}}X_4\}$
($a=1,2,3,4$).
Their Poincar\'e duals are
\beqa
&&[C]=[X_3=0]\cup [X_4=0]=H^2,
\nn\\
&&\sum_{a=1}^4[\ell_a]=[X_5=0]\cup[X_6=0]=H(H-2L).
\nn
\eeqa
Thus, if the four O-planes at $\ell_a$ are of the same type, the
consistency condition is
\beqa
{\rm ch}(E)\e^{-B}
&=&2\{\pm H^2\pm H(H-2L)\}
\nn\\
&=&\left\{
\begin{array}{l}
\pm 4(H^2-HL)\\
\mbox{or}\\
\mp 4HL
\end{array}
\right.
\eeqa
The first line of the RHS 
is when $C$ and $\ell_a$ contributes to the O-plane charge 
with $\epsilon[C]=\pm [C]$ and $\sum_{a=1}^4\epsilon[\ell_a]
=\pm \sum_{a=1}^4[\ell_a]$, while the second line is when
they contributes with
$\epsilon[C]=\mp [C]$ and $\sum_{a=1}^4\epsilon[\ell_a]
=\pm \sum_{a=1}^4[\ell_a]$.

\subsubsection*{\underline{$(++---)$}}

For $\tau:(X_1,...,X_6)\mapsto 
(X_1,X_2,-X_3,-X_4,-X_5,X_6)$,
the fixed point sets are the divisor
$D=\{X_6=0\}$
and eight points
$p_a=\{X_3=X_4=X_5=0, X_1=\e^{\pi i({1\over 8}+{a\over 4})}X_2\}$
($a=1,2,...,8$).
Their Poincar\'e duals are
\beqa
&&[D]=[X_6=0]=H-2L,\nn\\
&&\sum_{a=1}^8[p_a]=[X_3=X_4=X_5=0]=H^3.\nn
\eeqa
We have
\beqa
&&\int_DH^2=\int_M(H-2L)H^2=0 \,\,(\mbox{thus $H^2=0$ on $D$}),\nn\\
&&\int_DHL=\int_M(H-2L)HL=4.\nn
\eeqa
and
\beqa
c(N_D)&=&1+H-2L,\nn\\
c(T_D)&=&c(M)|_Dc(N_D)^{-1}
=1+2L-H-2HL,\nn\\
p_1(TD)&=&-c_2(T_D\oplus\overline{T}_D)=0,\nn\\
p_1(ND)&=&-c_2(N_D\oplus\overline{N}_D)=-4HL,
\nn\\
L({1\over 4}TD)&=&0,\nn\\
L({1\over 4}ND)&=&1-{1\over 12}HL,\nn\\
{\rm td}(M)|_D&=&1+{1\over 6}HL,\nn\\
{L({1\over 4}TD)\over L({1\over 4}ND)}
{\rm td}(M)|_D^{-1}&=&1-{1\over 12}HL.
\nn
\eeqa
If all the eight O3-planes are of the same type,
the consistency condition is
\beqa
{\rm ch}(E)\e^{-B}
&=&\pm 8(H-2L)(1-{1\over 24}HL)\pm {1\over 2}H^3
\nn\\
&=&\pm \left\{ 8H-16L-{1\over 6}H^3 \right\}
\pm {1\over 2}H^3\nn\\
&=&\pm\left\{
\begin{array}{l}
8H-16L+{1\over 3}H^3
\\
\mbox{or}\\
8H-16L-{2\over 3}H^3
\end{array}
\right.
\eeqa
The first line of RHS is when $D$ and the eight points $p_a$
contributes to the O-plane charge with
$\epsilon[D]=\pm [D]$ and
$\sum_{a=1}^8\epsilon[p_a]=\pm \sum_{a=1}^8[p_a]$,
while the second line is when they contribute with
$\epsilon[D]=\pm [D]$ and
$\sum_{a=1}^8\epsilon[p_a]=\mp \sum_{a=1}^8[p_a]$.

\subsubsection*{\underline{$(+-***)$}}

In all the cases with $\epsilon_1=-\epsilon_2$,
we have seen that the fixed point set
consists of a pair of homologous components.
Thus, one possible consistency condition is
\beq
{\rm ch}(E)\e^{-B}=0.
\eeq
There are of course other possibilities as well.

\subsubsection{Gepner model to the large volume with
$B={1\over 2}H+{1\over 2}L$}

Let us now see whether the set of Cardy
branes obeying the tadpole cancellation condition, when transported in
the orientifold moduli space, obey the condition at large volume.

For the parities $P_{1;+-***}^B$, we have seen that the O-plane has no
charge and therefore the tadpole canceling set of branes must have zero
total RR-charge. This is indeed one of the possibilities as we have just seen
--- the case where the two (set of) O-planes have the opposite
RR-charge. In particular, this is realized by the supersymmetric
O-plane configurations, where one of them is of $SO$-type and the
other is of $Sp$-type.

For the parities $P_{1;++***}^B$, the O-plane has non-zero RR-charge
and the check is non-trivial.
For the comparison,
we need to know the relation of the RR-charge of the Cardy branes
at the Gepner model
and the charge associated with the vector bundles at the large volume.
One relation is found in \cite{DiaDou}
\beqa
&&{\rm ch}(V_1)=1-H+L+2\ell+{2\over 3}v
\nn\\
&&{\rm ch}(V_2)=-1+H-2L+4h-2\ell-{8\over 3}v
\nn\\
&&{\rm ch}(V_3)=-3+2H-L-{4\over 3}v
\nn\\
&&{\rm ch}(V_4)=3-2H+4L-8h+{4\over 3}v
\nn\\
&&{\rm ch}(V_5)=3-H-L-2\ell+{2\over 3}v
\nn\\
&&{\rm ch}(V_6)=-3+H-2L+4h+2\ell+{4\over 3}v
\nn\\
&&{\rm ch}(V_7)=-1+L
\nn\\
&&{\rm ch}(V_8)=1
\nn\\
\label{DiDo}
\eeqa
where
$$
\ell:={H^2-2HL\over 4},\quad
h:={HL\over 4},\quad
v:={H^3\over 8}={H^2L\over 4}.
$$
Up to cyclic permutation, $V_1,...,V_8$ are identified as a certain
analytic continuation of the Cardy branes with
$L=(00000)$ and $M=0,2,4,6,8,10,12,14$.
We would first like to see which cyclic permutation is the relevant one.
To find it, we compute the rank (D9-brane charge) of the tadpole canceling
D-brane for the case $(+++++)$.
We need it to be 32.
A tadpole canceling D-brane has charge
$(-20,-8,-12,12,8,20)$ with respect to the first six of the
$L=(00000)$ Cardy branes.
We find
\beqa
V_m\leftrightarrow M=2m&\Longrightarrow& {\rm rank}=32,
\nn\\
V_m\leftrightarrow M=2m-2&\Longrightarrow& {\rm rank}=24,
\nn\\
V_m\leftrightarrow M=2m-4&\Longrightarrow& {\rm rank}=0,
\nn\\
V_m\leftrightarrow M=2m-6&\Longrightarrow& {\rm rank}=-24,
\nn\\
V_m\leftrightarrow M=2m-8&\Longrightarrow& {\rm rank}=-32,
\nn\\
V_m\leftrightarrow M=2m+6&\Longrightarrow& {\rm rank}=-24,
\nn\\
V_m\leftrightarrow M=2m+4&\Longrightarrow& {\rm rank}=0,
\nn\\
V_m\leftrightarrow M=2m+2&\Longrightarrow& {\rm rank}=24,
\nn
\eeqa
Thus, the identification $V_m\leftrightarrow M=2m$ is the correct one.
($V_m\leftrightarrow M=2m+8$ may also have a chance,
but it is simply the sign flip of $V_m\leftrightarrow M=2m$.)
Under this identification,
the tadpole cancellation condition
at the Gepner point continues to the condition at the large volume
with $B=-{1\over 2}H+{1\over 2}L$, as we now see.

\subsubsection*{\underline{$(+++++)$}}

The charge of a tadpole canceling D-brane $E$ is
\beqa
{\rm ch}(E)&=&
-20 {\rm ch}(V_8)
-8{\rm ch}(V_1)
-12{\rm ch}(V_2)
+12{\rm ch}(V_3)
+8{\rm ch}(V_4)
+20{\rm ch}(V_5)
\nn\\
&=&32-16H+16L-8H^2-12HL+{13\over 3}H^3.
\nn
\eeqa
If we choose $B=-{1\over 2}H+{1\over 2}L$, we find
\beq
{\rm ch}(E)\e^{-B}=32-12H^2-4HL.
\eeq
This is nothing but the tadpole cancellation condition in the large volume
regime.

\subsubsection*{\underline{$(++-++)$ etc}}

\beqa
{\rm ch}(E)&=&
-4 {\rm ch}(V_8)
-8{\rm ch}(V_1)
-12{\rm ch}(V_2)
-12{\rm ch}(V_3)
-8{\rm ch}(V_4)
-4{\rm ch}(V_5)
\nn\\
&=&-8H+4(H^2-HL)+{11\over 3}H^3
\nn
\eeqa
and, for $B=-{1\over 2}H+{1\over 2}L$,
\beq
{\rm ch}(E)\e^{-B}
=-8H+{11\over 3}H^3.
\eeq
This agrees with the large volume condition with $\epsilon[M^{\tau}]
=-[M^{\tau}]$.

\subsubsection*{\underline{$(++--+)$ etc}}

\beqa
{\rm ch}(E)&=&
4 {\rm ch}(V_8)
+0{\rm ch}(V_1)
+4{\rm ch}(V_2)
-4{\rm ch}(V_3)
+0{\rm ch}(V_4)
-4{\rm ch}(V_5)
\nn\\
&=&4HL-H^3
\nn
\eeqa
and, for $B=-{1\over 2}H+{1\over 2}L$,
\beq
{\rm ch}(E)\e^{-B}
=4HL.
\eeq
This agrees with the condition in the large volume regime where 
$C$ and $\ell_a$ contribute to the O-plane charge with
$\epsilon[C]=[C]$ and 
$\epsilon[\ell_a]=-[\ell_a]$.

\subsubsection*{\underline{$(++---)$}}

\beqa
{\rm ch}(E)&=&
4 {\rm ch}(V_8)
+0{\rm ch}(V_1)
+4{\rm ch}(V_2)
+4{\rm ch}(V_3)
+0{\rm ch}(V_4)
+4{\rm ch}(V_5)
\nn\\
&=&8H-16L-4(H^2-3HL)-{5\over 3}H^3
\nn
\eeqa
and, for $B=-{1\over 2}H+{1\over 2}L$,
\beq
{\rm ch}(E)\e^{-B}
=8H-16L-{2\over 3}H^3.
\eeq
This agrees with the condition in the large volume regime where 
$D$ and $p_a$ contribute to the O-plane charge with
$\epsilon[D]=[D]$ and
$\epsilon[p_a]=-[p_a]$.

\subsubsection{Gepner Model to the large volume with
$B={1\over 2}L$}

For the parities $P^B_{1;\epsilon_1...\epsilon_5}$
we are considering, the orientifold
moduli space contains another large volume region
--- the region with $B={L\over 2}$ (mod $\Z H+\Z L$).
Since this region is not separated from
the Gepner point in the moduli space,
the tadpole cancellation condition
at the Gepner point should match with the one at this large volume.
Let us confirm this.

\begin{figure}[htb]
\centerline{\includegraphics{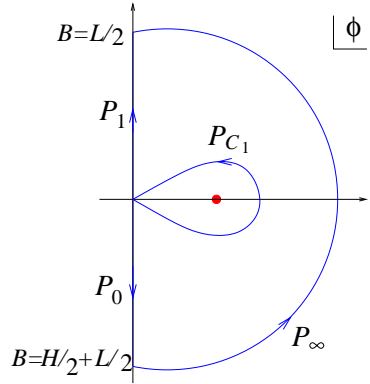}}
\caption{The paths}
\label{paths}
\end{figure}

The main task is to find
the transformation rule of the D-brane charge
--- from the Gepner point to the large volume.
Let $(\phi,\psi)$ be the coordinate
of the cover of the moduli space (before orientifold) that
are used in \cite{Candelas}. These are the natural parameters of the
superpotential of the mirror LG model (\ref{dualW2para}),
$\widetilde{W}=\widetilde{W}_G-8\psi \widetilde{X}_1\cdots\widetilde{X}_5
-2\phi\widetilde{X}_1^4\widetilde{X}_2^4$,
and is related to the linear sigma model parameters as
\beqa
&&\e^{t_1}=-2^{11}\psi^4\phi^{-1},\nn\\
&&\e^{t_2}=4\phi^2.\nn
\eeqa
The singular loci are described as
$$
C_1=\Bigl\{\,\phi^2=1\,\Bigr\},\quad
C_{\rm con}=\Bigl\{\,(\phi+8\psi^4)^2=1\,\Bigr\}.
$$
The $\omega=\e^{2\pi i/8}$ orientifolds impose constraints
$\e^{2\pi i /8}\psi=\overline{\psi}$ and
$-\phi=\overline{\phi}$, or
\beq
\psi\in \e^{-{\pi i\over 8}}\R\quad
\phi\in i\R.
\eeq
Let us consider
a path in this moduli space, ${\mathcal P}_0$:
$\psi=\e^{-\pi i/8}t^{3\over 8}$,
$\phi=\e^{-\pi i/2}\sqrt{t}$, $0\leq t<+\infty$.
(In the $(t_1,t_2)$ coordinates, it is
$\e^{t_1}=-2^{11}t$, $\e^{t_2}=-4t$.) 
It goes from the Gepner point to the large volume region with
$B={H\over 2}+{L\over 2}$.
The identification (\ref{DiDo}) for
$\Pi^{\rm Cardy}=(B_{{\bf 0},0},B_{{\bf 0},2},B_{{\bf 0},4},
B_{{\bf 0},6},B_{{\bf 0},8},B_{{\bf 0},10})^T$
and $\Pi^{\rm LV}=(1,H,L,H^2,HL,H^3)^T$: 
$$
\Pi^{\rm Cardy}=M_{{\mathcal P}_0}\Pi^{\rm LV};\qquad
M_{{\mathcal P}_0}=\left(\begin{array}{cccccc}
1&0&0&0&0&0\\
1&-1&1&{1\over 2}&-1&{2\over 3}\\
-1&1&-2&-{1\over 2}&2&-{8\over 3}\\
-3&2&-1&0&0&-{4\over 3}\\
3&-2&4&0&-2&{4\over 3}\\
3&-1&-1&-{1\over 2}&1&{2\over 3}
\end{array}\right)
$$
can be regarded as the
transformation of charges for this choice of path.
We would like to find the transformation with respect to
the other path,
${\mathcal P}_1$:
$\psi=\e^{-\pi i/8}t^{3\over 8}$,
$\phi=-\e^{-\pi i/2}\sqrt{t}$
($\e^{t_1}=2^{11}t$, $\e^{t_2}=-4t$), 
that goes to the large volume with
$B={L\over 2}$. In order to find it,
let us find a homotopy of paths from
the Gepner point to large volume,
that deforms
${\mathcal P}_0$ to ${\mathcal P}_1$.
The following does the job:
\beq
{\mathcal P}_s
:\,\left\{\begin{array}{l}
\psi=\e^{-{\pi i\over 8}}t^{3\over 8}\\
\phi=\e^{-{\pi i\over 2}}\e^{\pi i s}\sqrt{t},
\end{array}\right.\quad 0\leq s\leq 1.
\eeq
\begin{figure}[htb]
\centerline{\includegraphics{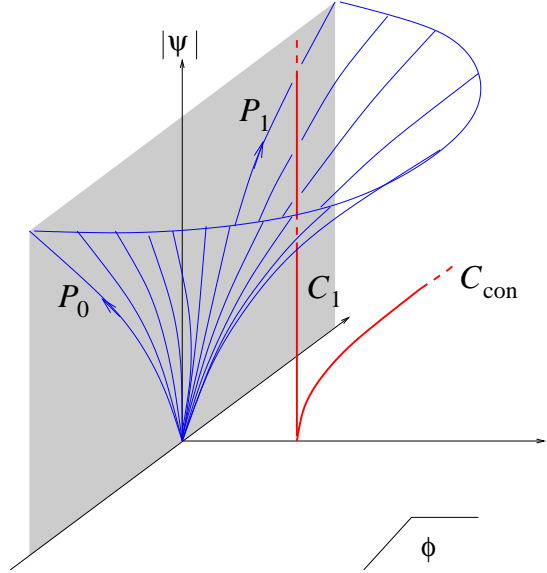}}
\caption{The homotopy ${\mathcal P}_s$. The shaded region is the
orientifold moduli space.}
\label{homotopyF}
\end{figure}
It intersects only with $C_1$ of the singular locus at
$t=1$ and $s={1\over 2}$. 
See Figure~\ref{homotopyF}.
Thus, we find that ${\mathcal P}_1$ is homotopic to
$-{\mathcal P}_{C_1}+{\mathcal P}_0+{\mathcal P}_{\infty}$
where ${\mathcal P}_{C_1}$ is the contour
that goes once around the singular locus $C_1$ and
${\mathcal P}_{\infty}$ is a contour that stays in the large volume limit.
See another figure, Fig.~\ref{paths}.

In \cite{Candelas}, the monodromy of the RR-charge for the contour
${\mathcal P}_{C_1}$ is computed
in a basis $\Pi^{\rm G}$ as
$$
\Pi^{\rm G}\to B\Pi^{\rm G};\qquad
B=\left(\begin{array}{cccccc}
1&0&0&0&0&0\\
0&1&-1&1&-1&1\\
0&0&2&-1&1&-1\\
3&-3&4&-3&3&-3\\
-3&3&-4&4&-2&3\\
-3&3&-3&3&-2&3
\end{array}\right)
$$
We also know that the intersection matrices with respect to the two bases are
related as $I^{\rm Cardy}=(1-A)I^{\rm G}(1-A)^T$
where
$$
A=\left(\begin{array}{cccccc}
0&1&0&0&0&0\\
0&0&1&0&0&0\\
0&0&0&1&0&0\\
0&0&0&0&1&0\\
0&0&0&0&0&1\\
-1&0&-1&0&-1&0
\end{array}\right)
$$
This implies that the two bases are related by
$$
\Pi^{\rm Cardy}=U\Pi^{\rm G};\quad
U=\pm (1-A)A^n,
$$
for some $n$.
In the Cardy basis, the monodromy along the contour $-{\mathcal P}_{C_1}$
is given by $\Pi^{\rm Cardy}\to UB^{-1}U^{-1}\Pi^{\rm Cardy}$.
Thus, the transformation of the charge basis along the path
${\mathcal P}_1$ is given by
\beq
\Pi^{\rm Cardy}=M_{{\mathcal P}_1}\Pi^{\rm LV};\quad
M_{{\mathcal P}_1}=UB^{-1}U^{-1}M_{{\mathcal P}_0}.
\eeq
It turns out that $U=(1-A)A^6$ is the right choice so that the Cardy branes 
canceling the tadpole at the Gepner point obey
the condition at the large volume with $B=-H+{3\over 2}L$:\\
\underline{$(+++++)$}
\beq
{\rm ch}(E)\e^{-B}
=32-12 H^2-4HL.
\eeq
This agrees with the large volume condition.\\
\underline{$(++-++)$ etc}
\beq
{\rm ch}(E)\e^{-B}
=-8H+{11\over 3}H^3
\eeq
This agrees with the large volume condition with
$\epsilon[M^{\tau}]=-[M^{\tau}]$.
\\
\underline{$(++--+)$ etc}
\beq
{\rm ch}(E)\e^{-B}
=4H^2-4HL
\eeq
This agrees with the large volume condition with
$\epsilon[C]=[C]$ and
$\epsilon[\ell_a]=[\ell_a]$.
\\
\underline{$(++---)$}
\beq
{\rm ch}(E)\e^{-B}
=-8H+16L-{1\over 3}H^3.
\eeq
The is agrees with the large volume condition
with $\epsilon[D]=-[D]$ and
$\epsilon[p_a]=-[p_a]$.

\subsubsection{Type of the O-planes}

We have determined the orientation $\epsilon[\sfX^{\tau}]$
of the O-plane in the two large volume regions, for
the $P_{1;\epsilon_1...\epsilon_5}$-orientifolds.
We would now like to know the type of each component of the O-plane.
We recall that there are roughly two types of O-planes
--- O$^-$ and O$^+$:
$N$ D$p$-brane on top of an O$^-p$-plane support $O(N)$ gauge
group while $N$ D$p$-brane on top of an O$^+p$-plane support
$Sp(N/2)$.\footnote{There is actually a finer classification
labeled by the discrete RR-flux \cite{WHWSH}, which we do not
discuss in the present paper.}
O$^-$-plane has a negative tension and O$^+$-plane has a positive tension.
Thus, in a supersymmetric configuration where the NSNS tadple
is also cancelled, it is impossible to have O-planes of type O$^+$ only.
In particular, if the O-plane consists of a single component, that must be
an O$^-$-plane.
Thus, the O9-plane for the $P_{*;+++++}$-orientifold
and the O7-plane for the $P_{*;++-++}$-orientifold must be both O$^-$,
as long as there are supersymmetric brane configurations at large volume.
For the $P_{1;+-***}$-orientifolds,
the O-plane has two (sets of) components
which are homologous to each other.
At the Gepner point,
we found that a configuration without a brane is supersymmetric
and tadpole canceling. Thus the two (sets of) O-plane components
are of the opposite types, as we have mentioned.
In what follows, we discuss the remaining cases, $P_{1;++--+}$
and $P_{1;++---}$.

Let us consider Type II orientifold on
a ten-dimensional manifold $\sfX$ with respect to an involution
$\tau$ of $\sfX$.
Let
$$
\sfX^{\tau}=\bigcup_iW_i
$$
be the decomposition of the O-plane into connected components.
Let $o_i=\pm 1$ be the type of the O-plane at $W_i$ ---
it is $\pm 1$ if $W_i$ is an O$^{\pm}$-plane.
As before we denote the orientation of the O-plane
by $\epsilon[W_i]$.
Then, the D-brane wrapped on $W_i$ preserves the same spacetime
supersymmetry as this O-plane at $W_i$ if its orientation
is $o_i\epsilon[W_i]$.
This also tells us the phase of the supersymmetry in the large volume limit.
To be specific, we consider the Type IIB orientifolds on a Calabi-Yau
three-fold. The overlap of the boundary state for a brane
wrapped on $W_i$ and the RR-ground state
$|0\rangle_{\RR}$ of the lowest R-charge
is given by
\begin{eqnarray}
{}_{\RR}\langle 0|B_{W_i}\rangle
&=&
\int_{[W_i]}\e^{-i\omega}+\cdots
\nonumber\\
&=&
(-i)^{\dim W_i}\int_{[W_i]}{\omega^{\dim W_i}\over
\dim W_i!}+\cdots
\nonumber
\end{eqnarray}
where $+\cdots$ are small in the limit $\omega\gg 1$.
Note that if $W_i$ is a complex submanifold and $[W_i]$ is the
standard orientation, $\int_{[W_i]}\omega^{\dim W_i}$ is positive.
Thus, in the large volume limit,
the phase of the supersymmetry preserved by the O-plane of type $o_i$
and orientation $\epsilon[W_i]=\epsilon_i\cdot [W_i]$ is given by
$$
\e^{i\theta_i}=o_i\epsilon_i(-i)^{\dim W_i}.
$$
In particular, in a supersymmetric orientifold, all the components
$W_i$ must have the same $o_i\epsilon_i(-i)^{\dim W_i}$.

This can be used to find the type of O-plane components
(up to an overall sign)
for the $P_{1;*****}$-orientifolds.
We recall the O-plane orientations for the four relevant cases:
$$
\begin{array}{l|l|l}
&B={H\over 2}+{L\over 2}&B={L\over 2}\\[0.1cm]
\hline
P_{1;++--+}&
\begin{array}{l}
\epsilon[C]=-[C]\\
\epsilon[\ell_a]=[\ell_a]
\end{array}
&
\begin{array}{l}
\epsilon[C]=-[C]\\
\epsilon[\ell_a]=-[\ell_a]
\end{array}
\\[0.2cm]\hline
P_{1;++---}&
\begin{array}{l}
\epsilon[D]=-[D]\\
\epsilon[p_a]=[p_a]
\end{array}
&
\begin{array}{l}
\epsilon[D]=[D]\\
\epsilon[p_a]=[p_a]
\end{array}
\end{array}
$$
Thus, for a common supersymmetry to be conserved, we need that
the types must be related as follows
$$
\begin{array}{l|l|l}
&B={H\over 2}+{L\over 2}&B={L\over 2}\\[0.1cm]\hline
P_{1;++--+}&
o_C=-o_{\ell_a}
&
o_C=o_{\ell_a}
\\[0.2cm]\hline
P_{1;++---}&
o_D=o_{p_a}
&
o_D=-o_{p_a}
\end{array}
$$
To fix the overall sign, we need to look at the tension of the O-plane.
Since we need some branes to cancel the tadpole in all cases,
we need the total tension to be negative.
Let us first discuss the $P_{1;++--+}$ orientifolds.
For a K\"ahler form $\omega=r_1H+r_2L$, we find
\begin{eqnarray}
&&
\int_{[C]}\omega+\sum_{a=1}^4\int_{[\ell_a]}\omega
=8r_1+8r_2
\nonumber\\
&&\int_{[C]}\omega-\sum_{a=1}^4\int_{[\ell_a]}\omega
=8r_1
\nonumber
\end{eqnarray}
Thus, the type of the O-plane at $C$ determines the sign of the total
tension, and it must always be O$^-$, $o_C=-1$.
Next, let us consider the $P_{1;++---}$ orientifolds.
In this case, the O-plane at $D$ has clearly
larger tension than the ones at $p_a$ in the large volume limit.
Thus, O-plane at $D$ must always be O$^-$, $o_D=-1$.
To summarize, we find that the type of the O-plane
components are given by 
$$
\begin{array}{|l|l|l|}
\hline
&B={H\over 2}+{L\over 2}&B={L\over 2}\\[0.1cm]\hline
P_{1;++--+}&
\begin{array}{l}
C:O^-\\
\ell_a:O^+
\end{array}
&
\begin{array}{l}
C:O^-\\
\ell_a:O^-
\end{array}
\\[0.2cm]\hline
P_{1;++---}&
\begin{array}{l}
D:O^-\\
p_a:O^-
\end{array}
&
\begin{array}{l}
D:O^-\\
p_a:O^+
\end{array}
\\\hline
\end{array}
$$

We found an interesting phenomenon.
{\it As we move from one large volume region to another,
through the non-geometric region of the constrained K\"ahler moduli space,
the type of O-plane changes}:
For the $P_{1;++--+}$-orientifold, the O5-plane at the rational curves
$\ell_a$ change from O$5^+$
to O$5^-$ while the O5-plane at the genus 9 curve
$C$ stays as O$5^-$.
For the $P_{1;++---}$-orientifold,
the O3-planes at $p_a$ change from O$3^-$ to O$3^+$
while the O7-plane at the divisor $D$ remains to be O$7^-$.
In this discussion, we have assumed that the sign of the total
O-plane tension remains the same as the Gepner point.
This can be justified by showing that the overlap
$\Pi^{\tau\Omega}_0$ does not vanish on a path from the Gepner point
to the large volume. (However, even if this assumption turns out
to be wrong, the change in the type of O-plane we have just discussed
remains true.)

This provide a challenge in finding
the classification scheme of D-brane charges
using K-theory that is valid uniformly in the moduli space.
For flat tori, it is known that
the D-brane charges in Type II orientifolds
are classified by using KR group \cite{WK,HT,BGH}
(see also \cite{Gukov,Stefanski}):
For $T^{9-p}/\Z_2$-orientifold
with all O$p^-$-planes (resp. all O$p^+$-planes),
the D-brane charges is classified by
KR$^{-(9-p)}$-group (resp. KR$^{-(5-p)}$-group).
One may guess that a similar rule applies when the space is
curved.
But the above phenomenon tells us that we need something very different
to describe the D-brane charges uniformly on the moduli space.

\subsection{Comments on Type IIA Orientifolds}

Let us consider
Type IIA orientifold on a large volume Calabi-Yau manifold
$M$ with respect to an antiholomorphic involution $\tau$ of $M$.
The O-plane is the fixed point set $M^{\tau}$ which is
a special Lagrangian submanifold, and 
the RR-flux generated by this is determined by the homology class
$[M^{\tau}]\in H_3(M;\Z)$.
The tadpole cancellation condition $n_i$ D6-branes wrapped on special
Lagrangian submanifolds $L_i$ are given by\footnote{We
assume that $M$ is simply connected, in which case
$K^1(M)=H_3(M,\Z)$.}
\beq
\sum_{i=1}^Nn_i[L_i]=4[M^{\tau}],
\eeq
where $[L_i]$ is the homology class of the submanifold
$L_i$.
One obvious solution to this condition is the configuration of
four D6-branes wrapped on $M^{\tau}$,
but other solutions may exist as well.
There can be a spacetime superpotential
depending on the $b_1(L_i)$ open string fields as well as
the K\"ahler moduli. There could even be open string fields
which are heavy at large volume limit but become light
in some interior regions of the K\"ahler moduli.
We would like to compare this with some results obtained at the Gepner
point.

Let us first consider the odd $H$ cases.
In each of such cases, we found a supersymmetric and tadpole canceling
configuration consisting of four copies of one brane
$B_{\bf {k-1\over2},{k-1\over 2}}$.
By comparison with the large volume condition,
it is natural to identify the brane $B_{\bf {k-1\over2},{k-1\over 2}}$
as the brane wrapped on the fixed point set $M^{\tau}$.
For example, in the case of the quintic,
$B_{\bf 1,1}$ is identified as the D6-brane wrapped on
the real quintic which has a topology of $\RP^3$.
In any of the odd $H$ cases,
the open string spectrum at the Gepner point includes
massless chiral multiplets charged under
$O(4)$ --- one or more symmetric tensors and in some cases
antisymmetric tensors as well (see Section~\ref{subsubsec:oddH}).
Let us consider the quintic case, where
there is a single massless matter in the symmetric
representation.
Can it be consistent with the large volume result?
At large volume, there is no open string moduli since
the D-brane is wrapped on a simply connected submanifold $\RP^3$.
However, as noted in \cite{BDLR}, there are choices in specifying
the supersymmetric configuration --- the choice of the flat gauge
connection on the brane.
In the present case where the gauge group is $O(4)$, this is given by the
$\pi_1(\RP^3)=\Z_2$ holonomy, namely, an element of
$O(4)$ which squares to $1$. This is up to gauge transformation, and thus the
vacuum manifold is
\beq
{\it V}_{\rm LV}=\Bigl\{\,g\in O(4)\,\Bigl|\,
g^2={\bf 1}_4\,\Bigl\}/{\rm ad} O(4),
\label{LVVac}
\eeq
where $/{\rm ad} O(4)$ is quotient by the action
$g\mapsto hgh^{-1}$, $h\in O(4)$.
Is there a field theory model consistent with this and
the massless spectrum at the Gepner point?
This problem is encountered in \cite{BDLR} in the context of a single
brane in a theory without orientifold in which
there are two vacua at large volume ${\it V}^{\rm single}_{\rm LV}=\{\pm 1\}$:
The model is
given by the superpotential $W=\phi \psi+\psi^3$ where $\phi$ is
a closed string field representing the K\"ahler
class and $\psi$ is the open string field.
If we identify $\phi=0$ as the Gepner point, $\psi$ is massless at the Gepner
point and at large volume there are two vacua
$\psi=\pm \sqrt{-\phi/3}$, which is consistent with
${\it V}^{\rm single}_{\rm LV}=\{\pm 1\}$.
A natural extension to the current situation is 
the theory with superpotential
\beq
W=\phi\Tr\psi +\Tr\psi^3,
\label{Wphipsi}
\eeq
where $\phi$ is the closed string field
representing the K\"ahler class and
and $\psi$ is the symmetric tensor for $O(4)$,
$\psi_{ij}=\psi_{ji}$, $i,j=1,2,3,4$.
The Gepner point is identified as $\phi=0$ where $\psi$ is massless.
Away from that point, say $\phi=-3$, the vacuum equations for
$\psi$ are
$$
[\psi,\psi^{\dag}]=0,\quad
\psi^2={\bf 1}_4,
$$
where the first equation is the D-term equation
(with $\psi^T=\psi$ taken into account) and the second equation
is the F-term equation, $\partial_{\psi}W=0$.
The vacuum manifold is obtained by moding out the solution space
by the adjoint $O(4)$ action.
By the D-term equation, $\psi$ can be diagonalized by $U(4)$ matrix
and it then follows from $\psi^T=\psi$ that $\psi$ is a real matrix.
Thus, $\psi$ is a four-by-four real matrix with
the constraint $\psi^T\psi=\psi^2={\bf 1}_4$.
Namely, the vacuum manifold
agrees with one at the large volume (\ref{LVVac}).
In this discussion, we have treated the K\"ahler modulus
$\phi$ as a parameter. Of course, 
in the full string theory,  we must treat $\phi$ as a dynamical field and
include $\partial_{\phi}W=0$ into the
vacuum equations. Then, we obtain an extra constraint
$\Tr \psi=0$ which means that $\psi$ has the same number of
$+1$ or $-1$ eigenvalues.
It would be interesting to find a similar field theory model for the odd $H$
cases other than the quintic.

Let us next consider an even $H$ case, the two parameter model
with $k_i=(6,6,2,2,2)$.
There is a freedom to dress by quantum symmetry, but
we only consider those without dressing for which
the large volume region is included in the moduli space.
There are six such cases $P^A_{+;00000}$,
$P^A_{+;00001}$, $P^A_{+;00011}$, $P^A_{+;00111}$
$P^A_{+;01000}$. $P^A_{+;01001}$.
At the Gepner point,
only one of them, $P^A_{+;01000}$, admits a tadpole canceling
supersymmetric solution
with exactly four copies of an elementary brane.
This brane, $B_{(30111),(30111)}$, is identified in the large volume limit
as the O-plane which has topology of $S^3$.
Since this is simply connected, there is a unique supersymmetric
configuration at the large volume.
At the Gepner point, we found no massless matter field.
A theory consisting of all these is the one
with only $O(4)$ super-Yang-Mills without matter and
exactly flat K\"ahler moduli space.
In all of the five other cases, we found that
there is no consistent supersymmetric configuration
with only four branes at the Gepner point
(Section~\ref{subsec:general}),
while ``four D6-branes wrapped
on the O-plane'' is always a solution at large volume.
Note that the O-plane has $b_1\geq 1$ in these cases,
and there are massless open string fields that correspond
to moving pairs of D6-branes away from the O-plane,
breaking $O(4)$ to $U(2)$ or further to $U(1)^2$.
One can expect a non-trivial superpotential depending on such open string
fields as well as K\"ahler moduli, and it is conceivable that the
supersymmetric vacua with unbroken $O(4)$ misses the Gepner point.
It is an interesting problem to verify it by explicit computation of
superpotential. Another interesting problem is
to analyze the interaction of
the supersymmetric solutions we found at the Gepner point and try to
connect to the large volume limit.

\section*{Acknowledgement}

We would like to thank M. Cvetic, C. Doran,
M. Douglas, J. Giedt, J. Gomis, M. Haack, M. Kapranov,
G. Mikhalkin, E. Poppitz,
R. Rabadan, A. Uranga for discussions. K.H. thanks Banff 
International Research Centre, Stanford University, and KITP 
Santa Barbara for hospitality during various stages of this work.
K.H. and K.H. were supported in part by Natural Sciences and Engineering
Research Council of Canada. K.H. is also supported by the Alfred P. Sloan
Foundation. The research is also supported in part by the National Science
Foundation under Grant No. PHY99-07949 and the PPARC grant
PPA/G/O/2000/00451.

\appendix{More General Gepner Models}\label{app:Gepner}

A Gepner model is defined as the orbifold of the product of
minimal models $\prod_{i=1}^rM_{k_i}$ 
by the group $\Gamma\simeq \Z_H$ ($H:={\rm l.c.m}\{k_i+2\}$)
generated by $\gamma=(\g,...,\g)$ with
$\g=\e^{-2\pi i J_0}(-1)^{\widehat{F}}$.
It can be used to define a fine compactification to
$3+1$ dimensions under the central charge condition
\beq
c=\sum_{i=1}^r{3k_i\over k_i+2}=9,
\label{ccondition}
\eeq
and the condition on the number of factors
\beq
r:\,\,\,{\rm odd}.
\label{oddc}
\eeq
The second condition is needed in order for
the RR-charge of the lowest R-charge (corresponding to the holomorphic
volume-form of the corresponding Calabi-Yau) to survive the orbifold
projection.

Gepner models coming from the
linear sigma models of the type described in Section~\ref{subsec:CYLG}
always have $r=5$.
But there are other models as well.
The equation (\ref{ccondition}) has
solutions with various number of factors, starting with $r=4$.
$r=4$ solutions can be made into $r=5$ by adding a single $k=0$ factor.
But there are solutions with $r>5$:
According to \cite{Gepnerclassification},
there are twenty-one models with minimal $r\geq 6$
 --- fourteen with $r_{\rm min}=6$,
four with $r_{\rm min}=7$, two with
$r_{\rm min}=8$ and one with $r_{\rm min}=9$.
(A solution with even $r$ {\it must be}
 added an odd number of $k=0$ factors
so that the condition (\ref{oddc}) is obeyed.)
To be complete we consider these more general cases in this Appendix.

One important identity is
\beq
\mu:=\sum_{i=1}^r\left(1-{1\over k_i+2}\right)
={r+3\over 2}
\label{mudef}
\eeq
where we have used the central charge condition (\ref{ccondition}).
If we use the second condition (\ref{oddc}),
we find that $\mu$ is an integer. Namely $\sum_i{1\over k_i+2}$
is an integer. We have implicitly assumed this
in the construction of the crosscap states: Look at the
the sign factor $(-1)^{\sum_i{\nu\over k_i+2}}$
in the RR part of the crosscap state (\ref{CAAA}).
This does not make sense unless $\sum_i{1\over k_i+2}$
is an integer.

In the main text of the paper starting Section~\ref{subsec:total}
we have assumed that $r=5$. Here we would like to present some formula
that is valid for all cases (with odd $r$).
The relation between
$\e^{\pi i J_0}|\Scr{C}_{P}\rangle$ and $|\Scr{C}_{(-1)^FP}\rangle$
changes (for both A and B types) by the sign factor $(-1)^{\mu}$.
Thus, the formulae for the total crosscap states are modified as
\beqa
(\ref{CARtotal})\longrightarrow&&
|C_{\omega;{\bf m}}\rangle_{\RR}
=|\Scr{C}_{P^A_{\omega;{\bf m}}}\rangle\otimes
|\Scr{C}_+^{\rm \,st}\rangle_{\RR}
-(-1)^{\mu}\omega
|\Scr{C}_{(-1)^FP^A_{\omega;{\bf m}}}\rangle\otimes
|\Scr{C}_-^{\rm \,st}\rangle_{\RR},
\nn\\
(\ref{CBRtotal})\longrightarrow&&
|C^B_{\omega;{\bf m}}\rangle_{\RR}
=|\Scr{C}_{P^B_{\omega;{\bf m}}}\rangle\otimes
|\Scr{C}_+^{\rm \,st}\rangle_{\RR}
-(-1)^{\mu}\tilomega^{-1}
|\Scr{C}_{(-1)^FP^B_{\omega;{\bf m}}}\rangle\otimes
|\Scr{C}_-^{\rm \,st}\rangle_{\RR},
\nn
\eeqa
If $\mu$ is even (note that $\mu=4$ (even) if $r=5$),
there is no difference in the discussion after 
Section~\ref{subsec:total}.
But there is some change if $\mu$ is odd.
The largest effect is in the action of parities on the branes.
The transformation rules (\ref{PAonD})-(\ref{PAonDshort})
 and (\ref{PBonBb})-(\ref{PBonBshort})
changes by sign (orientation).
As a consequence, this affects the set of parity invariant D-branes.
The analysis of the structure of
Chan-Paton factor goes through as in the discussion of
$r=5$, with an obvious modification of the result.

Addition of two $k=0$ factors shifts even $\mu$ to odd $\mu$ and vice versa.
What we have seen is that
this has a non-trivial effect on the physics involving branes in the
orientifold model.
In fact, without orientifold, addition of
even number of $k=0$ factors makes no difference since
the orbifold action is trivial on such pair of $k=0$ factors.
This is also true for the case involving D-branes (before orientifold):
for A-branes, the orbifold group flips the orientation of
the brane in a $W=X^2$ factor
but a pair of such flips cancel against each other.
For B-branes the same can be said (this is known as
the Kn\"orrer periodicity \cite{Knorrer}).
However, with an orientifold, this step 2 periodicity
is doubled to step 4.
This reminds us of the Bott periodicity:
complex K-theory has periodicity 2 but
Real K-theory has periodicity 8.

\appendix{Some Detail}\label{app:detail1}

We explain the projection factors (\ref{AshProj}) and (\ref{AshPA})
that are used to read the parity action on short-orbit branes.

We first compute the $\Z_2$ projection factor for the open
string stretched from a short-orbit brane
$\widehat{B}^{(\varepsilon)}_{\bf \bar{L},\bar{M}}$
to another $\widehat{B}^{(\varepsilon')}_{\bf L,M}$.
To this end, let us consider the loop-channel expansion of
the relevant overlaps in the minimal model with even $k$,
\beqa
 && \langle\Scr{B}_{\frac{k}{2},\bar{M},\bar{S}}|
  q_t^{H}|\Scr{B}_{\frac{k}{2},M,S}\rangle_{\NSNS\atop \RR}
 = \sum_{l\atop s=0,1}N_{\frac{k}{2}\frac{k}{2}}^l
     (\pm1)^s\chi_{l,\bar{M}-M,\bar{S}-S+2s}(\tau_l),
 \nonumber \\
 && \langle\Scr{B}_{\frac{k}{2},\bar{M},\bar{S}}|
  q_t^{H}|\Scr{B}_{\frac{k}{2},M,S}\rangle_{(\mp1)^Fa^{k+2}}
 = \sum_{l\atop s=0,1}N_{\frac{k}{2}\frac{k}{2}}^l
     (\pm1)^s(-1)^{\frac{1}{2}(l+\bar{M}-M-\bar{S}^2+S^2)}
\chi_{l,\bar{M}-M,\bar{S}-S+2s}(\tau_l)
 \nonumber\\
\eeqa
This is enough to see 
the open string states labeled by
$\otimes_{i=1}^r(l_i,n_i,s_i)$ are subject to the projection
\beq
  \frac12\left(1+\gamma^{H/2}\right) ~=~
  \frac12\left(1+\varepsilon\varepsilon'
 \prod_{w_i\,{\rm odd}}(-1)^{\frac{1}{2}(l_i+n_i-s)}\right),
%\label{AshProj}
\eeq
where $s=0$ for NS states and $1$ for R ones.
This is nothing but (\ref{AshProj}).

Let us next find the projection factor that appears in the
parity twisted partition function.
To do this, let us consider the minimal model with even $k$,
and take the M\"{o}bius strip amplitude
\beqa
\lefteqn{
  _{\NSNS}\langle\Scr{B}_{\frac{k}{2},M,S}|q_t^H
  |\Scr{C}^A_{(\pm1)^Fg^m\widetilde{P}_A}\rangle
} \nonumber \\ &=&
   \sum_l\delta^{(2)}_l(-1)^{m+\frac{l}{2}}
   \left\{e^{\mp\frac{\pi i}{4}}\hat{\chi}_{l,2M-2m,2S}
         -e^{\pm\frac{\pi i}{4}}\hat{\chi}_{l,2M-2m,2S+2}\right\}(\tau)
\label{mobminev}
\eeqa
where
\beq
 \hat{\chi}_{l,n,s}(\tau) =
 e^{-\pi i(\frac{l(l+2)-n^2}{4k+8}+\frac{s^2}{8}-\frac{c}{24})}
 \chi_{l,n,s}(\tau+1/2).
\eeq
Let us see how it changes under the shift
$m\to m+\frac{k+2}{2}$.
This tell us that the open string state $(l,n,s)$ on the M\"{o}bius
strip (\ref{mobminev}) has the following eigenvalue of $g^{\frac{k+2}{2}}$:
\beq
 g^{\frac{k+2}{2}} = \pm i(-1)^{\frac{l+n}{2}+S}.
\label{gkact}
\eeq
One can also perform a similar analysis on the RR sector states.
Let us then take a short orbit A-brane
$\widehat{B}^{\pm,A}_{{\bf L,M}}$ and an A-parity
$P^A_{\omega,{\bf m}}$.
Using (\ref{gkact}) for each minimal model, we can now easily
find the eigenvalue of $\gamma^{H/2}$ for the open string state
$\otimes_{i=1}^r(l_i,n_i,s_i)$ on the M\"{o}bius strip.
The result is independent of the choice of branes and
reads
\beq
  \gamma^{H/2} = \omega^{\frac{H}{2}}(-1)^{\frac{\sigma}{2}}
                 \prod_{w_i~{\rm odd}}(-1)^{\frac{1}{2}(l_i+n_i-s)},
\eeq
where $s=0$ for NS states and $1$ for R ones as before, and
$\sigma$ is the number of $i$'s such that $w_i$ is odd.
This leads to the projection factor (\ref{AshPA}).

\appendix{Integral bases of three-cycles in the quintic}
\label{integral}

In the main text, we have seen that the most convenient
way to write down and solve the tadpole conditions is to
find an integral basis of the charge lattice, so that the
coefficients in the equations (\ref{tadcan}) are manifestly
integer. In this appendix, we describe how such a basis
can be found for A-type branes on the quintic. We will also
describe an alternative way of solving the tadpole
conditions using $g_i$ polynomials.

\subsubsection*{Integral basis}

In a single minimal model with $k=3$, the charges of
A-type branes span a 4-dimensional lattice which is
generated by the $5$ branes $\Scr{B}^A_{0,M,1}$ with
$M=1,3,5,7,9$. As is easiest to see in the Landau-Ginzburg
picture, these $5$ charges satisfy one linear relation:
Their sum is zero. We can fit the set $\Lambda$ of 
five charges modulo this one relation into an
exact sequence
\beq
0\exact \{R\}\exact \A_5\exact \Lambda
\exact 0
\label{exactsingle}
\eeq 
where $\A_5$ stands for the set $\{n=0,1,2,3,4\}$
representing the charges $Q_n$ of the brane
$\Scr{B}^A_{0,2n+1,1}$, and $R$ is the relation
$$
R \mapsto \sum_{n} Q_n
$$
There is an obvious $\Z_5$ action on $\A_5$ and on
(\ref{exactsingle}) which cyclically
permutes the five elements, and leaves $R$ invariant.

If we now take the tensor product of $5$ such minimal models,
the charge lattice has dimension $4^5=1024$. It is generated
by the tensor products $Q({\bf n}) = \prod_i Q_{n_i}$ modulo
the relations
$$
R_1(i;{\bf n}) = \sum_{n_i} Q(n_1,\ldots,n_5)
$$
where ${\bf n}=(n_1,\ldots,\widehat{n_i},\ldots n_5)$. Thus
we have $5^5$ charges with $5^5$ relations, but these 
relations are not all independent. Namely, we have the
relations between relations
$$
R_2(i,j;{\bf n}) = \sum_{n_i} R_1(j;({\bf n},n_i))
- \sum_{n_j} R_1(i;({\bf n},n_j))
$$
where now ${\bf n}=(n_1,\ldots,\widehat{n_i},
\ldots,\widehat{n_j},\ldots, n_5)$.
Continuing this way, we obtain the long-exact sequence for
the set of charges $\Lambda^{\rm ten}$ of the tensor product
\beq
0
\exact \{R_5\}
\exact 5 \A_5
\exact 10 (\A_5)^2
\exact 10 (\A_5)^3
\exact 5 (\A_5)^4
\exact (\A_5)^5
\exact \Lambda^{\rm ten}
\exact 0
\label{exacttensor}
\eeq
from which we see that the dimension of the charge lattice
is indeed $4^5$. The advantage of this representation is that
it is now trivial to take the $\Z_5$ orbifold. All
relations $R_s$ with $s<5$ are related to one another under the
diagonal $\Z_5$ action, while $R_5$ is invariant. Thus,
the untwisted charges can be represented by the sequence
\beq
0\exact \{R_5\}
\exact 5 (\A_5)^0
\exact 10 \A_5
\exact 10 (\A_5)^2
\exact 5 (\A_5)^3
\exact (\A_5)^4
\exact \Lambda^{\rm Gep}
\exact 0
\label{exactgepner}
\eeq
from which we read off the dimensions of the charge lattice
of the Gepner model to be $204$, as expected. To obtain a basis 
of $\Lambda^{\rm Gep}$, we take a section of (\ref{exactgepner}). 
In view of solving the tadpole conditions, it is most useful 
to do this in such a way that respects the action of the parity. 
The parity acts in a single minimal model on $Q_n$ as $0\mapsto 4$,
$1\mapsto 3$, and $2\mapsto 2$, and similarly on all the
relations. We will show how this language simplifies finding
the explicit form of (\ref{tadcan}) in appendix
\ref{orbifold}.

\subsubsection*{Solving the tadpole conditions with
$g_i$ polynomials}

The tadpole cancellation condition can also be written in a
simple form by using the $g_i$-polynomials.
Here we again restrict our attention to the RR-charges and
tadpoles sitting in the untwisted sector.
Let us introduce the $(k_i+2)$-dimensional shift matrices
$g_i$ satisfying $g_i^{k_i+2}=1$,
and associate the following polynomial to each D-brane
\beq
 Q_{\bf L,M}(g_i)=\prod_{i=1}^5
 \left(\sum_{n_i=0}^{L_i}g_i^{n_i+(M_i-L_i)/2}\right),
\label{QLM}
\eeq
representing its RR-charge.
Let us also associate similar $g_i$-polynomials to the orientifolds
by first expressing their RR-charges in terms of D-branes and
then using the above formula.
These polynomials of $g_i$ are useful in computing the (twisted)
Witten indices between D-branes and O-planes.
The index between the branes $B$ and $B'$ is given by
the diagonal element(more precisely, the product of diagonal elements)
of the matrix
\begin{equation}
  I = Q_B(g_i)Q_{B'}(g_i^{-1})\prod_{i=1}^5(1-g_i^{-1})
  \sum_{\nu=1}^H(g_1g_2g_3g_4g_5)^\nu
\end{equation}
Since the last factor in the right hand side is the projection
onto the states on which $g_5^{-1}=g_1g_2g_3g_4$, we can eliminate
$g_5$ and obtain
\begin{eqnarray}
  I &=& Q_B(g_i)Q_{B'}(g_i^{-1})
 (1-g_1^{-1})(1-g_2^{-1})(1-g_3^{-1})(1-g_4^{-1})(1-g_1g_2g_3g_4)
 \nn\\ && \times
 \sum_{\nu=1}^{w_5}(g_1g_2g_3g_4)^{\nu(k_5+2)},
\end{eqnarray}
where the index is read off as the diagonal elements.
This agrees with the formula for quintic given in \cite{BDLR}.
The RR-charges of any configurations of branes and the O-plane
are therefore expressed as polynomials of $g_i$,
\beq
  (\mbox{RR-charge})=
  \sum_{m_i=0}^{k_i+1}
  N_{m_1m_2m_3m_4m_5}g_1^{m_1}g_2^{m_2}g_3^{m_3}g_4^{m_4}g_5^{m_5}.
\label{gpol}
\eeq
The configuration is free of tadpoles when the sum of $g_i$-polynomials
of the constituent D-branes and the O-plane vanishes up to
\beq
 1+g_i+\cdots+g_i^{k_i+1}=0,~~~
 g_1g_2g_3g_4g_5=1.
\label{defgi}
\eeq
It is cumbersome to have these equivalence relations in
analyzing the polynomial.
Therefore, it is more convenient to use the relations
(\ref{defgi}) to bring the polynomial into the following gauge
\beq
 \sum_{m_a=0}^{k_a+1}N_{m_1\cdots m_a\cdots m_5}=0,~~~
 N_{m_1m_2m_3m_4m_5}=N_{m_1+1,m_2+1,m_3+1,m_4+1,m_5+1}
\label{gauge}
\eeq
and see whether each coefficient is vanishing or not.
This is certainly possible because each of the polynomials
$Q_{\bf L,M}(g_i)$ can be brought to this gauge (\ref{gauge})
in the following way.
\beqa
 Q_{\bf L,M}(g_i) &=&
 \frac1H\sum_{\nu=0}^{H-1}
 \sum_{n_i=0}^{k_i+1}\prod_{i=1}^5
 (f_{L_i,n_i}g_i^{n_i+\nu+(M_i-L_i)/2)}),\nn\\
 f_{L_i,n_i} &=& \left\{
 \begin{array}{ll}
  1-{\textstyle\frac{L_i+1}{k_i+2}}&(0\le n_i\le L_i)\\
   -{\textstyle\frac{L_i+1}{k_i+2}}&({\rm otherwise})
 \end{array}\right.
\eeqa

As was noted before, these $g_i$-polynomials can only express the
RR-charges corresponding to polynomial deformations of hypersurfaces
defining the target space.
In the two parameter model with $(k_i+2)=(8,8,4,4,4)$ there are
six missing RR charges sitting in the twisted sector.
One can develop a similar argument using polynomials
for those RR-charges, too.

We can again see that the number of independent components of 
$N_{m_1m_2m_3m_4m_5}$ agrees with the number of RR ground states in 
the untwisted sector, which take the form (\ref{ground}),
\beq
  |l_i\rangle = \prod_{i=1}^5|l_i,l_i+1,1\rangle\times|l_i,-l_i-1,-1\rangle~~~
  (1\le l_i+1\le k_i+1,~\sum_{i}{\textstyle\frac{l_i+1}{k_i+2}}\in\Z)
\eeq
To see this, let us take the Fourier transform of $N_{m_1m_2m_3m_4m_5}$:
\beq
  \widetilde{N}_{n_1n_2n_3n_4n_5}
 = \sum_{m_i\in\Z_{k_i+2}}
   \exp(\sum_i\frac{2\pi im_in_i}{k_i+2})N_{m_1m_2m_3m_4m_5}
\eeq
Then $\widetilde{N}_{n_1n_2n_3n_4n_5}$ are nonzero only when
$n_i$ are all nonzero mod $k_i+2$ and $\sum_i\frac{n_i}{k_i+2}\in\Z$,
which is the same condition as the RR ground states satisfy
under the identification $n_i\leftrightarrow l_i+1$ (mod $k_i+2$).
So $\widetilde{N}_{n_1n_2n_3n_4n_5}$ has as many independent
components as there are untwisted RR vacua.

Of all the linear equations, there are some equations
among the tadpole cancellation conditions in which
all the D-branes appear with positive definite coefficients.
These essentially come from the overlaps with the RR ground states
$|\nu\rangle$, namely, those with $l_i=\nu$ mod $k_i+2$ for all $i$.
The reason for the positivity is that the overlaps of D-branes
or O-planes with any of these states have the same phases
if they preserve the same spacetime supersymmetry.
These equations are particularly important,
because they ensure that there are only finite number of
tadpole canceling configurations.
These equations also contain the condition that the sum of D-brane
tensions must cancel the O-plane tension.
One can obtain these special equations from
$g_i$-polynomials by setting $g_i=g^{w_i}$, where $g$ is
a $H$-dimensional shift matrix.

\appendix{Tadpole conditions for $\Z_5$ orbifold of quintic}
\label{orbifold}

In this appendix, we discuss the tadpole conditions for Type
IIB orientifolds of the orbifold of the quintic by the $\Z_5$
symmetry called $[1,4,0,0,00]$ in the notation of \cite{GreenePlesser}.
By mirror symmetry, this is equivalent to Type IIA orientifold
of $(\Z_5)^2$ orbifold of the quintic. In B-type language, the
model has $2h^{1,1}+2=12$ RR charges to cancel (the fact
that $h^{2,1}=49$ will not be important). The analogs of RS
branes in such orbifolds have been discussed for instance in 
\cite{BD}, and are also straightforward to obtain in the 
Landau-Ginzburg picture. It is easy to see that the branes 
are labeled as $\Scr{B}_{{\bf L},{\bf M},S}$ with ${\bf L}$ 
as before and ${\bf M}=(M_1,M_2,M_3,M_4,M_5)$ modulo a $(\Z_5)^3$ 
identification.

We can present an integral basis of the charge lattice similarly
to (\ref{exactgepner}) via
\beq
0\exact {R_3}
\exact 3 (\A_5)^0
\exact 3 \A_5
\exact (\A_5)^2
\exact \Lambda^{\rm orbi}
\exact 0
\label{exactorbi}
\eeq
such that $\Lambda^{\rm orbi}$ indeed has dimension $25-15+3-1=12$.
We now take a section through (\ref{exactorbi}) and specify an
integral basis as the charges of the branes with ${\bf L}={\bf 0}$
and ${\bf M}=2 {\bf n}+1$ with
$$
\begin{array}{l}
{\bf n}\in\Bigl\{
[0,0,2,2,2],
[0,4,2,2,2], 
[1,0,2,2,2], 
[1,1,2,2,2], 
[1,3,2,2,2], 
[1,4,2,2,2],\\
\qquad
[4,4,2,2,2], 
[4,0,2,2,2], 
[3,4,2,2,2], 
[3,3,2,2,2], 
[3,1,2,2,2], 
[3,0,2,2,2]\Bigr\}
\end{array}
$$
where $[\cdots]$ denotes $(\Z_5)^3$ orbits and where the second 
line is obviously the parity image of the first.

As before, branes preserving the same supersymmetry as the O-plane have 
arbitrary ${\bf L}$ and $\sum M_i=0\bmod 5$ and appropriate $S$ label 
depending on the parity of $\sum L_i$. Obviously, the charge of
such branes does not depend on the permutations of $L_3,L_4,L_5$, 
so we have the following representatives of ${\bf L}$ labels
\beq
\begin{array}{ll}
{\bf L}= 
&(0,0,0,0,0), \\
&(1,0,0,0,0),(0,1,0,0,0),(0,0,1,0,0),\\
&(1,1,0,0,0),(1,0,1,0,0),(0,1,1,0,0),(0,0,1,1,0),\\
&(1,1,1,0,0),(1,0,1,1,0),(0,1,1,1,0),(0,0,1,1,1),\\
&(1,1,1,1,0),(1,0,1,1,1),(0,1,1,1,1)\\
&(1,1,1,1,1)
\end{array}
\eeq
where we have ordered the branes according to increasing tension.
The possible ${\bf M}$ labels are ($\bmod 5$ and modulo $(\Z_5)^3$):
\beq
{\bf M}=
[0,0,0,0,0],
[1,4,0,0,0], 
[4,1,0,0,0],
[2,3,0,0,0],
[3,2,0,0,0] 
\eeq
The first of these is obviously invariant under the parity,
while the others are each others image. Thus, we have a total
of $16\times 3=48$ different charges to consider in the
tadpole cancellation. We will denote by $n_1,\ldots
n_{48}$ the number of times a given charge appears.

By utilizing the well-known expressions for the charges of
branes in the minimal model (see (\ref{QLM}) or (\ref{branec})),
we can compute the RR charges of all branes on the list in
the basis described above. This leads to the following
$6$ equations on the $48$ $n_i$'s
\beq
(n_1,\ldots,n_{48})
\left(
\hbox{\scriptsize
$\begin{array}{cccccc}
2&2&0&0&0&0\\
2&2&1&0&0&0\\
1&1&1&-1&0&0\\
2&0&0&1&0&-1\\
2&0&0&1&-1&1\\
1&0&1&0&1&0\\
2&0&0&1&0&1\\
2&0&1&1&1&-1\\
1&0&0&0&-1&0\\
2&0&-1&2&0&0\\
2&0&1&0&0&0\\
1&0&1&0&0&0\\
4&2&2&0&0&0\\
4&2&1&0&0&0\\
2&1&0&1&0&0\\
4&2&0&1&-1&-1\\
4&2&2&0&1&0\\
2&1&1&0&0&1\\
4&2&1&1&1&1\\
4&2&1&0&-1&0\\
2&1&1&0&0&-1\\
4&2&-1&2&0&0\\
4&2&2&0&0&0\\
2&1&2&-1&0&0\\
6&2&3&0&0&0\\
6&2&1&2&0&0\\
3&1&0&1&0&0\\
6&2&0&2&-1&-2\\
6&2&2&1&0&1\\
3&1&2&0&1&1\\
6&2&1&2&1&2\\
6&2&2&1&0&-1\\
3&1&1&0&-1&-1\\
6&2&-2&4&0&0\\
6&2&3&0&0&0\\
3&1&3&-1&0&0\\
10&4&5&0&0&0\\
10&4&2&2&0&0\\
5&2&0&2&0&0\\
10&4&0&3&-2&-3\\
10&4&4&1&1&1\\
5&2&3&0&1&2\\
10&4&2&3&2&3\\
10&4&3&1&-1&-1\\
5&2&2&0&-1&-2\\
16&6&8&0&0&0\\
16&6&3&4&0&0\\
8&3&0&3&0&0
\end{array}$}
\right)
=(32, 12, 0, 12, 0, 0)
\label{orbicond}
\eeq
Inspection reveals that the first two of these equations are nothing
but the equations (\ref{condition}) that we have solved in the context
of B-type orientifold of quintic. To see this, one has to take into 
account that for branes not invariant under parity, $n_i$ denotes the 
number of times the brane and its image under parity appear, and the 
fact that the mass depends only on the number of $1$'s in the ${\bf L}$ 
label. Thus, to find solutions of (\ref{orbicond}), we can take some
solution of (\ref{condition}) and scan through all ways of distributing
this mass among the branes on the list of $48$ with the same tension.

Here are a few examples of tadpole canceling brane configurations for the
$\Z_5$ orbifold of the quintic obtained in this way.
\\
\underline{Example 1}
$$
(B_{(00111), (20111)} + \mbox{\footnotesize image})
+(\overline{B_{(11110), (39116)}} + \mbox{\footnotesize image}
)
+2 B_{(11111), (11111)}
$$
\underline{Example 2}
\beqa
&4 (\overline{B_{(00000), (20666)}} +\mbox{\footnotesize image})
+2 (B_{(00100), (20166)} +\mbox{\footnotesize image}) 
+B_{(00100), (66166)} 
\nn\\ 
&+(B_{(01110), (25116)}+\mbox{\footnotesize image}) + 
(B_{(10000), (34666)}+\mbox{\footnotesize image}) + 
3(B_{(10000), (70666)}+\mbox{\footnotesize image})
\nn\\
& + 
\overline{B_{(11110), (11116)}}
\nn
\eeqa
\underline{Example 3}
\beqa
&
3 (\overline{B_{(00000), (20666)}}+\mbox{\footnotesize image})
 + 
\overline{B_{(00000), (66666)}} + 
(\overline{B_{(00000), (84666)}}+\mbox{\footnotesize image})
\nn\\
& + 
5 (B_{(00100), (20166)}+\mbox{\footnotesize image}) + 
B_{(00100), (66166)} + 
2B_{(01000), (61666)} + 
2 B_{(10000), (16666)} \nn\\
& + B_{(11111), (11111)}
\nn
\eeqa

\subsubsection*{Particle spectrum}

The spectrum of massless matters for these brane configurations
is analyzed in a similar way as in the case of ordinary quintic.
Here we only present the results.
{\footnotesize
\begin{center}
\begin{tabular}{|c|cc||cc|cc|c|}
\hline
${}^{\rm gauge}_{\rm group}$ &
$\sharp$&${\bf L},~{\bf M},~S$ &
1&2&3&4&5 \\\hline\hline
\lw{$U(1)$}
&1&(00111),(20111),0 & 12 &  0 &  0 &  3 &  0 \\
&2&(00111),(02111),0 &  0 & 12 &  3 &  0 &  0 \\ \hline
\lw{$U(1)$}
&3&(11110),(39116),2 &  0 &  3 & 12 & 7+3& 10 \\
&4&(11110),(93116),2 &  3 &  0 & 7+3& 12 & 10 \\ \hline
$O(2)$
&5&(11111),(11111),0 &  0 &  0 & 10 & 10 & 13+12 \\
\hline
\end{tabular}
\vskip2mm
Example 1
\end{center}
}

The first example consists of five kinds of branes,
$B_1+B_2+B_3+B_4+2B_5$, where $B_5$ is parity invariant
and $B_2,B_4$ are parity images of $B_1,B_3$.
The spectrum of chiral matters is summarized in the table above.

The gauge group is $U(1)^2\times O(2)$, and the labels
$({\bf L,M},S)$ of five D-branes are presented in the
second column.
The $5\times5$ numbers give the multiplicities of chiral primary
states on $i\!-\!j$ string ($i,j=1,\cdots,5$).
$3\!-\!4, 4\!-\!3$ and $5\!-\!5$ strings are parity invariant,
and they belong to symmetric or antisymmetric tensor representations
of gauge group according to their parity eigenvalues.
The numbers $7+3$ or $13+12$ represent the multiplicities of
symmetric and antisymmetric representations.

The table contains nine blocks.
Upper off-diagonal blocks are related with lower
off-diagonal ones by parity, namely,
the multiplicity of matters on $i\!-\!j$ string is the same
as that of $P(j)\!-\!P(i)$ string.
As was explained in section \ref{sec:FI}, the spectrum is chiral
if there is a block with the following property
\begin{itemize}
\item An off-diagonal block corresponding to one
      unitary and one non-unitary groups, with numbers
$$
  \left(\begin{array}{c} a\\b\end{array}\right)~~{\rm or}~~
  (a~b),~~~
  a\ne b
$$
\item An off-diagonal block corresponding to two different
      unitary groups, with numbers
$$
  \left(\begin{array}{cc} a&b\\c&d\end{array}\right), ~~~
  a\ne d ~~{\rm or}~~ b\ne c.
$$
\item A diagonal block for a unitary group, with numbers
$$
  \left(\begin{array}{cc} a&b_s+b_a\\c_s+c_a&a\end{array}\right), ~~~
  b_s\ne c_s~~{\rm or}~~b_a\ne c_a
$$
\end{itemize}
The table shows that the first example is non-chiral.

The other two examples are chiral, as can be seen
from the tables below.
Note that there are no antisymmetric tensor representations of $U(1)$
or $O(1)$.
\newpage
{\footnotesize
\begin{center}
\begin{tabular}{|c|cc||cc|cc|c|cc|cc|cc|c|}
\hline
${}^{\rm gauge}_{\rm group}$ &
$\sharp$&${\bf L},~{\bf M},~S$ &
1&2&3&4&5&6&7&8&9&10&11&12 \\\hline\hline
\lw{$U(4)$}
&1&(00000),(20666),2 &  0&0&2&0&1&0&1&0&0&0&1&0 \\
&2&(00000),(02666),2 &  0&0&0&2&1&0&0&1&3&0&0&0 \\ \hline
\lw{$U(2)$}
&3&(00100),(20166),0 &  2&0&2&0&1&0&3&1&2&0&0&0 \\
&4&(00100),(02166),0 &  0&2&0&2&1&0&0&0&0&2&0&0 \\ \hline
$O(1)$
&5&(00100),(66166),0 &  1&1&1&1&2&2&1&2&0&1&0&3 \\ \hline
\lw{$U(1)$}
&6&(01110),(25116),0 &  0&1&0&3&1&5&1&0&1&2&1&5 \\
&7&(01110),(07116),0 &  0&0&0&0&2&1&5&1&0&1&2&2 \\ \hline
\lw{$U(1)$}
&8&(10000),(34666),0 &  3&0&0&2&0&0&1&0&0&3&0&2 \\
&9&(10000),(98666),0 &  1&0&0&1&2&1&0&3&0&1&0&1 \\ \hline
\lw{$U(3)$}
&10&(10000),(70666),0 &  0&1&0&0&0&2&1&0&0&0&1&0 \\
&11&(10000),(52666),0 &  0&0&2&0&1&1&2&1&3&0&0&2 \\ \hline
$O(1)$
&12&(11110),(11116),2 &  0&0&0&0&3&2&5&1&2&2&0&6 \\
\hline
\end{tabular}
\vskip0mm
Example 2
\vskip5mm
\begin{tabular}{|c|cc||cc|c|cc|cc|c|c|c|c|}
\hline
${}^{\rm gauge}_{\rm group}$ &
$\sharp$&${\bf L},~{\bf M},~S$ &
1&2&3&4&5&6&7&8&9&10&11 \\\hline\hline
\lw{$U(3)$}
&1&(00000),(20666),2 &0&0&0&0&0&2&0&1&0&3&0 \\
&2&(00000),(02666),2 &0&0&0&0&0&0&2&1&3&0&0 \\ \hline
$O(1)$
&3&(00000),(66666),2 &0&0&0&0&0&1&1&2&0&0&1 \\ \hline
\lw{$U(1)$}
&4&(00000),(84666),2 &0&0&0&0&0&1&0&0&0&1&0 \\
&5&(00000),(48666),2 &0&0&0&0&0&0&1&0&1&0&0 \\ \hline
\lw{$U(5)$}
&6&(00100),(20166),0 &2&0&1&1&0&2&0&1&1&0&0 \\
&7&(00100),(02166),0 &0&2&1&0&1&0&2&1&0&1&0 \\ \hline
$O(1)$
&8&(00100),(66166),0 &1&1&2&0&0&1&1&2&0&0&3 \\ \hline
$O(2)$
&9&(01000),(61666),0 &3&0&0&1&0&0&1&0&0&3&1 \\ \hline
$O(2)$
&10&(10000),(16666),0 &0&3&0&0&1&1&0&0&3&0&1 \\ \hline
$O(1)$
&11&(11111),(11111),0 &0&0&1&0&0&0&0&3&1&1&13 \\
\hline
\end{tabular}
\vskip1mm
Example 3
\end{center}
}

\end{document}